\documentclass[12pt]{article}
\usepackage{amssymb,epsf,color}
\def\lsim{\mathrel{\rlap {\raise.5ex\hbox{$ < $}}
{\lower.5ex\hbox{$\sim$}}}}
\def\gsim{\mathrel{\rlap {\raise.5ex\hbox{$ > $}}
{\lower.5ex\hbox{$\sim$}}}} 
\def\sqr#1#2{{\vcenter{\vbox{\hrule height.#2pt

        \hbox{\vrule width.#2pt height#1pt \kern#1pt

           \vrule width.#2pt}

        \hrule height.#2pt}}}}

\def\lsim{{\displaystyle
{{\raise-8pt\hbox{$ <$}}
\atop{\raise5pt\hbox{$\sim$}}}}}
\def\gsim{{\displaystyle
{{\raise-8pt\hbox{$ >$}}
\atop{\raise5pt\hbox{$\sim$}}}}}
\def\slsim{{\displaystyle
{{\raise-8pt\hbox{$\scriptstyle <$}}
\atop{\raise5pt\hbox{$\scriptstyle \sim$}}}}}
\def\sgsim{{\displaystyle
{{\raise-8pt\hbox{$\scriptstyle  >$}}

\atop{\raise5pt\hbox{$\scriptstyle \sim$}}}}}

\newskip\humongous \humongous=0pt plus 1000pt minus 1000pt

\newcommand{\sumpf}[0]{\sum_{(H^{\rm f},G^{\rm f})}\! \! \! \!
{\raise
4pt
\hbox{$'$}}\,}

\newcommand{\sump}[0]{\sum_{(H,G)}\! \! {\raise 4pt \hbox{$'$}}\,}

\def\bs{\begin{subequations}}
\def\es{\end{subequations}}

\catcode`\@=11
\newcount\hour
\newcount\minute
\newtoks\amorpm

\hour=\time\divide\hour by60
\minute=\time{\multiply\hour by60 \global\advance\minute by-\hour}
\edef\standardtime{{\ifnum\hour<12 \global\amorpm={am}%
        \else\global\amorpm={pm}\advance\hour by-12 \fi

        \ifnum\hour=0 \hour=12 \fi
        \number\hour:\ifnum\minute<10 0\fi\number\minute\the\amorpm}}
\edef\militarytime{\number\hour:\ifnum\minute<10 0\fi\number\minute}
\def\draftlabel#1{{\@bsphack\if@filesw {\let\thepage\relax
   \xdef\@gtempa{\write\@auxout{\string
      \newlabel{#1}{{\@currentlabel}{\thepage}}}}}\@gtempa
   \if@nobreak \ifvmode\nobreak\fi\fi\fi\@esphack}
        \gdef\@eqnlabel{#1}}
\def\@eqnlabel{}
\def\@vacuum{}
\def\draftmarginnote#1{\marginpar{\raggedright\scriptsize\tt#1}}
\def\draft{\oddsidemargin -.2truein
        \def\@oddfoot{\sl preliminary draft \hfil
        \rm\thepage\hfil\sl\today\quad\militarytime}
        \let\@evenfoot\@oddfoot \overfullrule 3pt
        \let\label=\draftlabel
        \let\marginnote=\draftmarginnote
   \def\@eqnnum{(\theequation)\rlap{\kern\marginparsep\tt\@eqnlabel}%
\global\let\@eqnlabel\@vacuum}  }
\def\subequations{\refstepcounter{equation}%
  \edef\@savedequation{\the\c@equation}%
  \@stequation=\expandafter{\theequation}
  \edef\@savedtheequation{\the\@stequation}
  \edef\oldtheequation{\theequation}%
  \setcounter{equation}{0}%
  \def\theequation{\oldtheequation\alph{equation}}}

\def\endsubequations{\setcounter{equation}{\@savedequation}%
  \@stequation=\expandafter{\@savedtheequation}%
  \edef\theequation{\the\@stequation}\global\@ignoretrue
  \vspace*{-12pt} \\}

\def\bs{\begin{subequations}}
\def\es{\end{subequations}}
\relax
\def\tr{\,{\rm tr}\, }
\def\Im{\,{\rm Im}\, }

\def\thefootnote{\fnsymbol{footnote}}
\def\be{\begin{equation}}
\def\ee{\end{equation}}
\def\ba{\begin{eqnarray}}
\def\ea{\end{eqnarray}}

\newcommand{\ar}[2]{{#1\atopwithdelims[]#2}}

\def\ee{\end{equation}}
\def\bea{\begin{eqnarray}}
\def\eea{\end{eqnarray}}
\def\nn{\nonumber}

\newcommand{\uarrw}[0]{\mathrel{
{\raise.5ex\vbox{\hrule width 1cm}\hskip-6pt\rightarrow}}}
\def\thebibliography#1{%
\vskip 0.5cm \centerline{\bf References}
\list{%
[\arabic{enumi}]}{\settowidth\labelwidth{[#1]}
\leftmargin\labelwidth
\advance\leftmargin\labelsep
\usecounter{enumi}}
\def\newblock{\hskip .11em plus .33em minus .07em}
\sloppy\clubpenalty4000\widowpenalty4000
\sfcode`\.=1000\relax}

\renewcommand{\theequation}{\arabic{section}.\arabic{equation}}

\renewcommand{\section}{\setcounter{equation}{0}\@startsection%
{section}{1}{0mm}{-\baselineskip}{0.5\baselineskip}%
{\normalfont\normalsize\bfseries}}

\renewcommand{\subsection}{\@startsection%
{subsection}{2}{0mm}{-\baselineskip}{0.5\baselineskip}%
{\normalfont\normalsize\slshape}}

\renewcommand{\subsubsection}{\@startsection%
{subsubsection}{2}{0mm}{-\baselineskip}{0.5\baselineskip}%
{\normalfont\normalsize\slshape}}

\begin{document}
%
%
\renewcommand{\theequation}{\arabic{section}.\arabic{equation}}
\begin{titlepage}
\begin{flushright}
\end{flushright}
\begin{centering}
\vspace{1.0in}
\boldmath

{ \large \bf Combinatorics, observables, and String Theory: part II} 

\unboldmath
\vspace{1.5 cm}

{\bf Andrea Gregori}$^{\dagger}$ \\
\medskip
\vspace{1.2cm}
{\bf Abstract} \\
\end{centering} 
\vspace{.2in}
We investigate the string configuration that, in the framework of
the theoretical scenario introduced in \cite{assiom-2011},
corresponds to the most entropic configuration in the phase space
of all the configurations of the universe. This describes a universe with
four space-time dimensions, and the physical content is
phenomenologically compatible with the experimental observations and 
measurements. Everything is determined  
in terms of the age of the universe, 
with no room for freely-adjustable parameters.
We discuss how one obtains the known spectrum of particles and 
interactions, with massive neutrinos,
no Higgs boson, and supersymmetry broken at the Planck scale. 
Besides the computation of masses and couplings, 
CKM matrix elements, cosmological constant, 
expansion parameters of the universe etc..., all resulting, 
within the degree of the approximation we used,
in agreement with the experimental observations,
we also discuss how this scenario passes the tests provided by cosmology
and the constraints imposed by the physics of the primordial universe.

\vspace{6cm}

\hrule width 6.7cm
\noindent
$^{\dagger}$e-mail: agregori@libero.it

\end{titlepage}
\newpage
\setcounter{footnote}{0}
\renewcommand{\thefootnote}{\arabic{footnote}}

\tableofcontents

\vspace{1.5cm}

\noindent

\section{Introduction}
\label{intro}

In the \cite{assiom-2011} we have discussed a theoretical scenario
in which the universe is given by the superposition of all
possible configurations which describe the assignment of a certain amount of energy
along a vector space, of any possible dimension. The time ordering of the history
of the universe is given by the inclusion of the sets containing
all the configurations at a certain total energy. All the information about the
universe is encoded in a "partition function" which can be expressed as:
\be
{\cal Z}_{\cal E} ~ = ~ \sum_{\psi(E \leq {\cal E})} {\rm e}^{S (\psi)} \, ,
\label{zsum}
\ee
where $\psi (E)$ indicates a configuration (i.e. a distribution of $E$ energy units
along space),
and $S (\psi)$ is the entropy, i.e. the logarithm of the volume of occupation
of the configuration $\psi$ in the phase space of all the possible configurations.
After discussing how \ref{zsum} implies a quantum scenario which also embeds
special and general relativity, and therefore quantum gravity, we also identified in String Theory the theory which realizes a representation of this scenario, i.e.
the evolution of the superposition of geometries, in terms of propagating fields. 
The analogous of \ref{zsum} on the continuum is:
\be
{\cal Z}_V ~ = ~ \int_V {\cal D} \psi \, {\rm e}^{S(\psi)} \, ,
\label{zint}
\ee
where now $\psi$ indicates a string configuration, and $S(\psi)$ its volume in
the phase space of all string configurations at finite volume $V$, 
measured in the duality-invariant Einstein's frame.
The dominant configuration of the universe is the one
of highest entropy. As we discussed, the contribution of all
the neglected configurations falls under the error accounted for by the Heisenberg Uncertainty, and therefore is already considered by quantization.
The spectrum and the interactions of the
elementary excitations of the universe and their interactions are therefore basically 
the ones described by the string configuration
of highest entropy.

In this paper we want precisely to investigate the properties of this
configuration.
In \cite{assiom-2011} we have discussed how the 
absolute generality of the scenario described by \ref{zsum}, together with
the fact of being the properties of a string representation uniquely identified,
imply the uniqueness of string theory, namely the fact that
all perturbative string realizations are part (dual slices) of a unique
underlying theory. This means that we can not only
speak of volume  in a frame-independent way, but also of the string configuration
of highest entropy independently
on the particular representation, or perturbative string realization
we may find convenient to use in order to represent it.
Indeed, as this configuration is characterized by its having the
lowest amount of symmetry, a condition which in particular implies strong
coupling, it is not possible to realize it explicitly through a perturbative construction.   
In order to investigate its properties, we will make heavy use of string-string 
non-perturbative dualities.
We will find that, as already expected from the analysis of \cite{assiom-2011}, 
in this configuration
only a four-dimensional subspace is allowed to expand, 
and indeed, owing to the presence in the spectrum of massless fields, 
it expands. Along these four coordinates, T-duality is broken. 
At large volumes
a time-like coordinates can therefore be identified with what we ordinary
call ``time''. 
All the remaining coordinates
are twisted and therefore stuck, at the Planck scale. This scenario
describes a universe with, on the large scale, the geometry of a three sphere,
in which supersymmetry is broken at the Planck scale.
The ``volume ordering'', namely what corresponds to the time ordering
of the combinatorial approach, is equivalent to
an ordering according to the radius $R$ of the sphere. This implies
a non-accelerated expansion of the universe.  
Nevertheless, it can be shown that to an observer
the expansion of the universe appears to be accelerated as a consequence of
the time-dependence of red-shifts, in turn due to the
time-dependence of masses and energy levels.
The dominance of the contribution of the
configuration of highest entropy to the mean value of 
any observable increases the more and more 
``as time goes by'', and the universe expands and cools down. We have in this way     
a realization, at a non-field theory level, of the idea of
spontaneous breaking of symmetry, because the mean value of observables
effectively shows a ``progress'' toward more broken configurations.

The spectrum of the elementary excitations
corresponds to the degrees of freedom of all the known elementary
particles, and their interactions. However, their identification, as well as
a correct computation of masses, couplings, and cosmological parameters
such as the so-called cosmological constant, involves a deep change of
perspective as compared to the usual string or field-theoretical approach.
All this is a consequence of compactness of the whole space:
a finite volume space is here not conceived as a formal artifact introduced
in order to work with
a regularization of the infinity: it is the basic formulation of classical
space. This in particular means that there is
no invariance under time translations, because any 
progress in time is a progress in the history
of the universe, and therefore a flow toward a configuration with
different total energy, etc. There is no invariance under
space translations either, because, unless special boundary conditions
are imposed, any displacement implies approaching, or going away, from the
horizon of space. 
The basic absence of invariance under space-time translations
implies, by construction,
a different normalization of string amplitudes, and therefore a
different interpretation of the computed mean values: owing to
the absence of a normalization factor $1/{\cal V}$, where ${\cal V}$, 
the four-volume of space,
corresponds to the volume of the group of translations, 
densities are now lifted
to global quantities. The cosmological
constant is correctly predicted 
without fine tuning, despite the lack of low-energy supersymmetry,
because the string vacuum
energy expectation value, in our case of order 
one \footnote{Indeed it is precisely 1 by choice of normalization
of the mean values.}, does not correspond to an energy density, but
to a quantity that, in order to be transformed into a density, must be
rescaled by a Jacobian accounting for the coordinate 
transformation from the string to the Einstein's frame; this introduces
a suppression corresponding to a two-volume, the square of the radius of space-time. 
The so produced true density is therefore
$\Lambda \sim 1 / R^2 \sim H^2$, where $H$ is the Hubble
constant.
The cosmological constant is correctly predicted in its present day
value, and it turns out to be not at all a constant, but it evolves
with the inverse square of the age of the universe.
The masses of the various particles arise as momenta of a space in which
no coordinate is infinitely extended. These momenta have typically the size
of some root of the inverse of the radius of the classical universe, i.e.
the radius of the horizon, or, equivalently, the age of the universe.
They originate from a pure
string mechanism, they are not introduced via a Higgs mechanism.
There is no Higgs field. 
Masses are produced by shifts along the space-time coordinates, that lift the 
ground energy of a particle. These shifts have a non-trivial effect because
space-time is compact.
The values of masses indeed correspond to the experimental 
masses of the elementary particles.

As discussed in \cite{assiom-2011},
a consequence of the missing space-time
translational invariance
is also that the energy of the universe in not
conserved: the energy density of the universe can be
seen to scale, like the cosmological constant, 
as the inverse square of its age. Indeed, here we can see that 
the values of the
energy, matter and cosmological densities are of the same order of magnitude
as a consequence of a symmetry of the string configuration, which is broken
in a mild way. 
A scaling of these quantities like the inverse square 
of the age of the universe implies that
the total energy of the universe scales as its radius.

Inserting the values of the three energy densities
in a FRW Ansatz for the
universe, we can then solve the equations and obtain 
the large scale geometry of space-time. 
As it could have been argued from the fact that
the horizon is ``stretched'' from the expanding light rays,
the expansion of the universe turns out to be not accelerated.
Nevertheless, to our observations it appears to be accelerated: 
what we observe is in fact
a time-dependent red-shift effect, whose time variation, that could be
interpreted as due to an accelerated expansion, is produced 
by a time-variation of the energy and matter scales.

The chiral nature of the weak interactions too is a consequence
of the shifts along the
space coordinates, which not only give origin to masses but at
the same time also break parity. The separation of the matter world
into weakly and strongly coupled is instead a consequence of the breaking
of a T-duality along one internal coordinate. From a low-energy
effective point of view, this appears as an S-duality. 
Indeed, the shifts that give rise to masses break not only parity
and time reversal, but also explicitly the group of space rotations.
This agrees on the other hand with our experience of everyday
in the macroscopic world: the presence of objects,
i.e. clusters of energy,
in the space breaks the invariance under a change in the 
direction of observation.  These symmetries,
which appear to be broken at a macroscopic level, are here broken also 
in the fundamental description: the two levels are therefore sewed together
without conceptual separation.

Matter is basically non-perturbative:  in Planck units, the coupling of the
matter sector is one. From this ground value
depart the electro-magnetic, weak and strong coupling; 
the first two running, as the space-time volume 
increases, toward lower values; the third one toward higher values.
This poses a fundamental problem to the investigation of the matter
degrees of freedom, due to the fact that there are particles which feel
both strong and weak interactions.
An explicit, perturbative representation of particles as elementary states
can only be realized through an expansion around a vanishing 
ground coupling. Since couplings unify only at the Planck scale,
i.e. when we can no more speak of ``low energy world'', as a matter of fact 
there is no scale at which all the matter degrees of freedom
appear all at the same time as perturbative.  If we want to
see the matter degrees of freedom explicitly represented in a
perturbative construction, as they appear in usual field theory models, 
we must go to a picture in which the internal coordinates are
``decompactified''. 
In general, owing to the presence of some amount of T-duality, 
through decompactification we can have
access only to a part of the full theory. 
A decompactification is non-singular and preserves
all the properties of the physical vacuum only if the space under
question is flat. Otherwise, as in the cases of interest for us,
it is only an approximation, corresponding
to considering just the tangent space around a certain 
point~\footnote{This is for instance the case of the
so-called non-compact orbifolds.}. 
We may call this a "logarithmic representation" of the physical space,
in which a coupling of order one is mapped into its logarithm, therefore
of order zero.
In a logarithmic representation of space, i.e. on the
tangent space, the vanishing of the ground
coupling implies  also the vanishing of the tuning
parameter of the supersymmetry breaking. As a result, the linearized,
perturbative representation, in which particles show up 
as elementary states, appears to be supersymmetric, as is the case of many 
perturbative approaches to string and field theory.
On the other hand, in the real world there is no regime in which both 
leptons and quarks
appear at the same time as elementary, weakly coupled, free,
asymptotic states~\footnote{An artificial linearization 
of space-time is the cause of another 
false appearance of the string vacuum, namely the fact that 
from some respects string theory seems to require for its complete
description more than 11 coordinates. Indeed, the 12-th coordinate
should be better viewed as
a curvature. As is known, we can represent an $n$-dimensional
curved space in $n$ dimensions, with an ``intrinsic'' curvature, or we can 
embed it in a $n+1$-dimensional flat space. The degrees of freedom are 
in any case the same, because in the second case we don't consider
the full $n+1$-dimensional space, but an $n$-dimensional sub-manifold.
A perturbative representation of M-theory is something of this kind:
when the vacuum corresponds to a curved space,
by patching dual representations we have the impression that more than
eleven coordinates are required in order to describe the
full content, because these representations are necessarily perturbative
and therefore built on a flattened, ``tangent space''.}.

For the computation of masses and couplings,
we use the fact that, in order to be a representation of the combinatorial
scenario of \ref{zsum},the entire dynamics of the system is determined by
the entropy of the various processes in the phase space of all the configurations.
Since couplings and masses determine 
the interaction and decay probability of particles and fields,
masses and couplings
must be related to the volume occupied in the phase space
by the corresponding matter and field degrees of freedom.
The problem of computing these parameters in therefore translated 
to the one of computing the fraction of phase space these
particles and interactions correspond to.
This allows us to determine the ``bare'' mass and coupling values,
i.e. the parameters which are usually considered as external inputs 
in any effective action. Our approach is somehow
reversed with respect to the traditional one. 
Usually, the parameters and terms of the 
effective action are used in order to compute the full bunch of
interactions. From a general point of view, these can be seen
as ``paths'' coming out from, or leading to, a particle, 
or in general a physical state. Their number and strength
can be considered therefore a measure of the entropy of the state:
the higher is the mass of a particle, the higher is its 
interaction/decay probability, because higher is the number of final
states it can decay to. Entropy is then computed as a function of the 
interaction/decay probability, in turn determined by the dynamics.
In our approach, things go the other way around:
it is the dynamics which, consistently 
with~\ref{zsum}, is viewed as being determined by entropy.

Couplings and masses turn out to scale as powers
of the inverse of the age of the universe.  
They naturally unify at the Planck scale. There is no much to be surprised  
by the fact that, in usual field theory models,
supersymmetry seems to improve the scaling behavior
of couplings, making possible their unification.
If they correspond to a ``logarithmic'' representation of the
physical vacuum, couplings unify because, roughly speaking, 
they are logarithms of functions which unify. In our framework, 
a logarithmic scaling shows up if we want to compare the ``bare'' values
we obtain, with the parameters of a low-energy effective action,
in which space-time is considered as infinitely extended.
In order to make contact with the literature, this step in unavoidable: 
it happens in fact quite often that  
data of experimental observations are given as the result of elaborations
carried out within a certain type of theoretical scheme.
In particular, this is the case of effective couplings and masses
run to the typical scale of a physical process.
In this case, passing from a large but anyway finite
space-time volume to an infinitely extended one results in
a ``mild'', logarithmic correction to the ``bare'' mass, or coupling. 
Logarithmic corrections work in this case for small displacements
in the ``tangent space''. The bare parameters are instead 
derived in the full space, and their running is exponential with respect
to the one on the tangent space.

In general, any contact between our computations and the data found in the 
literature must be established at the level of experimental observations,
rather than on effective action parameters, whose derivation always
depends on a specific theoretical scheme.
Therefore, to be rigorous, better than effective couplings one should directly
consider scattering amplitudes and decay ratios; one should 
explain the emitted frequency spectra rather than
trying to match given acceleration parameters of galaxies, and so on.
This requires a deep change of perspective and a thorough re-examination
of any known result. On the other hand, 
in all the cases a prejudice-free re-analysis of
already known results is carried out, we find that
our theoretical framework provides us with a consistent scheme. Although
almost any physically observable quantity receives 
a different explanation than in traditional field theory or cosmology 
approaches, it is nevertheless consistent with what is experimentally
measured. Indeed, precisely the high predictive power of this theoretical
scenario, due to the fact
that there are no free parameters that can be adjusted in order to fit
data, enhances the strength of any matching with
experimental results: any discrepancy could in fact 
rule out the entire construction. Because of this, a large part of
our investigation is devoted to re-analyzing 
the most important data and constraints, coming not only from
elementary particles physics but also from astrophysics and cosmology. 
Our predictions and results are compatible with any experimental datum we have
considered, within the degree of approximation introduced in our derivation. 

Among the highlights of the predictions, there is the absence of
so-called ``new-physics'' (i.e. new particles or interactions) below the Planck scale.
This does not mean that no stringent tests can come from future high energy
experiments: the quantum mechanics 
arising from \ref{zsum} implies departures from the quantum behavior
expected in the traditional models of electro-dynamics and weak interactions,
especially in relativistic systems characterized by a high geometric complexity.
These effects can be viewed as due to quantization of space, as
implied in a quantum gravity scenario. When one thinks at the contribution
of gravity, one usually has in mind the contribution of \emph{classical} gravity,
and thinks at quantization as something that involves only the
description of the graviton as a quantum field. Indeed,
quantum gravity implies much more: it concerns the quantization of the
geometry, and its effects show up in any complex system. For instance, as
discussed in Ref.~\cite{spc-gregori}, it allows to explain the
correlation between complexity of the lattice structure and critical temperature in
high-temperature superconductors. Similarly,
it predicts at high energy a higher degree of non-locality of wave-functions, 
something that
should show out in a higher degree of correlation among products of a 
high energy collision~\footnote{It indeed seems 
that effects of this kind are being detected at LHC.}.
Apart from the absence of a Higgs field and the fact of being
the acceleration of the expansion of the universe only apparent, 
other phenomenological implications of our scenario 
that depart from what commonly expected
are to be found in the so-called ``time variation
of the fine structure constant $\alpha$''; 
and in the non-existence of dark matter.
In this scenario, for each of these phenomena the explanation relies in the
particular evolution of mass scales and couplings, 
as functions of the age of the universe.

One of the things that characterize this scenario is its high predictive power,
due to the fact that there are no free parameters other than the age
of the universe itself. Any measurable quantity is determined 
as a function of the age of the universe. 
It is therefore not a trivial fact that, besides its success
in producing values of observables (masses, couplings, CP violation
parameters etc...) correct within the degree of approximation we used,
this scenario also survives to the stringent tests coming from
cosmology, which is well known for imposing severe constraints on model
building.

\subsection{The outline of the work}
\label{plan}

The work consists for a large part in an update of \cite{spi}, 
and is organized as follows.
We start in section \ref{nps} by investigating the string phase space 
through orbifold constructions. The configuration of highest entropy, which, as discussed
in \cite{assiom-2011}, is also the one of minimal symmetry, is in this
subclass of constructions the one with the highest amount of twists and shifts.
In particular, we re-derive the result of \cite{assiom-2011}
about the number of dimensions of space-time, here identified with the
number of coordinates which remain untwisted, and therefore free to expand. 
We review also the discussion of \cite{assiom-2011} about the
number of dimensions in which the non-perturbative string theory lives,
in the light of non-perturbative string-string dualities between
orbifold constructions (section~\ref{spectrumZ2}).
Knowing the whole number of dimensions will turn out useful in order to compute the neutron mass in section \ref{nmass}.
After a discussion of the origin
of masses and the observable spectrum of the theory,
we pass to
the breaking of the Lorentz invariance, in particular of
the subgroup of space rotations (section~\ref{breakL}), which 
explains the observed slight inhomogeneities of the universe when observed
in different directions. We comment then the issues related to the magnetic
monopoles, in particular the topological ones, which in this scenario are expected to not exist.
We conclude the section with a discussion (subsection~\ref{eth}) of the
meaning of effective action in our framework of compact space with
time-dependent volume. In principle it is not obvious that even approximately
an effective action for our theory can be written at all; it indeed turns out to
make sense, even though only as an approximation, because of the
breaking of T-duality, which allows to speak of "extended" space-time.

In section~\ref{ubh} we discuss the geometry of the universe.
Interestingly, the curvature is given
by the sum of all energy densities, namely the cosmological one and
those corresponding to matter and radiation. 
As discussed in \cite{assiom-2011}, 
the contribution to the curvature of space due to the configuration
of highest entropy (which can be viewed as the "classical" one)
is of the same order as the sum of the contributions
of all the other configurations.
According to the discussion of section 7.4 of \cite{assiom-2011}, 
the part of the curvature which can be seen to
correspond to the pure geometry as described by light rays 
intended in the classical sense amounts to one-third of the actual geometry of the
universe.
This means that the value of the curvature receives contribution also from
non purely "classical" objects. These were interpreted in \cite{assiom-2011}
as quantum corrections, corresponding to matter and radiation, i.e. the degrees
of freedom that propagate in the space. Here we finally see, in the light
of a string scenario, how this is related to a non completely broken
symmetry, which exchanges
radiation, matter and gravity. 
We discuss then
the type of expansion the universe undergoes: it results to be
non accelerated, in agreement with the properties of the
universe as derived through the combinatorial approach discussed
in \cite{assiom-2011}.

In section~\ref{masses} we pass then to the detailed derivation of
the couplings, the masses of the elementary particles, and of the massive bosons.
All these quantities are given as functions of the age of the universe.
We derive then the scaling of the mean mass scale of the universe,
a quantity that can be 
non-perturbatively computed in an exact way: 
it corresponds in fact to
the only eigenvalue of the Hamiltonian at any finite space-time volume.
This scale can be seen to basically correspond to the mass of stable
matter: if the matter present in the universe was constituted by 
particles all of the same kind, these would have a mass precisely
corresponding to this scale. This scale can be shown to roughly correspond to the
neutron mass. 
With the scaling of the average mass of the universe at hand
we can discuss the issue of the apparent acceleration of the universe, and show that
in this scenario of non-accelerated expansion the acceleration
of red-shifts one observes can be explained as due to the variation with time of 
the atomic energy levels, and consequently of emitted light. In particular, 
we comment this point of view in comparison to the usual approach to the acceleration of the expansion of the universe. 
We discuss then the running of couplings, and 
the issue of their unification. We conclude the section with some comments
about approaches which share something in common to our 
one, such as those based on the geometric probability.

In section~\ref{todaymass} we come to an explicit 
evaluation of masses and couplings at present time.
We discuss the degree of approximation under which these values
are obtained, and give a rough estimate of the corrections they 
would receive if the string vacuum was known with a better accuracy.
In particular, in section~\ref{fsc} we compute the fine
structure constant (indeed its present-day, value because in our case it
is not a constant), obtaining a value which falls
within an error of $\sim 5 \times 10^{-6}$
away from its most updated experimental value. 
In section~\ref{hmass} we briefly discuss also
baryon and meson masses. Also their values agree, 
within the approximations introduced 
in the computing procedure, with what experimentally measured (in the case
of neutrinos, our computation remains a not yet tested prediction). 
The investigation of the mass sector of the theory is completed in
section~\ref{massmatrix}, where we consider the mixing angles 
of weak decays (the Cabibbo-Kobayashi-Maskawa matrix) and the CP
violations. Since in our scenario neutrinos are massive,
mixing of generations and off-diagonal decays are 
expected to occur also among leptons.

If the present-time value is useful for a comparison of our
predictions with the values of masses and couplings experimentally measured
in accelerators or in general in a laboratory, knowing their behavior 
along the history of the universe allows
us to test the predictions of this theoretical framework also in the case of
astrophysical and cosmological observations. 
In section~\ref{astropred} we consider  
the ``Cosmic Microwave Background''
radiation, and discuss how the existence of a $\sim 3 \, ^0 $ Kelvin
radiation comes out as a prediction
in this framework. We discuss then also, in subsection~\ref{darkm},
the case of dark matter. In our scenario, this is expected to not exist.
We comment several cases which are usually considered to provide 
evidence for its existence, and propose how, within our framework,
in each of them the effects attributed
to dark matter receive an alternative explanation.
In section~\ref{exc} we discuss then 
the constraints on the evolution of masses and couplings coming
from the observation of ancient regions of the universe, or, as is the case
of the Oklo bound, from the history of our planet. We find out that the 
predicted behavior is compatible with all the constraints. Not only, but
in the case of the so-called ``time dependence of $\alpha$'', it turns out
to correctly predict the magnitude of the observed effect 
(section~\ref{talpha}).

\newpage

\newpage

\section{\bf The non-perturbative solution} 
\label{nps}

The integral \ref{zint} contains in principle all the informations about
our universe. As discussed in~\cite{assiom-2011}, section~6.2,
the volume $V$ can be put in relation with ${\cal T}$,
the age of the universe. At any
time ${\cal T}$ the main 
contribution to the appearance of the universe is given by
the configuration of minimal symmetry, $\psi^{\rm min}({\cal T})
\equiv \psi^{\rm min}(V)$, the configuration that dominates in
the integral~\ref{zint} because it has at any
time the highest entropy in the phase space at fixed volume. 
The contribution to the mean values coming
from \emph{all} the other configurations gives a
correction of the order of the Heisenberg's Uncertainty, and therefore,
in the whole, becomes comparably smaller and smaller as the universe 
expands: ${\cal T}, V \to \infty$.
$\psi^{\rm min}({\cal T})$ is therefore at any time the configuration
that best corresponds to the
universe as it is experimentally observed. 
We stress however that $\psi^{\rm min}({\cal T})$
is not a solution in a classical sense, but only
the (largely) dominant configuration, that dominates
the more and more at larger times/temperatures. 
Our aim is to investigate the physical
content of $\psi^{\rm min}({\cal T})$.
To this purpose, we will use a ``perturbative'' approach.
In ordinary quantum field theory one separates the time evolution 
into a free propagation and an interaction part.
The physical configurations are inspected via
the conceptual separation of a base of free states, eigenstates of the
free Hamiltonian, which are exact solutions of the free theory.
As long as the coupling of the interaction is small, the full solution can
be considered as a small perturbation of the free propagation, and
the perturbative approach makes sense.
In our case, we have a truly non-perturbative string system, in which
even the space-time
is mixed up,  and in general will not be
factorisable into an extended one, ``the'' space-time as we experience it,
and an internal space.
Moreover, we can access to the whole theory only through 
``slices'', the perturbative (string) constructions, to be treated 
as the patches, the ``projections'', 
which allow to shed light into the ``patchwork'', 
the whole theory. We will
get information about the true vacuum through heavy use of string-string duality,
and follow the process of symmetry reduction through the
spectrum of possible string constructions in 
the class of orbifold contructions.


Orbifolds are particular string constructions in which we have, at any 
energy level, full knowledge about the spectrum of the perturbative states.
We are therefore able to write the partition function, the 
``one loop partition function'', which in principle
encodes all the information about the construction. 
With the string orbifold
partition function it is possible to perform one-loop computations 
of scattering amplitudes and threshold corrections, and
therefore compare string duals through pure string computations. 
The orbifold point is however in general a point of enhanced symmetry.
As we will discuss, degeneracies are removed by mass 
differentiations among the states that possess
a sub-Planckian mass. The latter turns out to depend on the (average)
radius $R$ of the extended space,
by definition the only part of the string space of $\psi^{\rm min}({\cal T})$
whose coordinates result to be non-twisted.
Mass corrections cannot be investigated in detail
at the orbifold point, and
will be treated as a perturbation. 
These corrections will turn out to 
be of order  $\sim \, {\cal O} \left(1/R^{\beta} \right)$,
$0 < \beta < 1$. 
Masses become equal in any perturbative limit, which is a limit of 
decompactification of some coordinate. On the other hand,
a perturbative investigation of  $\psi^{\rm min}({\cal T})$
makes sense because full theory and 
perturbative limits differ by the value of sub-Planckian masses, not by the nature of the 
spectrum: no new particles or fields are generated or lost 
in taking such limits.

In the search for the orbifold vacuum
that corresponds to the highest entropy in the phase space,  
$Z_2$ orbifolds will be the most favoured ones,
because they mod-out the space by the group with the smallest volume
among all the orbifold operations. A product of $Z_2$ twist/shifts allows 
therefore to achieve a configuration whith
a smaller surviving symmetry group than those 
obtained through any other product of orbifold operations. Entropy
will therefore be the maximal we can obtain with orbifold operations.
The most entropic orbifold vacuum will be the one with the 
highest amount of freely and non-freely acting $Z_2$ shifts
and twists.
Unfortunately, this configuration can be constructed explicitly only
in a perturbative regime. This corresponds to the decompactification
of some coordinate, which serves as coupling of the theory.
As a consequence, it will never show explicitly all the properties
of the whole theory.
The latter can only be indirectly inferred through the comparison of dual
constructions. In particular, supersymmetry will appear in a different way, 
depending on whether the decompactified coordinate does also tune the
supersymmetry breaking, or not.

\subsection{Investigating orbifolds through string-string duality}
\label{spectrumZ2}

Investigating the non-perturbative properties of a string vacuum
by comparing dual constructions is 
neither an easy task, nor a straightforward one. In general, at a generic
point in the moduli space the full set of dual
constructions, enabling to ``cover'' the full content, is not known.
Some progress on its knowledge has been done in the case of supersymmetric
vacua with extended supersymmetry, where it is in general possible
to identify a subset of the spectrum made ``stable'' by the properties
of supersymmetry. The case of orbifolds turns out to be particularly suited
for the investigation of non-perturbative string-string dualities. In this  
case it is possible to make a non-trivial comparison of
the renormalization of terms that receive
contributions only from the so called BPS states, and this
not just on the ground of the properties of supersymmetry, but through
the computation of true string contributions. Fortunately, $Z_2$ orbifolds,
the case of our interest, are the easiest and therefore more investigated 
constructions~\footnote{See for instance 
Refs.~\cite{dkl,kk,kkpr,solving,gkp,gkp2,kkprnew,6auth,gkr,gk,kop,striality,agn1,agn2,
ap,hm,hm4,fhsv,hmFHSV,afiq,
sv,gj,adk,kv,bl,agnt,Faraggi:2003yd,Faraggi:2004rq,Faraggi:2006bc,am}.}.
Indeed, through the analysis of these constructions, it is possible to
get an insight into the properties which are typical of string theory in 
itself: most of the investigations performed at other points in the moduli 
space must in fact rely on geometrical properties of smooth surfaces,
and their singularities. Although for some respects rather powerful, these 
techniques don't allow to capture the presence of states related to
non-geometrical singularities, or even fail in general for the simple reason 
that, owing to T-duality, the full string space simply cannot be
reduced to a geometrical one\footnote{For examples, see for instance 
Ref.~\cite{striality}.}. 
Our starting point will be a maximally supersymmetric string vacuum
with flat background: in our approach, 
the curvature of space-time will come out as an output, 
it is not an external input of the theory.
The constraints of two-dimensional conformal field theory impose that
$Z_2$ orbifold twists must act on groups of four coordinates
at once. In any string construction, there is room for a maximum of 3 such
operations, one of which is however redundant, in that it leads,
once combined with the other ones, to the re-introduction in the
twisted sectors of the states projected out. Therefore, we can say that
only a maximum of two independent $Z_2$ twists act effectively. 
However, the amount of 
supersymmetry surviving to these projections, as well as the amount of
initial supersymmetry, is different, depending on whether we start with
heterotic, type I, or type II strings. This means that in any construction
not all the projections acting on the theory are visible. Indeed, one of them
is always non-perturbative. The reason is that, by definition, a perturbative
construction is an expansion around the zero value of a parameter, the
coupling of the theory, which is itself a coordinate in the whole 
theory. An orbifold operation acting on this coordinate is forcedly
non-perturbative \footnote{A first investigation of a non-perturbative orbifold,
which produces the heterotic string, has been carried out
in~\cite{hw1,hw2}.}.

Our aim now is to derive the structure of the 
vacuum of lowest symmetry. 
Dual constructions correspond in general
to different perturbative limits,
i.e. decompactifications of different coordinates.
Essential for our work is that these decompactifications
must be possible; for this to make sense, the involved coordinate(s)
must not be twisted. Even in the case they are shifted, 
under decompactification
the scenario gets ``trivialized'' and we loose information about the 
construction. This problem does not exist for the first steps of 
symmetry reduction, because it is always
possible to decompactify a coordinate not involved in
some orbifold operation. Things are different when we reach the
maximum of orbifold operations. 
As we will see, for the  
configuration of lowest symmetry, 
a decompactification limit is a trivialization
of the theory, under which some physical content is lost: a perturbative
treatment of the degrees of freedom does not correspond in this case 
anymore to a simple ``limit'' of the theory, but to a phase which
can be obtained only through a ``logarithmic mapping'' of the
coordinates of the physical vacuum. This will be a key point
of the entire discussion of masses and supersymmetry breaking.
In this section we are not concerned with this problem; in the
purpose of understanding what is the singularity structure, 
we proceed, as in Ref.~\cite{estring},
by first ``counting'' the $Z_2$ operations in the various decompactification
limits, as long as these can be taken. By the way, we remark that it is 
precisely thanks to this limiting procedure that it is possible
to construct supersymmetric string constructions: 
in a fully compact space-time supersymmetry is always broken.

In the following we will 
often make use of the language of string compactifications to four dimensions,
especially for what matters our reference to the moduli of the 
string orbifolds. This will turn out to be justified ``a posteriori'': 
we will see that indeed the final configuration is the one of a string space 
with all but four coordinates twisted and therefore ``frozen''. 
Only four coordinates remain un-twisted and
free to expand, while all the others remain stuck at the string/Planck scale.
Massless degrees of freedom move along these and expand
the horizon of space-time at the speed of light. Although not infinitely
extended, this ``large'' space is what in our scenario corresponds to the
ordinary space-time. The language of 
orbifold constructions in four dimensions is therefore just an approximation,
that works particularly well at large times. Only at a second stage, we will
also discuss how and where this picture must be corrected in order
to account also for compactness of the space-time coordinates.
Although somehow an abuse of language, this approximation   
allows us to take and use with little changes
many things already available in the literature. In particular,
for several preliminary results and a rediscussion of the previous literature, 
the reader is referred to Ref.~\cite{striality}.

Let's see what are in practice the 
steps of decreasing symmetry we encounter 
when approaching the most singular configuration.
Although at the end it will be irrelevant the order in which we
apply freely and non-freely acting orbifold operations, it is convenient
to organize the analysis by considering first non-freely acting
operations, i.e. pure twists with orbifold fixed points.
Starting from the M-theory configuration with 32 supercharges,
we come, through orbifold projections, to 
16 supercharges and a gauge group of rank 16.
Further orbifolding leads then to 8 supercharges (${\cal N}_4 = 2$) and 
introduces for the first time non-trivial matter states (hypermultiplets).
As we have seen in \cite{striality} through an analysis of
all the three dual string realizations of this vacuum
(type II, type I and heterotic), this orbifold possesses
three gauge sectors with maximal gauge group of rank 16 in each.
The matter states of interest for us are hypermultiplets
in bi-fundamental representations: these are in fact those which
at the end  will describe leptons and quarks (all the others are eventually 
projected out). As discussed in \cite{striality},
in the simplest formulation the theory has 256 such degrees of freedom.
The less symmetric configuration is however the one in which,
owing to the action of further $Z_2$ shifts, the 
rank is reduced to 4 in each of the three sectors. These operations,
acting as rank-reducing projections, have been extensively discussed 
in \cite{dvv,6auth,gkr,striality}. 
The presence of massless matter is in this case still
such that the gauge beta functions vanish.
In this case, the number of bi-charged matter states is also
reduced to $4 \times 4 = 16$. 
These states are indeed the twisted states associated
to the fixed points of the projection that reduces the amount of
supersymmetry from 16 to 8 supercharges.

Let's consider the situation as seen from the type II side. We indicate
the string coordinates as $\{x_0,\ldots, x_9 \}$, and consider $\{x_0, x_9 \}$
the two longitudinal degrees freedom of the light-cone gauge. The
transverse coordinates are $\{x_1, \ldots , x_8 \}$. Here all the projections
appear as left-right symmetric. The identification of the degrees of freedom,
via string-string duality, on the type~I and heterotic side depends much on the
role we decide to assign to the coordinates, as we will see in a moment.
By convention, we choose the first $Z_2$ to twist $\{x_5,x_6, x_7, x_8 \}$:
\be
Z_2^{(1)} : ~~~ (x_5,x_6,x_7,x_8 ) \, \to \, (- x_5, - x_6, - x_7, - x_8 ) \, 
, 
\label{z_21}
\ee
and the second $Z_2$ to twist $\{x_3, x_4 , x_5, x_6 \}$:
\be
Z_2^{(2)} : ~~~ (x_3,x_4,x_5,x_6 ) \, \to \, (- x_3, - x_4, - x_5, - x_6 ) \, 
.
\label{z_22}
\ee
These two projections induce a third one: $Z_2^{(1,2)} \equiv Z_2^{(1)} \times
Z_2^{(2)}$, that twists $\{x_3, x_4 , x_7, x_8 \}$:
\be
Z_2^{(1,2)} : ~~~ (x_3,x_4,x_7,x_8 ) \, \to \, (- x_3, - x_4, - x_7, - x_8 ) 
\, .
\label{z_212}
\ee
Altogether, they reduce supersymmetry from ${\cal N}_4 = 8$ to 
${\cal N}_4 = 2$, generating 3 twisted sectors. 
Depending on whether we consider
the type IIA or IIB construction, the twisted sectors give rise either
to matter states (hyper-multiplets) or to gauge bosons (vector-multiplets).
As we discussed in Ref.~\cite{striality}, a comparison with the
heterotic and type I duals shows that the underlying theory must be 
considered as the union of the two realizations:
owing to the lack of a representation of vertex operators at once 
perturbative for all of them, for technical reasons no one of the constructions
is able to explicitly show the full content of this vacuum.
The matter (and gauge) content in these sectors is then 
reduced by six $Z_2$ shifts acting, two by two, by pairing each of the three
twists of above with a shift along one of the two coordinates of
the set $\{x_1, \ldots , x_8 \}$ which are not twisted. Each shift
reduces the number of fixed points of a $Z_2$ twist by one-half;
two shifts reduce therefore the matter states of a twisted sector 
from 16 to 4. Altogether we have then, besides the ${\cal N}_4=2$ 
gravity supermultiplet, three twisted sectors giving rise each one to 4 
matter multiplets (and a rank 4 gauge group). 
On the type~I side, these three sectors appear as two perturbative D-brane
sectors, D9 and D5, while the third is non-perturbative. On the heterotic side,
two sectors are non-perturbative.
As it can be seen by investigating duality with the type~I
and heterotic string, the matter states from the twisted sectors 
are actually bi-charged (see Refs.~\cite{gp,ang}, and \cite{striality}), 
something that cannot be explicitly observed, the charges
being entirely non-perturbative from the type II point of view.
The moduli $T^{(1)}$, $T^{(2)}$, $T^{(3)}$ of the type II realization, 
associated respectively to the volume form of each one of the three tori 
$\{x_3,x_4  \}$, $\{x_5,x_6  \}$, $\{x_7,x_8  \}$, are indeed 
``coupling moduli'', and
correspond to the moduli ``$S$'', ``$T$'',  ``$U$'' of the theory. On the
heterotic side, $S$ is the field whose imaginary part parametrizes the
string coupling: $\Im S = {\rm e}^{- 2 \phi}$.
It is therefore the coupling of the sector that contains the gravity fields.
$T$ and $U$ are perturbative moduli, and correspond to the couplings of the
two non-perturbative sectors. On the type~I side, on the other hand,
two of them are non-perturbative, coupling moduli, respectively of
the D9 and D5 branes, while only one of them is a perturbative modulus,
corresponding to the coupling of a non-perturbative 
sector~\cite{gp,ads,adds,adk}.
Owing to the artifacts of the linearization of the string space provided
by the orbifold construction, gravity appears to be on a different footing 
on each of these three dual constructions.

\subsubsection{The maximal twist}
\label{breaksusy}

The configuration just discussed constitutes the last stage of orbifold twists
at which we can ``easily'' follow the pattern of projections on all the 
three types of string construction. It represents also the 
maximal degree of $Z_2$ twisting corresponding to a supersymmetric 
configuration. As we will see, a further projection necessarily breaks 
supersymmetry. The vacuum appears supersymmetric only in certain dual phases,
such as the perturbative heterotic representation.
Non-perturbatively, supersymmetry is on the other hand broken.
This means that, when further twisted, the theory is basically no more 
de-compactifiable: perturbative, i.e. decompactification, phases,
represent only approximations in which part of the theory content 
and properties are lost, or hidden. This is what usually happens
when one for instance pushes to infinity the size of a coordinate 
acted on by a $Z_2$ twist. The situation is the one
of a ``non-compact orbifold''.

The further $Z_2$ twist we are going to consider is also the last that
can be applied to this vacuum, which in this way attains its maximal degree of
$Z_2$ twisting. This operation, and the configuration it leads to,  
appears rather differently, depending on the type of string approach.
Let's see it first from the heterotic point of view. So far we 
are at the ${\cal N}_4 = 2$ level.
The next step appears as a further reduction
to four supercharges (corresponding to ${\cal N}_4=1$ supersymmetry).
Of the previous projections, $Z_2^{(1)}$ and $Z_2^{(2)}$, only one
was realized explicitly on the heterotic string,
as a twist of four coordinates, say $\{x_5,x_6,x_7,x_8  \}$. 
The further projection, $Z_2^{(3)}$, acts on another four coordinates,
for instance $\{x_3,x_4,x_7,x_8  \}$. 
In this way we generate a configuration in which
the previous situation is replicated three times. 
When considered alone, the new projection would in fact
behave like the previous one,
and produce two non-perturbative sectors, with coupling parametrized by
the moduli of a two-torus, in this case $\{ x_5,x_6 \}$: $T^{(5-6)}$,
$U^{(5-6)}$. The product $Z_2^{(1)} \times Z_2^{(3)}$ leaves instead
untwisted the torus $\{ x_7,x_8 \}$ and
generates two non-perturbative sectors with couplings parametrized by the
moduli $T^{(7-8)}$, $U^{(7-8)}$. Altogether,
apart from the projection of states 
implied by the reduction of supersymmetry, the structure of the  
${\cal N} = 2$ vacuum gets triplicated.

The symmetry of the action of the
additional projection with respect to the previous ones suggests that 
the basic structure of the configuration, namely its repartition into
three sectors, $S$, $T$, $U$, is preserved when passing to the less
supersymmetric configuration. This phenomenon 
can be observed in the type II dual, that we 
discuss in detail in Appendix~\ref{N=0}.
From the heterotic point of view, the states
of these sectors come replicated ($\{ T \} \to \{ T^{(3-4)},
T^{(5-6)}, T^{(7-8)} \}$, $\{ U \} \to \{ U^{(3-4)},
U^{(5-6)}, U^{(7-8)} \}$).
On the type II side we observe a triplication also of the ``$S$''
sector. However, as we discussed in Ref.~\cite{striality}, 
we are faced here to an artifact of the
orbifold constructions, that by definition are built over a linearization
of the string space into planes separated by the orbifold projections.
The matter states are indeed charged under three sectors, $S^{i}, T^{j},U^{k}$,
but we can at most observe a double 
charge, as it appears on the type~I dual side; from an analysis based
on string-string duality, we learn that the states are in fact
multi-charged for mutually non-perturbative sectors
When one of the $S^{i}, T^{j},U^{k}$ sectors is at the weak coupling,
the other two are at the strong coupling, and it doesn't make sense
to ask what is this sort of ``splitting'' of the non-perturbative 
charge of the states: we simply observe that they have a perturbative index
and one running on a strongly coupled part of the theory.

On the type I dual realization of this vacuum,
besides a D9 branes sector we have now three D5 branes sectors
and a replication of the non-perturbative sector into three sectors, 
whose couplings are parametrized by 
$U^{(3-4)}$, $U^{(5-6)}$, $U^{(7-8)}$. 

A result of the combined action of these projections
is that all the fields $S^{i}$, $T^{j}$ and
$U^{k}$ are now twisted. This means that their vacuum expectation value
is not anymore running, but fixed. We will see below, section~\ref{d=4},
that minimization of entropy selects this value to be of order one, thereby
implying also the identification of string and Planck scale 
(section~\ref{eth}).
Nevertheless, for convenience here we continue with the generic notation 
$S$, $T$, $U$ used so far, because it allows to better follow the 
functional structure of the configuration we are investigating.
Twisting of the ``coupling'' moduli indeed suggests the non-decompactifiability
of this vacuum. This, as discussed,
would imply the breaking of supersymmetry. However, this property is not 
so directly evident: each dual construction is in fact
by definition perturbatively constructed around a decompactification limit.
The point is to see, with the help of string-string duality, whether
this is a real decompactification, or just a singular, non-compact orbifold
limit.
An important argument in favour of this second situation
is that, after the $Z_2^{(3)}$ projection
is applied, the so-called ``${\cal N}=2$
gauge beta-functions'' are unavoidably non-vanishing. 
According to the analysis of 
Ref. \cite{striality}, this means that there are hidden sectors at the
strong coupling~\footnote{We refer the reader to the cited work for 
a detailed discussion of this issue.}. 
As a consequence, supersymmetry is actually non-perturbatively
broken, due to gaugino condensation. 
Inspection of the type II string dual shows explicitly the
instability of the ${\cal N}_4 = 1$ supersymmetric vacuum.

In order to construct the type II dual,
it is not possible to proceed as with
the heterotic and type I string, namely by keeping un-twisted some coordinates.
On the type II side the ``${\cal N}_4 = 1$'' vacuum looks rather
differently: the new projection twists all the transverse coordinates,
leaving no room for a ``space-time''. This however does not mean that
a space-time does not exist:
all non-twisted coordinates, therefore
the space-time indices, are non-perturbative. Their volume is precisely related
to the size of the coupling around which the perturbative vacuum is expanded. 
After $Z_2^{(1)}$ and $Z_2^{(2)}$, the only possibility for applying
a perturbative $Z_2$ twist is in fact to act on $\{x_1, x_2, \}$ 
and on two of the
$\{x_3 , \ldots, x_8 \}$ coordinates, already considered by the previous 
twists. These can be either the pair $\{x_3,x_4 \}$ or
$\{x_5,x_6 \}$, or $\{x_7,x_8 \}$. Which pair, is absolutely equivalent. We 
can chose $Z_2^{(3)}$ such that:
\be
Z_2^{(3)}: ~~~ (x_1,x_2,x_3,x_4) \; \to \; (- x_1, - x_2, - x_3, - x_4) \, .
\label{z_23}
\ee
The other choices are anyway generated as $Z_2^{(3)} \times Z_2^{(2)}$
and $Z_2^{(3)} \times Z_2^{(2)} \times Z_2^{(1)}$.
Assigning a twist to some coordinates is not enough in order to define
an orbifold operation: the specification must be completed by an appropriate
choice of ``torsion coefficients''. The analysis of this orbifold
turns out to be easier at the fermionic point, where the world-sheet bosons
of the conformal theory are realized through pairs of free fermions~\cite{abk}.
We leave to the appendix~\ref{N=0} a detailed discussion of the construction
of this vacuum. There we see
how the duality map with ${\cal N} = 2 \to
{\cal N} \to 1$ heterotic theory imposes a choice
of ``GSO coefficients'' that leads to the complete breaking of supersymmetry,
and discuss, in appendix~\ref{susybreaking}, how the breaking is tuned by
moduli which in this vacuum are frozen at the Planck scale.
This is also therefore the scale of the supersymmetry 
breaking~\footnote{Among the historical reasons for the search of low-energy 
supersymmetry are the related smallness of the cosmological constant and 
the stabilization
in the renormalization of mass scales produced by supersymmetry.
In our framework, the value of the cosmological constant will be
justified in a completely different way (section~\ref{ubh}). 
Also the issue of stabilization of scales in this 
framework must be considered in a different way: masses are no more
produced by a field-theory mechanism, and field theory is not the
environment in which to investigate their running.}.

The reason why the breaking of space-time supersymmetry can be observed
in a dual in which space-time is entirely non-perturbative relies on the
unambiguous identification of the supersymmetry generators.
More precisely, what on the type II side it is possible to see is the 
projection of the supersymmetry currents on the type II perturbative space.
Target space supersymmetry is in fact realized in string theory
through a set of currents whose representation
is built out of the world-sheet degrees of freedom.
For instance, in the case of free fermions in four dimensions, we have:
\be
G(z) \, = \, \partial_z X^{\mu} \psi_{\mu} \, + \, \sum_i x^{i} y^{i} z^{i}
\, ,  
\label{gz}
\ee 
and
\be
G(\bar{z}) \, = \, \partial_{\bar{z}} X^{\mu} \bar{\psi}_{\mu} \, 
+ \, \sum_i \bar{x}^{i} \bar{y}^{i} \bar{z}^{i}
\, , 
\label{gzbar}
\ee 
where the index $i$ runs over the internal dimensions. 
At the ${\cal N}_4 = 2$ level it is possible to construct both
the representations of the type II dual, namely the one in which space-time
is perturbative, and the one in which space-time is non-perturbative.,
Tracing the representation of the supersymmetry currents in both these
pictures allows us to identify them also when 
the $Z_2^{(3)}$ twist is applied.
Although, strictly speaking, there is no simple one-to-one linear mapping 
between coordinates of dual constructions, the fact
that the dual representations of the currents share a 
projection onto a subset of coordinates common to both,
enables us to follow the fate of space-time supersymmetry anyway.

The analysis of the type II dual confirms that
the matter states of this vacuum
are indeed three replicas of the chiral fermions of the theory 
before the supersymmetry-breaking, $Z_2^{(3)}$ projection. In the type~II
construction their space-time spinor index runs non-perturbatively; 
they appear therefore as scalars. In total, we have
three sets of bi-charged states in
a ${\bf 16} \times {\bf 16}$. In the minimal,
semi-freely acting configuration, they get reduced to
three sets of ${\bf 4} \times {\bf 4}$ by the 
further $Z_2$ shifts, acting on the twisted planes. 
As it was the case of the ${\cal N}_4 = 2$ theory,
on the type II side their charges are 
non-perturbative, and they misleadingly appear as $({\bf 16},{\bf 16},
{\bf 16})$, reduced to $({\bf 4},{\bf 4},{\bf 4})$. The impression is that we 
have three families of three-charged states.
However, this is  only an artifact of the orbifold
construction. From the heterotic point of view, 
namely, the vacuum in which gauge charges are visible,
two sectors of each family are non-perturbative and,
as previously mentioned, the structure
of their contribution to threshold corrections is an indirect signal that
they are at the strong coupling (see Ref.~\cite{striality}).
The situation is the following: \emph{either} 
we explicitly see all the gauge sectors,
on the type~II side, but we don't see the gauge charges, \emph{or}
where we can explicitly construct currents and see gauge charges 
(the heterotic realization), we see the gauge sector, and the currents,
corresponding to just one index born by the matter states: the other
are non-perturbative and strongly coupled.

The type II realization appears to be a different ``linearization'',
or linear representation, of the string space, in which the non-perturbative
curvature has been ``flattened'' through an embedding in a higher number
of (flat) coordinates, which goes together with a redundancy of
states due to an artificial replication of some degrees of freedom. 
On the type II string, twisted states can only be represented
as uncharged, free states. Their charges are in any case non-perturbative,
and we cannot observe a ``non-abelian gauge confinement''.
These gauge sectors appear as partially perturbative on the type I side. 
However, the type~I vacuum, like the heterotic one, corresponds to 
an unstable phase of the theory:
it appears as supersymmetric although it is not. Moreover, inspection
of the gauge beta-functions reveals that they are positive. Therefore,
although appearing as free states, the currents on the D-branes run to
the strong coupling and the apparent gauge symmetries are broken by
confinement.

\vspace{.3cm}

Let's {\bf summarize} the situation. The initial theory underwent 
three twists and now is essentially the following orbifold:
\be
Z_2^{(1)} \times Z_2^{(2)} \times Z_2^{(3)} \, . 
\label{zzz}
\ee
In terms of supercharges, the supersymmetry breaking pattern is:
\be
32 ~ \stackrel{Z_2^{(1)}}{\longrightarrow} ~ 16 ~ 
\stackrel{Z_2^{(2)}}{\longrightarrow} ~ 8 ~ 
\stackrel{Z_2^{(3)}}{\longrightarrow} ~ 0 ~~ {\rm (4~only~ 
perturbatively)} \, .
\label{sbpattern}
\ee
The ``twisted sector'' of the first projection gives rise to 
a non-trivial, rank 16 gauge group; the twisted sector of the second
leads to the ``creation'' of one matter family, while after the
third projection we have a replication by 3 of this family.
The rank of each sector is then reduced by $Z_2$ shifts of the type discussed
in Ref.~\cite{6auth,gkp,gkr}, two per each complex plane. 
As a result, each ${\bf 16}$ is reduced to ${\bf 4}$.
On the type II side  one can explicitly see, besides the shifts,
both the total breaking of
supersymmetry and the doubling of sectors under which the matter 
states are charged. The product of these operations
leads precisely to the spreading into sectors that at the end of the day
separate into weakly and strongly coupled,
allowing us to interpret the matter states as quarks
\footnote{As we will discuss, the leptons show up as singlets inside 
quark multiplets.}. On the type I side, the states appear in an unstable
phase, as free supersymmetric states of a confining gauge theory, while
on the heterotic side they appear on the twisted sectors, and their gauge 
charges are partly non-perturbative, partly perturbative.
The perturbative part is realized on the
currents. Like the type~I realization, also the 
heterotic vacuum appears to be an unstable phase, before flowing
to confinement; they are indeed non-perturbatively singular, non-compact
orbifolds. This reflects on the fact that,
as also discussed in Ref.~\cite{striality}, 
both on the heterotic and type I side,
perturbative and non-perturbative gauge sectors have opposite sign of
the beta-function. This signals that, as the visible phase is confining, the 
hidden one is non-confining.

\subsubsection{Origin of four dimensional space-time }
\label{d=4}

The product~(\ref{zzz}) 
represents the maximal number of independent twists the 
theory can accommodate: a further twist would in fact superpose to the
previous ones, and restore in some twisted sector the projected states.
Therefore, further projections are allowed, but no further
twists of coordinates. These twists allow us to distinguish between
``space-time'' and ``internal'' coordinates. While the first ones 
(the non-twisted) are free to expand, 
the twisted ones are ``frozen''.
The reason is that the graviton, and as we will see the photon, live in 
the non-twisted coordinates.
Precisely the fact that graviton and photon propagate along these coordinates,
and therefore ``stretch'', expand the horizon, allows us to perceive
these as our ``space-time''.
We get therefore ``a posteriori'' the justification of our choice to analyze
sectors and moduli from the point of view of a compactification
to four dimensions.

The radius at which the internal coordinates
are twisted is fixed by the requirement of minimization
of symmetry to be of the order of the string scale.
In order to understand this, let's consider an ordinary, bosonic lattice, that
for simplicity we restrict to just one coordinate. The partition
function consists of a sum of two terms: the un-twisted/projected and the 
twisted/projected part. Roughly, the un-twisted part reads:
\be
{\cal Z} \, \sim \, \int  \left\{ 1 \, + \, \sum_{m,n \neq 0,0} 
{\rm e}^{- \tau [(m/R)^2 + (nR)^2]} \right\} \, \times \, | \eta |^{-1} \, .
\label{zun-tw}
\ee  
The $\eta$ factor encodes the contribution of
the oscillators that build up a tower of states over the 
ground momentum and winding.
The term of interest for us, the one that varies, is the factor within 
brackets.
Even in the case a shift acts on this coordinate (as is in our case), 
what remains of the
(broken) T-duality is enough to state that 
the string scale $R_0 \sim {\cal O}(1)$ is a point of minimal
symmetry of the theory.
This may seem odd, because it is somehow a self-dual point,
therefore related to an ``enhanced symmetry''.
However, even though there is a $Z_2$ enhancement of symmetry
(at least if T-duality is not broken, something which is not really
the actual case), the whole spectrum has, modulo $Z_2$, the highest
differentiation.
As the radius increases (resp. decreases beyond the extremal point),
the probability of the states with non-zero winding (resp. momentum) number
in fact decreases the more and more rapidly, 
while the probability of the momentum
(resp. winding) states approaches the limit value:
\be
P_m ~ \sim ~ {\int {\rm e}^{- \tau \, m^2/R^2} \over {\cal Z}}
~ ~ \stackrel{R \to \infty}{\longrightarrow} 
~ ~{\int 1 \over {\cal Z}} \, .
\label{pm}
\ee
Therefore, in the limit $R \to \infty$, for any
fixed momentum number $m_R$, the tower 
of momentum states with momentum number $m < m_R$ ``collapses'' to nearly 
the same probability: many of the 
formerly well separated steps shrink to a ``continuum'' of states of
nearly identical, maximal probability. Similarly goes for $R \to 0$ and the 
winding instead of momentum states.
Therefore, as the radius departs from the  extremal value $R = R_0$, 
either by increasing or by decreasing, the distribution of probabilities
``collapses'' toward two kinds of possibilities: half of the states
tend to acquire the same, maximal probability, while the other half
tend to disappear. From a well differentiated configuration, in which
any energy level possessed a non-trivial probability, we move toward
a situation of higher symmetry.
These two decompactification limits lead either to a restoration of the 
broken symmetry, or to a configuration of purely
projected, i.e. a non-freely acting orbifold, but without the additional
states coming from the twisted sectors.

\vspace{.4cm}

\subsubsection{In how many dimensions does non-perturbative String Theory 
live?}
\label{dimString}

We have seen that, with the maximal twisting, supersymmetry is broken.
The string space is therefore necessarily curved.
For some respects, this may seem strange. As we also discuss in 
Appendix~\ref{N=0} and \ref{susybreaking}, even with the maximal twist,
not all the string coordinates are twisted. At most, we have twisted eight 
of them. Why should then not be possible to decompactify one of the
non-twisted ones, and obtain anyway a flat space? The fact that
the space gets curved at the maximal twisting, even though this does not
involve all the coordinates, is a property of the orbifold constructions; 
they are a kind of singular spaces, with a geometry which is flat everywhere
apart from some singular points. The curvature is somehow all
``concentrated'' on these points. Although from a global 
point of view orbifold spaces are curved, 
locally they are almost everywhere flat. By the way,
this is the reason why one can have the impression of being able to
decompactify them and build a consistent supergravity theory 
at the decompactification 
limit~\footnote{This is also the case of Refs.~\cite{hw1,hw2}:
the Ho\v{r}ava-Witten analysis indeed obtains
the Heterotic string in ten dimensions as a non-compact orbifold 
(see Appendix~\ref{susybreaking}).}. 
Supergravity is a part of field theory, 
it is related to differential geometry concepts, it is a local theory.
The signal that something illegal has been done comes from the investigation
of the \emph{pure stringy}, non-perturbative properties of the vacuum
under consideration.
The fact that with the $Z^{(1)}_2 \times Z^{(2)}_2 \times Z^{(3)}_2$ 
twisting the string orbifold space is curved can be clearly seen
by considering the type II point of view. There, all the transverse
coordinates are twisted. Not the coupling, which remains untwisted.
Nevertheless, we are in the presence of a perturbative series in which,
at any order in the expansion around the non-twisted coordinate(s),
we have a fully twisted, curved space. This sums up therefore to a curved 
space; the perturbative contribution to the curvature cannot in fact be
cancelled by a possible non-perturbative term: the terms should be cancelled 
order by order.

Besides the above mentioned twists/shifts, the only way left out 
to further minimize
symmetry is to apply further shifts along the non-twisted coordinates.
How many are they? From the type II point of view, there are no further,
un-twisted coordinates. But we know that they are there, ``hidden'' as 
longitudinal coordinates eaten in the light-cone gauge and in the 
coupling of the theory. Some of these coordinates
appear on the heterotic/type~I side as \emph{two} transverse coordinates.
If we count the total number of twisted coordinates by collecting the 
information coming from intersecting dual constructions, and the coordinates
which are ``hidden'' in a certain construction and are explicitly realized
in a dual construction, we get the impression that the underlying theory 
possesses 12 coordinates. For instance, on the heterotic side we have 
a four-dimensional space-time plus six internal, twisted coordinates,
and a coupling. On the type II side we see eight twisted coordinates.
We would therefore conclude that the two additional twisted coordinates
correspond to the coupling of the heterotic dual. On the other hand,
no supersymmetric 12-dimensional vacuum seems to exist, at 
least not in a flat space: the maximal dimension with these properties
is 11. This seems therefore to be the number of dimensions in which 
non-perturbative string theory is natively defined.

Let's have a better look at the properties of supersymmetry.
As is known, the supersymmetry algebra closes on the momentum operator.
When applied to the vacuum, we have:
\be
\left\{ Q, \bar{Q}  \right\} \, \approx \, 2 \, M \, .
\label{qqm}
\ee
From a dimensional point of view, a mass can be viewed as the inverse of a length,
so that we can also write:
\be
< \left\{ Q, \bar{Q}  \right\} > \, \cong \, { 1 \over R } \, .
\label{qqr}
\ee
The supersymmetry algebra
suggests that the mass on the right hand side of \ref{qqm}, 
in all respects an order parameter for the supersymmetry
breaking, could be interpreted as the inverse of the length of a coordinate
of the theory. This coordinate refers to an extra internal dimension, or,
perhaps more appropriately, to a curvature, i.e. a function collecting the
contribution of several coordinates, perturbative as well as
non-perturbative. 
Precisely the fact that, in the breaking of ${\cal N}_4=2$
supersymmetry to ${\cal N}_4=1$, the dilaton and the other ``coupling'' fields
get twisted, is a signal that a non-vanishing curvature of the string space
has been generated. As we discussed in section~\ref{breaksusy}, 
this means that, even in the case of infinite volume, 
we are in a situation of non-compact orbifold.
Simply assigning a non-vanishing value to the coupling does not imply the 
generation of a non-vanishing curvature
of the string space: the corresponding field must also be twisted. 
In the orbifold language, this is implemented 
by the fact that, whenever the coupling field is ``explicitated''  
by going to a dual construction, the corresponding 
perturbative geometric field appears as a volume of a two-dimensional space.
This phenomenon can be observed for reduced supersymmetry (for maximal
supersymmetry, there is just the type~II string construction).
Consider for instance the eleventh coordinate of M-theory, that should 
correspond to the dilaton of the heterotic string. In the type~II orbifold
constructions (K3 orbifold compactifications), 
the heterotic coupling corresponds to a two-torus volume. 
Considering that this two-dimensional space corresponds, from the heterotic
point of view, to ``extra-coordinates'', one would say
that, in order to realize all these degrees of freedom, 
the full underlying theory should be (at least) twelve-dimensional.
However, this is only
an artifact of the linearization implied by the orbifold construction, 
and it means that the simple compactification on a circle is not enough, 
we need also an additional ``curvature coordinate'' in order to parametrize
a truly curved space.

From the type~II dual we learn that 
supersymmetry is not restored by a simple decompactification: the string space
is twisted \footnote{In some type~II/heterotic duality identifications,
the heterotic coupling is said to correspond to un-twisted coordinates
of the type~II string. This however does not change the terms of the problem:
in the artifacts of the flattening implied by the orbifold constructions,
part of the curvature may be ``displaced'', or referred, to some or some other
coordinates. This ``rigid'' distribution of the twists, basically
dictated by the need of recovering a description in terms of supergravity 
fields referring to the same space-time dimensionality for both the dual
constructions, may induce to misleadingly conclusions. The intrinsic twisted
nature of the space has to be considered by looking at the string space in its
whole (for more details and discussion, see for instance 
Ref.~\cite{striality}).}. 
Flatness of the string space
is broken by a ``twist'' of coordinates that fixes them to the
Planck scale. As a consequence, the supersymmetric partners
of the low-energy states are boosted above the
Planck scale.
In a situation of supersymmetry restoration, they should come down to the same 
mass as the visible world, and space should become ``flat''. 
However, this is only possible when
the twist is ``unfrozen'' and we can take a decompactification limit,
such as for instance the M-theory limit. Otherwise, at the decompactification 
limit the space becomes only locally flat (non-compact orbifold). 
In order to get a true flat space we must take out the ``point at infinity'' 
(the ``twist''), which closes the geometry to a curved space.

Let's collect the informations
so far obtained:

\begin{enumerate}

\item As soon as the string space is sufficiently twisted, supersymmetry is 
broken.

\item Equations~\ref{qqm} and \ref{qqr} suggest in this case a non-vanishing
curvature of space.

\item In the class of orbifolds, the phenomenon of curving the string 
space can only be partially and indirectly seen, through the comparison of
dual constructions. 

\item These constructions are built on a (perturbatively) flat, supersymmetric
background: they provide therefore ``linearizations'' of the string space. 

\item The maximal dimension of a supersymmetric theory on a flat background
is 11.  

\end{enumerate}

All this suggests that, when supersymmetry is broken, we are in the presence 
of an eleven-dimensional \emph{curved} background. Any, forcedly 
perturbative, explicit orbifold realization requires for its construction 
a linearization of the background. Since a 11-dimensional curved space can be
embedded in a 12-dimensional flat space, we have the impression of an
underlying 12-dimensional theory. However, this is only an artifact; in fact,
we never see all these 12 flat coordinates at once: we infer their existence
only by putting together all the pieces we can explicitly see. But this turns
out to be misleading: the linearization is an artifact.

\emph{The 12 dimensional background is only fictitious, 
we need it only in order to describe the theory in terms of flat coordinates. 
At the perturbative string level, of these coordinates we see only a maximum of 10.}

These conclusions match with those of section~6.1.1 of Ref.~\cite{assiom-2011}.
The counting of twisted and un-twisted coordinates has to be considered from
this point of view. Trading the two ``space-time'' transverse coordinates
on the type II side for the coupling coordinates
of ``M-theory'', as we did in section \ref{d=4}, 
doesn't mean that we really have two such 
coordinates: the only thing we know is that, taking the union
of all the coordinates of the various dual constructions, in
which this curved space appears to be linearized, one has to the impression
to have up to 12 dimensions. Indeed, going to the type II picture is a trick
enabling to explicitly see the instability of the ${\cal N}_4 =1$ vacuum, 
by switching on the operation on the ``hidden'', non-perturbative part of 
the theory. As a matter of fact, we are however 
in the presence of a maximum of seven
``twisted'' coordinates, i.e. coordinates along which the degrees of freedom
don't propagate, and four un-twisted ones, along which the
degrees of freedom can propagate. 
By comparison of dual string vacua, we can see that there 
is room to accommodate two more ``perturbative'' $Z_2$ shifts:
through the heterotic and/or type I realization, we can explicitly see
only two transverse non-twisted coordinates,
plus two longitudinal ones, along which no shift can act.
It remains then one ``internal'', truly non-perturbative coordinate,
to which no shift has yet been applied. This can only 
be indirectly investigated: if we try to explicitly realize it, 
it will appear as a set of two coordinates, giving the fake impression 
to have room for two independent shifts.

Let's now count the number of degrees of freedom of the matter states.
We have three families, that for the moment are absolutely identical:
each one contains $16 = 4 \times 4$ chiral fermions.
These degrees of freedom are suitable to arrange in
two doublets of two different $SU(2)$ subgroups
of the symmetry group. Each doublet is therefore like 
the ``up'' and ``down'' of an $SU(2)$ doublet of the weak interactions,
but this time with a multiplicity
index 4. As we will see, this 4 will break into 3+1, the 3 
corresponding to the three quark flavours, and the 1 (the singlet) to
the lepton. The number of degrees of freedom is therefore the right one
to fit into three families of leptons and quarks. However, so far all these
fields are massless and charged under an $SU(2)$ symmetry. We will see
how precisely shifts along the space-time coordinates lead to the
breaking of parity of the weak interactions and to a non-vanishing mass
for these particles.

\subsubsection{The origin of masses for the matter states}
\label{massmatter}

Shifts applied to the ``internal'' coordinates reduce the symmetry group
through mass lifts that, owing to the fact that the coordinates are
also twisted, remain for ever fixed; in this specific case, 
at the Planck scale. Also the shifts acting on the space-time 
coordinates reduce further the rank of the symmetry group.
However, the breaking in this case is obtained through mass shifts that
depend on the length of the untwisted coordinates, i.e. on the size
of space-time. Therefore, the matter states ``projected out'' by these 
shifts are not thrown out from the spectrum of the low energy theory: 
they acquire a ``weak'' differentiation in their masses; 
the mass difference is inversely related to the scale of space-time.

We have seen that, before any shift in the space-time coordinates is applied,
each twisted sector gives rise to four chiral matter fermions transforming in 
the ${\bf 4}$ of a unitary gauge group.
The first $Z_2$-shift in the space-time breaks this symmetry, reducing 
the rank through a ``level doubling'' projection.
On the heterotic string, this operation acts by further doubling the level
of the gauge group realized on the currents~\footnote{Notice that, once
the four-dimensional space-time is included in the orbifold operations,
owing to the rank reduction produced by the further shift made in this 
way possible, the gauge group realized on the currents
becomes non-confining, inverting thereby the situation discussed at the end
of section~\ref{breaksusy}.}. The consequences  of this operation are that:
1) half of the gauge group becomes massive; 2)
half of the matter states become also massive. 
The initial symmetry is therefore 
broken to only one rank-2 unitary group, under which only half of matter 
transforms.
The remaining matter degrees of freedom become massive. 
Since this phenomenon takes place at a scale related to the inverse
of the space-time size, therefore lower than the Planck scale, 
field theory constitutes a good framework allowing to understand the fate of 
these degrees of freedom, leading to the creation of massive states.
In field theory massive matter is made up of four degrees of freedom, 
corresponding to two chiral massless fields. 
In order to build  up massive fields, the lifted matter 
degrees of freedom must combine with those that a priori were left
massless by the shift. Namely, we have that the ${\bf 4}$, corresponding
to $U(4)$, is broken by this shift to ${\bf 2}   + {\bf \not{\! 2}}$:
\be
U(4) \, \to \, U(2)_{(\rm L)} \otimes \, \, 
{ \not{\!   U(\not{\! 2})_{(\rm R)} }} \, ,
\label{4to2+2}
\ee 
where the second symmetry factor is the broken one, 
with the corresponding bosons lifted to a non-vanishing mass.
The two matter degrees of freedom charged under this group
acquire a mass below the Planck scale, and combine with the two
charged under $U(2)_{(\rm L)}$. Therefore, of the initial fourfold
degeneracy of massless matter degrees of freedom, we make up light
massive matter, of which only the left-handed part feels an 
$U(2)$ symmetry, namely what survives of the initial symmetry.
Indeed, the surviving group is the non-anomalous, traceless 
$SU(2)_{(\rm L)}$ subgroup. As we will see, this 
group can be identified with the group of weak interactions.
The chirality of weak interactions comes out therefore as a
consequence of a shift in space-time. This had to be expected:
the breaking of parity is in fact somehow like a free orbifold projection
on the space-time. Together with the generation of non-vanishing particle 
masses and the breaking of parity, this shift also breaks the rotation symmetry
of space-time, by separating the role of the two space-time transverse 
coordinates. We will come back to this issue in section~\ref{breakL}.

The spectrum does not contain the degrees of freedom of a possible Higgs boson.
On the other hand, here there is no need of such a field, because
masses are generated with a pure stringy mechanism, and
are basically related to the compactness of the whole space. As remarked
in~\cite{assiom-2011}, the Higgs boson of
ordinary field theory can in some way be thought as the parametrization
of a boundary term through a field propagating in the bulk of space.

It is legitimate to ask what is the mass scale of the gauge bosons 
of the ``missing'' $SU(2)$, namely whether
there is a scale at which we should expect to observe an enhancement of 
symmetry. The answer is: there is no such a scale.
The reason is that the scale of these bosons is simply T-dual, with 
respect to the Planck scale, to that of the masses
of particles. Let's consider this shift as seen from the heterotic side.
On the heterotic vacuum, matter states originate from the twisted sector,
while the gauge bosons (the visible gauge group, the one involved in this
operation) originate from the currents,
in the untwisted sector of the theory. Similarly, on the type~I side,
gauge bosons and the charged states we are considering originate
from D-branes sectors derived respectively from the untwisted, and the twisted
orbifold sectors of the starting type II theory 
\footnote{The type II vacua are on the other hand not appropriate
for the investigation of this phenomenon, because the gauge charges are 
non-perturbative. In any case, although in the form of just the Cartan 
subgroup of their symmetry group, gauge bosons and matter states
arise from mirror constructions, related each other by the type~II
dual of the heterotic T-duality under consideration \cite{striality}.}.
It is therefore clear that a shift on the string lattice 
lifts  the masses of gauge bosons and those of matter states in a T-dual way.
Since the scale of particle masses is below the Planck scale,
the mass of these bosons is above the Planck scale; at such a scale,
we are not anymore allowed to speak of ``gauge bosons'' or,
in general, fields, in the way we normally intend them.

\
\\

The shift just considered is the last 
``level-doubling, rank-reducing'' projection allowed by this perturbative
conformal approach. 
We are left however with two more coordinates which can accommodate
a shift. One is a further coordinate of the extended space-time, the other
is one of the twisted coordinates. From the heterotic point of view, this is
an internal non-perturbative coordinate; just for simplicity, we can identify
it with the 11-th coordinate of M-theory. 
A shift along the extended coordinate is somehow related to the breaking
of the last perturbative symmetry we are left with, the $SU(2)_{(\rm L)}$ 
symmetry. A shift along the 11-th coordinate
breaks instead the underlying S-duality of the theory. This symmetry
exchanges the role of strong and weak coupling.

With our approximation of $Z_2$ orbifolds
it is however not possible to investigate these effects in a complete way: 
with the first shift, we have reached the boundary of the
capabilities of the approximation we are using. 
$Z_2$ orbifolds are a good base on which to expand the 
string vacua of interest for us, but they 
constitute anyway an approximation. This results particularly clear if 
one considers that, in the orbifold description, the string target space
does not account for the whole detail of the extended space, namely for
its actual content, the full, real geometry of the universe, as implied
by \ref{zsum}. This space appears as
``factorized out''. Indeed, in the perturbative orbifold construction    
the whole target space appears factorized into orbifold 
planes, and also for the internal space
we get the fake impression of a symmetry between these planes.
As a consequence, the three matter families appear on the same footing. 
We have on the other hand an indirect indication that, non-perturbatively,
this symmetry is broken, and the extended space gets somehow ``embedded''
into the internal space.

\subsubsection{Breaking the symmetry between the three matter families}
\label{breakthree}

By comparing dual string constructions,
it is possible to indirectly see how the shifts acting on the various 
coordinates of the string space break 
the symmetry between orbifold planes. 
This holds both in the case of the symmetry between the three
orbifold twists that give rise to the three matter families, and
for the symmetry between these twists and the twist "along the
eleventh coordinate". In this second case, the breaking of the symmetry
will mean not only the breaking
of the strong-weak coupling symmetry (S-duality), and
the related equivalence of weakly and strongly coupled matter, 
but also the breaking of the $S-T-U$ symmetry exchanging
weakly with strongly coupled matter, and either of the two with radiation 
(the physical meaning 
of this symmetry will be discussed in section~\ref{ubh}).  

Let us consider the action of shifts
at the "${\cal N}=2$" level, the step at which the three sectors,
that we will shortly indicate as the "$S$", "$T$" and "$U$"
sector, originate.
The interpretation of them as the sectors corresponding
to the three matter families, or to the $S-T-U$ non-perturbative
symmetry, depends on the interpretation one gives to
the specific dual perturbative constructions, namely on
whether one decides to explicitate the "eleventh coordinate"
trading it with one of the perturbative coordinates of the
string target space, or not.
These sectors appear on a different footing on dual constructions:
what in a heterotic construction
appears as a perturbative shift on the momenta of a "T-U" lattice which affects the 
corrections to some coupling of the same sector (the "S" sector, by convention), 
on a type II dual construction shows out as an operation that lifts the mass of
states which were non-perturbative from the heterotic point of view.
From the heterotic point of view, the T and U sectors are related by a perturbative 
T-duality. In this realization one can explicitly see that
the shifts break this symmetry. In particular, the action of the
shifts on the internal coordinates results in a different
(T-dual) rescaling of the volume
of the string space.
This implies a rescaling of the coupling, which as a consequence affects 
the measure of masses in units of the Planck 
mass~\footnote{The coupling is related to
the ratio of Planck and string scale (see for instance
expression~\ref{mpms} or \ref{mpmstu} for the dependence
at the leading order). In the simple case
of a series of freely-acting orbifold projections
realized through lattice shifts on a two-torus, the loop corrections
to the coupling amount to a sum of logarithms of Jacobi $\theta$
functions, which distinguish between the dependence on the modulus T
and U. In more complicated cases, when also a dependence on the coordinates
of the extended space is introduced, the $\theta$ 
functions are substituted by more complicated expressions.
It remains however the differentiation in the contribution
to the volume factor, i.e. in the scaling of the 
coupling.}: as a consequence of the shift, that rescales
differently the coupling of the T and of the U sector, 
in each of these sectors 
the dependence of sub-Planckian masses on the radius of the
extended space enters multiplied by a different factor. 
As we will discuss in section~\ref{masses}, the orbifold construction corresponds
to a logarithmic realization, i.e. on the tangent space, of the full, non-perturbative
string configuration. Instead of 
the true coordinates of the physical space,
one works with their logarithm: $R \to \log R \equiv r$.
As a consequence, the rescaling of a dependence in the logarithmic realization on $1/r$
by some numerical factor, $1 / r \to k  / r$ reflects in a power-law dependence on the real
radius of space: $1/R \to (1 / R)^{1 / k} $. As we will see, the ratios of masses between
different families are indeed of the order of some power of the radius $R$:
$m_i / m_j \approx (1 / R)^{\alpha_{ij}}$.

Through a comparison of dual constructions we can see where 
the differentiation of sectors comes from, but we cannot
follow the whole pattern of symmetry breaking, because
we can only see the fate of
pairwise two sectors out of three: one of the three sectors is always
excluded from the investigation, because it involves the coupling
over which one of the dual realizations is perturbatively
constructed.
All this leads to expect that the configuration of highest entropy
lies a bit ``displaced'' from the orbifold point.
However, we expect the corrections to the orbifold 
approximation to be related to the scale of extended space-time.
More precisely, owing to the breaking of T-duality produced by the
shift along the coordinates of the extended space, we expect these corrections
to be of the order of some root of the inverse of the radius, or age,
of the universe. 
The mean value of the observables on the configuration of minimal symmetry
is expected to be of the type:
\be
< {\cal O} >_{\psi^{\rm min}} ~ \sim ~
\left. < {\cal O} >_{\psi^{\rm min}} \right|_{Z_2} \, + \,
\Delta < {\cal O} >_{\psi^{\rm min}}  \, , 
\label{psidel}
\ee
with
\be
\Delta  < {\cal O} >_{\psi^{\rm min}}  ~ \approx ~
{\cal O} \left( 1 / {\cal T}^p  \right) \, , ~~~~ p > 0 \, .
\label{delpsi}
\ee
The dependence on this scale as a root and not simply to
the first power depends on the action of the orbifold shifts, something we will
discuss more in detail.

\subsection{Origin of the $SU(3)$ of QCD and low-energy spectrum}
\label{les}

At the point we arrived,
the low energy world appears made out of light fermionic matter
(\emph{no scalar} fields are present!),
massless and massive gauge bosons. Matter states are
charged with respect to bi-fundamental representations.
More precisely, they transform in the 
$(({\bf 2} \oplus \, {\bf \not{\! 2})}, {\bf 4 })$, 
replicated in three families: $(({\bf 2} \oplus \, {\bf \not{\! 2}}), 
{\bf 4^{\prime}})$,
$(({\bf 2} \oplus \, {\bf \not{\! 2}}), {\bf 4^{\prime \prime}})$, 
$(({\bf 2} \oplus \, {\bf \not{\! 2}}), {\bf 4^{\prime \prime \prime}})$.
As we discussed, a perturbative shift
lifts the mass of the gauge bosons of the second ${\bf 2}$
above the Planck scale and gives at the same time a light mass
to the matter states. The ${\bf 4}$ (i.e. ${\bf 4^{\prime}}$,
${\bf 4^{\prime \prime}}$ and ${\bf 4^{\prime \prime \prime}}$)
are however at the strong coupling, and 
therefore broken symmetries. Indeed, outside 
of the orbifold point, we should 
better speak in terms of a larger symmetry, 
into which the matter degrees of freedom transform as 
${\bf 12} \supset {\bf 4^{\prime}} \oplus {\bf 4^{\prime \prime}}
\oplus {\bf 4^{\prime \prime \prime}}$, which is 
broken and at the strong coupling. The (perturbative)
orbifold point constitutes a 
simplification, in which the ${\bf 12}$ has been rigidly broken by 
the orbifold twists;
the real configuration is a perturbation of this simplified situation, in which
the bosons of the broken symmetry, which were acting among orbifold planes,
acquire a non-vanishing mass, of the order of some power of
the age of the universe. 
On the other hand, minimization of symmetry tells us that indeed
the ${\bf 12}$, besides being broken into 
${\bf 4^{\prime}} \oplus {\bf 4^{\prime \prime}} 
\oplus {\bf 4^{\prime \prime \prime}}$, are also further
broken to the minimal confining subgroup. Had we to represent
this in terms of field theory, this would imply as unique possibility
the breaking of ${\bf 4}$ into ${\bf 1} \oplus {\bf 3}$,
corresponding to $SU(4) \to U(1) \times SU(3)$.  
However, from the string point of view, strictly speaking the
$�SU(3)$ symmetry does not
exist: the vacuum appears to be already at the strong coupling, and the only
asymptotic states are $SU(3)$ singlets.
More than talking about gauge symmetries, what we can do is
to  \emph{count} the matter states/degrees
of freedom, and \emph{interpret} their multiplicities as due to 
symmetry transformation properties, 
as they would appear, in a field theory description,
``from above'', i.e. before flowing to the strong coupling.
This is an artificial situation, that does not
exist in the actual string realization. The string vacuum
indeed describes the world as we observe it, namely, with quarks
at the strong coupling. From: i) the counting of
the matter degrees of freedom, ii) knowing that this sector
is at the strong coupling as soon as we identify the 
``$({\bf 2} \oplus \, {\bf \not{\! 2}})$'' 
as the sector containing the group of
weak interactions, and iii) the requirement of consistency with
a field theory interpretation of the string states below the Planck 
scale~\footnote{For us, field theory is not something realized
below the string scale. It is rather an approximation (that locally works),
obtained by artificially considering the space-time coordinates as
infinitely extended. As a consequence, the ``field theory'' scale
is in our set up the scale of the compact, but non-twisted, string coordinates.
In order to keep contact with the usual approach and technology,
what we have to do is to send to infinity the size of space-time,
and consider the \underline{massless} spectrum of four
dimensional string vacua.},
we derive that the only possibility is that this representation
is broken to $U(1) \times SU(3)$. A further breaking would in fact
lead to a negative beta-function, in contradiction with point ii).
The operation that breaks the ${\bf 4}$ into ${\bf 3}$ and ${\bf 1}$
corresponds to the ``shift'' along the 11-th coordinate we mentioned
in section~\ref{massmatter}. This operation,
that cannot be represented at the orbifold point, doesn't work as a
``level doubling'', but rather as a ``Wilson line''.
Notice that, coming from the breaking of an $SU(4)$ symmetry, the
$U(1)$ factor is traceless. This means that it acts by transforming with
opposite phase states charged under $SU(3)$ and uncharged ones:
\ba
U(1)_{\beta} 
\, \varphi & = & {\rm e}^{{\rm i} \beta  } \, \varphi \, , \nn \\
& & \label{U1phi} \\
U(1)_{\beta} 
\, \varphi_{a} & = & {\rm e}^{- {\rm i} \, \beta / 3  } \, \varphi_a \, , 
~~~~~ a \, \in \, {\bf 3} \, {\rm of} \, SU(3) \, .
\nn 
\ea
At this point, we are left with one more extended
space-time direction where to accommodate a shift, and therefore further 
break the symmetry.

At the point of minimal symmetry, the gauge groups are broken to $U(1)$'s.
This is true for the groups which are at the weak coupling, 
namely those corresponding to
the $({\bf 2} \oplus \, {\bf \not{\! 2}})$. It doesn't hold for the
$SU(3)$ ``colour'' symmetry, which is broken and at the strong coupling.
Let's therefore concentrate on the $({\bf 2} \oplus \, {\bf \not{\! 2}})$.
This derives from an initial $U(4)$, then broken to $U(2) \otimes U(2)$
and ``patched'' to a single, massless $U(2)$, with the second $U(2)$ 
lifted at over-Planckian mass. Of this massless $U(2)$, acting on half
of the chiral matter states, indeed only an $SU(2)$ subgroup survives,
the traceless part. This can be identified with the group of the weak 
interactions, $SU(2)_{\rm w.i.}$, which acts only on the left moving part
of the matter states. A slight displacement away from the orbifold point,
driven by moduli of the size of the inverse of the coordinates of 
the extended space-time (${\cal O}(1 / {\cal T}^p)$, $p < 1$), 
breaks this group to $U(1)$. 
This displacement does not break parity: it commutes with the 
operation that lifted the second $U(2)$ (indeed, one can think to perform 
this displacement first, and then the level-doubling shift on the 
$U(1) \times U(1)$ group). 

Consider now the last shift operation, either at the $U(1)$ or at the 
extended $SU(2)_{\rm w.i.}$ gauge symmetry point. 
This shift cannot act as a level-doubling projection, as it
can be seen by looking from a heterotic 
point of view, in which the group from which the $SU(2)_{\rm w.i.}$
of the weak interactions originates
is realized on the currents. A further rank reduction through a level-doubling
orbifold projection is forbidden by the embedding of the spin connection 
into the gauge group. The effect of this shift is that of lifting the mass of
all the gauge bosons. This action cannot be followed in a heterotic 
construction, where the gauge fields arise from the currents, and cannot be
``shifted''. What happens can be observed on the type~II side, where the 
gauge fields originate in the twisted sectors. From the type~II point of view,
it is clear that such a shift corresponds to making the orbifold projection
to act freely. For a shift realized ``on the momenta'' of space-time,
i.e. a ``field theory shift'', as the one we are considering,
the mass of the bosons will be below the Planck mass:
\be
m_{W^{i}} \, {\to} \approx \, {1 \over 2} \times {1 \over R} \, ,
\label{mWshift}
\ee
where $R$ is the radius of the shifted coordinate. 
We will come back to discuss the mass of
the $SU(2)_{\rm w.i.}$ bosons after having gained some insight into
the relation between perturbative mass expressions and the perturbatively 
resummed ones, in sections~\ref{masses} and \ref{todaymass}.
Differently from the usual case, in which the matter
states are lifted in a T-dual way to the bosons, here the mass of the matter 
states receives a shift of the same order as the
gauge bosons. This phenomenon can be traced by looking at the type~II dual
realization. As discussed also in Ref.~\cite{striality}, on the type~II
side there are two mirror constructions: in one it is the (Cartan subgroup of 
the) gauge group which is explicitly realized on the twisted sectors;
in the mirror we see instead the matter states. The two constructions
are related by a T-duality in the internal coordinates, therefore not
involving a transformation of the space-time
coordinates in which this last shift acts. A shift that lifts gauge boson
masses by acting on the momenta of one such coordinate remains
a shift in the momenta also as viewed from the mirror construction, in which
it is seen to lift the mass of matter states. What we expect is therefore
that the mass difference between the up and the down of a broken   
$SU(2)_{\rm w.i.}$ doublet is of the order of the broken boson masses.
Namely, we expect the mass difference of the heaviest doublet,
the one that sets the scale of the breaking of this symmetry 
in the matter sector, to correspond to
the scale of the breaking of this symmetry in the gauge sector
($m_W^{i}$ mass). We will come back for a more detailed
discussion in sections~\ref{masses} and~\ref{todaymass}.  
At the point of minimal entropy, the mass lift \ref{mWshift} is ``combined'' 
with a further mass shift produced by the
breaking of the $SU(2)_{\rm w.i.}$ symmetry to $U(1)$. 
This is a ``second order'' deformation, driven by moduli as in \ref{psidel},
\ref{delpsi}. The $U(1)$ boson
will have therefore a mass in first approximation of the same order
of the one of the $W^{+}$ and $W^{-}$ bosons, plus a mass difference
produced by a parity-preserving, and therefore left-right symmetric,
displacement.

\vspace{.3cm}

Leptons and quarks originate from the splitting of the 
${\bf 4}$ of $SU(4)$ (replicated in three families,
$({\bf 4^{\prime}},{\bf 4^{\prime \prime}},{\bf 4^{\prime \prime \prime}})$),
as ${\bf 3} \oplus {\bf 1}$ of $SU(3)$. 
All in all, we have therefore two $U(1)$ groups, or, better, if we also 
consider the Cartan of the lifted $SU(2)$ ``Right'', that we know
to have acquired an over-Planckian mass, three $U(1)$s.
Let's indicate their generators as $T_3^{\rm L}$, $T_3^{\rm R}$ and 
$\tilde{Y}$.
The charge assignments along each of the three $({\bf 4} \otimes {\bf 4}^{i})$
are then:
\be
\left( Q(T_3^{\rm L}) \oplus Q(T_3^{\rm R}) \right)
 \otimes \left( Q (\tilde{Y}) \right) \,
= \, \left( {1 \over 2}, - {1 \over 2} \oplus {1 \over 2}, - {1 \over 2}  
\right)
\otimes \left( \beta, - \beta/3 , - \beta/3 , - \beta/3 \right)
\, .
\label{Q44}
\ee
$U(1)_{\tilde{Y}}$, the group \ref{U1phi}, can be identified with some kind of
hypercharge group. This is not exactly
the hypercharge of the Standard Model: the charge assignments are here 
left-right symmetric, and $SU(2)$ invariant, because the two
$SU(2)$, $SU(2)_{(\rm L)}$ and $SU(2)_{(\rm R)}$, of which the 
``Left'' gives rise to the weak interactions group, are here
factorized out. Since the hypercharge and the $SU(2)$ groups arise from
mutually non-perturbative sectors, the hypercharge eigenstates
are here singlets of the $SU(2)$, left \emph{and} right, symmetries.
This means that the phase refers to the sum of the up and down
of each family:
\ba
\beta & \propto & \tilde{Y}_e \, + \, \tilde{Y}_{\nu} \, , \label{QtildeYl} \\
{\beta / 3} & \propto & \tilde{Y}_{\rm up} \, + \, 
\tilde{Y}_{\rm down} \, ,
\label{QtildeYq}
\ea 
this for any family and colour. By using the standard normalization of 
Lie group generators, we set $\vert \beta \vert = 1$, and,
by pure convention, $\beta = -1$. This group, the only surviving gauge group,
whith a truly massless boson, can be identified with the electromagnetic
group, and its charges with the ``electric'' charge assignment of the 
elementary particles of this vacuum.
The problem is to see how the electric charge is distributed among the
$SU(2)$ doublets, namely, according to which fraction the decompositions
\ref{QtildeYl} and \ref{QtildeYq} are made. We know that, as a matter of fact,
in nature this is not done ``democratically'': 
the electron has charge -1, and the neutrino is uncharged. The same
charge difference applies also to colour triplets of quarks. This can be
interpreted as the result of having taken a linear combination, 
more precisely the sum, 
of the $\tilde{Y}$, $T^{\rm L}_3$ and $T^{\rm R}_3$ charges.
In our scenario, this distribution of the electric charge among the $SU(2)$ 
multiplets turns out to be required by minimization of the amount
of symmetry of the configuration. 
Among all the possible arrangements of this kind of hypercharge,
the one corresponding to
the most entropic string configuration
is the one for which one particle becomes uncharged for one of
the interactions, the electromagnetic one. In this case, highest is
the differentiation among the spectrum of particles,
i.e. the symmetry of the configuration is the lowest possible. 
Of course, in the whole phase space also
other choices are present, but they correspond to less entropic
configurations, which therefore give a minor contribution to the
mean value of observables.
The less interacting particle must also be
the lightest one.
As we will discuss in section~\ref{entropymass}, 
the mass of a particle is related to its weight
in the phase space. The latter is in turn
related to the number of paths leading to its configuration.
The number of paths is related to the strength of the interacting power
of the particle: the more are the interactions the particle possesses,
the more entropic is the physical configuration it corresponds 
to, because larger is the particle's 
``spread out'' in the phase space.
The neutral particle we obtain must be identified with the neutrino.

The inclusion of the right-moving quantum numbers
is not an ``ad hoc'' assignment, chosen in order to get the correct charge 
assignments: these are instead the consequence of the fact that, in this 
scenario, ${\cal T} \to 1$ is a limit of left-right symmetry restoration, where
the dual, over-Planckian masses, come down to match the sub-Planckian ones.
This is basically the meaning of having broken parity in a soft-way, by
a shift along the extended coordinates.
Therefore, no one of the charge assignments in \ref{Q44}
can explicitly break this symmetry \footnote{The case of the strong-weak 
coupling separation, related to the breaking of an $S$-duality, is different, 
because it involves a shift along
coordinates which are also twisted, and therefore there is no restoration
at the limit. The breaking is neither soft nor ``spontaneous''.}. 
Here is a point of difference
with respect to the ordinary approach to the building of the electromagnetic 
group: in the usual approach, the hypercharge assignments explicitly
break the $SU(2)$ and the parity symmetries.

Once the charge of the neutrino has been fixed, 
all other charges result correctly determined. Owing to the shift of the
``hypercharge'' with the $SU(2)$ charges, the electromagnetic current doesn't 
couple only with the ``long range'' force, the ``photon'', i.e. the
massless field associated to the $U(1)_{\tilde{Y}}$ symmetry (indeed
the sum of the up and down electric charges), but also with the massive 
neutral fields originating from the broken $SU(2)$s.
The ``Right'' one is very massive, over the Planck scale, and therefore 
the corresponding interaction channel can be neglected: from a field theory
point of view, ``it does not exist'' \footnote{This is a major
point of difference with respect to the field theoretical ``left-right''
extensions of the Standard Model.}. In section \ref{gmass}
we will discuss the masses of the broken $SU(2)$ and the ``Left''
$U(1)$ boson. The coupling of the electromagnetic current will
be derived and discussed in sections~\ref{U1beta} and \ref{a-fine}. The
$SU(2)_{\rm w.i.}$ coupling and the coupling of the neutral current
with the under-Planckian massive neutral boson, the ``$Z$'' boson,
will be discussed in sections~\ref{betaf} and \ref{gmass} respectively.

\begin{figure}
\centerline{
\epsfxsize=10cm
\epsfbox{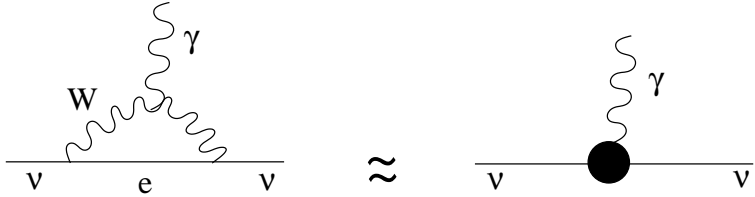}
}
\vspace{0.4cm}
\caption{An effective neutrino electric 
charge is generated through higher order corrections.}
\label{q-nu}
\vspace{.8cm}
\end{figure}

It may seem disappointing that relations that we are used to consider
associated to an exact symmetry, or concerning
the exact vanishing of a charge, in this case the electric charge
of the neutrino, appear in this scenario ``softened'', a consequence
of maximization of entropy: there are no
``conservation laws'' strictly forbidding other solutions, and
protecting these charges and masses. However, this is precisely
what we should expect in a quantum scenario, and is accounted for
in \ref{zsum} and \ref{zint}. 
In a way similar to what the well known path integral does,
here \emph{all} possible configurations contribute to build up the
universe as we observe it, although not all with equal weight.
Also in traditional quantum field theory
the electric charge of the neutrino
vanishes only as long as we consider the neutrino as a
free, asymptotic state. Interactions generate indeed an effective
non vanishing electric charge for this particle, 
in the sense that, as illustrated in figure~\ref{q-nu},
at higher orders, a neutrino effectively interacts
also with the photon. 
\noindent

\subsection{The breaking of Lorentz/rotation invariance}
\label{breakL}

We have seen that in our framework 
the origin of masses is related to the breaking of parity and time-reversal,
and that the same operation implies also the breaking of rotation invariance.
This is not a surprise, because \ref{zint} is expected to describe the universe
``on shell'', i.e. as it is, the space-time 
\emph{with} its energy and matter content at any scale, and the breaking of  
these symmetries at the macroscopical level is part of the common
experience of everyday.
Indeed, the entire analysis of the string orbifold vacua has been performed in
the light-cone gauge, where the Lorentz symmetry is explicitly broken.
However, the breaking due to a particular choice of gauge is not a real
breaking: it is just a convenient representation, through a particular
``slice'', of an invariant construction. 
On the other hand, in our framework, what happens is that
the Lorentz symmetry is really broken. The breaking is realized at two levels:
1) the breaking of the Lorentz boosts; 2) the breaking of the subgroup of space
rotations. The Lorentz boosts are not a symmetry
in a compact space-time, in which transformations in space and time
correspond to an evolution of the universe. Nevertheless, 
rotations in principle could remain a symmetry of space: they are not in 
contradiction with its compactness. However, here we discover that maximization
of entropy not only implies the breaking of parities, 
but also the breaking of the symmetry of space 
under rotations. And this precisely at the same time as masses are produced.
This does not come unexpected: the very fact of
placing a matter excitation somewhere in space breaks this invariance, 
selecting a preferred direction, in a way similar to the one a Higgs field 
breaks a gauge 
symmetry by selecting a position of minimum 
among the entire orbit of the symmetry group. 
In the average, owing to the presence of a lot of particles, existing as single
sources of curvature or grouped into larger objects almost homogeneously
distributed in the space, we can still say that
the universe is on a large scale invariant under rotation. 
Nevertheless, strictly
speaking this is not a ``pure'' invariance. From a field theory point of
view we would say that it is ``softly broken''.
It is not a surprise that precisely the mechanism that in our
framework substitutes the Higgs mechanism in giving rise to masses is
also responsible for the breaking of the space rotation symmetry, through 
shifts acting on the transverse space coordinates: 
the one associated to the breaking of 
parity, and the one associed to the breaking of $SU(2)_{\rm w.i.}$.
The amount of breaking of space rotations produced in our framework is
of the same order of the particle masses. 
There is no fracture 
between a fundamental, microscopical physics
possessing certain symmetries, and a macroscopical world in which for some
reasons these laws are violated. 
In our framework the microscopical
and macroscopical descriptions of the world are sewed together: 
time-reversal, space-parity, reversibility, and symmetry under rotations 
are broken at a very fundamental level.

\subsection{The fate of the magnetic monopoles}
\label{Mpoles}

Under the conditions of the scenario we are discussing,
namely of a universe ``enclosed'' within a finite, compact space, also 
the issue of the existence of magnetic monopoles changes dramatically. 
Magnetic monopoles can be of two kinds: the ``classical'' ones, namely
those associated to a non-vanishing ``bulk'' magnetic charge that
parallels the electric charge in a symmetric version of the Maxwell's 
equations, and the topological ones. In our scenario 
there are no ``classical'' monopoles: their
existence would be possible only in the absence of an electro-magnetic
vector potential, what we have called the ``photon'' $A_{\mu}$; their existence
has therefore been ruled out as soon as we have discussed 
the existence and the masslessness of this field. 

The first idea about the existence of 
magnetic monopoles in the classical sense (i.e. non-topological)
originated by a request of symmetry: were not 
for the absence of magnetic charges, the Maxwell equations would be 
completely symmetric in the electric and magnetic field. However, the symmetry 
of these equations, preserved in empty space, is precisely spoiled by the 
presence of matter states that are also electrically charged. In our scenario, 
the description of the universe is ``on-shell'' and the presence of matter
comes out as ``built-in'': it cannot be disentangled from the existence of 
space itself. 
As the most often realized configuration of the universe is the one with the 
lowest amount of symmetry, it is not so surprising that, 
for the same reason for which charged matter states are generated, 
in correspondence with a local breaking of the invariance of space under
$SO(3)$ rotations, also the symmetry of the Maxwell's equations is broken.

In this scenario there are no topological monopoles 
either. As all vector fields 
are twisted (i.e. massive at the Planck scale or above it) with the only 
exception of the photon $A_{\mu}$, propagating in the four-dimensional space 
time, and as this space-time dimensionality is electro-magnetically self-dual, 
the only possible topological monopoles would be those of the four-dimensional 
space coupled to the same photon field $A_{\mu}$, 
namely, configurations \`{a} la t'Hooft and Polyakov or 
similar~\footnote{for a review and references, 
see for instance~\cite{Coleman,Bais:2004sc}.}.
However, any such topological configuration is characterised by its
being living in an infinitely-extended space: only in this way it is in fact
possible to make compatible the existence of a $p$-form working as 
a ``potential'' $A_{(p)}$, defined as an analytic function 
in every point of the space, 
with the presence of a non-trivial magnetic flux.  
As is well known, the magnetic flux through a surface can be computed 
as a loop integral of the vector potential. 
In the case of a surface enclosing a 
finite volume, the total flux is the sum of the loop integral circulated 
in both the opposite directions, so that it always trivially vanishes.
However, things are different if the field has a non-trivial behaviour at 
infinity. At infinity we need just the circulation in one sense, because
there is no ``outside'' from which field lines can ``re-enter'' in the 
space: if there is a non-vanishing circulation, there is a non-vanishing
magnetic flux, and therefore also a non-vanishing magnetic charge. This however
also means that, provided it exists, such a magnetic monopole is a highly 
non-localised object, with a magnetic field/vector potential such that the
magnetic flux vanishes through any compact finite closed surface 
\footnote{Notice that the situation around the zero-dimensional
point is equivalent to the one around the surface at infinity: 
if on one side the Dirac string can be considered as somehow
the ``dual'' picture of the surface at infinity of the
t'Hooft and Polyakov construction, in our scenario both infinity and
the dimensionless point are excluded. Differential 
geometry and gauge theory are here only approximations.}.
As a consequence, also the magnetic charge density vanishes point-wise at any
place in the ``bulk''. 
Therefore, in our setup, where space is compact, these monopoles cannot
exist. Moreover, 
in our case we don't have a Higgs mechanism either, and,
since the surface at infinity does not belong
to any configuration of space-time, there is no smooth
limit with a true restoration of
the conditions at infinity allowing the existence of non-trivial topologies
and homotopy groups. Light states with topological magnetic
charges do not exist at all, not even approximately as the
time becomes very large \footnote{The situation is similar to
the case of the volume of the group of translations and its identification
with the regularized volume of space in the usual normalization of
operators and amplitudes, completely absent in our scenario, 
something that leads to a different interpretation of string amplitudes
as global quantities instead of densities, cfr. Ref~\cite{assiom-2011}.}.

\subsection{Effective action}
\label{eth}

The various string constructions have to be considered as slices,
at different points in the moduli space, of the same theory. 
Dual constructions corresponding
to slices of overlapping regions of the moduli space. 
The comparison of dual 
string constructions makes only sense once the expressions of the
physical quantities are converted into common units, i.e. when, 
instead of being expressed in terms of the proper length of each string 
construction, they are expressed in terms of the
Planck length or mass scale.
Whenever one can write an effective action for the string light modes,
this conversion corresponds to the passage from the string frame to the
duality-invariant, so called  Einstein's frame, also referred to as
the ``supergravity frame''. In the case of compact space-time,
the existence of an effective action
is however not obvious. Traditionally, i.e. in an infinitely extended space 
time, the limit of infinite volume selects as elementary excitations of the 
propagating modes one of the two T-dual worlds, or towers of states
whose quantum numbers correspond either to windings or to momenta. 
The other ones (therefore either the momenta or the windings)
are ``trivialized'' to a constant contribution. 
In the case of compact space, at any volume both T-dual degrees of freedom are 
instead non-trivially present, and the 
selection of a configuration close to the case of infinite volume is only
possible when T-duality along the space-time coordinates is broken.
The breaking of T-duality leads in fact to the choice of a ``direction'',
or scale, for a field theory representation: either below 
or above the Planck scale, two situations no more equivalent. 
As we have seen, in the configuration of minimal symmetry 
T-duality in the space-time coordinates is broken. 
It is then possible to talk about ``extended'' space-time, along which
degrees of freedom, representable as fields, can propagate.
We are therefore allowed to talk about an effective
action, in which the light modes of the theory, namely those whose mass
is below the Planck scale, move in a space-time frame 
of coordinates larger than the Planck length, while
heavier modes are integrated out and their
existence manifests itself only through their contribution to the 
parameters of the effective theory. Although 
the case of infinitely extended space-time is never realized,
at large space-time volumes it can constitute a useful approximation.
This justifies the use of the language of traditional
string compactifications, and an
analysis in terms of orbifold dualities and comparison of dual
constructions as we did in section~\ref{spectrumZ2}, namely 
in terms of the ``sectors'' $S$, $T$ and $U$ of the theory,
compared through the reduction of the effective actions
to the common, Einstein's frame.
At the ${\cal N}_4 = 2$ level, the last step at which we could still
explicitly follow the pattern of dual constructions, 
for the heterotic string the relation between the string and the Planck scale
is given by:
\be
{\rm M}^2_{\rm P} \, \equiv \, {\rm M}^2_{\rm Het} \Im S^{-1} \, , 
\label{mpms}
\ee 
where $S = \chi + {\rm i} {\rm e}^{- 2 \phi}$, $\phi$
is the dilaton field. 
On the type I side there are two such fields, called $S$ and 
$S^{\prime}$, that parametrize the coupling of different D-branes sectors
of the theory. Each sector has therefore its own ``string length'', to be
rescaled to the Planck length. 
By inspecting string-string duality with the type II string, 
we have on the other hand learned that
the vacuum of interest for us  possesses
at the ${\cal N}_4 = 2$ level three such ``coupling fields'', corresponding to
three sectors of the theory (see also Ref.~\cite{striality} 
for a more detailed discussion). 
This structure, namely the triplication of the string vacuum, is preserved 
when going to lower (super)symmetric, less entropic 
configurations, up to the one of minimal symmetry.
We can therefore talk of three ``waves'', the ``$S$'', ``$T$'' and ``$U$''
``wave'' of the theory. Correspondingly, there are three
``string scales'' to be related to the Planck scale. In terms of the notation
``$ S , T , U $'', these relations read:
\be
{\rm M}^2_{\rm P} \, = \, {\rm M}^2_{\rm S} \Im S^{-1} ~~ = ~
{\rm M}^2_{\rm T} \Im T^{-1} ~~ = ~ {\rm M}^2_{\rm U} \Im U^{-1} \, .
\label{mpmstu}
\ee 
As we have seen in section~\ref{d=4}, the internal space is frozen
at the self-dual radius, ${\cal O}(1)$ in \underline{string units}.
This implies that string and Planck scales are equivalent, as anticipated.
From this equivalence, we conclude that the masses
of the states lifted by the shifts acting along the internal coordinates
are of the order of the Planck mass.

Owing to the implied breaking of time reversal, the breaking of T-duality 
leads to the choice of an arrow for the time evolution: space-time
volume expansion is not equivalent to space-time contraction.
Indeed, in this scenario the effect of the boundary of a finite volume
space cannot be neglected: it reflects in a dependence of the parameters
of the effective action on the age of the universe. To be more precise, instead of one effective action, we should better speak in this case of a series of 
effective actions, $S_{\rm eff} \to S_{\rm eff}({\cal T})$, for each one of
which one can formally write a Lagrangian density
for propagating fields bearing a dependence on a "bulk" time coordinate $t$,
but this remains only an approximation.

\newpage

\section{\bf The geometry of the universe}
\label{ubh}

According to the results of section 7.2 of Ref.~\cite{assiom-2011},
the energy density scales as~\footnote{We recall that in our theoretical framework,
owing to the absence of symmetry under translations, string amplitudes
are normalized in such a way that densities scale as the inverse of the Jacobian of
the transformation between string world-sheet and two target space coordinates
(see Ref.~\cite{assiom-2011}, section~7.2).}:
\be
\rho(E) ~ \sim ~ {1 \over {\cal T}^2} \, ,
\label{rEscaling}
\ee
whereas the entropy scales as:
\be
S ~ \propto ~ {\cal T}^2  \, .
\label{Sscaling}
\ee
The total energy at a certain time ${\cal T}$ of the history 
of the universe, given by the integral 
of the energy density over the space volume of the universe at time ${\cal T}$,
scales then as:
\be
{\rm E} ({\cal T})  
 ~ \sim ~ \int_{\cal T} d^3 {1 \over {\cal T}^2} ~~ \approx ~  {\cal T}
\, .
\label{et}
\ee
We will now discuss in detail the evaluation
of the mass and energy content of the universe, and confirm 
that the metric of the 
universe is the one of a sphere.

\subsection{The energy density and the cosmological constant}
\label{cosmo}

Let us start by computing in detail the energy density of the universe.
When investigating the minimization of symmetry, in
section~\ref{spectrumZ2}, we have seen that ${\cal N}_4 = 2$ was the
last step at which the couplings of the theory appeared as 
moduli of the string space. There, the theory
consists of three ``sectors'', corresponding to couplings parametrized by the
moduli ``$S$'', ``$T$'' and ``$U$'' \footnote{As we already pointed
out, we use the language of compactifications to four dimensions, whereas 
strictly speaking in our framework all string coordinates are compactified.
Nevertheless, as we are interested in situations
in which the extended space is very large as compared 
to the string scale, this abuse of language is not completely
inappropriate, and allows to simplify the discussion.}.
When the space is further twisted,
these moduli are frozen. Nevertheless, as we already pointed out in 
section~\ref{breaksusy}, the less supersymmetric theory
inherits the structure from ${\cal N}_4 = 2$. As we have seen, 
at the ${\cal N}_4 = 2$ level the heterotic realization,
being constructed around a small/vanishing expectation value of 
the coupling parametrized by $S$ ($\Im S$),
corresponds to a perturbative realization
of the ``$S$-sector''. This is the ``slice'' of the theory which in natural way
describes the coupling with gravity. We will call it the 
``$S$-picture''. When going to vacuum of lowest symmetry, by ``$S$-picture'' 
we have to intend a non-perturbative slice, whose perturbative
limit would correspond to the heterotic construction.

By comparison of dual constructions at the extended supersymmetry level,
it is possible to see the symmetry of the theory under exchange of the three
sectors, symmetry which is then inherited by the ${\cal N}_4 = 0$ vacuum.
In each sector, the energy of the universe is 
given by the mean value of the identity 
operator \footnote{See Ref.~\cite{assiom-2011}, section~7.2.}; the total amount
is given by the sum of the contributions of the three sectors. 
In the configuration of minimal entropy, 
the $S \leftrightarrow T \leftrightarrow U$ symmetry
is broken. As discussed in section~\ref{breakthree}, the breaking is tuned by the
non-twisted coordinates, and, owing to the breaking of T-duality,
the corrections to mean values 
are of the order of some root of the inverse of the age of 
the universe, as in \ref{psidel}, \ref{delpsi}. 
As a consequence, we expect that also the contributions to
the energy of the universe differ by some power (indeed some root)
of the inverse of its age:
\be
{E_i \over E_j} ~ \approx ~ 1 \, + \, 
{\cal O} \left( 1 / {\cal T}^{p_{ij}} \right)\, , ~~~~ p_{ij} > 0 \, ,
\label{Ei/Ej}
\ee  
where $E_i, E_j$ stay for $E_{S}$, $E_{T}$, $E_{U}$. 
These three contributions to the energy of the universe
differ in their interpretation. 
A closer look at the situation through comparison of the heterotic and 
type~I pictures (see also Ref.~\cite{striality}) reveals that the $S$, $T$ and
$U$ pictures are related by S- and T-dualities, that exchange weakly and
strongly coupled sectors (i.e. confining matter and non-confining particles,
together with their gauge bosons, including the photon)
and the ``bulk'', i.e. the gravity sector.
Therefore, the mean value of the identity operator,
when inserted in the gravity sector, computes the vacuum energy of the
``empty space-time'' in the ``M-theory'', or ``String Theory'' frame.
In the other sectors, the mean value of the identity
corresponds to the energy of matter
(the confining part) and of ``relativistic particles'' respectively.
The basic equivalence of these three contributions to the
energy of the universe is therefore a consequence of the symmetry
between the ``$S$-''. ``$T$-'' and ``$U$- waves'' of the 
dominant configuration; symmetry which is eventually broken   
by the very minimization of symmetry, but the breaking is ``soft'' as compared
to the scale of the twisted space.

We want to pass now from energies to energy densities.
As discussed in \cite{assiom-2011}, this passage introduces a dependence
on the Jacobian of the transformation of two space-time coordinates
from the string/Planck scale to the actual scale of space-time.
Let's start by considering the contribution to the energy density in 
the ``$S$-wave''. This corresponds to the standard heterotic picture.
According to \ref{rEscaling}, it is given by:
\ba
E_{S}\vert_{\rm Einstein~frame} & = &
\, < {\bf 1} >\vert_{\rm Einstein~frame} \; =
 \nn \\
& & \nn \\
& = & < {\bf 1} >_{S} \vert_{\rm string~frame} \, 
\times \, {\kappa \over {\cal T}^2} \; = 
\; E_{S}\vert_{\rm string~frame} \times {\kappa \over {\cal T}^2 }\, ,  
\label{es}
\ea
where $\kappa  / {\cal T}^2$ is the Jacobian of the transformation
from the string frame to the Einstein's frame, and $\kappa$ a normalization 
constant. According to \ref{Ei/Ej}, $E_{S}\vert_{\rm string~frame}$
depends on time at the second power, and, owing to the fact that:
\be
< {\bf 1} >_{S} \vert_{\rm string~frame} \, + \,
< {\bf 1} >_{T} \vert_{\rm string~frame} \, + \,
< {\bf 1} >_{S} \vert_{\rm string~frame} \; = \, 3 \, ,
\label{1stu}
\ee 
we have also that:
\be
E_{S}\vert_{\rm string~frame} \, + \,
E_{T}\vert_{\rm string~frame} \, + \, 
E_{S}\vert_{\rm string~frame} 
\; = \, 3 \, .  
\label{Estu}
\ee 
The quantity \ref{es} corresponds to the so 
called cosmological constant.
In order to fix the normalization $\kappa$ we must take into account that 
in passing from the string picture to the effective action term, besides
the conversion from target space to world sheet of the two longitudinal
coordinates, responsible for the two-volume factor, there is
also a further space-time contraction due to the $Z_2$ orbifold projection.
On the four (two transverse and two longitudinal) space-time coordinates
act altogether two $Z_2$ shifts, which lead to a factor 4 in the conversion
of a four volume. In our case, we have a two-volume, and the factor is 2. 
We obtain:
\be
\Lambda ({\cal T}) ~ \equiv ~
E_{S}\vert_{\rm Einstein~frame} \,  =  \, {2 \over {\cal T}^2} 
\, \times \, \left[  1 \, + \, {\cal O} \left( { 1 \over {\cal T}^{q_S} }  
\right)    \right]
\,  , 
\label{cosmoc}
\ee
where the quantity within brackets accounts for the second order correction
mentioned in \ref{Ei/Ej}, which can be positive or negative, i.e. contribute
to slightly increase the value of the cosmological constant as compared to the
other energies, or decrease it. In any case, 
the quantity \ref{cosmoc} is here not a constant,
but, through the age of the universe, turns out to 
depend on time. For the sake of simplicity, we will anyway keep
the usual terminology, and refer to it as to the
``cosmological constant''. According to \ref{cosmoc}, at present time 
the value of this parameter is:
\be
\Lambda ({\cal T} = 10^{61}{\rm M}^{-1}_{\rm P} ) \; \approx \, 10^{-122} \,
{\rm M}^2_{\rm P} \, ,
\label{lambdatoday}
\ee
as it effectively seems to be suggested by the experimental observations 
\cite{p,debern,melch}. 

\
\\
Let's now introduce a ``cosmological density'', $\rho_{\Lambda}$,
defined through:
\be
8 \pi \, G_{\rm N} \, \rho_{\Lambda} \, \equiv \, \Lambda \, 
= \, {2 \over {\cal T}^2} 
\, \times \, \left[  1 \, + \, {\cal O} 
\left( { 1 \over {\cal T}^{q_{S} } }  
\right)    \right]
\, .
\label{rholambda}
\ee  
In a similar way, we introduce two other densities,
related to the energy density of the $T$- and $U$-''waves'',
$\rho_m$ and $\rho_r$:
\be
8 \pi \, G_{\rm N} \, \rho_{m} \, \equiv \, 
E_{T}\vert_{\rm Einstein~frame} \, 
= \, {2 \over {\cal T}^2} 
\, \times \, \left[  1 \, + \, {\cal O} \left( {1 \over {\cal T}^{q_T} }  
\right)    \right]
\, ,
\label{rhom}
\ee  
and
\be
8 \pi \, G_{\rm N} \, \rho_{r} \, \equiv \, 
E_{U}\vert_{\rm Einstein~frame} \, 
= \, {2 \over {\cal T}^2} 
\, \times \, \left[  1 \, + \, {\cal O} 
\left( {1 \over {\cal T}^{q_{U}} }  
\right)    \right]
\, .
\label{rhor}
\ee  
The correction terms in the brackets are of different signs, in order to
sum up to a constant, in agreement with \ref{Estu}.
Each one of the above equations can be written as:
\ba
\rho_i & = & 2 \times 
{< {\bf 1} >_{i} \vert_{\rm string~frame} \over 3} 
\, \times \,  {R \over 2 G_{\rm N} 
\left[  {4 \over 3} \pi R^3 \right]} \nn \\
&& \nn \\
& = & 2 \times {< {\bf 1} >_{i} \vert_{\rm string~frame} \over 3} 
\, \times \, {M_{\rm Schw.} \over V } \, ,
\label{rischw}
\ea
where $M_{\rm Schw.} \equiv R / 2 G_{\rm N}$ is the ``Schwarzschild mass'',
and $V$ is the volume of the universe up to the horizon, $R \, = \, {\cal T}$.
If we sum up the three densities, taking into account \ref{1stu} we obtain:
\be
\rho \; \equiv \, \rho_{\Lambda} \, + \, \rho_m \, + \, \rho_r \, = \, 
{2 M_{\rm Schw.} \over V } \, . 
\label{rschw}
\ee
This agrees with the interpretation of the universe as a
black hole: the mass/energy content corresponds to 
the Schwarzschild mass of a black hole of radius 
$R = {\cal T}$ (to be more precise,
the energy content is twice as much as the one of a black hole, because
quantum space-time is an orbifold,
and the volume is effectively reduced by a factor 2 with respect to the one
of an ordinary smooth space).
It may seem absurd that a space with the geometry of a sphere, therefore 
a geometrical object without boundary, behaves for what matters
energy, entropy, and the traveling of information, like a black hole, i.e.
like an object with boundary, a ball.
Indeed, the coexistence of these aspects is possible because 
the universe is not simply a classical object. As discussed 
in~\cite{assiom-2011}, the strange geometry of the universe implies
that geometric points
as entities without dimensions are substituted by "cells" of Planck-length size.
In such a scenario of quantum gravity / quantum geometry,
close to the horizon "points" are completely non-local (see Ref.~\cite{blackholes-g},
and the discussion in
section~5.5 of Ref.~\cite{assiom-2011}), and
they are topologically equivalent to two-spheres. At the origin of any such geometric
puzzle is basically the fact that any attempt to view the geometry of the universe
as something "really" existing, beyond the measurement that can make of it an 
observer, is not legitimate. The universe is not a sphere nor a ball: simply, the
local geometry around the observer is the one of a three-sphere, and
light rays are interpreted by him as coming from the boundary of a three-ball
(see the discussion in Ref.~\cite{assiom-2011}).

\
\\

The fact that \ref{zsum} receives contributions from any type
of configurations implies that
the relation~\ref{rschw},
the relation between 
radius and total energy, is modified by vacua of non-maximal entropy.
In this sense, we can say, something not at all surprising, 
that the universe is a ``quantum Black Hole''. The Schwarzschild relation,
a classical relation, is violated by ``quantum gravity'' fluctuations. 
Indeed, if we consider the whole universe
as a fluctuation out of the vacuum, from the Heisenberg's Uncertainty Relations
we get that its energy must satisfy a relation of the type:
\be
\Delta M \, \geq \,
{1 \over 2 {\cal T}} \, ,  
\label{delMbh}
\ee
where the equality is saturated by the ``classical'', Schwarzschild mass. 

There is nothing to worry about considering ourselves at the border, 
or, depending on the point of view, at the center, 
of a black hole, our universe:
although we are intuitively led to consider black holes as small, 
extremely dense objects, indeed the mass of a black hole scales linearly with 
its radius, while the mass density scales with the (inverse) volume, 
i.e. the (inverse) cubic power of the radius. Therefore, the larger the 
black hole, the lower is its density. 
For a radius equal to the age of the universe, 
matter inside is as rarefied as we see it to be in our universe.

Owing to the interpretation of the universe as a black hole, we can define
a temperature of the universe as the total energy divided by the entropy.
The temperature turns out therefore to be proportional to the inverse of the 
age of the universe:
\be
T \; \stackrel{\mathrm{def}}{=}  
\; k^{-1} {{\rm E}({\cal T}) \over {\cal S}({\cal T})} \; = \; (4 \pi k)^{-1}
{1 \over {\cal T}} \, ,
\label{tdeft}
\ee
where $k \approx \; 8,62 \times 10^{-5} {\rm eV}K^{-1}$
is the Boltzmann constant, and the normalization
has been fixed according to the usual black hole thermodynamics 
\cite{Bardeen:1973gs,Bekenstein:1973ur} 
(see also section~6.4 of \cite{assiom-2011}). 
The present value of the age of the universe, 
converted in mass units, is
${\cal T} \, \sim 10^{61} {\rm M}^{-1}_{\rm P}$. At present time,
the temperature of the universe is therefore:
\be
T \, \approx \, 1,1 \times 10^{-29} \; ^0 K \, . 
\label{temphodie}
\ee
This temperature, for any practical purpose indistinguishable from the absolute
zero \footnote{The temperature was around one 
Kelvin at a time in Planck units $10^{29}$ times earlier than
today, i.e. at ${\cal T} \, \sim \, 10^{33} {\rm M}_{\rm P}^{-1}$. Since
1 yr $\sim \, 10^{51} {\rm M}_{\rm P}^{-1}$, this temperature corresponds
to the age ${\cal T}_0 \, \sim \, 10^{-19}{\rm yr} \, \sim \, (3 \times) 
10^{-12} {\rm sec}$.}, 
has not to be confused with that of the CMB radiation ($\sim \, 3^0 K$),
that we will discuss in detail in section~\ref{CMB}.

\subsection{The solution of the FRW equations}
\label{frw}

In the previous paragraph we have seen
that the energy density and the scaling of entropy 
agree with the interpretation of the 
universe as a black hole.
As the universe evolves, the energy density and the curvature
of space-time decrease toward a flat limit, and the dominant
configuration  tends to a 
``classical'' description. At large ${\cal T}$ it is therefore 
reasonable to suppose that this configuration
admits a description in terms of Robertson-Walker metric,
i.e. a classical metric of the type:
\be
ds^2 \, = \, dt^2 \, - \, R^2 (t) \left[ {dr^2 \over 1 - kr^2} \, + \,
r^2 (\, d \theta^2 \, + \, \sin^2 \theta \, d \phi^2 )  \right] \, , 
\label{rw}
\ee
where for us $t \, \equiv \, {\cal T}$, and $r \, \leq \, 1$.
The metric should correspond to a closed universe,
$k = 1$. 
Under the assumption of perfect fluid for the energy-momentum tensor,
the Einstein's equations lead to:
\be
\left( { \dot{R} \over R } \right)^2 \, = \, - \, {k \over R^2} \, 
+ \, \left\{ {8 \pi \, G_{\rm N} \, \rho \over 3 } \, 
+ \, {\Lambda \over 3}  \right\} \, ,
\label{flm}
\ee 
where we have collected within brackets the contribution of the
stress-energy tensor and of the cosmological term. 
According to our previous 
discussion, the stress-energy contribution consists of two
terms ($\rho = \rho_m + \rho_r$), each one of the same order of the
$\Lambda$-term. Inserting the values given 
in \ref{rholambda}, \ref{rhom}, \ref{rhor} with the ``Ansatz'' 
$R \,= \, {\cal T}$, and summing up, we obtain:
\be
\left( { \dot{R} \over R } \right)^2 \, = \, - \, {(k = 1) \over R^2} \, 
+ \, \left\{ {2 \over R^2 }  \right\}  ~~ = \, 
 { 1 \over R^2  } \, .
\label{rwsphere}
\ee 
The equation is solved by $R \, = \, t $, consistently with our Ansatz.
This confirms that the dominant
configuration corresponds to a spherical Robertson-Walker metric,
describing a universe bounded by a horizon expanding at the speed of light.
If instead of the densities we introduce the usual quantities 
$\Omega_m$, $\Omega_r$, $\Omega_v$, i.e. the densities rescaled in units of the
Hubble parameter $H \, \equiv \, ( \dot{R} / R )$, the Hubble equation 
\ref{flm} becomes trivially:
\be
0 \, = \, {k \, (= 1) \over R^2} \, = \, H^2 \left( \Omega_m  + \, \Omega_r
\, + \, \Omega_v \,  \, - \, 1  \right) \, ,
\label{hubble}
\ee
where we dropped the label ``$_0$'', which usually indicates the current,
present-day value, from the Hubble parameter: this equation is now valid at 
any time, and is trivially solved by $\Omega_m  + \, \Omega_r
\, + \, \Omega_v \, = \, 2$ and $\dot{R} \, = \, 1$.

Besides the Hubble equation, one usually derives further
equations of motion, by imposing energy-momentum conservation
to the Roberts-Walker solution of the Einstein's equations. 
In the present case however no further equation can be derived: 
energy-momentum conservation remains
valid only as a ``local'' law. At the cosmological scale, energy is not conserved: 
$E_{tot} \, \propto \, R \, = \, {\cal T}$.

\begin{centering}

${\star \star \star}$

\end{centering}

The comparison of our results with astronomical data, as we did in 
eq.~\ref{lambdatoday}, contains a possible weak point. 
Experimental data are given as a result of a process of 
interpretation of certain measurements, for instance through a
series of interpolations  of parameters. All this is consistently
done within a well defined theoretical framework. Usually, one
takes a ``conservative'' attitude and lets the interpolations run
in a class of models. However, this is always done within
a finite class of models. In principle,
we are not allowed to compare theoretical predictions
with numbers obtained through the elaboration of measurements in a 
different theoretical framework: in general, this doesn't make any 
sense.
However, in the present case this comparison is not meaningless, and this
not on the base of theoretical grounds: the reason is that,
for what concerns the time dependence of cosmic parameters and energy 
densities, the solution we are proposing does not behave, at present time, much
differently from the ``classical'' cosmological models usually
considered in the theoretical extrapolations from the experimental 
measurements. The rate of variation of energy density is in fact:
$\dot{\rho} \, \sim \, \partial (1 / R^2) / \partial {\cal T} \, = 
\, 1 / {\cal T}^3 \, = \, 1 / R^3$. The values of the three kinds of 
densities can therefore be approximated by a constant within a wide range of 
time.
For instance, as long as the accuracy of measurements does not go beyond the 
order of magnitude, these densities can be assumed to be constant
within a range of several billions of years. For the purpose of testing 
the statements and conclusions of the present analysis, the use of the
known experimental data about the cosmological constant,
derived within the framework of a Robertson-Walker universe
with constant densities, is therefore justified.

\begin{centering}

${\star \star \star}$

\end{centering}

\
\\

A universe evolving according to eq.~\ref{rwsphere}  is not accelerated:
$\dot{R} \, = \, 1$ and $\ddot{R} \, = \, 0$. 
Owing to the existence of an effective Robertson-Walker description, 
the red-shift can be computed as usual. We have:
\be
1 \, + \, z \, = \, {\nu_1 \over \nu_2 } \, = \, {R_2 \over R_1} \, 
= \, {{\cal T}_2 \over {\cal T}_1 },
\label{rs}
\ee
where $\nu_1$ is the frequency of the emitted light, $\nu_2$ the
frequency which is observed, and $R_1$, $R_2$ are respectively the scale 
factor for the emitter and the observer. $R = {\cal T}$ is precisely the
statement that the expansion is not accelerated. 
Expression~\ref{rs} 
however accounts for just the ``bare'' red-shift, namely the part due to the
expansion of the universe: it does not account for the corrections
coming from the time dependence of masses, that we will
discuss in section~\ref{accel}. Usually, this effect is not
taken into account,
because in the standard scenarios masses are assumed to be constant.
As we will discuss, in our scenario they depend instead on the age of the 
universe. A change in the values of masses
reflects in a change of the atomic energy levels, and therefore in a change
of the emitted frequencies. In order to properly include this effect
in the computation of the red-shift, we must therefore discuss the time
dependence of masses. We will see
that, once the observed frequencies in expression~\ref{rs}
are corrected to include also the change in the scale of energies, 
the scaling of the emitted to observed frequency ratio is
not anymore proportional to the ratio of the corresponding
ages of the universe. Since the conclusions about the
rate of expansion are precisely derived
by comparing red-shift data of objects located at a certain 
space-time distance from each other, this explains why the expansion 
\emph{appears to be accelerated}.

\subsubsection{Speed of light versus speed of expansion}
\label{cexp}

\begin{figure}
\centerline{
\epsfxsize=7cm
\epsfbox{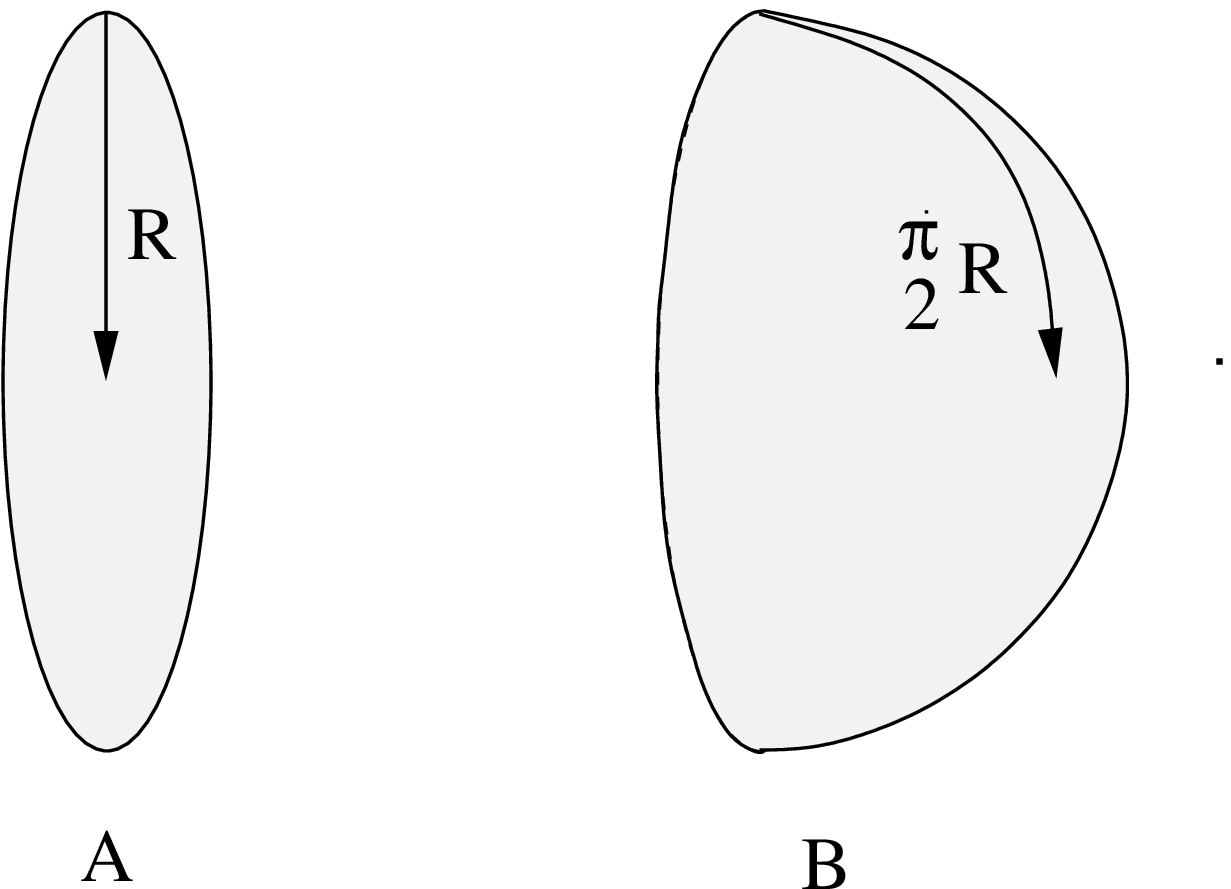}
}
\vspace{0.3cm}
\caption{A schematic representation of the universe in reduced dimension. 
A light ray seen by the observer as coming from the horizon, 
represented as the boundary of a flat space, the disc of picture A, 
follows in reality a curved path, a geodesic on the hemisphere depicted
in B. The real distance is longer: $R$ in (A) and ${\pi \over 2} R$ 
in (B).}     
\label{boundary2}
\end{figure}

As we discussed in section~7 of \cite{assiom-2011}, 
to an observer the universe appears in the average
as a ball bounded
by a two-sphere, with radius $R = c {\cal T}$, where ${\cal T}$
is the age of the universe. 
However, the geometry of space-time is curved, and corresponds to the one of a
three-sphere, this too of radius proportional to $R = c {\cal T}$. 
Light paths correspond
therefore to geodesics in this curved space. A light beam that appears
to the observer to be long as much as $R = c$ times the age of the universe
${\cal T}$, i.e. coming from the origin of time, corresponding to the horizon
of the ball,
in reality follows a longer, curved path, because light travels through
a curved space (see figure~\ref{boundary2} for an illustration of the
situation in lower dimension). 
The ``true'' speed of light is 
therefore higher than the speed of expansion, or equivalently 
the ``scale factor'' $R$ of the universe is shorter, and expands 
at a lower speed, than real distances.
However, this is of no physical relevance: as long as the two speeds
are proportional to each other, with a constant proportionality factor,
it does in fact make no sense to ask what is the ``real'' speed of expansion,
the ``real'' age of the universe, in a world in which everything looks
like if we were living in a flat space bounded by a horizon at distance
$R = c {\cal T}$. The only signal of the existence of a curvature
comes from indirect experiments, in which cosmological data are
interpreted within a theoretical framework. Namely, for what matters
parameters such as the energy density of the universe, the matter content,
the speed of expansion of the horizon, and consequently the age of the 
universe, everything works consistently with the identification of the
speed of light with the speed of expansion, the total energy with
the integral of the energy density
over the (hyper)disk, i.e. the ball enclosed by the horizon, and the
universe itself with a black hole.
For what matters the cosmological evolution too, densities
and equations of motion are normalized for a space with boundary, the
``ball'' enclosed by the horizon. The two scales,
either of times or of lengths, are proportional, and
any computation of red-shifts and similar parameters is insensitive
to this detail.
Things are in this interpretation rather different from the usual 
cosmological scenario, of a universe described by a Roberts-Walker metric
which undergoes an accelerated expansion. In
section~\ref{accel} we will make a comparison of the two points of view.

\newpage

\section{\bf Masses and couplings}
\label{masses}

\subsection{the mass of a particle}
\label{entropymass}

We consider now the masses of the elementary particles.
In our approach, a particle is a certain localized amount of energy that
rests or moves in space (indeed, the difference between a massive particle
and a massless one relies precisely in the fact that the massive
particle can stay at rest, whereas the massless one can not). When this
scenario is translated into the string representation, masses arise
as ground momenta associated to the states of the string spectrum.
Since through~\ref{zint} the string
scenario is a representation of the combinatorial one, even in the string 
space a mass is related to the weight of a certain state in the phase space. 
In the ordinary perturbative approach to field theory (no matter whether it
is string-inspired, string-derived, or not) masses, after they
have been introduced via some mechanism (Higgs mechanism),  
are attributes which in general
receive corrections at various perturbative orders.
The corrections 
appear as the sum of a series of insertions in the free propagator: 

\vspace{1cm}
\be
\centerline{
\epsfxsize=11cm
\epsfbox{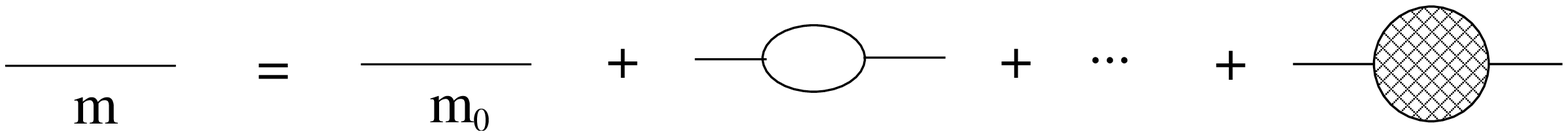}
}
\label{mdiagr}
\ee
\vspace{0.3cm}

\begin{figure}
\centerline{
\epsfxsize=6cm
\epsfbox{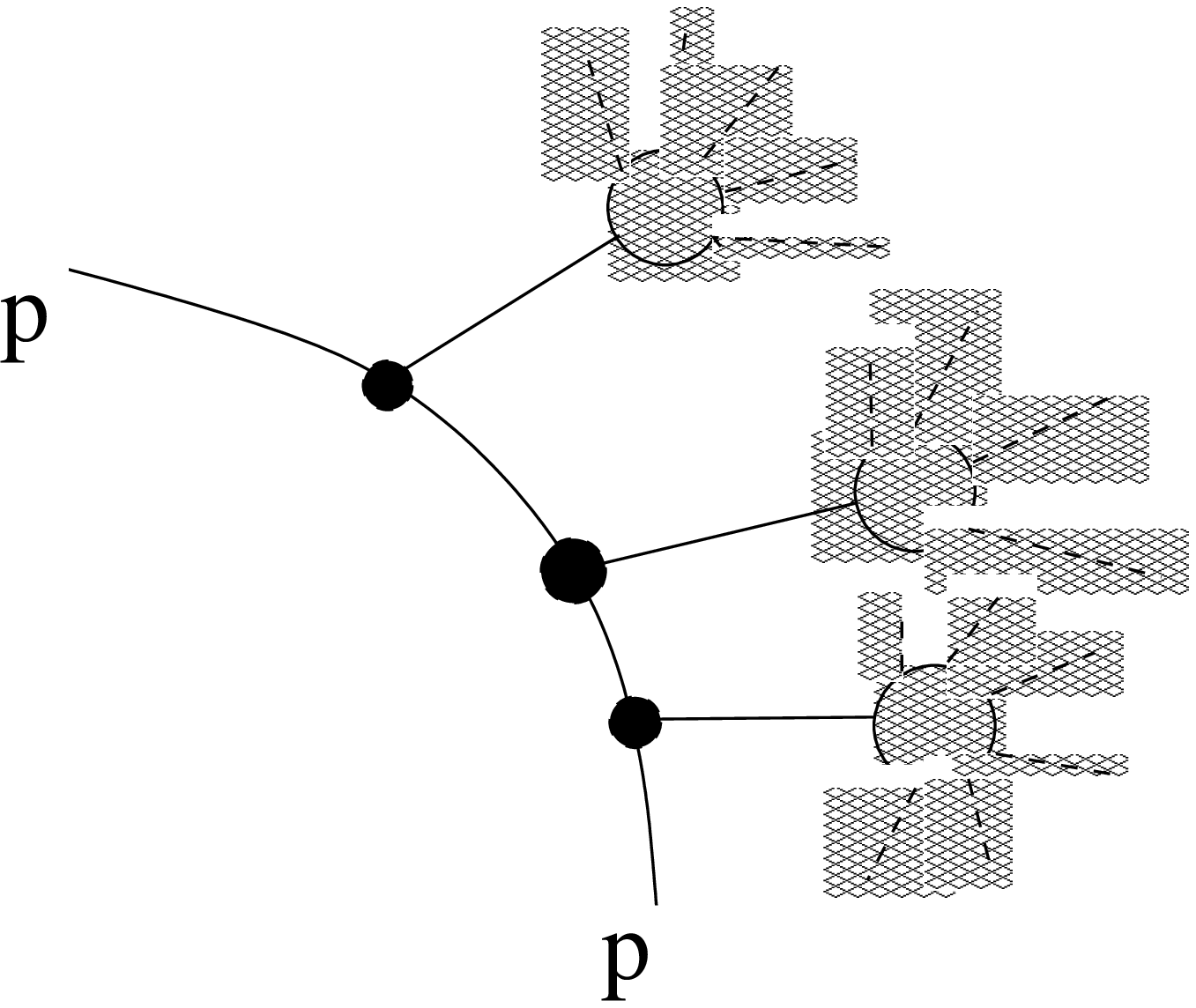}
}
\vspace{0.4cm}
\caption{The more and stronger are the interactions of the particle p,
the higher its entropy.}
\label{p-spread}
\vspace{.8cm}
\end{figure}

\noindent
Mass and volume in the phase space are related by the fact that
the more are the decay channels of a particle, the larger is its 
entropy, and also the correction to the mass,
because higher is the number of virtual lines (Feynman diagrams) 
contributing to the mass renormalization
(all this is illustrated in 
figure~\ref{p-spread} of page \pageref{p-spread}). 
\noindent
Heavier particles possess a huger
decay phase space: quarks are heavier than leptons, and among leptons neutrinos
are the lightest particles. Inside each family of particles,
the heavier (for instance the top as compared to the bottom of an
$SU(2)$ doublet) has the larger absolute value of the electroweak charge.
In each family, the lightest particle is the one which has less
interactions, or less charge (and therefore a lower interaction probability).
For instance, $|Q_{\nu}| \, < \, |Q_{e}|$,  
$|Q_{b}= -1/3| \, < \, |Q_{t} = + 2/3|$, and quarks,
that feel also the $SU(3)$ interactions, are heavier than 
leptons~\footnote{The first quark family makes an apparent exception: 
the down quark is heavier, 
although less charged, than the up quark. 
This issue will be discussed in detail in section~\ref{1family}.}.
Along this line, we can view the lightest particle as
the end-point of a chain of projections that reduce the symmetries
of the internal space. 
In string theory, this corresponds to the fact that heavier momenta
are given by inverse of mean radii obtained 
by taking the geometric average over a higher number
of internal coordinates.
Masses are the lowest
momenta of the string space, corresponding to inverses of radii.
Owing to the existence of internal dimensions,
not all the radii correspond to the radius of
the extended space, $R$, and therefore there are masses which are higher
than $1 / R$. In general, we have: 
\be
m_{(p)}^{-1} = 
\langle  R \rangle = \sqrt[p]{R \times r_1 \times r_2 \cdots \times r_{p-1}}
\label{mprod}
\, .
\ee
Since all internal radii are of Planck size,
that is, in our conventions, $r_i = 1$,
this relation becomes:
\be 
m_{(p)}^{-1} ~ = ~ \sqrt[p]{R \times 1 \times 1 \cdots 
\times 1} \, . 
\label{mrri}
\ee
In terms of string phase space, this expression
says that masses are in relation with the "weight", 
the volume of an internal space:  
$\langle R \rangle_p = \sqrt[p]{R}$ and $\langle R \rangle_q = \sqrt[q]{R}$  
stay in ratio related to the number of "1"s, i.e. of internal dimensions, they 
involve. Larger masses correspond to larger internal space, i.e.
radii of spaces with a larger number of internal dimensions.
Heavier particles are therefore
those which ``occupy'' a larger space; they
correspond to a larger internal symmetry than lighter particles.
Lighter particles correspond to sub-volumes, sub-spaces of those
of the heavier particles:
the phase space of lighter particles is contained in the phase space of
heavier ones. To figure out this point,
think for instance at the case of a heavy particle that decays
into lighter ones: the physics of these latter is ``contained'', in the sense
that it is produced, derived, by the physics of the heavier one. In terms
of combinatorials of distribution of energies, this simply means that
the ways of distributing an amount of energy $E$ along space contain
the ways we can distribute an amount $E^{\prime} < E$.

Although in the most entropic string configuration the string
space is not simply factorized into a product of radii as in the
above expression of masses, \ref{mrri}, from \ref{mrri}
one can see that the volumes corresponding to the various masses
are not simply those of the internal 
space, but they realize a kind of "embedding" of the extended space into the 
internal one, as indicated by their dependence on $R$. 
Mass ratios turn out to be of the type 
$m_{(p)}/m_{(q)} = \sqrt[q-p]{R}$, with $p$ and $q$ however not simply
integer, but rational numbers.

\subsection{The couplings}

In this scenario, also couplings
are related to weights in the phase space. Indeed, in the
combinatorial formulation of the physical world, that, we recall, is the
fundamental one, a configuration
is representated by a group of symmetry, and its weight corresponds
to the volume of the symmetry group. Also ratios of masses correspond
therefore to ratios of symmetries, i.e. to cosets. When we pass from
a particle to another, lighter one, we reduce the symmetry
of the subset of the configuration of the universe
(i.e. the subgroup of the whole symmetry group representing the 
configuration of the universe) which represents the particle under consideration. 
We say we break a certain group to a subgroup. 
The inverse of the volume of the coset is what
we call ``coupling''. Indeed, in the case we start from a particle and
follow a decay process leading to some products (particles and fields with
a certain energy), the ratio of the volume of the initial particle to
the volume of its decay products gives the coupling of the interaction.
Interactions are therefore in a natural way related to groups and
symmetries. We will see that this point of view includes both the symmetries
which, in the representation in terms of fields living on the continuum,
correspond to ordinary gauge symmetries, and the symmetries that in such
a representation are going to be ``hardly broken'', i.e. appear as rigid 
symmetries. This second type does not give rise to interactions propagating
through gauge bosons, but describe for instance the leading
contribution to ``transitions'' such as the passage from a family of
quarks or leptons to another one. In this interpretation, they
are viewed as an appropriate type of rotations in the phase space.   
In both cases, these processes, and their relative couplings, follow
the rules of composite probabilities typical of the weights in the phase
space. A composite transition/decay process: $A \to B \to C$,
corresponds to a rotation with an element of the
group ${\bf G}_{(AC)}$, given by a product ${\bf G}_{(AC)} \, = \, 
{\bf G}_{(AB)} \times {\bf G}_{(BC)}$. This transition corresponds therefore
to: 1) first a rotation with an element of the group
${\bf G}_{(AB)}$ and then: 2) a rotation with an element of ${\bf G}_{(BC)}$. 
Therefore, we expect the probability of the decay $A \to C$ to be the 
product of the decay probability of $A \to B$ and of $B \to C$:
\be
P(A \to C) \, =  \, P(A \to B) \times P(B \to C) \, .
\label{pabc}
\ee
The effective coupling determines the probability amplitude:
\be
P (A \to B) \, \sim \, \alpha_{(AB)} \, ,
\label{pab}
\ee
where, as usual, $\alpha_{(AB)} \equiv g^2_{(AB)} / 4 \pi$.
The effective coupling for the transition from A to C is given by the product
of the effective couplings of the single steps:
\be
\alpha_{(AC)} \, \propto \, \alpha_{(AB)} \times \alpha_{(BC)} \, .
\label{alphaABC}
\ee
The square coupling of the group  
${\bf G} \, = \, {\bf G_1} \otimes {\bf G_2} \otimes \ldots
\otimes {\bf G_n}$ is therefore:
\be
\alpha_{\bf G} \, = \, \alpha_{\bf G_1} \times \alpha_{\bf G_2} \times \cdots
\times \alpha_{\bf G_n} \, .
\label{alphaG}
\ee
In the usual renormalization group approach one works 
in the algebra $\bf{ \cal G}$ of the group ${\bf G}$.
If ${\bf G} \, = \, {\bf G_1} \otimes {\bf G_2} \otimes \ldots
\otimes {\bf G_n}$, the algebra decomposes as 
$\bf{\cal G} \, = \,
\bf{{\cal G}_1} \oplus \bf{{\cal G}_2} \oplus \ldots \oplus \bf{{\cal G}_n}$,
and the inverse of the effective coupling seems to renormalize additively.
For instance, the one-loop beta-function of $SU(N)$ with gauge bosons in the
adjoint and (massless) matter states in the fundamental representation is
one half of the beta-function of $SU(2N)$ with an analogous content of matter
and gauge states. From the point of view of our approach,
what seems to behave additively is just the 
logarithmic derivative of the inverse of the coupling.

\subsection{Entropy and mass}
\label{entromass}

From a technical point of view,
subplanckian masses are introduced by a shift
along the coordinates of the extended space, required
in order to reach the configuration of lowest symmetry. 
This shift lifts to a non-zero
mass all matter states. Their mass is the lowest energy excitation of
these states, and, corresponding to the lowest momentum of space, is of the
order of the inverse of its radius. 
As discussed, perturbatively the $Z_2$ orbifold appears
as a point of enhanced symmetry
of the theory, and it doesn't allow to see the fact
that such an operation doesn't produce an equal mass for all the matter
sectors. On the other hand, as we discuss in section~\ref{breakthree}, 
the symmetry between the three matter families
(u,d,e, $\nu_e$), (c,b,$\mu$,$\nu_{\mu}$) and (t,b,$\tau$,$\nu_{\tau}$)
is non-perturbatively broken.
The fact that both masses and couplings are related to volumes of symmetries
implies that masses  are related to couplings.
The hierarchy of masses, as implied by the hierarchy of the corresponding 
volumes in the string target space, reflects into a hierarchy of the interactions
experienced by the particles. We will now see in detail how the ratios of masses
are in relation to the ratios of the volumes of these groups, in
turn related to the ratios of the strengths of the corresponding couplings.

Let us first see how masses are in general generated
by shifts. From the inspection
of the generic action of orbifold shifts that produce masses
we will also gain a further indication that
the perturbative construction corresponds to
a logarithmic representation of space.  
Let us consider the simple example of the orbifold vacua analysed in
Ref.~\cite{striality}. One starts with a construction
with matter in the $N \oplus N$ of $SU(N) \otimes SU(N)$. With a 
level-doubling, rank-reducing $Z_2$ shift one derives a configuration
with matter in the $N$ of $SU(N)$, whereas the other $N$ has been lifted up,
and acquired a mass, in the case of a shift acting on the momenta of a circle
contained in an internal torus. The mass of the lifted states is given by:
\be
m ~ = ~ \sim {1 \over R_1} \, ,
\ee
where $R_1$ is the radius of the circle. Let us now suppose to apply a second
shift, acting on another circle, that we assume to be of radius $R_2$,
in a way to lift all matter states. We have therefore two kinds
of massive states: those which already acquired a mass with the first shift,
that now have mass:
\be
m_{1+2} ~ = ~ \sim \left[ {1 \over R_1} \, + \, {1 \over R_2} \right] \, ,
\ee
and those which after the first shift were left massless, and now have mass:
\be
m_{2} ~ = ~ \sim {1 \over R_2} \, .
\ee
Let us suppose that $R_1 = R_2$; $m_{1+2}$ results in this case to be exactly
twice $m_2$. Generalizing, 
we can say that the mass is proportional to the gauge beta-function:
\be
m ~ \propto ~ {\beta_0 \over R} \, ,
\ee
where $\beta_0$ is the gauge beta-function coefficient (the beta-function is
$\beta_0 \log \mu$, $\mu$ an appropriate scale). 
This can also be expressed as:
\be
\Delta \left(  {1 \over m} \right) \, \approx \, {\Delta m \over m^2} ~ \sim ~ \Delta \beta_0 \times R \, ,
\label{dmr}
\ee
to be compared with a similar expression one can write for the couplings:
\be
\Delta \left(  {1 \over \alpha} \right) ~ \sim ~ \Delta \beta_0 \times \log \mu \, ,
\label{damu}
\ee
that we derive from  the one-loop renormalization of the inverse of
a gauge coupling: ${4 \pi \over g^2} \, \sim \, {4 \pi \over g_0^2} \, 
+ \, \beta_0 \ln \mu$.
According to our previous discussion about the couplings,
the strength $\alpha(G)$ is
by definition proportional to the volume of the group, $|| G ||$
(not to be confused with the
volume of the Lie algebra $|| \mathfrak{g}  ||$), and we can write:
\be
{ \alpha (G_i) \over \alpha (G_j)} ~ = ~
{|| G_i  || \over || G_j  ||  } \, .
\label{gvol}
\ee
On the other hand, also masses are related to volumes of symmetries,
so that we can write a similar expression:
\be
{ m_k \over m_{\ell}} ~ = ~
{|| G_k || \over || G_{\ell}  ||  } \, .
\label{mvol}
\ee  
The comparison of these two expression with \ref{dmr} and \ref{damu}
suggests that:
\be
{m_k \over m_{\ell}} ~ = ~ { \alpha (G_{k}) \over \alpha (G_{\ell})} \, ,
\label{magkell}
\ee
a relation which implies the identification of the radius $R$ entering
in the perturbative expression of masses with $\log \mu$,
as it is to be expected if \emph{in the perturbative construction
the space coordinates are a logarithmic representation of the physical 
space}~\footnote{More precisely, the comparison suggests the identification
up to a proportionality constant $\kappa$ on the tangent space, i.e.
$R \sim \mu^{\kappa}$.} \label{logpict}.
Expression~\ref{magkell} can also be written as:
\be
{m_i \over m_j} ~ = ~ \alpha (G_{ij}) ~ = ~
|| G_{ij}  || \, ,
\label{mgij}
\ee
where $G_{ij}$ too is a symmetry group. In terms of familiar Feynman diagrams,
this can be undersood as follows.
Let's consider the following diagram for the vacuum renormalization
in the case of a two-particle theory with a
broken symmetry relating particle 1 and particle~2:  
\be
\centerline{
\epsfxsize=6cm
\epsfbox{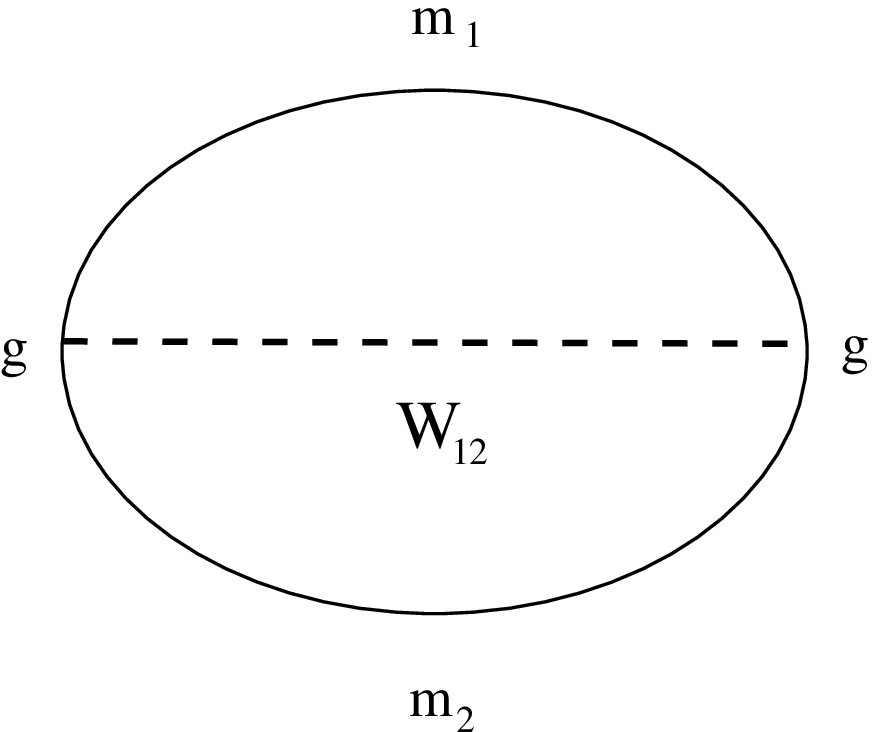}
}
\label{m1m2}
\ee
This has to be thought not as a field theory diagram, but as a diagram
for our effective non-perturbative theory. 
There is no question about non-renormalizability
of a theory without a Higgs particle: here masses have to be treated as 
effective parameters, more or less like what one does with the electric field 
in a semiclassical approximation of a quantum relativistic scattering by
an external field. 
The theory is finite, and we know from above that this diagram,
intended as the non-perturbative resummation of the terms
corresponding to this transition process, should correct the vacuum 
amplitude \emph{multiplicatively}, by a factor 1~\footnote{We recall that
for us the additive representation corresponds to a logarithmic representation
of the physical string vacuum.}: this means in fact that there is no 
renormalization of the vacuum, when for $g$, $m_1$, $m_2$, $m_{W_{12}}$ we
take the ``on-shell'', non-perturbatively resummed physical parameters. 
By computing the contribution of this diagram, we find:
\be
1 \; \approx \; g^2  \, \times \, 
\int {d^4p_1 d^4p_2 \over (\not{\! p}_1 + m_1) (\not{\! p}_2 + m_2) 
\left((p+ p_1 - p_2)^2 - M_W^2 \right)} ~~
\sim ~ g^2 \, \int^{< p > } { d^8 p \over p^4 } \, , 
\label{gm1m2int}
\ee
where the integration has to be performed up to a cut-off $< p >$.
This corresponds to the
typical mass scale of the (sub)space in which the process takes place.
In practice, we can consider that the subsystem consisting of the
particles 1, 2 and the boson $W_{12}$ ``lives'' in a space with
``temperature'' given by the average of the temperatures of the three
states. Equivalently, we can think that the typical length of the process
is the (multiplicative) average of the typical lengths associated to the
three states. In any case, 
\be
< p > ~ \sim ~ \left( m_1 m_2 M_W^2  \right)^{1 / 4} \, .  
\label{p-cutoff}
\ee  
Therefore, the integral pops out a factor $[< p >]^4 \, = \, m_1 m_2 M^2_W$, 
where the $W_{12}$ mass is T-dual to the lightest particle mass:
\be
M_W^2 \, = \, { 1 \over m_2^2} \, ,
\label{mWm2}
\ee
where by convention we have chosen $m_2$ to be the lower mass,
and T-duality is performed with respect to the Planck mass. 
From this we derive that:
\be
g^2 m_1 m_2 M_W^2 \; = \; g^2 {m_1 \over m_2 } \; \approx \; 1   
\label{gm1m2}
\ee
The ratio of the two masses is therefore given by the inverse square
coupling of the broken symmetry. We stress that these arguments make only 
sense in our scenario, not in a generic field theory.
In this example, the boson
mass, T-dual to the mass of the particle, is higher than the Planck scale,  
and there is no boson line really coming out from the line of the particle
transition: the latter looks like a rigid symmetry. 
When the transition is ``off-diagonal'' with
respect to the traditional classification of particles into families, 
we interpret this phenomenon as a ``family mixing'', whose effect is collected
in the off-diagonal entries of the CKM mass matrix.

Expressions~\ref{magkell} and \ref{mgij}, together with 
the identification of the logarithm of the
scale of perturbative renormalization of the couplings, $\log \mu$,
with the scale $R$ of the extended space,
tell us that not only masses, but
also the couplings scale as powers of the radius, or 
equivalently the age, of the universe.
Indeed, the idea of renormalization group 
is based on the possibility of letting the parameters run as functions of a scale,
in such a way that the sum of higher order terms can be re-written so as
to reproduce the functional expression of the first non-trivial
correction. This means that there is always an appropriate scale $\tilde{\mu}$
and an effective ``beta-function'' $\tilde{\beta}$ for
which $\tilde{\beta} \ln \tilde{\mu}$ constitutes a good approximation
to the expression of the coupling, at least for its perturbative part. 
Since the logarithmic picture is for us
precisely the representation in which the coupling of the theory vanishes,
this could be taken somehow as the definition of the running of a 
coupling in this picture. And 
indeed, a series like the usual perturbative expansion of the corrections
to a coupling:
\be
 \partial \ln \alpha / \partial \ln \mu 
\, \approx \,  \sum_n \beta_n \alpha^n  \, , 
\label{dadm}
\ee 
somehow looks like the expansion around a small value of the coupling
of an exponential expression, of the type:
\be
\alpha ( \mu ) \, \sim \, \exp \left[ \beta \ln \mu \right] \, ,
\label{alphaexp}
\ee
over $\ln \alpha$:
\be
\ln \alpha \, \sim \, \beta \ln \mu \, .
\label{logalpha}
\ee
In our approach, \ref{alphaexp} expresses the full, non-perturbative
behaviour of a coupling, whereas a perturbative series~\ref{dadm}
constitutes only an approximation. 
Since any field-theoretical approach is in itself only 
an approximate representation of the full physical content, encoded 
in~\ref{zsum}, in this framework it does not make sense to look 
for a rigorously proof that the first perturbative corrections to a gauge 
coupling are the first terms of the expansion of \ref{alphaexp}.

Since the perturbative slices of the string configuration we want to consider
correspond to logarithmic realizations (i.e. on the tangent space), 
in order to evaluate the volumes in the phase space associated to the masses
it is convenient to think in terms of entropy.
The mass is the ground energy of a system, and corresponds
to a ``ground entropy'': we may think that
an object with mass possesses a ``ground entropy''.
In the case of a black hole, entropy can be seen as somehow related
to a counting of the elementary-state 
paths going from outside into the horizon of the 
black hole \cite{'tHooft:1984re,Susskind:1993ws,Susskind:1994sm,Kabat:1995eq}. 
In a similar way, here the relation between mass corrections and 
entropy, established in \ref{mdiagr} and \ref{dqde},
suggests that the ``ground entropy'' of a particle can be seen as a way of 
counting the paths leading to the particle. The ``bubbles'' of the terms
in \ref{mdiagr} represent in fact paths that come out of the particle and end 
up back into the particle itself. The equivalence with the usual
interpretation of the entropy of a black hole is more evident if
we ideally view these terms as made up of two mirror cuts: 
\be
\epsfxsize=7cm
\epsfbox{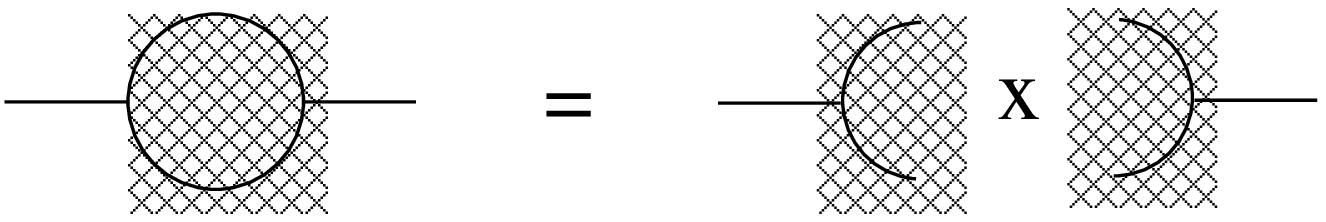}
\label{mdiagr12}
\ee
\vspace{0.3cm}

\noindent
Each half looks like a decay process, or the other way around as 
the process of formation of a black hole.
The mass renormalization can be seen as a physical process
in which the final state is the same particle as the initial state: no
lines are propagating out of the bubble. It corresponds therefore to 
an adiabatic process. The first law of thermodynamics tells us that
in this case, 
\be
d Q \, = \, 0 \, = \, d E \, - \, T d S \, .
\label{dqde}
\ee
If the temperature were constant, we would conclude that
the variation of energy along this process corresponds
to the variation of entropy. However, the temperature of the system is 
related to the energy, and varies with energy.
Consider the case of the universe
itself. We have seen that its mass, i.e. its rest energy, 
scales with time, ${\rm M}_{\rm Univ.} \,
\sim \, {\cal T}$, while the temperature scales as the inverse of its age:
$T \, \sim \, {\cal T}^{-1}$. Finally, entropy scales as 
$S \, \sim \, {\cal T}^2$. In this case, $d S \sim d {\cal T}^2 \sim (2) {\cal T}$
and the relation \ref{dqde} reads:
$$ 
dE \, \equiv \, d{\rm M}_{\rm Univ.} ~~
\sim ~~  {1 \over {\cal T}} \times dS  \, d {\cal T}
~ \approx ~ d {\cal T} \, .
$$
Let's now consider the case of a 
``universe'' consisting of just one particle. In other words, let's restrict
these considerations to the system illustrated in fig.~\ref{mdiagr}.
In this case, the ``universe'' behaves in itself like a quantum black hole.
It possesses a ``temperature'', which is related to the inverse of its ``age'',
and is therefore proportional to its minimal energy gap, given by the
lower bound of the Heisenberg inequality: $\Delta E \sim 1 / {\cal T}$. 
This energy gap is the mass of the particle itself: $m \sim \Delta E $,
and therefore the mass of the particle is proportional to the temperature:
$m \sim T$. In this case,
equation~\ref{dqde} reads:
\be
{dm \over T} \, \sim \, {dm \over m} \, = \, dS \, .
\label{dmTds}
\ee  
Once integrated, it gives:
\be
\ln m \, = \, S \, + \, {\rm const.} \, ,
\label{lnmS}
\ee
which is again the relation $m \sim \exp S = W$, stating that
the mass is proportional to the
weight in the phase space.

In order to evaluate entropy, we proceed as follows.
If we think the 
string space as fibered over the ``space-time'' base, we 
can say that for any coordinate of the fiber there is a  phase space 
of space-time size. Since the size of the internal coordinates, those of
the fiber, is one in Planck units, the volume of the phase space will be:
\be
{\cal V}_P \, = \, \mu^{\beta} \, , 
\label{vpmubeta}
\ee
where $\mu$ is the length
(the volume) of space-time, and $\beta$ a coefficient. 
For instance, if we start with a phase space
of volume $\mu^{\beta_0}$ and act on the states with a $Z_2$ projection 
that halves the internal space, the un-projected states will
have a phase space at disposal for their decays given by $\mu^{\beta_0/2}$.
The probability density is proportional to the inverse
of the volume \ref{vpmubeta}:
\be
P (x) \, = \, {1 \over {\cal V}_P } \, ,
\label{px}
\ee
and entropy, given by definition as $S = - \sum P \ln P$,
is then:
\be
S \, = \, - \int_{{\cal V}_P} dx \; {1 \over {\cal V}_P} \, 
\ln { 1 \over {\cal V}_P } \, = \, \beta \ln \mu \, ,
\label{smulog}
\ee
where the result follows immediately from the fact that ${\cal V}_P$ is
a constant in the domain of integration.
After a projection like the one of above, entropy will be reduced by one half:
$S \, \to \, (\beta /2) \ln \mu$. The mass renormalization
reads:
\be
\ln m  \,  = \, \beta_0 \ln \mu  
 \, + \, \beta \ln \mu  \, + \, {\rm const.} \, ,
\label{lnmbeta0}
\ee 
that we can also write as:
\be
\ln m  \,  = \, \ln m_0 
 \, + \, \beta \ln \mu  \, + \, {\rm const.} \, .
\label{lnmm0}
\ee
This expression suggests that the term $\beta \ln \mu$ can be calculated in
a logarithmic representation, and the coefficient $\beta$, corresponding to the
``exponent'' of the volume of the subspace of the phase-space of the particle,
can be obtained by computing the amount of projections acting on the sector
under consideration.
We have separated a term $\beta_0$, because we must allow for a minimal value
of entropy. At finite space-time volume, the minimal
mass gap is not arbitrarily small. This sets also the minimal entropy
allowed for the process, and results in a ``minimal subtraction''
in the phase-space volume.
After exponentiation, we obtain:
\be
m \, = \, ({\rm Const.}) \, \times \, m_0 \, \exp (\beta \ln \mu) \, 
= \, ({\rm Const.}) \, \times \, m_0 \, \mu^{\beta} \, .
\label{mexp}
\ee
The ``bare'' mass $m_0$ is given by the inverse square root of the age  of the
universe: $m_0 \, = \, 1 / {\cal T}^{1/2}$. This value,
common to all particles, is produced by the shift acting along 
the space-time coordinates, that we described in section~\ref{massmatter}. 
In the logarithmic picture, where, owing to the linearization of space, shifts
along the coordinates contribute additively, it appears as an additive term:
\be
\ln m_0 \,  = \, {1 \over 2} \ln {1 \over {\cal T}} \;
\equiv  \, \beta_0 \ln \mu  \, ,
\label{lnm0T}
\ee
as in \ref{lnmm0}.
In the two pictures the mass renormalization reads therefore respectively:
\ba
({\rm log \, picture})~~~ &
\tilde{m}  \, [ = \, \ln m ] &  = ~~ {1 \over 2} \ln {1 \over {\cal T}} ~ 
 \, + \, \beta \ln \mu \nn \\
&& \nn \\
&& \equiv ~~  \tilde{m}_0 ~ \,  + ~ \, \beta \ln \mu  \nn \\
&& \nn \\
& ~~
\stackrel{\mathrm{exp}}{\longrightarrow} ~~ & \label{mlogmreal} \\
&& \nn \\
({\rm real \, picture})~~
& m & = ~~ {\cal T}^{ - {1 \over 2}} ~ \times ~
 \mu^{\beta}  \, .
\nn
\ea 
In these expressions, $\mu$ is the age of the universe raised to some power:  
$\mu \, = \, {\cal T}^{p}$.
It is clear that any change in $p$ can be reabsorbed by a change in
$\beta$ and $\beta_0$. We set by convention 
$\mu \, \equiv \, {\cal T}$, so that $\beta_0 \, = \, -1/2$.  
By inserting these values in \ref{mexp}, we obtain that masses scale as:
\be
m \, = \, ({\rm Const})  \times \,  {\cal T}^{\beta - 1 / 2} \, ,
\label{mTbb0}
\ee
With the identification of $\mu^{\kappa}$ 
with the age of the universe, ${\cal T}$,
the ratios \ref{mvol}--\ref{mgij} can be written as:
\be
{m_i \over m_j} ~ = ~ \alpha (G_{ij}) ~ = ~ {\cal T}^{\beta_i - \beta_j} \, .
\label{mTbeta}
\ee 
Notice that, although the mass of a particle decreases with time, the
``ground entropy'' increases. This is due to the fact that entropy is related
to the statistics of the physical system, i.e. to the structure
of the phase space. As it can be seen from the Uncertainty Relations
($ \Delta E \Delta t \geq 1/2 $), in this space the volume of the minimal 
cell goes like $\sim 1 / {\cal T}$. It decreases therefore much faster than 
any mass scale, which is bounded by the square root scale: 
$m \geq 1 / {\cal T}^{1/2}$ for any mass $m$.  
The ``ground'' phase space volume of a particle is:
\be
{\cal V}_0 \, = \, {m \over \Delta E} ~ \leadsto ~ 
{ {\cal T} \over {\cal T}^{1 \over 2} } \, \sim \, {\cal T}^{1 / 2} \, .
\label{v0T}
\ee
As time goes by, the phase space of a particle increases, and therefore
also its entropy.
In section \ref{GeomP} we will comment on the relation of this approach 
to the one based on the ``geometric probability'' techniques.

\
\\

In order to obtain the masses,
we must first obtain the ``beta-functions'' $\beta_i$, 
$\beta_j$. According to our discussion, we cannot compute them using the rules
of ordinary field theory: here we are interested in the full, 
non-perturbative beta functions. 
We can proceed as follows: we can determine the ratios 
of these beta-functions if we know the ratios of the phase-space
volumes. For instance, if a projection reduces by one half the phase space,
the beta-function will be one half of the initial one. On the other hand,
the phase space volumes can be determined if we know the spectrum of 
interactions of the various particles (i.e., the pattern of 
figure~\ref{p-spread}), 
but the important point is that, in this framework, this
in principle is equivalent to knowing the chain of projections,
$\equiv$ symmetry reductions, giving origin to the sector a certain massive
state belongs to. 

In order to recover the correct dependence on the radius of the
extended space, which as we have seen does not show out correctly
in perturbative orbifold constructions, 
we will have to combine several considerations
in subsequent steps, that somehow constitute a sequence of 
perturbative approximations, in which the fine details of the structure
of the physical configuration of highest entropy are the better and better
investigated. This way of proceeding makes sense, because 
these corrections, depending on inverse powers (roots) of the age of the 
universe, at present time are sufficiently small. 

Once obtained the ratios of beta-functions, in order to get all their values
we must fix one of them. To this purpose, we must
consider that the mass of the state with maximal, unbroken symmetry, does
not change with time, it is a constant. 
Maximal symmetry, and therefore also supersymmetry, implies
in fact that masses either vanish (as also the cosmological constant does), 
or they do not renormalize out of the initial value at the Planck scale:
among the preserved symmetries, there is in fact also time reversal,
so that masses do not run. 
In this case, we have:
\be
m_{\rm max.\, symm.} \, = \,  {1 \over 2} \, 
{\cal T}^{\beta_{\rm max} - {1 \over 2}}  ~ 
= \, {\rm Constant}  \, ,
\label{mmax}
\ee 
where we have set the normalization of the mass as a function of the
inverse of a proper time to be $ 1 / 2$, as according to the Heisenberg's
Uncertainty Principle. 
The condition \ref{mmax} implies $\beta_{\rm max} = {1 \over 2}$.
With this normalization, at the Planck time
the maximal mass is the one of a black hole, in agreement with
our discussion of section \ref{ubh}, namely it is  
given by the relation: $2 M \, = \, R$, 
with the identification $R = {\cal T}$.
Therefore, at the Planck time the minimal mass 
excitation is forced by the Uncertainty Principle to be 1/2. 
On the other hand, being the universe a Planck size black hole,
this is also the maximal allowed mass excitation. If we allow one
coefficient to be greater than 1, there exists a certain size
of the universe, \emph{larger} than the Planck length, 
at which the mass is larger than the black hole's Schwarzschild mass.
If we run back in time, starting from the present age, where this state
appears among those of the sub-Planckian spectrum, this excitation
drops out from the spectrum, by definition formed by
the degrees of freedom which are below the black hole threshold,
\emph{before} reaching the Planck scale.
Therefore, it cannot correspond to a perturbation over the massless spectrum
corresponding to some momentum (= inverse of radius) of the string space,
of the kind of those giving rise to the
effective, ``low energy'' theory. From~\ref{mTbeta}
we see that the overall normalization is the same for all the states
which are $SU(3)$, $SU(2)_{\rm w.i.}$ and $U(1)_{\rm e.m.}$ singlets:
\be
m_i \, = \, \kappa_i {\cal T}^{\beta_i} \, , ~~~~ \kappa_i = \kappa ~~ 
\forall \, i \, .
\label{mkappaT}
\ee
Therefore all these masses are expressed as:
\be
m \, = \,  {1 \over 2} \; {\cal T}^{\beta - {1 \over 2}}  ~ 
\,  , ~~~~~~~ \beta \in \{0, {1 \over 2} \} \, .
\label{mparticles}
\ee

\subsubsection{Elementary masses}
\label{Emass}

Let's indicate with $p$ the ``beta'' function coefficient, i.e.
the exponent of the phase-space volume, ${\cal V} = \mu^{\beta}$, 
of the maximally symmetric configuration, and with $q$ the one of the 
end-point, minimal symmetry
(minimal entropy) configuration. The couplings of the various subgroups
in which the initial symmetry group gets divided are then given by:
\be
\alpha^{-1}_i \, = \, \mu^{n_i - n_{ i+1} \over p } \, ,
\label{alphapq}
\ee
where the ratios ${ n_i \over p}$, with $n_i : p = n_0   \geq  n_{\rm i}  
\geq  q  $, are the volume exponents
of the various steps. $n_0$ corresponds to the last step (the one which
selects the lightest neutrino). 
An accurate evaluation of masses is then equivalent to an exact determination
of these coefficients. 
Our ``perturbative'' approach starts with a first degree of
approximation, consisting of a ``rough'' determination of the volume of
the phase space of each elementary particle, as seen ``at the Planck scale''.
This allows us to map to a logarithmic picture, where all couplings are
perturbative, because they are of order one in the original picture.
Further corrections of the weak coupling scale are to be 
expected at the present age of the universe. This leads us out of the domain
of a logarithmic picture; the problem can in principle be 
treated as an ordinary perturbative correction, whose complete evaluation
must take into account the details of every decay channel. In any case,
these corrections should be of second order, with a relative magnitude 
proportional to the inverse ratio of the phase volume of the particle under 
consideration and the one of its decay products.
Namely, we expect: 
\be
m \, = \, m_0 \, + \, \delta m  \, ,
\label{m0delta}
\ee
where
\be
{ \delta m \over  m } \,  
\sim \, {\cal O} 
\left[ { m_{\rm final} \big/  m_{\rm initial} (\equiv m) } \right] \, .
\label{deltam}
\ee
We will consider these corrections in sections \ref{corrm}--\ref{pi-k}.

\
\\

Let's consider the first step of the analysis, the one in which we can
map to the logarithmic picture. If the hierarchy of the three families
were ordered in such a way that all the particles of the higher family were
heavier than those of the neighbouring lighter family, we could immediately 
conclude that the phase space is divided by the particle's families into three
parts, with volumes staying in ratios given by 3:2:1.
Namely, in passing from one family to the other one, 
the phase space volume ${\cal V}$ should undergo a contraction  
${\cal V} \to {\cal V}^{2 \over 3} \to {\cal V}^{1 \over 3}$. 
However, this sequence is valid only as long as \emph{all} the particles
of the ``heavier'' family are indeed all heavier than the particles of the
following family. Otherwise, the pattern changes. 
In order to understand the implications of this condition, we must consider 
that relating the mass to the corresponding volume in the phase space
roughly means that heavier particles are also the more interacting ones.
On the other hand,   
saying that all the particles of the heaviest family are heavier than those
of the lighter families means in particular that even the tau-neutrino,
$\nu_{\tau}$, is heavier than the charm and strange quarks, something
not expected for a particle definitely less interacting. Indeed, we expect
the three neutrinos to be the lightest among all particles.  
This means that the phase space is \emph{first} separated into the block of
neutral and charged particles, and \emph{then} each block is contracted
according to the 3:2:1 rule. 
Apart from the separation of the $SU(4)$ symmetry into 
${\bf 1} \oplus {\bf 3}$ of leptons and quarks, 
all the other separations are built on $SU(2)$ steps, as
a consequence of the $Z_2$ structure of the projections. We expect therefore
the $SU(2)$ coupling to play a key role in the mass ratios. 
In principle, an $SU(2)$ coupling factor separates also the up and down, both
lepton and quark, inside each family. However, the coupling, and the group,
of interest for us is not the left-moving $SU(2)$ of the weak interactions,
which at the string scale remains unbroken: mass separations are in relation
with the breaking of the $SU(2)_{(L)} \leftrightarrow SU(2)_{(R)}$
symmetry. Requiring that all neutrinos are lighter than any charged
particle implies a ``flip'' in the action of the up/down
separation of the $SU(2)$ doublets, such that the minimal block, the
$SU(2)$-coupling, instead of acting ``diagonally'', by separating the 
tau lepton from its neutrino, acts off-diagonally, being interposed between the
$\nu_{\tau}$ and the lightest charged particle, the electron.

Let's start by considering the lightest steps: the mass ratios of neutrinos. 
They should be separated by ``minimal blocks'' consisting of $SU(2)$ coupling
factors:
\be
{ m_{\nu_{\tau}} \over m_{\nu_{\mu}} } \, \sim \,
{ m_{\nu_{\mu}} \over m_{\nu_{e}}} \, \sim \, x \, , ~~~~~~
{ m_{\nu_{\tau}} \over m_{\nu_{e}} } \, \sim \, x^2 \, ,
~~~~~~ x \, = \, \alpha^{ -1}_{SU(2)} \, .
\label{nuratios}
\ee
According to the hypothesis of the 3:2:1 separation of the neutrino subspace
of the phase space, a $x = \alpha^{ -1}_{SU(2)}$ factor should also separate 
the mass of the lightest neutrino, $\nu_{\rm e}$, from the pure ``vacuum''
of the mass sector, namely the square-root energy scale corresponding to the
basic shift: ${\cal T}^{- 1/2} / 2$, the scale which, in the language 
of~\ref{mmax}, corresponds to $\beta = 0$. 
A further $\alpha^{-1}_{SU(2)}$ should then separate the heaviest neutrino
from the lightest charged particle, the electron:
\be
{ m_{e} \over m_{\nu_{\tau}}} \, \sim \, \alpha^{-1}_{SU(2)} \, .
\label{memnutau}
\ee  
As we discussed, from now on we should expect a 3:2:1 relation between
the phase space sub-volumes of the three families in the logarithmic picture:
\be
\ln {\cal V}(t,b,\tau) \, : \, \ln {\cal V}(c,s,\mu) \, : \, 
\ln {\cal V}(u,d,e) 
~~ \approx ~~ 3:2:1 \, .
\label{lnV321}
\ee
In order to determine one of these volumes, what we need now is to know 
the down-quark-to-lepton mass ratio. From the down to the up quark the
separation should be again an $SU(2)$ coupling factor. There are however 
subtleties. First of all, we must notice that it is ${\cal V}(t,b,\tau)$
that contains these factors at their ``ground'' level.
The reason is that, as we have seen in section~\ref{les}, 
the shift which, acting along the space-time coordinates, realizes
a further ``$SU(2)$ step'' in the reduction of symmetry, breaking the group
of the weak interactions, also introduces a mass differentiation in the
matter states of the same order of the scale of the breaking of the gauge 
group.
The full width of the phase space separation corresponds therefore to the
scale of the breaking, i.e. the heaviest scale at which a 
mass separation between matter states related to this broken symmetry appears. 
We will derive ${\cal V}(c,s,\mu)$
and ${\cal V}(u,d,e)$ as fractional powers: 
\be
{\cal V}(c,s,\mu) \, \sim \, \left[ {\cal V}(t,b,\tau) \right]^{2 \over 3} \, ;
~~~~
{\cal V}(u,d,e) \, \sim \, \left[ {\cal V}(t,b,\tau) \right]^{1 \over 3}
\, ,
\label{V321}
\ee
A second subtlety is that, in the case of quarks, we expect
the $\alpha_{SU(2)}$ factor between the top and bottom quark to separate
not the masses of the single quarks, but $SU(3)$ triplets, 
i.e. singlets of the confining symmetry.
In other words, the equivalence of 
leptons and quarks is established at the level of asymptotic states, which are
singlets for the confining theory. We expect therefore a factor $1/3$
correcting the mass ratio between the top and bottom quark.
This normalization is due to the requirement that, once run back to the
Planck scale, only $SU(3)$ singlets (in this case quark triplets)
go to the same limit mass value as the leptons~\footnote{We will discuss below
how further normalization factors are needed in order to take into account
the fact that at the Planck scale also the electromagnetic and weak 
interactions are ``strong'', so that masses must be normalized 
with respect to singlets of these interactions too.}. 
The top-to-bottom mass ratio should therefore be:
\be
{m_t \over m_b} \, \sim \, {1 \over 3} \, \alpha^{-1}_{SU(2)} \, .
\label{mt/mb}
\ee
The mass separation between quarks and 
leptons is the consequence of the breaking of 
the ${\bf 4}$ of each family into ${\bf 3} \oplus {\bf 1}$.
This separates the phase space in two parts of unequal volumes. 
In first approximation, this separation corresponds to
disentangling ``one quarter'' of $SU(4)$, 
and therefore we expect the ``up'' of 
the ${\bf 1}$ part to lie a $\sqrt{\alpha_{SU(2)}}$ factor below the
``down'' of the ${\bf 3}$ part. This is the separation factor between
bottom quark and $\tau$ lepton:
\be
{m_b \over m_{\tau}} \, \approx \, {1 \over 3} 
{1 \over \sqrt{\alpha_{SU(2)}}} \, .
\label{mb/mtau}
\ee  
Again a $1/3$ factor is needed in order to account for the passage
from $SU(3)$ singlets to free quarks.   
Altogether, the top-tau mass ratio is:
\be
{ m_t \over m_{\tau}} \, = \, 
{ m_t \over m_b } \times { m_b \over m_{\tau} }
~ \sim ~ 
{1 \over 3} \; \alpha^{-1}_{SU(2)} 
\, \times \,
{1 \over 3} \, {1 \over \sqrt{\alpha_{SU(2)}}} \, .
\label{mt/mtau}
\ee
According to~\ref{V321}, the analogous separation for the first family
should read:
\be
{m_u \over m_{e} } ~ \sim ~ {1 \over 9} \, 
\left( 9 { m_t \over m_{\tau}} \right)^{1 \over 3} \, ,
\label{mu/me}
\ee
where we have first removed the ${1 \over 3} \times {1 \over 3}$ factor
from the $m_t / m_{\tau}$ ratio, and then reintroduced it after having
taken the third root. This was required by the fact that these normalization 
factors, accounting for the passage from free quarks to $SU(3)$ singlets, 
don't enter in the contraction of phase sub-spaces. 
Putting all the informations together, 
we conclude that the phase-space sub-volume of the charged particles
of the first family, ${\cal V}(u,d,{e})$, should be given by:
\be
{\cal V}(u,d,{e}) \, = \, 9 \; {m_u \over m_{\nu_{\tau}}} ~
\sim ~  \; \alpha^{- 1 / 3}_{SU(2)} \; 
\left( \sqrt{\alpha_{SU(2)}} \right)^{- 1/3} \, \times
\alpha^{-1}_{SU(2)} \, .
\label{V1nutau}
\ee 
The second and third power of this volume give finally ${\cal V}(c,s,\mu)$
and ${\cal V}(t,b,\tau)$. To summarize, the mass ratios are:
\ba
m_{\mu} \over m_{\nu_{\tau}} & \sim & \alpha^{-2}_{SU(2)} \, ; 
\label{ratio-mu}\\
& & \nn \\ 
m_s \over m_{\nu_{\tau}} & \sim & {1 \over 3} \; 
\left( \sqrt{\alpha_{SU(2)}} \right)^{- 2/3} 
\; \alpha^{-2}_{SU(2)}  \, ; \label{ratio-s} \\
& & \nn \\ 
m_c \over m_{\nu_{\tau}} & \sim & {1 \over 9} \; 
\alpha^{- 2 / 3}_{SU(2)} \; 
\left( \sqrt{\alpha_{SU(2)}} \right)^{- 2/3} \; \alpha^{-2}_{SU(2)} 
\, ; \label{ratio-c} \\
& & \nn \\ 
m_{\tau} \over m_{\nu_{\tau}} & \sim & \alpha^{-3}_{SU(2)} \, ; 
\label{ratio-tau} \\
& & \nn \\
m_b \over m_{\nu_{\tau}} & \sim & {1 \over 3} \; 
\left( \sqrt{\alpha_{SU(2)}} \right)^{- 1}
\; \alpha^{-3}_{SU(2)} \, ; \label{ratio-b}  \\
& & \nn \\
m_t \over m_{\nu_{\tau}} & \sim & {1 \over 9} \; 
\alpha^{- 1}_{SU(2)} \; 
\left( \sqrt{\alpha_{SU(2)}} \right)^{- 1} \; \alpha^{-3}_{SU(2)} 
\, . \label{ratio-t}
\ea
These relations are completed by:
\ba
m_{\nu_{\tau}} & \sim & {1 \over 2} \; 
{\cal T}_{({\rm string})}^{- {1 \over 2}} 
\, \times \,
\left[ \alpha^{-1}_{SU(2)} \right]^3 \, .
\label{mnutau} \\
& & \nn \\
m_{\nu_{\mu}} & \sim & {1 \over 2} \; 
{\cal T}_{({\rm string})}^{- {1 \over 2}} 
\, \times \,
\left[ \alpha^{-1}_{SU(2)} \right]^2 \, .
\label{mnumu} \\
& & \nn \\
m_{\nu_{e}} & \sim & {1 \over 2} \; 
{\cal T}_{({\rm string})}^{- {1 \over 2}} 
\, \times \,
\alpha^{-1}_{SU(2)}  \, .
\label{mnue} 
\ea 
What we expect to be able to normalize is 
not the single electron's and neutrino mass, but the 
$SU(2)_{\rm w.i.}$-neutral combination $(e , \nu_{e})$.
Furthermore, what we should be able to normalize is not the
pure electron's mass but that of the electrically neutral ``compound''
$(e , \bar{e})$. In the first case, we can say that
$m_{(e, \nu_e)} \sim m_e + m_{\nu_e} \sim m_e$. In the second case, however,
similarly to what happens with $SU(3)$ and the quarks, 
we expect $m_{(e, \bar{e})} \sim 2 m_e$. Of course, analogous 
considerations apply to all families, and to quarks as well, because they are 
all charged also under $SU(2)_{\rm w.i.}$ and $U(1)_{\rm e.m.}$.
If an almost logarithmic sequence of masses in passing from ups and downs
of $SU(2)$ doublets allows in general 
to neglect the correction due to the lighter particle of the pair,
what we cannot neglect is the factor 2 due to the fact that, as it
was for the case of the non-perturbative mean scale, section \ref{nmass},
we are calculating the mass of a particle-antiparticle pair.
As a consequence, we expect that with the formulae obtained in this section
what we get is twice the mass of any state.

\begin{figure}
\centerline{
\epsfxsize=12cm
\epsfbox{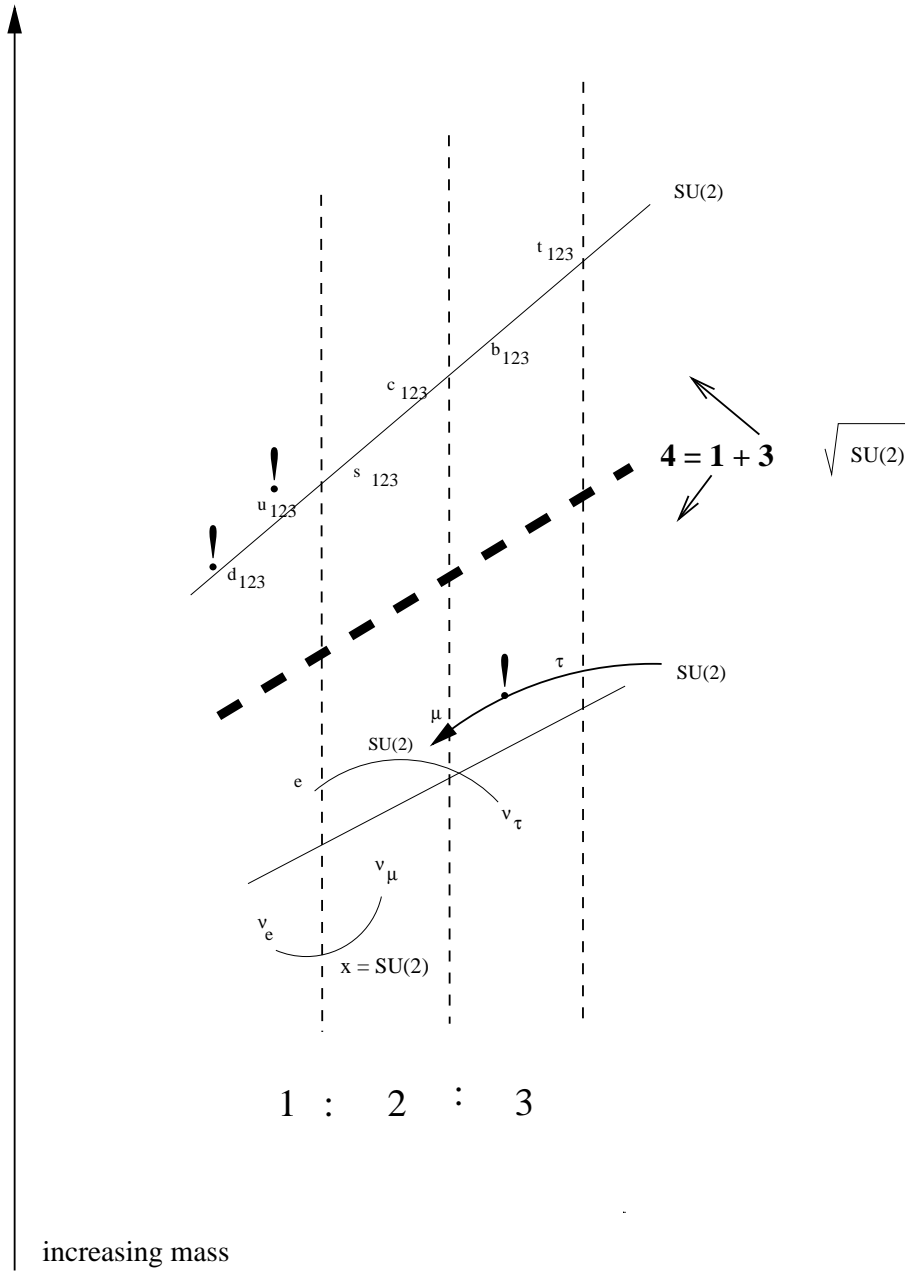}
}
\vspace{0.5cm}
\caption{The diagram of elementary masses. Notice that up and down quarks
are flipped. For leptons, the $SU(2) = SU(2)_{\Delta m}$ 
coupling factor separates the 
heaviest neutrino, $\nu_{\tau}$, from the lightest charged particle, 
the electron. This is due to the rearrangement of the ${\bf 1}$ in
the ${\bf 4} = {\bf 1} \oplus {\bf 3}$,
so that the three neutral particles are also the lightest ones, as required
from entropy considerations.}
\label{Mparticles}
\end{figure}

The values we obtain in this way are just the ``bare'' values 
of the mass ratios, the first step in the approximation, which must be 
improved by ``actual time'' corrections, in order to account for finer details
of the phase spaces.
We didn't consider yet the quark masses of the first family. As it appears
from our discussion, the up quark seems to be heavier than the down quark,
as it is reasonable to expect by analogy with the other families.
However, this is wrong, as is also clearly indicated by the experimental
observations. In the next sections we will pass to the explicit
evaluation of all the mass values. We will there discuss also
the corrections to the bare expressions, required by an improved description
of the details of the string configuration. This is particularly necessary
in order to discuss the masses of the second family, strongly affected by 
the ``stable'' mass scale of the universe, the mean scale discussed in
section~\ref{mms}, and the quarks of the first family. As we will see
in section~\ref{1family}, what happens in this case is that consistency
of the vacuum implies an exchange in the role of the up and down quark. 
The mass relations are shown in the table~\ref{Mparticles}.

\subsubsection{The $SU(2) \equiv SU(2)_{\Delta m}$ coupling}
\label{betaf}

In order to compute masses, what remains to know is the
beta-function of the broken $SU(2)$ group which constitutes the basic 
ingredient of mass ratios.
In principle, this is not the $SU(2)_{\rm w.i.}$ of 
weak interactions, which acts only on the ``left-moving'' part of
the particles. 
We indicate it as ``$SU(2)_{\Delta m}$'', to distinguish it from the group
of the weak interactions.
In order to determine the $SU(2)_{\Delta m}$ beta-function, we cannot proceed
as in the traditional approach, through a (perturbative) analysis
of the spectrum and the corrections to the gauge coupling. 
Any perturbative computation, therefore performed in a logarithmic 
representation, does not simply account for the ``logarithm of'' the 
true beta-function: in any explicit string realization, 
part of the spectrum is perturbative and weakly
coupled, but there is also a part which is ``hidden'', either because entirely
non-perturbative, or because at least part of its interactions
are non-perturbative. 
Therefore, it makes no sense to count the states 
of the spectrum, compute the beta-function as is usual in field theory, and
then trade this quantity for the logarithm of the real beta-function.
In order to get this quantity, we proceed in another way.
For the computation of the beta-function
a passage to the logarithmic picture is dangerous:
we don't know what are all the states, perturbative and non-perturbative,
of the spectrum, and we don't know how do they exactly appear in
the logarithmic picture~\footnote{Indeed, we will discuss in section~\ref{rlp}
how in this picture the spectrum of the configuration of highest entropy
appears to be supersymmetric.}. 
The analysis of the spectrum and the interactions carried out in 
section~\ref{nps} was based on a comparison of several, dual ``logarithmic
pictures'', or perturbative constructions, no one of them accounting
for the full content of the theory. The gauge charges
appeared in a rather different way in dual constructions. As long as it is a 
matter of counting the degrees of freedom, and listing their transformation 
properties, gathering together information obtained by ``patching''
dual pictures proved to be sufficient. However,
in order to derive the strength of the couplings, the question is:
how do we ``patch beta-functions''?  On the other hand,
we have seen that, with a good degree of approximation, we can
analyze the pattern of projections leading to the minimal entropy
vacuum through the counting of $Z_2$ projections in orbifold representations.
Through this procedure, non-perturbative sectors are automatically taken into 
account, being generated and/or projected out by operations that we can easily
trace in some picture. 
In order to be sure to not forget, in the counting of the beta
function, some states or misunderstand their role, we will therefore
derive the volume occupied by the broken $SU(2)_{\Delta m}$ 
by counting the volume reductions produced by
the various projections we have applied in order to reach the configuration
of minimal symmetry.

The analysis of section~\ref{nps} tells us that, within the conformal theory,
we have at disposal~7 internal and 2 extended transverse coordinates
for the projections. In total, we can apply 7 + 2 = 9
projections \footnote{Said differently: we have room for shifts along 9
coordinates.}. We must however consider that,
for what concerns our problem, two of them don't reduce
the volume of the phase space:  with a first 
projection, supersymmetry is reduced from ${\cal N}_4 = 8$ to ${\cal N}_4 = 4$.
With a second projection, supersymmetry is further reduced to ${\cal N}_4 = 2$,
and new sectors of the spectrum are generated.
However, in our specific case also at the ${\cal N}_4 = 2$ level the 
gauge beta functions vanish. This is not a general property of any 
${\cal N}_4 = 2$ vacuum,
but it is precisely what happens in the case of a configuration obtained
with non-freely acting projections, even when coupled with freely acting,
rank-reducing shifts, as is our case~\footnote{See the discussion of 
section~\ref{spectrumZ2} and Ref.~\cite{striality}.}. Indeed, ${\cal N}_4 = 2$
is the step at which matter is generated in its full content, with
the maximal amount of twisted sectors. For what
concerns the gauge interactions, it is 
therefore the point of largest symmetry group. It is only through
a further projection that, owing to the reduction to ${\cal N}_4 = 0$
\footnote{We have seen that, although in some representation this 
configuration may appear as perturbatively supersymmetric with 
${\cal N}_4 = 1$, supersymmetry is indeed broken.}, 
the gauge beta-function do not vanish anymore. Starting from this point,
any reduction of the spectrum of matter states results in a corresponding
reduction of the volume of the symmetry group. 
By counting the projections with the exclusion of the first two, we obtain that
the last step, the one corresponding to the breaking to the minimal symmetry,
corresponds, in the logarithmic picture, 
to a reduction of the volume of the phase space, with respect
to the ${\cal N}_4 = 2$ configuration, by a factor $2 \times 7 = 14$.
Through these projections the full initial symmetry group has 
\emph{effectively} been reduced into a product of 14 equivalent factors,
which correspond to $SU(2)$ enhancements of $U(1)$ symmetries. 
In other words, we can consider that the full symmetry of the phase space is
a group $G$ such that $G \supset U(2)^{\otimes 7}$. Each $U(2)$ factor
rotates a subspace of dimension 2. Namely, on the tangent space 
(logarithmic representation) the ``fundamental'' representation is
a direct sum of ${\bf 2}$:
\be
\underbrace{{\bf 2} \oplus {\bf 2} \oplus \ldots \oplus {\bf 2}}_{7 \; 
{\rm times}}
\, ,
\label{2-7}
\ee
or, after the last step,
in which also the $U(2)$ symmetry is broken and we remain with $U(1)$,
a sum of 14 $U(1)$ representations:
\be
\underbrace{{\bf 1} \oplus {\bf 1} \oplus \ldots \oplus {\bf 1}}_{14
\; {\rm times}}
\, .
\label{1-14}
\ee 
The ``beta-function coefficient'' (or better ``exponent'') of $SU(2)$  
is then ${1 \over 14}$ of the full exponent, which is fixed as follows.
For ${\cal N}_4 = 2$ the beta function must vanish: no renormalization
at all. The range of values is therefore $[0 - 1/2]$. According to 
expression~\ref{mparticles}, the beta-function exponent of 
$SU(2)_{(\Delta m)}$ is:
\be
\beta_{SU(2)} \, = \, {1 \over 14} \times {1 \over 2} = {1 \over 28} \, .
\label{bSU2}
\ee
The coupling of $SU(2)_{(\Delta m)}$ is therefore:
\be
\alpha_{SU(2)} \, = \, {\cal T}^{- {1 \over 28}} \, .
\label{betaSU2}
\ee 
Using the value of the age of the universe given in appendix \ref{A1},
we obtain that, at the present day, $\alpha^{-1}_{SU(2)} \, \sim \, 147$. 
If more precisely we use the age of the universe suggested by the agreement
with neutron's mass, eq.~\ref{calt0} 
(i.e. $\sim 5,038816199 \times 10^{60}{\rm M}_{\rm P}^{-1}$, 
see Appendix~\ref{A1}), we obtain:
\be
\alpha^{-1}_{SU(2)} \, \sim \, 147,2 ~(147,211014) \, .
\label{aSU2}
\ee
Being obtained through a counting of projections in the
$Z_2$ orbifold approximation, \ref{betaSU2} probably
constitutes only an approximation
of the real value of the beta function. However, we expect
the relative correction to $\beta^{-1} = 28$ to be small, of the order
of the relative magnitude of an inverse root of the age
of the universe, as compared to this integer value:
\be
\beta^{-1} ~ \approx ~  28 \, + 
\, {\cal O} \left( \alpha^{-1} {\cal T}^{- 1/p_{\beta}} \right) \, ,
\label{bcorr}
\ee   
which should reflect in a similar correction also for the
coupling $\alpha$ 
($\alpha^{-1} \to \alpha^{-1}(1 + {\cal O} ( 1 / {\cal T}^{1 / p_{\beta}}) $).

\subsubsection{The $U(1)_{\gamma}$ coupling}
\label{U1beta}

In order to obtain the coupling of $U(1)_{\gamma}$, the
electromagnetic group, we don't need  
to determine the absolute fraction of a group factor within the full
symmetry group: we can determine the ratio of
the $U(1)_{\gamma}$ and $SU(2)$ phase spaces, or equivalently
the ratio of the two exponents, by counting the charged matter states,
and subtracting the number of gauge bosons. We can justify this if we
consider that the latter
contribute somehow ``in opposite way'' to the matter-to-matter
scattering probability amplitudes. Consider a diagram corresponding
to a matter-to-matter transition:
\vspace{1cm}
\be
\epsfxsize=6cm
\epsfbox{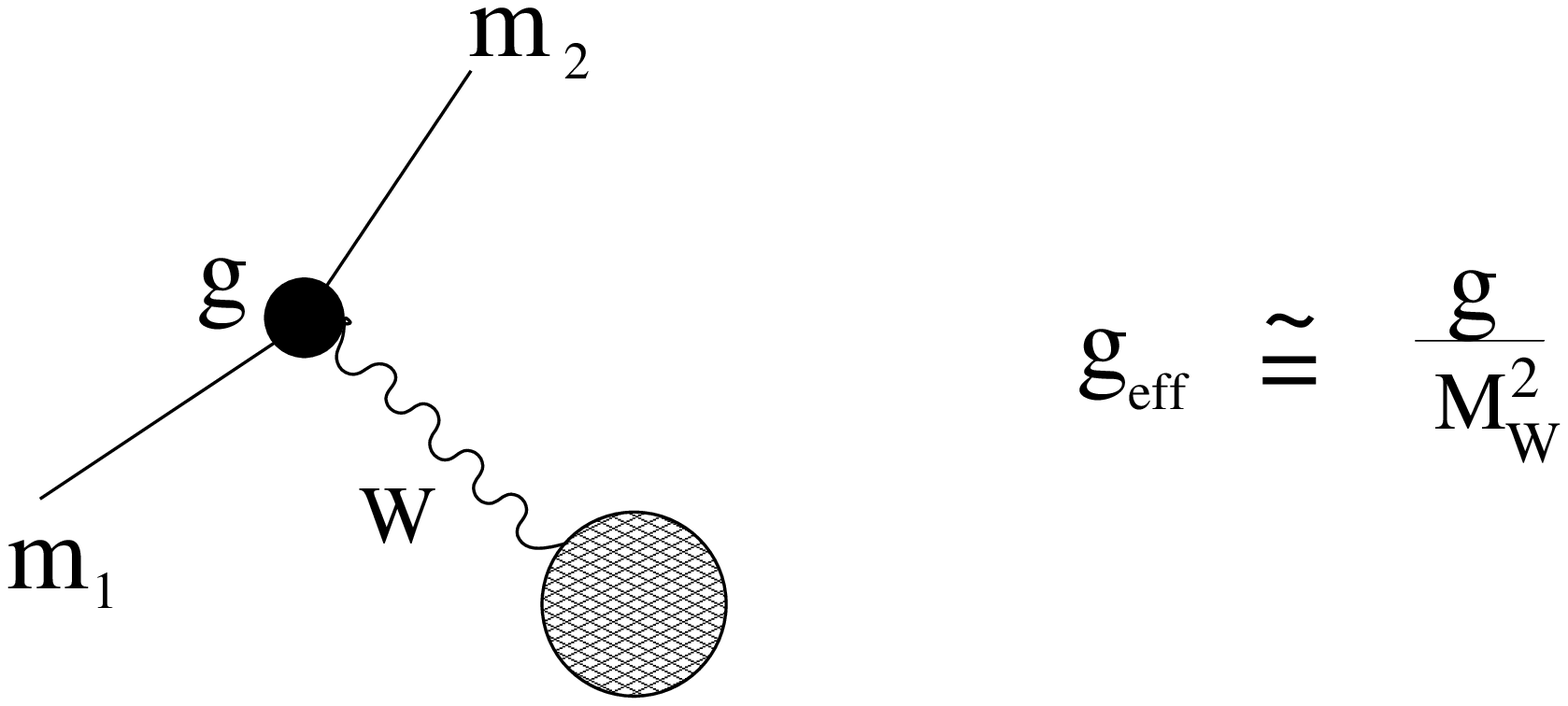}
\label{geffW}
\ee
\vspace{0.3cm}

\noindent
For what concerns the initial and final 
matter states, we have that the larger the mass ratio
between initial and final state, the larger is the decay amplitude.
The boson mass appears instead at the
denominator in the expression of the effective coupling, and
suppresses the process.

A better way to see this is to consider that, as
we will discuss in section~\ref{rlp}, in
the logarithmic picture the most entropic vacuum 
appears as effectively supersymmetric, with ${\cal N}_4 = 2$
extended supersymmetry. As seen from the logarithmic picture, 
the beta-function exponent is a ${\cal N}_4 = 2$ beta function coefficient.
In this case $b \, = \, T(R) - C(G)$. 
An equal number of matter states and gauge bosons, transforming in the
same representation, corresponds to an effective ${\cal N}_4 = 4$ restoration, 
a situation of non-renormalization, with vanishing beta-function 
exponent \footnote{In the philosophy of this analysis,
when the number of bosons matches the number of matter states, this also
means that the representation of the symmetry group is the same: a 
differentiation in the symmetries of gauge and matter states
would correspond to a less symmetric/more entropic phase produced by some 
operation, even in the case this is not explicitly related to broken gauge 
states. This is for instance
the case of the separations into different planes introduced
by orbifold projections, where these mechanisms appear in a ``frozen phase''.}.
The phase space coefficient of $U(1)_{\gamma}$ is proportional to:
$3({\rm families}) \, \times \,
2 (SU(2){\rm doublets}) \, \times \, ({\bf 1}+ {\bf 3}) ({\rm leptons ~+ ~
quarks}) \, \times \, 2 ({\rm left+right~chirality}) \, [\, = \, 48] \, - \, 
1 ({\rm gauge~boson}) \, = \, 47$. 
Notice that, in the counting, we have considered
that \emph{all} the matter states are charged under $U(1)_{\gamma}$. Three
states, the three neutrinos, are however uncharged. However,
the electromagnetic charge is simply ``shifted'' from the central value
$({1 \over 2}, - {1 \over 2})$, but the traceless condition is preserved.
As a result, the charge is only ``rearranged'' among the states: 
some states result more charged, some less.
In total, the strength of the renormalization is the same as
with a traceless $U(1)$ with a charge equally distributed among all the 
states. This is true in first approximation: from a field theory
point of view, this would be strictly true if all the masses of the matter 
states were the same, i.e. vanishing. Otherwise,
the diagrams corresponding to the contribution
of different particles have different amplitudes. From the point of view
of this work, the corrections to this first order approximation
are kept to a ``minimal'' degree by the requirement of entropy maximization,
which implies a minimization of the total strength of the electro-magnetic 
interaction.

The beta-function coefficient of $SU(2)$ is proportional to 
$48$ (the same effective number of states as for $U(1)_{\gamma}$) minus 3 
(the number of gauge bosons),
i.e. 45, where the coefficient of proportionality is the same
as for $U(1)_{\gamma}$. The ratio of the two coefficients is therefore:
\be
{ \beta_{U(1)} \over \beta_{SU(2)} } \, = \, { 47 \over 45 } \, .
\label{bU1/bU2}
\ee
Using~\ref{betaSU2} and~\ref{bU1/bU2}, 
and the scale $\mu \, = \, {\cal T} \, \sim \, 5,038816199 \times 10^{60}
{\rm M}_{\rm P}^{-1}$, the present age of the universe \ref{calt0},
adjusted on the neutron mass, we get:
\be
\alpha^{-1}_{\gamma} \, \sim \, 183,777867 \, .  
\label{bU1}
\ee
The has to be considered 
as a ``bare'' value of the
coupling, not an effective coupling in the field theory sense.
We will discuss in sections~\ref{a-fine} and~\ref{fsc} how
this value should be ``run back'' to obtain the effective coupling 
to be compared with the value experimentally measured at a certain scale.

\subsubsection{The $SU(2)_{\rm w.i.}$ coupling}
\label{alphaweak}

Determing the coupling of the $SU(2)$ of the weak interactions is
even more problematic than determining $\alpha_{\gamma}$.
The point is that for us this symmetry is not spontaneously broken in
the classical sense, and we cannot 
compute the beta-function coefficient in an effective
theory with unbroken gauge symmetry.
In the usual field theory approach, the $SU(2)$ acting on just 
one of the two helicities 
transforms only half of the matter degrees of freedom,
and therefore, if we neglect the contribution of the gauge bosons, its 
beta-function coefficient turns out to be one half of that of a ``full'' 
gauge group, namely, with a vectorial coupling to the matter currents.
This is however true as long as the matter states are massless (on the other 
hand, once they acquire a mass, the gauge symmetry is broken).
Massive states consist of both left and right degrees of freedom.
From the point of view of the volume occupied in the phase space, although
interacting with just their left-handed part, massive matter degrees of freedom
count as much as left + right chiral states. The volume occupied by
$SU(2)_{\rm w.i.}$ is therefore something ``intermediate'' between
the situation of pure chiral gauge symmetry acting on massless states, 
therefore on half the space of the degrees of freedom, and a full vectorial
interaction. We don't know a rigorous way of counting the volume 
of this interaction in the phase space.
If we want just to give a rough estimate, we can approximately consider
that our coupling lies somehow ``in between'' the two situations:
since in first approximation
the matter states acquire a mass through a shift that reduces
by half the logarithmic volume of the space (resulting therefore in a
square-root scaling law), we can expect that the logarithmic volume occupied by
$SU(2)_{\rm w.i.}$ is the mean value between
the one of the vectorial interaction (acting therefore on the same number
of degrees of freedom as the massive matter states), and the one of the
pure chiral interaction, viewed as acting on massless states:
\be
\beta_{SU(2)_{\rm w.i.}} \, \approx \, 
{1 \over 2} \left( 1 \, + \, {1 \over 2}  \right) \, \times \, {1 \over 28}
\, .
\label{bSU2w}
\ee
The present-day value of the inverse of the $SU(2)_{\rm w.i.}$  
coupling should therefore be:
\be
\alpha^{-1}_{w} \, \approx \, 
{\cal T}_0^{- \left( \beta_{SU(2)_{\rm w.i.}} \right) } \, \sim \, 42,26 \, ,
\label{awToday}
\ee 
where we have used the estimate of the age of the universe \ref{calt0}.
The value \ref{awToday} is roughly a factor 4,4 smaller than the inverse 
electromagnetic 
coupling, given in \ref{bU1}. Also this number has to be considered 
a ``bare'' value, to be corrected in the way we will discuss in 
section~\ref{a-fine}.

\subsubsection{The strong coupling}
\label{alphastrong}

In our framework, the $SU(3)$ colour symmetry is always broken, and in 
principle there is no phase in which the strong interactions can be treated
as gauge field interactions at the same time as the electromagnetic ones. 
In particular, there is no (under-Planckian) phase in which
the strongly coupled sector comes down to a ``weak'' coupling, which  
merges with the other couplings of the theory to build up a
unified model with a unique coupling, taking up the running 
up to the Planck scale. For us, the strongly coupled sector
is strongly coupled at any sub-Planckian, i.e. field theory, scale.
The coupling $\alpha_s$ will always be larger than one:
\be
\alpha_s \, \sim \, {\cal T}^{-  \beta_s} \, , ~~~~~ \beta_s \, < \, 0 \, .
\label{asrunning}
\ee
Indeed, the representation in terms of an $SU(3)$ gauge symmetry is
something that belongs more to an effective field theory realization than
to the non-perturbative string configuration we are considering.
Namely, in our case we just know that, as soon as
the space is sufficiently curved (i.e. symmetry sufficiently reduced),
we have the splitting into a weakly and a strongly coupled sector,
mutually non-perturbative with respect to each other.

In order to derive the exponent $\beta_s$, we must proceed as in section 
\ref{betaf}, by computing the amount of symmetry reduction, this time however 
in the ``S-dual'' representation. As we discussed in section~\ref{nps},
when seen from the point of view of the full space, this duality is indeed a 
T-duality . This is basically the reason why the coupling increases
as the temperature of the universe decreases (or equivalently its volume
increases). We expect therefore that, when seen from the point of view of a 
dual picture, the coupling arises in a vacuum which underwent the same amount
of symmetry reduction as in the case of the dual $SU(2)$ case of 
section~\ref{betaf}. However, the space-time coordinates feel a 
``contraction'' which is T-dual to the one experienced in the picture
of the electro-weak interactions. Therefore, when referred to
the time scale of the ``electroweak picture'', the ``beta-function''
exponent, the coefficient $\beta_s$, should be 1/4 of its analogous 
given in \ref{bSU2}.
Of course, as seen from the electroweak picture, the sign is also
the opposite (an inversion in the exponential picture reflects
in a change of sign of the logarithm). We expect therefore:
\be
\beta_s \, = \, - \, {1 \over 4} \, \times \, {1 \over 28} \, .
\label{bStr}
\ee
In other words, the strong coupling in itself should run as:
\be
\alpha_s \, \sim \, \left( {\cal T}_{\rm dual} \right)^{1 \over 28} \, ,
\label{asdual}
\ee 
but the time scale ${\cal T}_{\rm dual}$ is related to ${\cal T}$
by an inversion \emph{times} a rescaling. As the value \ref{betaSU2}
can be seen as the ``on-shell'' value at the matter scale $1/ \sqrt{\cal T}$,
logarithmically rescaled by a factor $1/2$ with respect to the
un-projected time scale ${\cal T}$, the scale ${\cal T}_{\rm dual}$
feels an inverse logarithmic rescaling, $(1/2)^{-1}$. In total, as compared 
to the square-root scale, it has a logarithmic rescaling by a factor $4$.
In order to refer the value of the strong coupling to the square-root scale
of the electroweak picture, we must therefore take its fourth root. 
The present-day ``bare'' value of the strong coupling is 
therefore \footnote{Also in this case we don't need a high precision in
the estimate of the age of the universe, whose value appears here rather
suppressed.}:
\be
\alpha_s \vert_{\rm today} ~ \sim \, 
\left[ \left( {\cal T}_0 \right)^{1 \over 4} \right]^{1 \over 28} 
\, = \,
{\cal T}_0^{- \beta_s} \, \sim \, 3,48 \, .
\label{alphastoday}
\ee 
As in the case of $\alpha_{\gamma}$ and $\alpha_w$, in order to be
compared with the coupling currently inserted in scattering amplitudes
also this one has to be ``run back''
in the way we will discuss in  section~\ref{a-fine} .

\subsection{The mean mass scale} 
\label{mms}

The mean mass scale corresponds to the inverse of the mean radius
of space. In order to have an estimate of this value, it is enough
to consider all the coordinates of unit size, apart the three
which are of the order of the radius of the classical horizon, $R$,
proportional to the age of the universe ${\cal T}$. Indeed, 
as we will see, the real spectrum of momenta is more complicated
than just the list of inverse of radii of $3 + n$ dimensional subspaces
of the string target space, $n$ running from zero to the dimension of
the string target space $D$ minus three.
Nevertheless, since in the configuration of maximal entropy
T-duality is broken we expect
the corrections to the geometry of space
to be weak, of the order of some (positive)
power of the inverse of the age of the universe.

We have seen that three space coordinates 
are extended up to ${\cal T}$, while the 
other ones are frozen at the Planck scale. Once this is taken into account,
expression \ref{mprod} reads:
\be
< \, m \, > ~ \sim \, {1 \over {\cal T}^{3 / D}} \, .
\label{mTD}
\ee
(here $D$ is the dimension of space, so that the
full space-time has dimension $D + 1$).
In order to obtain the value of the mean mass, it is now a
matter of inserting the correct value for $D$.
Until now we have considered, as is usually done, \emph{linearized}
realizations of the string space. 
We know however that from a non-perturbative 
point of view space-time is not so simply factorized. As we discussed,
it seems that we are in the presence of 11 dimensional curved space-time, that,
owing to the linearization introduced in the various dual 
``slices'' of the theory, gives the impression to be twelve dimensional. 
Twelve dimensions
are precisely what is required in order to embed in flat space
an 11-dimensional space-time, of which 10 are coordinates of
the curved space-like part. The ``true'', intrinsic space-time dimension is
therefore 11, not 12, and the
mean value of the mass scales as:
\be
< \, m \, >  ~ \sim \, \left[ \sqrt[10]{\left( \prod_i^{10} R_i = 
{\cal T}^3 \times 1^7 \right) }  \right]^{-1}  ~ = ~
{1 \over {\cal T}^{3/10}} \, .
\label{mT3}
\ee 
If we include also the correct normalization of the mass, which, according to
the Heisenberg's Uncertainty relation, $\Delta E \Delta t \geq {1 \over 2}$, 
should be proportional to 1/2 the inverse of the space-time length,
we conclude that the true value of the mean mass is:
\be
< \, m \, >  ~ = ~ 
{1 \over 2} ~ {\cal T}^{- {3 \over 10}}  \, .  
\label{maverage}
\ee 
In this expression, the time ${\cal T}$ is the age of the universe as seen
from the string frame. The age of the universe, as derived
with interpolations based on the usual Big Bang cosmology, 
is supposed to range from 11,5 and 14 billions years. Its central value
is therefore $\sim 12,75 \, \times \, 10^{9}$ yrs,  
($\sim \, 5 \, \times \, 10^{60} {\rm M}^{-1}_{\rm P}$, 
see Appendix~\ref{A1}). Inserting it in~\ref{maverage}, we obtain:
\be
< \, m \, >  ~ \sim ~ 7,49 \, {\rm GeV} \, .
\label{mval}
\ee
It may seem surprising that the mean value of the mass,
by definition given by:
\be 
\langle  m \rangle \, = \, \sum_i^n \langle i | m |  i \rangle \, , 
\label{mi}
\ee
is here obtained as a multiplicative average:
\be
\langle  m \rangle \, = \, \sqrt[n]{\prod_i^n  {1 \over R_i}  } \, . 
\ee
This is due to the multiplicative nature of the phase space, and
to the fact that any perturbative construction, in which elementary states
can be observed as free states, is a representation ``on the tangent space'',
a logarithmic realization ($g=1 \to \log g = 0$)
in which products are transformed into sums.

Let's see how this works in the case of the mean mass. 
Mean values of observables can be viewed as obtained by inserting
an appropriate operator in \ref{zint}:
\be
\langle A \rangle ~ = ~ {1 \over \cal Z} \int {\cal D} \psi \, A \,
{\rm e}^S{\psi}\, . 
\label{meanA}
\ee
Similarly to what is
usually done in the ordinary path integral, we can imagine to produce the
insertion of $A$ in expression \ref{meanA} by switching on, in the exponential
of entropy, currents $J$ that couple to the operator. 
In the case of the mass, these
are radii deformations. Since entropy scales as the square of the ``radius'' 
(i.e. the age of the universe), eq.~\ref{Sscaling}, we have:
\be
\exp - S ~ \approx ~ \exp - \left( \prod^n R_i \right)^{2 / n} 
~ \to ~  \exp - \left( \prod^n (R_i + R_i J) \right)^{2 / n}    \, .
\label{SprodR}
\ee
Notice that the integral deformation is $R_i \to R_i (1 + J)$
and not $R_i \to R_i + J$: the latter would be an infinitesimal deformation
on the tangent space.
The mean mass is therefore given by:
\be
< m > ~ \approx ~ \left[ {\delta \over \delta J} \ln {\cal Z} \right]_{J=0} ~ 
\approx ~ \left( \prod^n R_i \right)^{1 / n} \, . 
\label{mddJ}
\ee
Had we instead used as deformation the one of the tangent space,
$R_i + J$, we would have obtained for the mean mass the additive formula
$m \sim \sum {1 / R_i}$, typical of the traditional perturbative string 
approach.

\subsubsection{The neutron mass}
\label{nmass}

We want now to discuss the physical meaning of the mean mass scale just
considered. According to its definition,
eq.~\ref{mi}, the contribution to the mean value should be provided by 
the asymptotic stable mass eigenstate(s) of the theory. 
These are not necessarily elementary mass/interaction 
eigenstates: in general they will be
compounds. Usually,
one thinks at the singlets of the strong interactions, because 
the theory is constructed as a perturbative vacuum around the
zero value of the electromagnetic and weak couplings. 
Here however the situation is
different: a finite, non-perturbative functional mass expression,
valid at any value of the space-time volume, corresponds to a regime
in which not only the strong interactions are non-perturbative, but
also the electroweak interactions cannot be considered weak:
the perturbative description of electro-weak interactions is an approximation,
whose degree of accuracy increases with the age of the universe 
\footnote{The behaviour of these couplings will be discussed in 
sections~\ref{betaf} and~\ref{U1beta}.}. 
The true free mass eigenstates are 
neutral to \underline{both} strong and electroweak
interactions. The mean value \ref{mT3} corresponds therefore to the
average value of the mass of stable matter in the universe.
Since the time dependence of gauge couplings is much milder than
that of masses:
\be
\alpha \, \sim \, {1 \over {\cal T}^{1 / p}} \, , ~~~~
m \, \sim \, {1 \over {\cal T}^{1 / q}} \, , ~~~~ p \gg q \, ,
\label{alphamTpq}
\ee 
already outside of a close neighborhood of the Planck scale
we rapidly fall into a regime in which
the gravitational interaction is weak, while all other interactions
are still strong. This is the regime of interest for our problem
(at precisely the Planck scale the configuration becomes trivial).
In this phase, the only state neutral under strong, electromagnetic and weak 
interactions is a compound made out of a neutron-antineutron 
pair at rest, and their decay 
products, i.e. the proton-electron-neutrino/antiproton-positron-antineutrino 
system. At the ``strong'' limit of
the weak coupling, family mixings can be neglected because one can assume
that all heavier particles have rapidly decayed to the ground family.
As it happens for stable matter,
the decay probability of the neutron is compensated by an equal
probability of the inverse process of neutrino capture, and the system is 
stable under weak interactions. 
It is invariant under charge reversal,
and stable under electromagnetic interactions as well.
This is the \emph{only} singlet under all the above interactions,
and therefore the only mass eigenstate at finite volume.
At the present age of the universe, the volume of space-time
is anyway large enough to assure weakness of the electro-weak interactions.
This compound is therefore not necessarily a ``bound state'', as
it has presumably been at earlier times. We expect
expression~\ref{maverage} to account for the mass of
the ``composite bound state'', i.e. roughly
twice as much as the mass of the neutron-antineutron pair. Therefore:
\be
m_{\rm n} \, = \, {1 \over 4} < m > \, = \, {1 \over 8} 
{\cal T}^{-{3 \over 10}}
\, .
\label{mneutron}
\ee
By inserting in \ref{mneutron} the current value for the age of the
universe, as obtained by extrapolating data of experimental observations
within the theoretical framework of Big Bang cosmology, we obtain
a value quite close to the neutron mass. 
Namely, from \ref{mval} and~\ref{mneutron} we obtain:
\be
m_{\rm n} \, \approx \, 937 \, {\rm MeV} \, ,
\ee
quite in good agreement with the value experimentally measured,
$939,56563 \, \pm \, 0,00028$ MeV~\cite{pdb2006}.
On the other hand, we recall that expression~\ref{mT3} is given up to
orders ${\cal O} (1 / {\cal T}^p)$, $p > 0$, and
the age of the universe itself is not known with great accuracy.
Therefore, the only thing we can say here is that our expression
is \emph{compatible} with the current experimental extrapolations. 
A more correct analysis would require a new derivation of the value of the age
of the universe completely \emph{within our framework}. Anyway, 
owing to the degree of approximation applied to the usual computations, 
we expect the data about the age of the universe, obtained by integrating the
equations of motion for the expansion of the universe, 
to catch at least the correct order of magnitude.

On the other hand, within our theoretical scheme one can reverse the argument,
and take the mass of the neutron as the
most precise measurement of the age of the universe. In this case, we 
obtain as its actual value:
\be
{\cal T}_0 \, = \, 12,62028271 \, \times \, 10^9 \, {\rm yr} \, . 
\label{calt0}
\ee 
The fact that our mass formula gives as average mass the mass of the neutron
is nicely in agreement with what we would expect from a universe
behaving as a black hole. 
According to the common astrophysical models, a black hole
is in fact the step just following the ``neutron star'' phase of a star at the
end of its life. Our considerations of above suggest
that the universe, as ``seen from outside'', can be roughly thought as a 
kind of neutron star at the point of transition to a black hole.

\subsubsection{The apparent acceleration of the universe} 
\label{accel}

We are now in the position to come back to the issue of the
apparent acceleration of the universe.
We have seen that the average mass of the stable matter scales with time as:
\be
m \, \sim \, \, {\cal T}^{- 3 / 10} \, .
\ee 
If we take this mass as the reference for the atomic mass
scale, we derive that the above behaviour induces an apparent shift in the 
frequencies of the light emitted at different distances from the observer,
i.e. at different ages of the universe, 
due to the different scale of the atomic energy levels:
\be
{\tilde{\nu}_1 \over \tilde{\nu}_2} \, 
= \, \left( { {\cal T}_2 \over {\cal T}_1 } \right)^{3 \over 10} \, .
\ee  
Once ``subtracted'' from the bare red-shift \ref{rs}, 
this gives an apparent, effective red-shift $z_{\rm app.}$:
\be
1 \, + \, z_{\rm app.} \, = \,
\left( {\nu_1 \over \nu_2 } \right)_{\rm observed} \, = 
\, \left( { {\cal T}_2 \over {\cal T}_1} \right)^{7 \over 10} \, ,
\ee
as if the universe were expanding with rate 
$\tilde{R} \, \sim \, {\cal T}^{7 / 10}$,
normally expected for a matter dominated era.

At the base of what is considered an experimental evidence of the accelerated
expansion of the universe is
the observed acceleration in the time variation of the red-shift 
effect. Here, this effect receives a different explanation, in terms of
accelerated variation of ratios of mass scales, and therefore of 
observed emitted frequencies. 
Indeed, one may question whether a pure expansion of the metric
is observable. In the classical approach,
the expansion occurs at the level of the 
overall scale factor of the space part of the Robertson-Walker metric:
\be
ds^2 ~ = ~ dt^2 \, - \, R^2(t) \left[ d \vec{x}^2  \right] \, .
\ee
From a physical point of view, the scale factor $R(t)$ precisely \underline{defines} 
the speed of light, obtained from the condition $ds^2 = 0$, therefore $dx/dt = 1/R$. 
The classical argument is that the Robertson-Walker metric is 
the metric of the cosmological evolution,
not the metric of local physics. This saves things from a formal point of view, but is 
not satisfactory from a \underline{physical} point of view: 
saying that there is an expansion of the overall scale of the 
metric is equivalent to saying that there is an expansion of the scale in 
which space lengths are measured in terms of time length.
In other words, saying that there is such an expansion means that 
there is an expansion (more precisely a contraction) of the speed of light. 
Suppose we want to compare wavelengths between present time and a time 
at which the scale was 1/2 of the present one. 
From a physical point of view, what we observe is radiation produced by 
atomic transitions, and we compare wavelength
keeping fixed the period of the light wave. 
Since in the past time lengths 
are contracted by 1/2 with respect to today, during each period 
of the wave light travels twice as much as today. 
Therefore, the same atomic transition generates a 
photon with twice the wavelength as today. 
However, if the space scale is contracted, also energies are different. 
Energies scale in fact as 
inverse of lengths (consider for instance the electric potential, $V = e^2 /R$). 
In our specific example, this means that
energies were doubled, and, according to $E = h \nu$, 
also frequencies were doubled, or equivalently periods were halved. 
The same atomic transition produced therefore photons with twice the frequency, 
or half the period, as compared to today. 
This fact, combined with the fact that the speed was doubled,
implies that, for the same physica phenomenon,
the effective wavelength was the same as today: any
such an overall scale of the metric is therefore physically
unobservable.

\subsubsection{The ``unification'' of couplings}
\label{unialpha}

As one can see, in our framework the fundamental scaling of couplings
as a power of the age of the universe does not
involve the ``field theory gauge strength'' $g$, the strength of
the gauge covariant derivative, which couples the gauge field to the
matter kinetic term, but the quantity $\alpha = g^2 / 4 \pi $. 
On the other hand, it is $\alpha$ the physical
coupling entering in any expansion, always given as a series
of powers in $g^2 / 4 \pi$. 
As a consequence, for us the quantities which go to 1 and unify, 
in the specific case at the
Planck scale, are not the three couplings $g_s, g_1, g_2$
introduced through a gauge mechanism, but the effective strengths
entering in scattering and decay amplitudes, namely the three
couplings $\alpha_s$, $\alpha_w$ and $\alpha_{\gamma}$:
\be
\alpha_i \, = \, {\cal T}^{- \beta_i} ~~ \Longrightarrow ~~
\lim_{{\cal T} \to 1} \alpha_i \, = \, 1 \, .
\label{limalpha}
\ee
This in particular means that, at a certain scale, close to but
below the Planck scale, the couplings $g$ of the electro-weak interactions 
will become ``strong'': $g > 1$. This however doesn't mean that 
the corresponding interactions are going out of the weak coupling regime.
In our framework, strictly speaking there are no gauge interactions,
the gauge representation being only a useful approximation, and
the gauge connection $g$ doesn't have a particular physical meaning,
besides being a useful tool in order to arrive to $\alpha$. 
On the other hand, even at the ``$m_Z$'' mass scale, 
$\alpha_s$ is in our case larger than one. What is then the meaning of
a $\sim 0,2$ value for this coupling, as predicted by the usual $SU(3)$
colour analysis, and ``confirmed'' by experiments? In our case,
such a value would mean that the strong interactions are not strong at all,
but well perturbative, as are the electromagnetic and the weak one.
In the next section, we will discuss how the values we have obtained 
for the three couplings do 
compare with the parameters of an effective action, and therefore
with the data one finds in the literature.

\subsubsection{The effective couplings: part 1}
\label{a-fine}

The couplings $\alpha_{\gamma}$, 
$\alpha_{\rm w.i.}$ and $\alpha_s$ derived
in section~\ref{betaf} and~\ref{U1beta} and \ref{alphastrong} run with time,
and therefore with an energy scale, but not in the usual sense of the
renormalization group. Namely,
they are the \emph{fixed} couplings at a specific age of the universe.
The ``electron mass scale'', or the ``Z-boson scale'', here would mean
a different age, and size, of the universe. The usual running
according to the equations of a renormalization group refers instead
to the ``effective'' rescaling in a universe in which the
fundamental parameters remain fixed. This is due to the fact that
space-time, the space in which the effective action is framed,
is normally assumed to be of infinite extension.
The infinities which are in this way produced in the effective parameters
must be regularized according to certain rules; in practice,
by keeping as reference points certain ``on shell'', ``physical'' values.
If we want to compare our results with the parameters of such
an effective action, we must take into account this mismatch in the
interpretation of space-time. Namely, we must correct for a 
finite extension of the universe. 
This is necessary in order to compare with the experimental data
as they are quoted in the literature.
Any experimental value is derived in fact through comparison 
of a certain ``scattering amplitude'' with an effective formula, expressed in 
terms of couplings, masses, momenta
etc... For instance, the value of the fine structure constant is obtained
from a process taking place at the electron's scale. The amplitude is computed
by integrating over the momentum/space-time coordinates. In the traditional
field theory approach, this is done in an infinitely extended space-time. 
In our case, instead, the volume of space-time is finite; the fraction of
phase space occupied by the process is therefore relatively higher
as compared to the case of infinite volume, and the probability of the 
transition too. 
We understand therefore how it is possible that, in our framework, 
the same amplitude is obtained
with a smaller electromagnetic coupling as in the usual approach, 
where the volume of space-time is infinite. 

The correction due to the finiteness of space-time involves not only the
couplings, such as those of the electro-magnetic and strong interactions, but
also, as a consequence, the masses. Indeed, when we say that an elementary 
particle has a certain mass, it is always intended that this is the 
``on shell'' mass of the particle ``at rest'', and considered as an
asymptotic, free state. For instance, in the case of the electron this means
that it is considered ``at the electron's scale'', and in a phase in which
it can be assumed to be decoupled from the other particles, i.e. in a weak 
coupling regime. From our point of view, this means in a decompactification
phase. However, in our scenario the true, physical decompactification
occurs only at the infinite future. The decompactification implied in these
arguments is therefore an artifact: one takes a limit of weak coupling
of some string coordinate, 
a linearization that corresponds to working on the tangent space, while curing
the so produced infinities/zeros and trivializations by imposing a 
regularization procedure, a renormalization prescription.
In our case, we keep on imposing that the neutron's mass is the one 
given as in \ref{mneutron}. This means: we treat the neutron's mass
as an already~\emph{renormalized} value, and consider the 
relation~\ref{mneutron} as
an ``on shell prescription'' which we use in order to fix the regularization. 
The finiteness of the space volume can then be taken into account
by considering any mass and coupling in a renormalization group analysis 
in which we use a finite-volume regularization 
scheme. The physical masses and couplings become therefore scale dependent:
not only they depend on the age of the universe, by prescription
fixed by the value of the neutron's mass, but also on the scale
at which the process they correspond to takes place. Couplings and masses
have therefore a ground time-dependence, as a power of the age of the universe,
as a consequence of the time-dependence of the renormalization prescription,
and a milder, logarithmic dependence on the cut-off scale of the 
renormalization.

\subsubsection*{The electromagnetic and weak couplings}

To start with, in this section we consider the correction to the weak
``gauge'' couplings~\footnote{The strong coupling $\alpha_s$ requires  
a separate discussion.}. The correction to the $SU(2)_{\Delta m}$ couplings, 
and to the masses, will be considered in a further section~\ref{inftyV}. 
In the representation at infinite volume,
the effective gauge couplings are corrected according to:
\be
\alpha^{-1}_i \, \approx  \, \alpha^{-1}_i \vert_0 \, + \, b_i \ln \mu / \mu_0 
\, , 
\label{alphamu}
\ee
where $b_i$ are appropriate beta-function coefficients, and 
$\mu$ is the scale of the process of interest
(this can be the electron mass in the case of the fine structure constant).
In principle, in this vacuum there is no Higgs field, however it is not
clear what should be the best linearized representation of the physical
configuration: is it non-supersymmetric, as in the ``real'' picture, or
does the process of linearization lift down some supersymmetry?
How does one correctly \emph{approximate}
the physical situation, which in itself escapes the rules of field theory,
with a field theory in which to consistently perform computations?
It could be that the introduction of a Higgs field
\emph{mimics} with a certain accuracy the effect of masses, in the practical
purpose of computing the running of effective parameters in the neighbourhood
of a certain scale of the universe. There is therefore an uncertainty
in the definition of the running. However, as a consequence of~\ref{limalpha},
in first approximation we can assume that, in the effective
representation of the physical configuration, couplings run
logarithmically with an effective beta-function such that, starting from
their ``bare'' value at the actual ${\cal T}^{- 1/2}$ scale, they meet
at zero at the Planck scale: 
\be
\alpha^{-1}_i \, \approx  \, \alpha^{-1}_i \vert_0 \, + 
\, b^{({\rm eff.})}_i \ln \mu / \mu_0 
\, , 
\label{alphamueff}
\ee
with $b^{({\rm eff.})}_i $ such that:
\be
b^{({\rm eff.})}_i \ln \mu_0 \, = \, \alpha^{-1}_i \vert_0 \, . 
\label{bieff}
\ee
The scale $\mu_0$ is
fixed to be the end scale of our symmetry reduction 
process, the one
corresponding to the lowest level attained with the projections, the
square root scale:
\be
\mu_0 \, =\, {1 \over 2} \, {\cal T}^{- {1 \over 2}} \, ,
\label{mu0T}
\ee 
where ${\cal T}$ is the age of the universe as fixed by the
neutron's formula \ref{mneutron}. The choice of the 
square root scale \ref{mu0T} as the starting scale is dictated by the fact that
this is the fundamental scale of matter states, and their interactions.
Matter consists of spinors and their compounds, and a spinor feels a 
square-root space, in that twice a spinor rotation corresponds to a true
vectorial space rotation. From a technical point of view, the square root
scale is the one produced by the shift giving rise to masses for the matter 
states in this string scenario. As we will discuss in section~\ref{inftyV},
the exact normalization of the end scale for elementary states
is 1/2 of \ref{mu0T}.

Let's consider the electromagnetic coupling.
The value of $\alpha_{\gamma}$ given in section~\ref{U1beta}
must be considered as a bare value at the scale $\mu_0$.
The fine structure constant, which for us is not
really a constant, but just the present-day value of this coupling,
will correspond to the value of $\alpha_{\gamma}$ run 
from~\ref{bU1} at the scale~\ref{mu0T} to
a scale $\mu_{\gamma}$, typical of some process related to the electric 
charge. As it is experimentally given, this quantity refers to
the scale of the electron at rest. This is on the other hand
the original scale at which historically
the electric charge has been referred to.
Although in general modern experiments are not performed at the electron's 
scale, through renormalization techniques their measurements are anyway always
reduced to the electron's scale. From the point of view of our theoretical
framework, this is the scale at which the ``charged world'' starts.
Below this scale, there are the un-charged particles, and,
from a classical point of view,
the electric charge effectively ceases to exist.   
Once recalculated on the electron's mass scale,
\ref{bU1} gets corrected to:
\be
\alpha^{-1}_{\gamma} \, : \, 
\alpha^{-1}_{\gamma} \vert_{\mu_0} \ = \, 183,78 
\, \to \, \alpha^{(0) \; -1}_{\gamma} \vert_{m_e} \, \approx \, 132,85 \, ,
\label{a-fine-rough}
\ee
where we used the value~\ref{melex} for the electron's mass. 
The result~\ref{a-fine-rough} 
is definitely closer to the experimental value, nevertheless 
still quite not right, being out for an amount higher than the error
in our approximations. The reason is that
the value \ref{a-fine-rough} has been calculated
by assuming a perfect logarithmic running,
without taking into account for
an important modification in the volume of the phase space of the charged
matter particles around the electron and up quark mass scale, 
something we will do in section~\ref{1family}.
We postpone therefore a detailed evaluation of the fine structure
constant to section~\ref{fsc}.

For what matters the weak coupling, the contact with experiment is
made through the Fermi coupling constant $G_F$, basically the weak coupling
divided by the $W$-boson mass squared. Any discussion about this must
therefore be postponed after we have obtained this mass. However, the
$W$ mass too, in order to be calculated, requires a first order
estimation of the weak coupling. Indeed, proceeding as in \ref{a-fine-rough},
we can see that also this coupling undergoes relative corrections
of the right magnitude. We will come back to this coupling in 
section~\ref{GF}.

\subsubsection*{The strong coupling}

In the case of the strong coupling, things are, for obvious reasons,
more involved, being more model-dependent
also the theoretical framework in which
its effective experimental value is obtained. 
A possible ``contact with the experiment'' 
is the value $\alpha_s$ at the scale of some typical quark process, 
for instance the $Z$-boson mass in a $e^{+} e^{-} \to 4 {\rm J}$ event:
$\alpha_s (M_Z) = 0,119$ \cite{pdb2006}.
As it is usually given, $g_s$ runs logarithmically with the scale.
It seems therefore impossible to think that
the ``on shell'' value \ref{alphastoday} can be effectively
corrected to a current value lower than 1 at around 100 GeV.
However, not necessarily $\alpha_s$ must admit
an effective representation in terms of a logarithmic running
\emph{at the same time}, i.e. \emph{in the same picture} 
as the electromagnetic and weak couplings.
Namely, although strongly and weakly coupled sectors are usually
described in an effective action that accounts for all of them
at the same time, attributing a logarithmic running to all of them in
a unified picture, there are good reasons to believe that, especially for
low energies, in the case of the
effective $\alpha_s$ the logarithmic behaviour 
is only a first approximation. Indeed,
electro-weak and strong coupling are mutually non-perturbative
with respect to each other, and, although 
we don't know what the correct resummed running of $\alpha_s$
should be, and we can only make some speculation, we may expect that
its logarithmic behaviour is only the first order approximation
of a running that, in the representations in which
the electro-weak couplings are linearized, is exponential.  
If we suppose that the amount of change computed in a certain
scale interval should be seen as the first step of an exponential correction,
namely, if we suppose that it increases/decreases by a factor $\sim 15$
for each  $\Delta \mu \approx 10^{12 \sim 13}  \, 
{\rm M}_{\rm P}$, then it is not impossible that,
in passing from the scale $\mu_0 \sim 10^{-30}\, {\rm M}_{\rm P}$ to 
$\sim 10^{-17}\, {\rm M}_{\rm P}$ ($\sim 100$ GeV) the value of the strong
coupling passes from \ref{alphastoday} to $\sim 0,2$. 
It appears therefore that an effective value of $\alpha_s$ lower than 1,
as it is usually obtained, is not a signal of weakness of the interaction,
but the result of working in a ``fictitious'', infinitely extended
space-time.

\subsubsection{Running in the ``logarithmic picture''}
\label{rlp}

In our analysis, we have used several times the mapping to an artificial
logarithmic representation of space-time, in order to investigate
properties related to a vacuum that, in its correct, ``physical''
representation, appears to be strongly coupled.
This proves to be particularly useful in the investigation
of some properties of leptons and quarks as free states. 
In the logarithmic picture, a coupling of order one
becomes weak, and therefore strongly coupled elementary states become
weakly coupled. When considered at early times, namely at a stage in 
which even the weak interactions are ``strong'', this representation
allows to deal at the same time with leptons and quarks. This is not
something so unfamiliar:
although not stated in these terms, all the perturbative, 
either field theoretical or stringy, constructions of a spectrum such as that
of the Standard Model of electro-weak interactions are based on
the assumption of working in such a kind of representation.

What we did not yet discuss is how does the spectrum of the
low-energy theory look in such a representation, as compared
to the ``exponential'' picture, the physical one.
In section \ref{breaksusy} we discussed how
supersymmetry breaking is related to the appearance of
a non-vanishing curvature of space-time. This effect can therefore be viewed
as ``tuned'' by a ``coordinate'' of the theory, that 
remains twisted, and therefore frozen, at the Planck scale.
Since the strength of the coupling of the theory is related to the volume 
of the ``internal'', i.e. non-extended, coordinates,
the breaking of supersymmetry is a phenomenon related also to the
appearance of strongly coupled sectors, and strongly coupled matter states.    
Roughly speaking, we could say that strong coupling and supersymmetry 
breaking go together and are tuned by the same parameter.
This is however a ``composite'' parameter (such as a function of the product
of coordinates).

We have seen that logarithmic pictures, and more in general, perturbative 
constructions, correspond to ``decompactifications'' of the theory.
The decompactification can be either a true flat limit,
or be just a singular, non-compact orbifold,
which appears flat only locally and perturbatively. 
In both cases, the theory doesn't appear higher dimensional, because this
operation involves an ``internal'' coordinate, the coupling of the theory.
Owing to the flattening, the ``trivialization'' of this parameter,
from a perturbative point of view it appears as
a limit of partial restoration of supersymmetry. 
However, strictly speaking this is not a smooth limit of the
string space, which remains basically twisted. That's why we prefer to 
speak in terms of logarithmic mapping rather than of ``limit''.  
The amount of restored
supersymmetries depends on the details of the way the ``decompactification''
is obtained. Indeed, a full bunch of over-Planckian states could 
in principle come down to a light mass: extended
supersymmetries could appear, as well as the extra states lifted by the 
rank reductions. However,
for the investigation of the properties of matter states,
the logarithmic mapping of interest for us is the  
``minimal'' one, such that just one parameter, 
the one responsible for the separation of weakly and strongly coupled sectors, 
is mapped to zero, while all the other internal coordinates remain
``twisted''. Depending on the ``dual'' representation we
want to consider, the amount of supersymmetry we are going to 
recover in the logarithmic picture is therefore either ${\cal N}_4 = 1$
or ${\cal N}_4 = 2$. We already discussed that
${\cal N}_4 = 1$ is a fake, unstable configuration consisting of the
projection onto just the perturbative part of the spectrum of a theory
which, non-perturbatively, is non-supersymmetric. 
In the case we are interested in a correct understanding of the 
beta-functions, mapping to a ${\cal N}_4 = 2$ logarithmic
representation is more appropriate than to a ``fake'' ${\cal N}_4 = 1$.
This is what we have done in section~\ref{U1beta}, in order to
understand the role played by matter states and gauge bosons in the
evaluation of the $U(1)_{\gamma}$ beta-function as compared to the
$SU(2)$ beta-function. In this representation, there is no
``parity restoration'' in the sense of the $SU(2)_{(\rm R)}$ bosons coming
to zero mass. The fact that ${\cal N}_4 = 2$ supersymmetry doesn't have
a chiral matter spectrum (hypermultiplets include
the conjugate states of fermions) simply means that we must expect 
a doubling of the matter states, due to the fact that both the left and right
moving part of a matter state get paired to a conjugate. On the other hand,
this is not a problem, because
this picture is just a useful representation, in which we can understand 
certain properties, that must however be
appropriately pulled back to the physical picture.
In the computation of the beta-function coefficients we don't need to consider
this doubling of degrees of freedom, because this is also related to an
effective disappearance of one of the projections.

\subsubsection{Recovering the ``Geometric Probability'' tools.}
\label{GeomP}

Our method of deriving masses and couplings through an analysis of the
associated entropy in the phase space can be viewed as a ``lift up'' 
of the methods of deriving these quantities through
a computation of the geometric probability of the interaction processes.
The idea goes somehow back to the work of Armand Wyler \cite{Wyler1}, in which
the value of the fine structure constant is given as a ratio
of volumes, which can be interpreted as phase space volumes.
Further developments have shown how, through an appropriate representation
of the phase space of particles and interactions, it is possible to obtain
couplings and masses which are extremely close to the experimental ones 
\cite{Smith:1997uw,Castro:2005nb,Castro:2006,Smilga:2004uu}.
However, these are given as pure numbers, no dependence on the
fundamental scale of the universe seeming to be implied.
In order to understand this point, we must consider that, from the point
of view of our work, all these
computations are performed in \emph{linearized} representations
of the physical space. Even in the case
space-time, and the associated phase space,
is seen as the ``tangent space'' of a curved embedding space, this
higher space is somehow just the ``minimal'' embedding space, a ``first order''
departure out of the base, the flat four dimensional space. 
From our point of view, these analyses are 
a kind of ``logarithmic picture analyses''. Let's consider what usually happens
to the coupling. In our framework, it scales as a power of the age of the 
universe; in a logarithmic representation, as a logarithm of the age. 
Nevertheless,
up to a certain extent it is possible, and it does make sense, 
to reparametrize the scale dependence by approximating a power-law dependence
with a logarithmic one, in such a way that:
\be
\alpha^{-1} \, = \,
\left( {1 \over {\cal T}^{p_{\alpha}}} \right)^{-1} 
~ \leftrightarrow ~ \approx \, {1 \over \alpha_0} \, + \, \beta \log \mu \, ,
~~~~~ \alpha_0^{-1} \equiv \beta \log \mu_0 \, , ~ ~ \mu  \equiv  {\cal T} \, ,
\label{exp/log}
\ee   
for some value of $\alpha_0$  and $\beta$.
On the left hand side, we have the real running in the physical picture
(we can call it the ``cosmological picture''), on the right hand side
we have the usual running in a perturbative, effective field theory 
representation. We have discussed in a previous section (\ref{a-fine})
how and why this basically works. 
In sections \ref{entropymass}--\ref{alphastrong} 
we have also mentioned the problem of computing the exponents to which the 
age of the universe must be raised, alluding at the possibility of
calculating their ratios in a logarithmic representation.
Of course, in such an effective representation, the coefficient
$\beta$ is not the same number as the exponent $p_{\alpha}$ on the l.h.s. of
\ref{exp/log}. For instance, we have seen in \ref{betaSU2}
that the $U(1)$ exponent is $\approx 1/28$, which is quite far from the 
electro-weak beta-function coefficients of ordinary field theory.
This because the r.h.s. of expression \ref{exp/log} is an effective 
reparametrization of the physical problem, 
not simply the logarithm of the l.h.s. 
From this point of view, the Wyler's formula for the electromagnetic coupling
is the measure, in a linearized, logarithmic representation, of the
volume corresponding to this coupling in units of the volume of a 
five-dimensional space. To be concrete, using the value \ref{bcorr} we obtain:
\be
{ \left[ \alpha^{-1} \, \approx \, 
\left( {\cal T}^5 \right)^{\beta} \right] 
\over V({\cal T}^5 ) }
~~ \stackrel{\log}{\leadsto} ~
\simeq \, 
{ \beta \log (\mu / \mu_0) \over \log (\mu / \mu_0)} \, ,
~~~~~
\beta \, \approx {1 \over 5} \times 
{1 \over 28} \, = \, {1 \over 140}   \, ,
\label{awyler}
\ee
where $\mu / \mu_0$ corresponds to ${\cal T}^5$ 
and $\beta \approx {1 \over 140}$ is here our rough approximation
of the fine structure constant, what in Wyler's formula indeed comes out 
closer to the experimental value. This would probably happen also in
the present case, after a more accurate inclusion of the corrections 
discussed in section~\ref{a-fine}. 
The fact that with this procedure one gets a number that
corresponds to the electromagnetic coupling, which can be interpreted
as a geometric probability in a five-dimensional embedding of
the four-dimensional physics \cite{Smilga:2004uu,Castro:2006} is then here
no more than a matter of coincidence: the coupling evolves with time,
and ``measuring'' the exponent \ref{betaSU2} in a five-dimensional space
is just a lucky choice, which works because, at present time, 
${\cal T} = {\cal T}_0 \sim 5 \times 10^{60} \, {\rm M}^{-1}_{\rm p}$,
indeed ${\cal T}^{- (1 / 28)}_0 \sim {1 \over 5} \times {1 \over 28}$.
However, all this acquires a deep meaning when the so derived couplings and 
masses are instead seen as ratios of volumes, normalized to a specific 
scale (\cite{Smith:1997uw,Castro:2005nb,Castro:2006}). If we consider ratios
of couplings in a logarithmic representation, it is easy to see that
any scale dependence drops out. 
At the ground of the disappearance of any scale dependence is the fact
that all the quantities we deal with in field-theoretical effective 
representations
are regularized quantities. Namely, one performs the analysis around a 
starting point, such as $\alpha_0$ on the r.h.s. of expression~\ref{exp/log}, 
a value obtained 
after regularization of the infinities, ``endemic'' of a representation in
an infinitely extended space-time. 
One deals then with constant, regularized values,
and perturbations around the regularization point. The scale dependence 
appears only as a correction to a scale-independent bare value.
Although corresponding to an artificial reparametrization at a certain
fixed point of the cosmological evolution, 
such a representation of the physical space
can anyway be very useful. In principle, with our approach one catches the
full behaviour of masses and couplings. In particular, 
we get the running along the history of the universe, something
essential in order to understand astronomical experimental observations, 
or more in general the physics of much earlier
times of the history \footnote{See sections \ref{darkm},
\ref{talpha}, \ref{oklo} and \ref{nsyn} for
a discussion of these issues.}.
In practice however, in order to perform fine computations limited to
our present time, it may turn out convenient to map to an appropriate
``linearized'' representation, such as those considered in 
Refs.~\cite{Castro:2005nb,Castro:2006,Smith:1997uw}, in order to
carry out a refined calculation of masses and couplings.

\newpage

\section{\bf Present-time values of masses and couplings}
\label{todaymass}

Now that we have at hand
the value of the $SU(2)_{\Delta m}$ coupling, we can proceed to
an explicit evaluation of the masses of the elementary particles, listed in 
table~\ref{Mparticles}. 
Free elementary particles correspond to our conceptual classification
more than to the real world. 
As we mentioned in section~\ref{entromass}, asymptotic running mass formulae
are naturally given for states which are neutral
under the three elementary interactions, all strong at the Planck scale.
At present time, leptons are weakly coupled, while quarks feel a coupling
which is even stronger than it was at the Planck time. 
In order to obtain the mass of the elementary particles, intended as free
states, we must therefore identify what are the ``minimal composite states'',
i.e. the lightest singlets they can form. 
We will in the following proceed by discussing the situation case by case.

\subsection{``Bare'' mass values}
\label{baremass}

We consider here the mass values corresponding to the elementary particles,
as they can be computed using the mass formulae given in section~\ref{Emass}.
These can be considered the ``bare'' values, to be corrected in various ways,
in order to account for the mismatch between the finite-volume and the 
usual infinite-volume approach, and for the fact that at any finite time
completely free states are an ideal representation, but don't really exist. 
Mass scales can therefore be perturbed by the ``stable''
mass scale \ref{mneutron}.
In section~\ref{corrm} we will discuss these corrections, and how, 
in some cases, it is even more appropriate to consider these
values themselves as ``corrections'' of a ``bare'' mass scale.

\subsubsection{Neutrino masses}
\label{neumass}

We start with the less interacting, and therefore lightest, particles.
According to the considerations of sections~\ref{entropymass} and~\ref{Emass},
the lightest mass level must correspond to the lightest electrically
neutral particle, the electron's neutrino. 
Using the value of the present-day age of the
universe derived from the neutron's mass, expression~\ref{calt0}, 
we obtain the following value for the ``square root scale'' :
\be
{1 \over {\cal T}^{1/2}} 
\, \approx \, 4,454877246 \times 10^{-31} {\rm M}_{\rm P} \, .
\label{rootscale}
\ee 
Following~\ref{nuratios} and~\ref{mparticles}, 
the first neutrino mass should be a $\alpha^{-1}_{SU(2)}$ 
factor above 1/2 this scale. Furthermore, as discussed
in section~\ref{Emass} after the expression~\ref{mnue}, 
this procedure, being related by
a chain of symmetry reduction factors to the mass of an electrically neutral
electron-positron pair, gives twice the mass of the neutrino, or, better,
the $\nu \bar{\nu}$ mass.
Using the value~\ref{aSU2}, we obtain therefore:
\be
2 \, 
m_{\nu_e} \, \approx \, 3,279 \, 
\times 10^{-29} {\rm M}_{\rm P} \, \sim \,
4,0037 \times 10^{-10} {\rm GeV} \, = \, 0,40 \, {\rm eV} \, .
\label{mnueV}
\ee
After multiplication by a further $\alpha^{-1}_{SU(2)}$ factor, 
we obtain the second neutrino mass:
\be
2 \, m_{\nu_{\mu}} \, \approx \, 5,89 \times 10^{-8} {\rm GeV} \, = \,
58,9 \, {\rm eV} \, .
\label{mnumuV}
\ee
Finally, multiplication by a further $\alpha^{-1}_{SU(2)}$ factor
leads us to the tau neutrino:
\be
2 \, m_{\nu_{\tau}} \, \approx \, 8,677 \, {\rm KeV} \, .
\label{mnutauV}
\ee
These values agree with the experimental indications of 
possible neutrino oscillation effects at the electronvolt scale.

\subsubsection{The charged particles of the first family}
\label{1family}

An $\alpha^{-1}_{SU(2)}$ factor above the mass of the
tau-neutrino there is the electron's mass:
\be
m_e \, \sim \, \alpha^{-1}_{SU(2)} \, \times \, m_{\nu_{\tau}} ~
\sim ~ 0,639 \, {\rm MeV} \, .
\label{m-e}
\ee
As discussed in section~\ref{Emass}, 
this should be the mass of an electron-neutrino compound.
However, as we have seen, neutrino masses are negligible in comparison to
lepton masses, and with a good approximation the mass of such a compound
coincides with the lepton's mass. 

Continuing along the lines of section~\ref{Emass}, from~\ref{V1nutau}
we should be able to derive then the down and up quark masses, obtaining
$m_d \sim 0,48 \, {\rm MeV}$ and $m_u \sim 0,87 \, {\rm MeV}$\label{up-bare}.
However, this is not correct, and is contradicted by the experimental
observations.
The explanation has to do with the way the symmetry breaking is realized
in our framework. At low energy, the $SU(2)_{\rm w.i.}$ symmetry appears 
as a broken gauge symmetry, with the breaking tuned by a 
parameter of the order of a negative power of the age of the universe. 
As we will see in section~\ref{gmass}, the $SU(2)_{\rm w.i.}$
gauge boson masses scale in such a way that ${\cal T} \to \infty$
is a limit of approximate restoration of the $SU(2)_{\rm w.i.}$ symmetry. 
Moreover, remember that
the weak force in itself is stronger than the electromagnetic force:
$\alpha_w > \alpha_{\gamma}$ (it is called weak because for low transferred
momenta, $p/M_W \ll 1$, effective scattering/decay amplitudes are suppressed by
the boson mass: $\alpha^{\rm eff}_w \approx \alpha_w / M_W$). 
Therefore the ``hierarchy'' of matter is prioritarily determined by the 
$SU(2)_{\rm w.i.}$ charge, more than by the electric charge. As a consequence,
the matter spectrum can be thought as made of two subspaces, the ``up''
and the ``down'' subspace, and the trace of the electric charge
can be viewed as:
\be
< Q_{e.m.} > \, = \, \sum_{\ell, {\rm q} } <{\rm up }| Q_{e.m.} | {\rm up}> 
\, + \, \sum_{\ell, {\rm q} } <{\rm down }| Q_{e.m.} | {\rm down}> \, ,
\label{Qup/down}
\ee  
where $\sum_{\ell, {\rm q} }$ indicates the sum over leptons and quarks.
As we discussed in section~\ref{les}, minimization of entropy requires
to choose a particular distribution of the electric charge among the
$SU(2)_{\rm w.i.}$ singlets (up + down), 
so that we can have an electrically neutral 
particle. The condition of approximate restoration of the $SU(2)_{\rm w.i.}$
symmetry, and the dominance of the weak force with respect to the 
electromagnetic one, require that the two terms of the r.h.s. 
of \ref{Qup/down} give an
equal contribution to the total mean value of the electric charge.
Otherwise, this would explicitly break the $SU(2)_{\rm w.i.}$ invariance.
This imposes that the trace of the electric charge has to
vanish separately on the ``up'' and ``down'' multiplets. In practice,
both of them must vanish.

For the validity of this argument it is essential that the weak force
ends up by dominating the more and more over the electric one, and that
the symmetry is restored at infinitely extended space-time; therefore,
the full space must be
essentially thought as separated in two $SU(2)_{\rm w.i.}$ eigenspaces. 
Compatibility of the theory at any finite time with the situation at the
limit tells us that:
\be
\tr (\nu, d) \, = \, 0  \, .
\label{tr0}
\ee
Since the $\nu$ charge vanishes, we have that:
\be
\tr (d) \, = \, 0 \, .
\label{trd}
\ee   
This is only possible if, for one family, the roles of the up and down quarks,
for what matters the electric charge,
are exchanged, so that we have $\tr (d) \, = \, 3 \, \times \, \left(
{2 \over 3} - {1 \over 3} - {1 \over 3} \right) \, = \, 0 $. 
Correspondingly, the trace of the ``ups'' is also vanishing:
\be
\tr({\rm e},\mu, \tau, {\rm u}) \, = \, -1 \, -1 \, -1 \, + \, 
3 \times \left( - {1 \over 3} + {2 \over 3} + {2 \over 3}  \right) \, = \, 0
\, .
\label{trup}
\ee
Therefore, in one of the three quark families the role of up and down is 
interchanged: the quark with electric charge +2/3 is indeed the ``down'', 
while the one with charge -1/3 is the ``up''. 
In the ordinary field theory approach, this argument does not apply
because the symmetry remains broken 
also at infinitely extended space-time~\footnote{Notice that the usual charge 
assignment breaks the $SU(2)$ symmetry explicitly.}.

Simple entropy considerations
allow us to identify in which family the flip occurs. Let's consider the
$SU(3)_c$-singlet made out of the three quarks, one per each family,
with higher electric charge, and the one made in a similar way out
of the three quarks with the lower electric charge. Clearly, the first one is
the most interacting singlet we can form by picking one quark from each family,
and conversely the other one is the less interacting one we can form.
The first must therefore also be the most massive out of all
the possible $SU(3)$-singlets formed by one quark per each family, while the
second one must be the lightest. The only possibility we have to achieve
this condition is when the flip between charge +2/3 and -1/3 quarks
occurs in the lightest family, i.e., for the quarks we usually
call the up quark and the down quark. 
Therefore, approximately the value of the 
mass of the up quark is the one we computed for the lightest ``down'' quark 
states, and conversely the mass of the down quark
is the one we assigned to the lightest ``up''. However, now 
the lightest quark has a higher electric charge.
Namely, from charge $|Q| = {1 \over 3}$ we pass to
$|Q |= {2 \over 3} $. This transformation is \emph{not} a rotation
of the group $SU(2)_{\rm w.i.}$, but a pure electromagnetic charge shift. 
Therefore, here it does not matter that the former down
had negative charge, so that the charge \emph{difference} is $\Delta Q
= {2 \over 3} - \left( - {1 \over 3} \right) = 1$: a charge conjugation
is a symmetry for what matters the occupation in the phase space, or
equivalently the mass. 
What counts is the pure increase in the absolute value of the charge,
which implies an increasing of the strength of the interaction of a
particle, therefore the probability of interaction, and as a consequence
also its volume of occupation in the phase space, that is, the mass.  
Indeed, doubling the charge means logarithmically doubling, i.e. 
squaring, the interaction probability, $P \propto \alpha \propto g^2$. 
Since in the present case we increase $|Q|$ by ${1 \over 3}$ of the unit
electric charge, we expect that,
in passing from the electron to the lightest quark, 
besides the factor \ref{mu/me}, we approximately gain an extra 
$(\alpha_{\gamma})^{- 1/3}$ factor~\footnote{No further 1/3 normalization 
factors are needed, because in this operation we are leaving unchanged the 
$SU(3)$ indices.}\label{e-up-flip}. 
The upper quark of the $SU(2)$ pair passes on the other hand from
$|Q| = {2 \over 3} $ to $|Q| = {1 \over 3}$, but it does not
acquire mass shifts (in the sense either of expansion or of contraction
of its volume in the phase space) other than what already inherited by the
expansion in the phase space of the lower partner quark. The two
are in fact separated by an $SU(2)$ rotation, and the absolute
value of their mass difference
remains the same: the electric charge modification $| Q | \, : \, {2 \over 3}
\to {1 \over 3}$ has to be seen as the result of an $SU(2)_{\rm w.i.}$
rotation from the lower member of the pair, therefore
a $| \Delta Q | = 1 $ rotation, not as a charge shift by 
$| Q | = {1 \over 3}$. If, in order to better compare
with experimental data, instead of using
the inverse of \ref{bU1}, we consider the current value of
the fine structure constant at the MeV scale we will obtain 
in section~\ref{fsc}, putting everything together we get:
\ba
m_d \, & \approx & \, 4,39 \; {\rm MeV} \, , \label{mdown} \\
m_u \, & \approx & \, 2,50 \, {\rm MeV} \, , \label{mup}
\ea
so that:
\be
\delta m_{u/d} \, = \, m_u - m_d \approx 1,89
\, {\rm MeV} \, .
\label{dmu/d}
\ee

\subsubsection{The charged particles of the second family}
\label{2family}

The masses of the charged particles of the second
family are obtained from \ref{ratio-mu}, \ref{ratio-s}, \ref{ratio-c}.
At present time, they are:
\ba
m_{\mu} & \approx &  94 \, {\rm MeV} \, ;
\label{m-mu} \\
m_s & \approx &
167 \, {\rm MeV} \, ;
\label{mstrange} \\
m_c & \approx & 1,539 \, {\rm GeV} \, . 
\label{mcharm} 
\ea

\subsubsection{The charged particles of the third family}
\label{3family}

The masses of the charged particles of the third
family are obtained from \ref{ratio-tau}, \ref{ratio-b}, \ref{ratio-t}:
\ba
m_{\tau} & \approx &  13,85 \, {\rm GeV} \, ;
\label{m-tau} \\
m_b & \approx &
56 \, {\rm GeV} \, ;
\label{mbottom} \\
m_t & \approx & 2749 \, {\rm GeV} \, . 
\label{mtop} 
\ea
One can see that, up to the second family,
the mass values, although all more or less slightly differing from
those experimentally measured, are anyway of the correct order of magnitude.
The values obtained for the third family, instead, seem to be hopelessly 
wrong. In the next sections we will discuss how these ``bare'' values 
get corrected by a refinement in our approximation.

\subsection{Corrections to masses}
\label{corrm}

Some of the mass values we have obtained
are close to the
experimental ones. Other masses, in particular those of the charged
particles of the third family, are decidedly out of their experimental
value by almost one order of magnitude.
In any case, no one really coincides with its known experimental value.
Indeed, as we already pointed out, free elementary particles correspond
to a conceptual classification of the real world, that makes sense
only in the case of weakly coupled states. In our scenario, for the leptons
this condition is better and better satisfied as the universe expands.
Quarks are instead strongly coupled, and for us 
their coupling will become stronger and stronger as time goes by.
In order to disentangle the properties of the elementary states as 
``free states'', we have mapped to a logarithmic representation of the 
string vacuum. In this picture, owing to the linearization of the string 
space, it was easier to consider the ratios of the volumes occupied 
in the phase space by the various particles, and their interactions. 
However, as we pointed out, this mapping works only at times close to the 
Planck scale, where the ``logarithmic world'' becomes weakly coupled. 
At a generic finite time,
the entire spectrum of particles is ``strongly coupled'': not
only the ``colour'' interactions are strong, but even the electro-weak
symmetry can hardly be expanded around a vanishing value of the coupling.
As we discussed in section~\ref{nmass}, in such a world, 
the only true ``asymptotic'' state is neutral to all the interactions. 
We have identified this
as a bound state made of neutron, proton, electron and its neutrino and
their antiparticles.
The corresponding mass scales as ${\cal T}^{ - 3 / 10} / 2$.
Strictly speaking, at finite time
this is the only true ``bare'' state of our theory, and its mass scale 
can be used in order to set the scale of the universe. The problem of
correctly computing masses is then twofold:

1) first of all there is the fact that masses, as they are experimentally 
derived,
correspond to a theoretical framework of scale-running, finite volume
regularization of a theory basically defined in an infinitely extended
space-time. The values we gave in the previous sections must be
corrected in such a scheme, 
taking as starting point the ``regularized'' value of
the fundamental scale, related to the neutron's mass;

2) besides this correction, we must also consider that
the masses below the ${\cal T}^{-{3/10}}$ scale should be treated as 
perturbations of this scale. This is the case of the proton and the neutron, 
which are made of up and down quarks of the first family,
but have a mass much closer to the GeV scale than to the one 
of the quarks they are made of. 
By consistency, we should apply the same argument also to the electron.
The electron's mass too should be considered as a
perturbation of the mean scale. And indeed, strictly speaking it is:
free electrons exist for a short time, until they ``recombine'' into atoms
or anyway they bound into some materials. However, since the strength
of their interaction (as well as that of neutrinos) decreases with time
and at present is sufficiently small, we can safely speak of electrons
as free states; 

3) for masses above the ${\cal T}^{- 1 / 3}$ scale, things are reversed:
it is ${\cal T}^{- 1 / 3}$ which is rather 
a perturbation of the bare mass scale.

When we consider the corrections
to the masses we want to compute, it is therefore 
of primary importance to distinguish,
at least from an ideal point of view, what are the corrections with
respect to what: with respect to the mass of a stable state, or to that
of an unstable phase, whose existence can anyway be indirectly detected?
This problem is of particular relevance when we talk about quarks.
For instance, the ``bottom'' or ``top'' mass. In this case,
what are indeed measured are the masses of the quark compounds, mesons that
exist for a short time: transitory states, that we mostly know through
their decay products. It is currently assumed that the mass of the
compound reflects with a good approximation the mass of the heavy quark.
However, for us the question is: how is this mass related to the ``bare''
mass quoted in~\ref{mbottom}, \ref{mtop}?

\subsection{Converting to an infinite-volume framework}
\label{inftyV}

As discussed in section~\ref{a-fine} for the couplings, also 
for elementary masses,
in order to convert their values
to on-shell values at the appropriate scale, in
a framework of infinitely extended space-time, we must treat them
in a renormalization scheme based on a finite-volume regularization.
In this picture
they are converted to running values, fixed to reduce to the
``bare'' values at a certain age ${\cal T}$ at the fundamental mass
scale for spinors, the ${\cal T}^{-1 /2}$ scale. 
The ``regularization
prescription'' is that the effective value
of the age of the universe ${\cal T}$ is adjusted 
on the neutron mass through its relation to the
${\cal T}^{-3 /10}$  scale~\ref{maverage}.

The evaluation of masses proceeds therefore along a sequence of perturbative
steps: at first we roughly determine, as in 
sections~\ref{neumass}--\ref{3family}, 
the energy scale ``at rest'' of the a certain
particle. Then we improve the computation by letting the
mass to run from the fundamental ${\cal T}^{- 1/2}$ scale to the
specific scale, obtaining thereby an improvement in the perturbation
process. Exactly knowing the running of masses entails a detailed
knowledge of the interaction and decay processes the particle is involved in:
these in fact decide what is the weight of a particle in the phase space.
This investigation too can be viewed as part of a sequence of perturbative 
steps. At the first step,
the logarithmic running of masses can be inferred from \ref{gm1m2}:
by differentiation of this equation one obtains a renormalization group equation 
in which the running of mass ratios results to be the opposite 
of the running of couplings. In this case,
the coupling concerned is $\alpha_{SU(2)_{\Delta m}}$, that, according to
\ref{bU1/bU2}, runs more or less like the electromagnetic 
coupling $\alpha_{\gamma}$, just a bit slower. It is a kind of ``non-chiral
weak coupling'', and the correct evaluation of its beta function
suffers of the same theoretical problems of the other gauge couplings,
namely of the lack of exact knowledge of the most appropriate effective
theory (non-supersymmetric? minimally supersymmetric? with an effective
Higgs field?...). In section~\ref{a-fine} we assumed that,
in first approximation, in the effective
representation of the physical configuration, couplings run
logarithmically with an effective beta-function such that, starting from
their ``bare'' value at the actual ${\cal T}^{- 1/2}$ scale, they meet
at zero at the Planck scale. We can here assume that this holds for
the $\alpha_{SU(2)_{\Delta m}}$ coupling too.  
From \ref{gm1m2} we derive then that the relative variation of 
a mass along a certain scale variation
is opposite to the one of the $SU(2)_{\Delta m}$ coupling:
\be
{\Delta m \over m } \, = \, - {\Delta \alpha_{SU(2)_{\Delta m}} 
\over \alpha_{SU(2)_{\Delta m}}} \, . 
\label{DmDa}
\ee
Notice that, while the inverse couplings decrease to zero, 
and therefore
couplings increase when going toward the Planck scale, 
masses instead decrease. This is correct,
because what we are giving here are relative
corrections to mass ratios, not masses in themselves. It must be
kept in mind that this linearized representation makes only sense 
reasonably away from the Planck scale.
In first approximation the mass corrections are of order:
\be
{\Delta m_{i} \over m_{\rm i}} \, \approx \, 
{\ln \mu_0 - \ln \mu_{i} \over \ln \mu_0} \, ,
\label{Dlogm}
\ee
where $\mu_{i}$ are the mass scales given in 
sections~\ref{neumass}--\ref{3family},
$\mu_0 = \left( {1 \over 2} \right)^2
{\cal T}^{-1 /2}$, $\mu$ and $\mu_{i}$ are 
expressed in reduced Planck units, an appropriate Planck mass rescaling
in the argument of each logarithm being implicitly understood.
Indeed, since masses are obtained from the expressions of mass ratios,
the higher mass of a pair is obtained
as a function of an inverse coupling times the lower mass, which
sets the scale of the process. Effectively, expression~\ref{Dlogm}
is therefore shifted to: \label{improvmass}
\be
{\Delta m_{\rm i} \over m_{\rm i}} \, \approx \, 
{\ln \mu_0 - \ln \mu_{\rm i} + \ln \mu_{\nu_e} \over \ln \mu_0} \, .
\label{DlogmShift}
\ee
The first neutrino mass remains unvaried. For the
other masses, we obtain:
\ba
m_{\nu_{\mu}} \, : && 2,945 \, {\rm eV}  \, \to \, 2,739 \, {\rm eV} \, ; 
\label{massnumu}\\ 
m_{\nu_{\tau}} \, : && 4,3385 \, {\rm KeV}  \, \to \, 3,731 \, {\rm KeV} 
\, ; \label{mnutau1} \\
m_{e} \, : && 0,639 \, {\rm MeV}  \, \to \, 0,505 \, {\rm MeV}  \, ; 
\label{electrmass}\\
m_{\mu} \, : && 94 \, {\rm MeV}  \, \to \, 67,7 \, {\rm MeV} \, ; \\
m_{\tau} \, : && 13,85 \, {\rm GeV}  \, \to \, 8,99 \, {\rm GeV} \, ; 
\label{masstau1}\\
m_{u} \, : && 2,50 \, {\rm MeV}  \, \to \, 1,93 \, {\rm MeV} \, ; 
\label{massup} \\
m_{d} \, : && 4,39 \, {\rm MeV}  \, \to \, 3,35 \, {\rm MeV} \, ; 
\label{massdown} \\
\delta m_{u/d} \, : && 1,89 \, {\rm MeV}  \, \to \, 1,42 \, {\rm MeV}  \, ; 
\label{muddelta}\\
m_{c} \, : && 1,539 \, {\rm GeV}  \, \to \, 1,048 \, {\rm GeV} \, ; 
\label{mcharm1} \\
m_{s} \, : && 167 \, {\rm MeV}  \, \to \, 118,9 \, {\rm MeV} \, ; \\
m_{t} \, : && 2749 \, {\rm GeV}  \, \to \, 1582 \, {\rm GeV} \, ; 
\label{mtop1}\\
m_{b} \, : && 56 \, {\rm GeV}  \, \to \, 35,3 \, {\rm GeV} \, . 
\label{massbott}
\ea 
The only elementary particle mass we
can here use for a precise comparison with experimental data is the one
of the electron: neutrino masses are not yet known, and the other
masses will undergo further corrections (see next sections).

The correction~\ref{electrmass} must be considered as a first order
correction: once determined at ``order zero'' the bare mass, 
$0,639 \, {\rm MeV}$, we have rescaled it according to~\ref{DlogmShift},
by recalculating the effective coupling on the zero order electron's scale.
Now that we have the first order electron's mass scale, 
$\sim 0,505 \, {\rm MeV}$, we can improve our approximation by recalculating
the effective coupling on this new scale, and using this newly obtained
relative mass correction in order to correct the scale of the 
$0,639 \, {\rm MeV}$. We obtain in this case:
\be
m_e \vert_{2^{nd}} \, ; ~ 0,505 ~ \to ~ 0,5069397 \ldots \, {\rm MeV} \, .
\label{electrmass1}
\ee 
This is still about 1\% lower than the
experimental value. Indeed, in order to get the \emph{physical} mass of the 
electron, to the ``bare'' mass~\ref{electrmass} we must add also the masses 
of the lighter states. The reason is the following. 
In the derivation of\label{addmass}
the mass ratios of section~\ref{Emass}, namely proceeding 
from~\ref{gm1m2}, there is the implicit assumption that all particles
lighter than a certain one belong to a subspace of its phase space.
Suppose we have just two particles, particle $A$ with mass $m_A$, and
particle $B$, with mass $m_B = \alpha \, m_A$, 
$\alpha < 1$. When we say that
$\alpha$ is the ratio of the two volumes in the phase space, we also
imply that particle $A$ is heavier than particle $B$ in that
the space of $B$ has been obtained by a process of symmetry
reduction, by truncating the space of $A$. Particle $A$ has more 
interaction/decay
channels than $B$, because the space of $A$ contains the space of $B$.
Let's now consider the full phase space of a sub-universe consisting of
$A$ and $B$. The full volume is:
\be
V(A) \, + \, V(B) \; = \; V(A) \, + \, \alpha \, V(A) \, .
\label{shiftAB}
\ee
Now, in our specific case $A$ is the electron, and $B$ is basically
the $\tau$-neutrino (we neglect here the other neutrinos,
that give corrections of order ${\cal O} (\alpha^2)$).
When we measure the mass of the physical electron, what we look at is
the modification to the geometry of the space-time produced by the
existence of the electron. For what we just said, 
deriving the electron's mass from~\ref{gm1m2} implies considering that,
when generating the electron, we generate also the $\tau$-neutrino
and the lighter particles. They also interact, and the modification to
the whole phase space produced by the existence of the electron is
indeed the full $V(A) + V(B) = V(A) + \alpha \, V(A)$. This implies that
what we call the physical electron mass is the sum of the  
bare electron mass~\ref{electrmass} \emph{plus}, in first
approximation, the mass of the $\tau$-neutrino.
Summing to~\ref{electrmass1} the $\nu_{\tau}$ mass \ref{mnutau1}, we obtain 
then:
\be
m^{\prime}_e \vert_{2^{nd}} \, ; ~ 0,50694 ~ \to ~ 0,51057 \ldots \, 
{\rm MeV} \, .
\label{melex}
\ee 
Of course, we can correct to the second order also the $\nu_{\tau}$
mass, and further refine our evaluation. At this order the $\nu_{\tau}$
mass gets increased, thereby increasing also the estimate of the
electron's mass. A further recalculation of the coupling at the new
scales leads on the other hand to a subsequent lowering of all masses.
The approximation of the electron's mass proceeds through 
a converging series of ``zigzag'' steps of decreasing size,
below and above the final value. One can easily see that
in this way we better and better approximate the experimental value
of the electron's mass (see \cite{pdb2006}).  
However, we don't want here to go into a detailed fine evaluation
of mass values, because $\sim 1 \% \div 0,1 \%$ is our best precision in
many steps of our analysis of masses. 

In general, accounting for the shifting 
of phase space~\ref{shiftAB}
amounts in a small (${\cal O} (\alpha^{-1})$)
correction to mass values, but for the quarks of the first and second family
the relative change is much higher (${\cal O} (\sqrt[3]{\alpha^{-1}})$
and ${\cal O}(\sqrt{\alpha^{-1}})$ respectively).
Once this is taken into account, the
masses of the up and down quarks get further corrected to\label{addmassq}:
\ba
m_{u} \, : && 1,93  \, {\rm MeV}  \, \to \, 2,435 \, {\rm MeV} \, ; 
\label{massup1} \\
m_{d} \, : && 3,35 \, {\rm MeV}  \, \to \, 5,785 \, {\rm MeV} \, ; 
\label{massdown1} \\
\delta m_{u/d} \, : && 1,42 \, {\rm MeV}  \, \to \, 3,35 \, {\rm MeV}  \, .
\ea

\subsection{The fine structure constant: part 2}
\label{fsc}

Let's now come back to a more precise determination of the fine structure
constant. As discussed in section~\ref{a-fine}, the fine structure constant
is the value of $\alpha^{-1}_{\gamma}$ at the electron's scale,
the scale that can be considered
as the reference for the operational definition of the electric charge.
According to our analysis of section~\ref{Emass},
summarized in the diagram~\ref{Mparticles},
the phase space of the elementary particles divides 
into two subspaces, the electrically uncharged and the charged space,
the latter being the upper one in the sense that all charged particles
are heavier than the uncharged ones. This second subspace 
starts at the electron's scale. As we discussed 
at page~\pageref{e-up-flip}, section~\ref{1family}, 
after the up-down flip in the quarks of the first family,
the phase space gets further expanded by a 
$\sqrt[3]{\alpha^{-1}_{\gamma}}$ factor. 
This shift modifies the effective strength of the projections
applied in order to get the mass hierarchy of section~\ref{Emass}
in the sub-volume of the phase space corresponding to the first
charged family.
As a consequence, it modifies also the effective weight of the 
corresponding states, and the ratio of the effective
$U(1)_{\gamma}$ and the $SU(2)_{\Delta m}$
beta-functions \emph{around this scale}. The effect is that, as the states
weight more, the effective running of the coupling is faster, or,
equivalently, the one of its inverse slower. Namely, as
the volumes of the matter phase space are expanded (or, logarithmically,
shifted), the value of the
electromagnetic coupling at the scale $m_e$ effectively
corresponds to the value of the coupling \emph{without correction} at a 
run-back scale, $m^{\rm eff.}_e$. The amount of running-back in the scale
of the logarithmic effective coupling is equivalent to the
amount of the forward shift in the logarithmic representation 
of the volumes of particles in the phase space. If volumes get
multiplied by a factor, their logarithm gets shifted, and so 
gets shifted back the scale at which the coupling in its
logarithmic representation is effectively evaluated. 
From an effective point of view, we can therefore
derive the value of the fine structure constant by evaluating
the electromagnetic coupling proceeding as in~\ref{alphamueff}, but at
a scale a factor $\sqrt[3]{\alpha^{-1}_{\gamma}}$ below the electron's scale,
rather than precisely at the electron's scale as we did in~\ref{a-fine-rough}
(see also Appendix~\ref{Ashift}). To have a first
rough estimate, we can use~\ref{a-fine-rough}) to calculate
that the effective scale $\mu_{\gamma}$ is
lower than $0,511 \, {\rm MeV}$ by a factor $\sim 5,102549027 \dots$.  
In this case we obtain:
\be
\alpha^{(1) \; -1}_{\gamma} \vert_{m_e} ~ = ~ 137,0700548 \, .
\ee
In order to improve our evaluation, 
we need a better approximation of the shape and size of the effective shift
of the phase space of the first family. If we consider that the
$\sqrt[3]{\alpha^{-1}_{\gamma}}$ shift on the up quark translates
also to the down quark, the heavier in this case,
we should conclude that the scale at which to evaluate
$\sqrt[3]{\alpha^{-1}_{\gamma}}$ is around the down quark mass scale.
Using the value~\ref{massdown} for the point of evaluation, we obtain:
\be
\alpha^{(2) \; -1}_{\gamma} \vert_{m_e} ~ = ~ 137,0366167 \, .
\label{a-fine-corr2}
\ee
In order to further improve the estimation, one should then proceed
as we did for the electron, by iterated steps of corrections of 
the down and electron scale, recalculating the $\alpha_{SU(2)}$ factors
at the new scales to obtain improved estimations of $m_d$ and
of $\alpha^{(0)}_{\gamma}$ at the down mass scale, and so on,
obtaining a series of converging ``zigzag'' steps.
The first step corresponds to 
a slight increasing of the effective down mass,
thereby lowering the factor $\sqrt[3]{\alpha^{(0) \; -1}_{\gamma}}$,
eventually resulting in a slight, higher order decrease of 
the value of the inverse of the fine structure constant.
The value~\ref{a-fine-corr2} is around $0,0005 \, \%$ above the
current experimental one,
$\alpha^{-1} \sim 137,035999 \ldots $ \cite{pdb2010}, and these considerations
induce to expect that the further steps
of the approximation do improve the convergence toward the
experimental value. However, one should not forget the major point
of uncertainty, namely that we are here attempting to
parametrize the effective modification
of the size of the projections applied to the phase space, 
and therefore the value of the fine structure constant,
due to a local dilatation
of the phase space. A true fine evaluation
of $\alpha^{-1}_{\gamma}$ requires first of all
a better approximation of this effect. Last but not least,
there is the question whether, and at which extent, a convergence
toward the official ``experimental value'' of this parameter should be
expected and desired. Beyond a certain order of approximation,
current evaluations heavily rely on QED techniques, and are
extrapolated within a theoretical scheme that only at the first orders
corresponds to the one discussed in this paper .
A mismatch beyond this regime of approximate correspondence
does not necessarily implies and indicates that the values here obtained
are wrong, provided the effective computation of
physical amplitudes nevertheless produces correct results.

Finally,
we repeat and stress that, in our framework, the electric charge is 
time-dependent, and \ref{a-fine-corr2}, possibly corrected
at any desired order, only represents the present-day
value of this parameter. The rate of the time variation at present time
can be easily derived from the very definition. From~\ref{betaSU2} 
and~\ref{bU1/bU2} we obtain:
\be
{1 \over \alpha } {d \, \alpha \over d \, t} ~ = ~
{1 \over 28} \times { 47 \over 45} \times {1 \over {\cal T}} \, .
\ee
In one year, the expected
relative variation is therefore of order $\approx 3 \times
10^{- 12}$. This is a rather small variation, however not so small when
compared with the supposed precision with which $\alpha$ is
obtained. Indeed, the most recent measurements give for its inverse
a number with precisely 12 digits, a number whose variation could be
observed by repeating the measurement at a distance of some years.
Since however a fine experimental determination of $\alpha$ depends,
through the theoretical framework within which it is derived, on 
time-varying parameters such as lepton masses etc..., it would not be
an easy task to disentangle all these effects to get the ``pure $\alpha$
time-variation''. This kind of effects can be better detected when expanded
on a cosmological scale, as we will discuss in section~\ref{talpha}.

\subsection{The Heavy Mass Corrections}
\label{hmass}

Any perturbative approach is based on the identification of a ``bare''
configuration, a set of states, which serve as starting point for a
series of corrections. For the sake of consistency, these must be ``small''
as compared to the main contribution, the ``bare'' quantity.
The selection of a ``bare'' set is in general not uniquely determined:
precisely in string theory, a complete investigation requires a full
bunch of ``dual constructions'', built as perturbations around different
sets of bare states. In our case specific case, several physical quantities
can be viewed both from the  
the physical (``exponential'') picture, 
and from the point of view of a logarithmic 
representation of the vacuum they are mapped to.
As we have discussed, a comparison with experimental data presented
within a framework of infinite space volume and free particles may require
to map quantities to this picture.
The advantage of a logarithmic representation
resides in that, in the cases of interest,
the logarithm maps large quantities into small ones, and vice-versa.
In particular, in the case of masses which are below the Planck scale, 
since everything is measured in units of the Planck scale, higher
masses are mapped into smaller ones:
\be
m_1 \, > \, m_2 ~~~~~ \Rightarrow ~~~| \ln m_1| \,
<  \, | \ln m_2 | \, , ~~~~ \left( |m_1| \, , \, |m_2| \, < \, 1 \, \right) . 
\label{m-logm}
\ee 
When we evaluate the corrections to the masses,
we must therefore consider the perturbations
to the bare values as seen from both the point of view of
the ``exponential picture'',
where for instance the mass of the up quark is smaller than the
mean mass scale $m_{3/10}$,  which has then 
to be considered as the ``bare'' scale around which to perturb, and 
the point of view of the ``logarithmic picture'', in which,
according to \ref{m-logm}, this hierarchy is reversed, and it is the
quark mass which is going to be perturbed. Once resummed, depending
on the picture one considers, the mass
corrections may therefore be quite large, larger than the scale they are going
to correct.

We will in the following consider two types of
corrections to the bare quark masses,
depending on whether they refer to
a perturbation of the stable non-perturbative mass scale, the $m_{3/10}$
(the case of the proton mass), or to unstable particles, whose existence 
in itself is a perturbation of the mean configuration of space-time.

\subsubsection{Stable particles}
\label{n-p}

At large volume-age of the universe, $1 \ll {\cal T} \to \infty$,
electro-weak interactions are very weak, while strong interactions
become stronger and stronger. Asymptotically, the universe settles
to a configuration in which only the lightest particles of the decay
chain are normally present, 
the probability of producing higher mass, unstable ones
becoming lower and lower. 
The particles charged under the strong interaction tend to form bound
states, ``attracted'' by the mean mass scale ``$m_{3/10}$'',
defined in eq.~\ref{maverage}. In practice, this means that the universe
tends toward a world with matter made out of up and down quarks, electrons
and neutrinos~\footnote{We will comment later about muon- and tau- neutrinos.}.
The masses of these objects as \emph{free} particles are those computed
in sections~\ref{neumass}, \ref{1family},
and corrected in \ref{massnumu}--\ref{massbott}. Quarks however
are not present as free particles, but form the bounds we call proton and 
neutron, which are stable in the sense that the neutron's decay into 
proton+electron+neutrino is balanced by the inverse process of 
electron and neutrino capture by the proton.  
As discussed, the stable mass scale $m_{3/10}$ corresponds to the rest
energy of this system, namely the
neutron + proton-electron-neutrino plus their antiparticles.
If we look at the masses of the quarks and leptons constituting
this system, we see that the neutron is made of an heavier quark
set than the proton, and that the quark mass difference between neutron
and proton is higher than the sum of the electron and neutrino masses. 
This means that the neutron decay into proton+electron+neutrino leaves
some amount of energy. As far as the $m_{3/10}$ compound at equilibrium
is concerned, this energy must be included in the account. This allows us to 
deduce that ``at equilibrium'' $(1/2) \, m_{3/10}$
corresponds to four times the mass of the neutron.
Furthermore, neutron and proton differ for their electromagnetic
charge, i.e. for ``weak'' interaction properties as compared
to the strong coupling; it is therefore
reasonable to expect that their mass difference is basically 
due to the mass difference between the up and down quark. 
On the other hand, the $m_{3/10}$ scale is much higher 
than the down-up mass difference, 
and, as it corresponds to a stable scale, it can be considered weakly 
coupled. This implies that the quark mass difference can be treated as a small
perturbation of the $m_{3/10}$ scale. We can therefore write:   
\be
m_{\rm p} \, \approx \, {1 \over 4} \,
m_{3/10} \, + \, {\cal O}(\delta m_{u,d}) \, .
\label{mMq}
\ee
Indeed, since the proton is a stable particle, the difference in
the volume occupied in the moduli space by neutron and proton
is entirely due to the difference of the volumes occupied
by the quarks they are formed of.
Differently from what discussed at pages~\ref{addmass}
and~\ref{addmassq}, a correction to the
down quark mass does not require summing the mass of the lighter
states, as it was the case for instance of the electron, whose
bare mass had to be corrected by adding the $\nu_{\tau}$ mass. 
The reason is that, in this case, 
once bound to form the heavy, strong-coupling-singlet compound, the
lighter particle, the quark up, does not interact anymore: it does not
have an independent phase space, as it was the case of the
$\tau$-neutrino. For what concerns the physical mass of the proton and the 
neutron, namely, as long as, like for the case of the electron, we look at the 
modification caused to the geometry of space-time by the existence
of the proton and the neutron, the phase space of the up quark
does not add to the phase space of the ``bare'' down quark. 
The quark mass difference entering in this game is therefore
the (corrected) bare quark mass difference~\ref{muddelta}.
\emph{At the quark scale}, this is
$\delta m_{u/d} = 1,42 \, {\rm MeV}$. However, for what matters the
mass difference between neutron and proton, the scale at which the
quark masses have to be run is the proton/neutron scale. At this scale,
once recalculated according to~\ref{DlogmShift}, the quark masses are:
\ba
m_u \vert_{E = m_n} & = & 1,7189 \, {\rm MeV} \, \\
m_d \vert_{E = m_n} & = & 3,0183 \, {\rm MeV} \, 
\ea 
that imply:
\be
m_d - m_u \vert_{E = m_n} \, = \, 1,299 \, {\rm MeV} ~ \approx \,
m_{\rm n} \, - \, m_{\rm p} \, , 
\ee
quite in good agreement, apart the usual ${\cal O} (1 \, \%)$ mismatch,
with the experimental value of the neutron-proton mass difference 
\cite{pdb2006}.
Notice that, in their logarithmic running, masses, and mass differences,
decrease when increasing the scale. This is the opposite of
what happens in the real, cosmological scaling.
The point is that in the real scaling they all tend to 1 in Planck
units. As it happens for the couplings, also masses tend to the
logarithm of 1, namely, to zero, at the Planck scale.

Expression~\ref{mMq} accounts
with good approximation for the behaviour of the proton-to-neutron
mass relation far away (i.e. well below)
the Planck scale. As we get close to this scale, this approximation looses
its validity.

\subsubsection{Unstable particles}
\label{pi-k}

Let's now consider the particles that exist only for a short time: the leptons
$\mu$, $\tau$ and the mesons. As seen from the point of view
of the universe along the running of its history, at a late stage,
as it is today, the existence, i.e.
the production and decay, of these particles can be seen as a 
fluctuation out of a ``vacuum'' characterized by the mass scale $m_{3/10}$, 
of which they are a perturbation. Indeed,
since we are going to compare masses with values given in an
infinite-volume theoretical
frame corresponding to a logarithmic picture, 
in practice it is the lower mass what is going to
be seen as the ``bare'' value to be corrected. This may be the
$m_{3/10}$ mass itself 
in the case of particles with a bare mass higher than the $m_{3/10}$ scale,
as is the case of the particles of the third family 
(the quarks top, bottom and the $\tau$). Or it can be the mass
of the particle, as is the case of the second family (charm, strange and
muon). In any case, the correction is of the form:  
\be
M^2_0 ~ \leadsto ~ M^2 ~ \approx ~ M^2_0 \left(1 + \, \alpha \, 
{m^2 \over M^2_0} \right) \, ,
\label{Mam}
\ee
where $M_0$ and $M$ are the bare and the corrected mass, and $m$ is the
perturbing mass, the mass scale with which the state of mass $M$ is in
contact through an interaction with strength $\alpha = g^2 / 4 \pi$.

In the case of particles of the third family, $M$ is the $m_{3/10}$ scale,
that from the point of view of a logarithmic picture is the higher scale,
and $m$ the bare quark or lepton mass.
For the second family, things work the other way around:
$M$ is the bare mass of the particle, and $m$ corresponds to
the $m_{3/10}$. When we say ``the bare quark mass'' here we intend
something different from the usual concept of bare quark mass.
As they are usually given, quark masses are in general
directly derived from the mass of mesons they form, possibly
subtracted of the mass of the partner quark they are bound with,
and corrected within the framework of an $SU(3)$-colour symmetry based
model of hadrons. Apart from the case of the up and down quarks,
the quark mass turns out to be, although
not really coinciding and sometimes considerably 
different~\footnote{See for instance the case of the strange quark and 
the $K$ mesons.}, anyway of the same order of magnitude of the meson mass.  
In our case, bare mass means instead the value given 
in~\ref{massup}--\ref{massbott}.
The coupling $\alpha$ is in general the electromagnetic coupling,
which provides the strongest interaction between the two mass scales.
Masses enter in expression~\ref{Mam} to the second power, because
this mass correction can be viewed as a propagator correction
of an effective boson, as here illustrated  
$$
\centerline{
\epsfxsize=6cm
\epsfbox{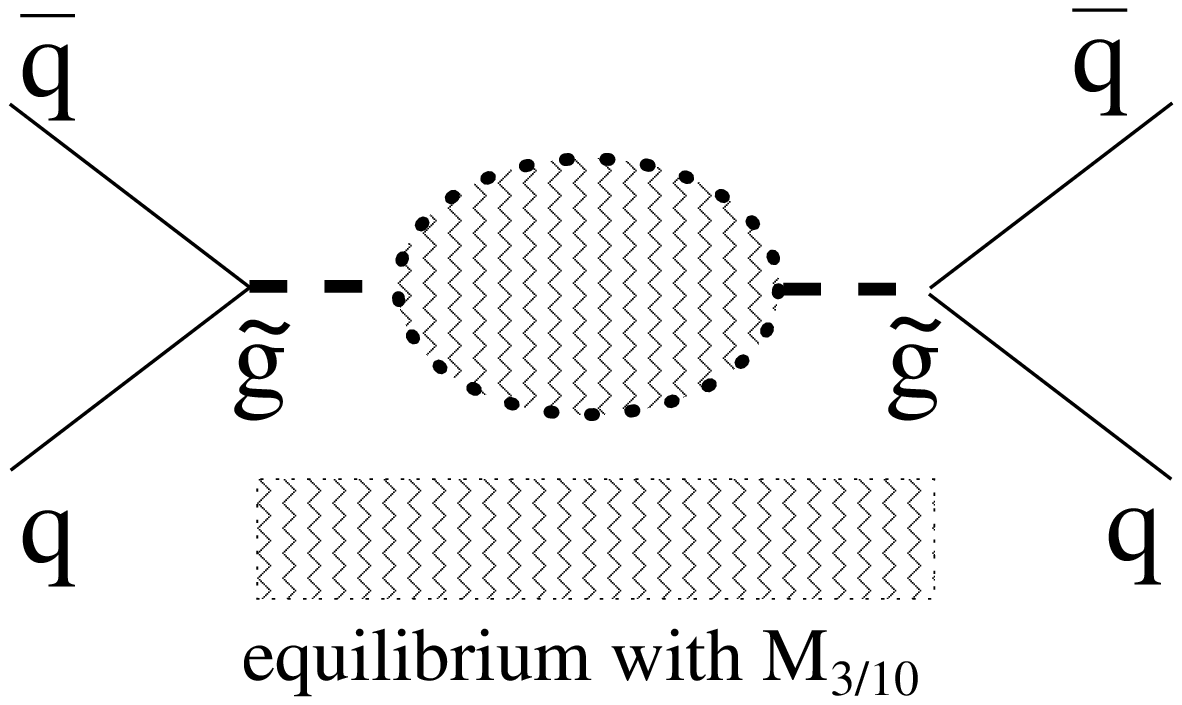}
}
$$
\vspace{.5cm}

\noindent
where $q$ and $\bar{q}$ stand for a quark-antiquark pair, in the
simplest case for instance in a $\pi$-meson. 
Indeed, an expression similar to \ref{Mam} could be 
considered also for the stable
barions considered in section~\ref{n-p}. In that case, the strongest
contact between the two scales, the up and down quark scale and the 
$m_{3/10}$ scale, is given by the strong coupling itself, of order one.
The correction ends up therefore to the $m_{3/10}$ scale itself.
For the $\pi$-mesons, or the other mesons, $K$, $C$, $B$ etc...
(these last ones
more or less ``by definition'' in direct relation to the mass of their 
heaviest quark), although their constituents interact
strongly, this interaction involves the quarks within each meson.
The strongest contact between the two scales
is however given by the electromagnetic interaction, and
$\alpha$ is basically the electromagnetic coupling.

In the case of neutrinos, their only contact with the $m_{3/10}$ scale
occurs through the weak coupling. In itself, $\alpha_{\rm w.i.}$ is 
even a bit stronger than the electromagnetic coupling. However, the
effective strength of the interaction is of order:
\be
\alpha^{\rm eff.} ~ \approx \, \alpha_{\rm w.i.} \times
{ m^2_{\nu} \over M^2_W} \, , 
\ee 
where $\alpha^{\rm eff.}$ already takes into account typical energies
of neutrino processes, and should not be confused with $G_F$, the Fermi
coupling constant. The neutrino mass corrections are therefore
extremely suppressed.

The correction \ref{Mam}
reduces the effective mass of the $t$ or $b$ quarks and the $\tau$
lepton by around one order of magnitude, producing 
values close to those experimentally measured. 
More precisely, the top mass gets corrected to:
\be
m_t \, \to \, \sim  164 \, {\rm GeV} \, ,
\label{mtcorr}
\ee
where, besides the value \ref{mtop1}, we have also used the
value of the electromagnetic coupling logarithmically
corrected to the bare top scale ($\alpha_{\gamma}^{-1} \, : \, 
183,78 \to 92,91$) \footnote{In principle, this value could be affected
by the shift in the effective beta function, centered to the electron's scale,
we discussed in section~\ref{fsc}. However, we don't have a recipe
in order to derive the full non-linear effective running of the
electromagnetic coupling. We suppose that the local modification has its peak
at around the electron/up/down scale, and tends to vanish both
toward the ${\cal T}^{- 1/2}$ and the $m_t$ scale. Therefore, we neglect
it in this and the following computations, already affected in themselves
by possibly larger uncertainties\label{alphashift}.}.
As in the case of the electron, this value too should be corrected at higher
orders, by recalculating the ``bare'' top mass, from the $1582 \, {\rm GeV}$
of the first order, to a second order value, to be used as starting point for 
the correction, to be plugged in \ref{Mam}. Then, as we did for the electron,
in order to catch the full phase space of the physical top particle,
we must add the lighter masses, the heaviest of which are
the bottom, tau, and charm masses. 
Then, here too one can easily see that
these higher order corrections better and better approximate the
experimental value of the top mass. 
Let's see the first steps of this correction.
First of all, we recalculate the relative mass correction, or
equivalently the relative coupling correction, run at the new corrected
top mass, $1582 \, {\rm GeV}$. We obtain:
\be
m^{(0)}_t \, = \, 2749 \, {\rm GeV} ~ \to ~ 
2749 - (2749 \times 0,4167) \, = \, 1603,5298 \, {\rm GeV} \, .
\ee 
To this, we must sum the non-negligible contributions of the
bottom, $\tau$ and charm bare masses, obtaining:
\be
m^{\prime}_t ~ \approx ~ 1603,53 \, + \, 35,3 \, + \, 8,99 \,
+ \, 1,048 = \, 
1648,87 \, {\rm GeV} \, .
\label{mprimet}
\ee
Of course, to be more precise we should re-correct at the second order
also the bottom, $\tau$ and charm masses, something we are not doing here.
Re-plugging~\ref{mprimet} in \ref{Mam}, we obtain:
\be
m^{\prime \prime}_t ~ \approx ~ 171,07 \, {\rm GeV} \, ,
\label{mtcorr2}
\ee
quite more in agreement with the experimental value, which is around
$\sim 171,4 \, \pm \, 1,7 \, {\rm GeV}$ \cite{topmass1}. 
We don't go further in the refinement
of~\ref{mtcorr2}, because, to start with, we should
recalculate also the bottom, $\tau$ and charm bare masses. 
Then, to be more precise,
we should also take into account the modifications to the effective
coupling and bare mass logarithmic scales, as due to the $SU(3)$ normalization
factors of the quark mass ratios, $1/3$ and $1/9$ for the bottom and the top
of each $SU(2)$ doublet. All these corrections contribute
for at most an order $\sim 1 \, \%$, therefore an uncertainty 
lower than the error in the experimental value of the top mass. More
importantly, we must
warn here that the agreement we obtain between our estimate and the
experimental value has to be taken more as the indication of
the plausibility of our analysis, rather than a real fine test.
We are trying to evaluate the ratios of the volumes in the phase space
of the particles in a rather complicated part of the spectrum, 
where the regions of validity of our simple perturbative dual approaches meet.
For instance, it is not completely clear whether the best approximation
is obtained by summing to the top phase space the lower masses \emph{before}
the correction through the $m_{3/10}$ scale, or \emph{after} it. 
Here and in the following we choose the first option. In the case of the
top quark, since the top scale
is well above all these scales, this does not make such a big difference.
Things become however more critical when looking at the corrections
to the lower masses, such as the one of the  bottom quark,
the $\tau$ or the charm quark.

For the bottom, the effective coupling we use is the inverse electromagnetic
at the bottom scale, $\alpha_{\gamma}^{-1} \vert_{b} \sim 102,95$. We obtain:
\be
m_b \, \to \, \sim 3,61 \, {\rm GeV} \, .
\label{mbcorr}
\ee
This scale too should then be corrected in a way similar to the top
mass. Adding the tau and charm masses, we obtain:
\be
m_b \, \to \, \sim 4,57 \, {\rm GeV} \, .
\label{mbcorr1}
\ee
This value is slightly above the average experimental estimate.
However, the latter is basically \emph{extrapolated} from the
$B$-meson width, and \ref{mbcorr1}, although above
the extrapolated value, is actually still compatible with
the mass of the $B$-meson. A serious comparison would require a better
understanding of the theoretical uncertainties underlying
the entire derivation, both on the side of our evaluation
of volumes in the phase space, and on the side of the experimental
derivation: for consistency, the extrapolation
from experimental data should be done entirely within the light of our 
theoretical scheme.

For the $\tau$ lepton we use a value of the electromagnetic coupling  
run to the lepton's scale, 
$\alpha_{\gamma}^{-1} \vert_{\tau} \sim 106,55$, and obtain:
\be
m_{\tau} \, \to \, \sim
1,28 \, {\rm GeV} \, .
\ee
For the further corrections to this value,
analogous arguments apply also here, with the difference that, being
the $\tau$ mass so close to the $m_{3/10}$ scale, the final result
is more sensitive to these corrections than in the top and bottom case.
For instance, at the second order the corrected bare $\tau$ mass, instead 
of \ref{masstau1}, is $m_{\tau} \vert_{2^{nd}} \sim 9,52 \, {\rm GeV}$,
that gives $1,32 \, {\rm GeV}$. Adding the charm mass, 
we get a further correction by some $5 \%$, leading to:
\be
m_{\tau}^{\prime} \, \sim \, 1,39 \, {\rm GeV} \, .
\label{mtaucorr}
\ee
As it is also the case of the quarks of this family,
in particular the bottom quark,
it doesn't make however sense to go on
with refinements of scale evaluations, as it is already clear that something
more fundamental is here missing, in order to explain the
gap between the values we obtain and the so-called experimental one
($\sim 1,78 \, {\rm GeV}$~\cite{pdb2006}). As we said, a better
understanding of the corrections to the volumes of
phase spaces around the $m_{3/10}$ scale for unstable particles is in order.
In the case of the bottom quark, the experimental value too is strongly
affected by model-dependent considerations, and things are even more 
complicated.

When we pass to the second family, analogous considerations
hold for the charm quark, whose mass is extremely close
to $m_{3/10}$. In first approximation,
by inserting the renormalized value of the electromagnetic coupling
at the bare charm mass scale, 
$\alpha_{\gamma}^{-1} \vert_{c} \sim 113,5$, we obtain a slight decrease
of the quark mass:
\be
m_c \, : \, 1,048 \to 0,946 \, {\rm GeV} \, .
\ee
However, as it is already evident from the $\tau$ mass evaluation of above,
as the bare scale approaches 
the $m_{3/10}$ scale, our perturbation method starts showing 
its limitations. Indeed, in the case of the charm quark, it would be
also possible to invert the role of bare mass and perturbing mass,
using the charm bare mass~\ref{mcharm1} as the mass $M$ in the
expression~\ref{Mam}, and, for $m$, the neutron mass, obtaining:
\be
m^{\prime}_c \, : \, 1,048 \to 1,051 \, {\rm GeV} \, .
\ee  
Including the strange-quark mass shift, we would obtain a light increase to:
\be
m^{\prime}_c \,  \sim \, 1,170 \, {\rm GeV} \, .
\ee
Similar considerations as for the bottom
and $\tau$ masses are in order here too, and we leave any further
analysis for the future.

For the strange quark and the $\mu$-lepton, they are below the $m_{3/10}$ 
scale, and, as we start to get far away from it, the reliability
of our estimate starts to improve again.  
For the strange quark, we use
$\alpha_{\gamma}^{-1} \vert_{s} \sim 117,94$, to obtain,
if we don't consider the $\mu$-mass shift: 
\be
m_s \, \to \, \sim 147 \, {\rm MeV} \, ,
\label{mscorr1}
\ee
and, when including the muon mass shift:
\be
m_s \, \to \, \sim 205,7 \, {\rm MeV} \, .
\label{mscorr2}
\ee
A comparison with
what is known as the experimental value of the strange quark mass
is affected by theoretical considerations. In itself, the
strange quark mass is extrapolated via $SU(3)_c$-related techniques
from the width of the $K$-mesons. Surely, in the space of the
$K$-mesons there is also the $\mu$- channel. However, when the
``bare'' $s$-quark mass is disentangled from the total width, does this
mean that also the $\mu$- shift gets decoupled? In this case,
the value to be considered for a comparison should not be the second one,
\ref{mscorr2}, but the $\mu$-unshifted one, \ref{mscorr1}.
The difficulties rely here also on the fact that we are comparing
\emph{extrapolated} values, not true ``experimental'' ones.  

\vspace{.5cm}
\noindent
Finally, for the muon we use
$\alpha_{\gamma}^{-1} \vert_{\mu} \sim 119,42$, that leads to:
\be
m_{\mu} \, \to \, \sim 109,4 \, {\rm MeV} \, ,
\label{mmucorr1}
\ee
and, when including the electron mass shift:
\be
m_{\mu} \, \to \, \sim 109,8 \, {\rm MeV} \, .
\label{mmucorr2}
\ee
One may notice that our mass corrections become the less and less precise
as we get closer to the $m_{3/10}$ mass scale. Indeed,
our approximation of the correction works better 
when the bare scale of the particle is far away from $m_{3/10}$, so that
we can either treat the particle's scale, or the $m_{3/10}$ scale, as the
perturbing or the perturbed scale. When they are close,
other ``non-linear'' effects become important, and with our approximation
we systematically obtain an overestimate for the particles 
with a mass below $m_{3/10}$ (muon and $s$-quark), and an underestimate 
for the particles that are above (charm, tau, (bottom ?)).

\subsubsection{The $\pi$ and $K$ mesons} 
\label{K-pi}

The $\pi^0$ mesons are bound states of the up and down quarks,
that, differently from the proton and the neutron,
``interact'' with the $m_{3/10}$ scale through the electroweak coupling
felt by their quarks,
instead than directly through the strong force.
As a consequence, the relation of the meson to the quark mass is given
as according to \ref{Mam}, with $\alpha$ the electromagnetic coupling. 
We expect therefore:
\ba
m^2_{\pi} & \sim & {\cal O}(m^2_{\rm q}) \, 
\times \, \left\{ \alpha_{e.m.} \, {\cal O}(m^2_{3/10})  \, + \, 
{\cal O}(1) \right\} \nn \\
&& \nn \\
&  \approx &
{\cal O}(m^2_{\rm q}) \times \left\{
\alpha_{e.m.} \, \left( 2 m_{\rm n}  \right)^2 \; + \,
{\cal O}(1) \right\}
\, .
\label{mpiK}
\ea
This leads to a $\sim \, 100$ MeV scale. 

As we already observed, in principle
the $s$-quark mass corrected by the $m_{3/10}$ scale as
given in~\ref{mscorr2} is somehow already the effective mass
``corresponding'' to the $K$ meson. It is not our scope here to enter
into the details of the relation between the effective 
quark and meson mass, that, according to the common framework in which
experimental data are interpreted, and therefore masses are derived,
are supposed to be linked through $SU(3)$-colour-splitting relations. 
We want here only point out that, for what matters 
the charged mesons $\pi^{\pm}$ and $K^{\pm}$, they
occupy a different phase space volume than the corresponding neutral ones; 
since the difference is due to the $U(1)_{\gamma}$
transformation properties, i.e. to the quark content, we expect
the mass difference between charged and neutral mesons 
to be of the order of the mass difference of the component
quarks. However, differently from the case of the neutron--proton
mass difference, here we don't have stable particles.
While for proton and neutron the phase space is basically the
same (in they sense that they stably transform the one into the other),
so that their differences simply reflect the differences
in the properties of the bare particles they are formed of,
for the $\pi$ and $K$ mesons charged and neutral ones have
access to completely different decay and interaction chains.
Their phase spaces are therefore really different. As a consequence,
although of the order of the mass difference of their quarks, 
the mass difference of the mesons are further modified by the
modifications of the volumes of their effective phase spaces,
and should be investigated as higher order corrections, after a 
recalculation of the phase spaces obtained 
by correcting the bare ones according to the meson interactions.

\subsection{Gauge boson masses}
\label{gmass}

We already discussed how,
in any perturbative realization of a string vacuum with gauge bosons and
matter states charged under a symmetry group,
gauge and matter originate from T-dual sectors. For instance, in heterotic
realizations the gauge bosons transforming in the adjoint
of the group originate from the currents, while the fermions transforming
in the fundamental representation originate from a twisted sector.
As a consequence, any projection 
producing a non-vanishing mass for the matter states
as the result of a shift on the windings, produces also a mass 
for the projected gauge bosons as a consequence of a shift on the momenta.
Therefore, the mass of the projected matter states scales in a T-dual way 
to that of the gauge bosons. 
After the projections involved in the breaking of the internal symmetry
into a set of separate families of particles,  matter acquires
a light mass, below the Planck scale, while the gauge bosons of the
broken symmetry acquire a mass above the Planck scale.

As anticipated in section~\ref{les},
the gauge bosons of the $SU(2)_{\rm w.i.}$ interaction don't follow
this rule: they are lifted by a shift ``on the momenta'', not on the
windings, and acquire an under-Planckian mass, of the same order
as the matter states. As discussed in section~\ref{les}, on the space-time
coordinates act two shifts. One of them produces the breaking of parity, and 
gives a light mass to the matter states ($m \sim {\cal T}^{- 1/2}$) 
while lifting in a T-dual way the mass of the gauge bosons coupled to the 
right-moving degrees of freedom. The other shift produces instead the lifting
of the left-moving gauge group, the group of weak interactions. 
As discussed in section~\ref{les},
this operation does not act, like the previous one, through
``rank-reducing level-doubling'', and its effect is not
simply a further, equivalent shift. However, from the
analysis of section~\ref{Emass}, we see that
more or less the effect of this `
`Wilson-line like'' operation is the one of producing a mass lift
whose typical length  is approximately 
a fourth root power, $\sim {\cal T}^{- 1/4}$. 
The space-time gets in fact doubly contracted,  
the root being the lift up of 
what in a perturbative, logarithmic picture appear as projection 
coefficients, ${1 \over 2}$ for the first shift and 
${1 \over 2} \times {1 \over 2}$ from the second operation.
Indeed, what we didn't say in section~\ref{les}, is
why parity should be broken by the first shift, and be related
to the ``square root'' scale, and the weak group by the second shift,
resulting approximately in a 
$\sim {\cal T}^{- 1/4} $ scale~\footnote{Indeed,
as it can be seen from the mass expressions given in section~\ref{Emass}, 
the highest mass scale is a $\sqrt{\alpha_{SU(2)}}^{-1}
= {\cal T}^{1/56}$ factor above the the ${\cal T}^{-1/4}$ scale.}.
Why not the opposite? In principle, there seems to be no ground for this
choice of ordering. Once again, what explains
this choice is the minimization of the
volume of the symmetry group, resulting in the maximization of the
entropy of the configuration~\footnote{We recall that the volume of
occupation of the whole configuration, in the phase space of configurations,
is dual to the volume of the symmetry group: the smaller the second one, the
larger the first one.}. 
In the case of the first shift the volume of the symmetry is reduced 
by the strength of the symmetry breaking driven by the mass of the right 
moving bosons, scaling as a positive power of the age of the universe. 
The higher this power,
the higher the reduction of the volume of the symmetry group ( $1/2 > 1/4$). 
In the case of the second shift,
the boson masses scale as negative powers of the age of the universe, and 
therefore the higher reduction of the volume of the symmetry, 
obtained with a higher boson
mass, is achieved with a lower exponent ($1/4 < 1/2$).

The scale
at which first the breaking of the $SU(2)_{\rm w.i.}$ symmetry takes place
approximately corresponds the scale of the
top-bottom mass difference, and, like the latter is higher than the 
experimental hadron mass scale, this one 
is around one-two orders of magnitude above the 
experimental mass scale of the $SU(2)_{\rm w.i.}$ bosons. 
It is reasonable to think that this one too is 
subjected to the same kind of renormalization
as the other scales which are above the $m_{3/10}$ scale.
However, as it is the case of the other masses, here too
thinking in terms of shifts and elementary
orbifold projections is a too simplified picture to get the
fine details of mass differencies; in order to understand the mass
of these bosons it is convenient to follow an approach similar to the one
we have used for the masses of elementary particles, and use in our formulae
the mass values already corrected according to section~\ref{pi-k}.

Let's first concentrate on what happens to the $W$ bosons.
The computation of the mass of these bosons proceeds similarly
to that of the elementary particles. The basic diagram for a broken symmetry is
again~\ref{m1m2} of section~\ref{entromass}, 
where, in this case, $m_1$ and $m_2$ refer to the masses
of the heaviest $SU(2)_{\rm w.i.}$ doublet, the top-bottom quark pair.
However, this time we are not interested in evaluating the ratio of
the volumes occupied in the phase space by the up and down particle of the
doublet. The relation between the two particles we want to consider is
therefore not a $SU(2) = SU(2)_{\Delta m}$ symmetry, but the one
established through the $SU(2)_{\rm w.i.}$ symmetry.
Being produced by a shift ``on the momenta'', 
differently from the case of the broken symmetries discussed in 
section~\ref{entromass}, here the $W$ mass is T-dual to the mass 
of the over-Planckian bosons .
Therefore, instead of~\ref{p-cutoff}, here the cut-off energy of the 
space of the process is:
\be
< p >_{\rm w.i.} ~ \sim ~  \left( { m_t m_b \over M^2_W  } \right)^{  1 / 4}
\, ,
\label{p-coffW}
\ee
and the relation \ref{gm1m2} is replaced by:
\be
\alpha \, { 3 m_t  m_b \over M_W^2 } ~ \approx ~ 1 \, ,
\label{mtmbmW}
\ee 
where $\alpha \equiv \alpha_{SU(2)_{\rm w.i.}}$ The factor 3 can be understood 
in this way: each $SU(2)_{\rm w.i.}$ transformation rotates one quark colour;
we need therefore three such rotations in order to pass from the bottom to the 
top quark.
Notice that the relation~\ref{mtmbmW} can be viewed as the integral form
of a renormalization group equation. Differentiated and mapped to a logarithmic
(and therefore also supersymmetric) representation, it roughly corresponds
to the usual expressions of the beta-function:
\be
\alpha \, {  m_t \,  m_b \over M_W^2 } ~ \approx ~ 1 ~~~~
\stackrel{\partial , \, \log}{\leadsto}  
~~~ b \, \approx \, T(R) - C(G) \, ,
\label{logbCT}
\ee
where $b$ is the gauge beta-function coefficient and
$T(R)$, $C(G)$ are the contributions of matter and gauge, entering
with opposite sign.
Inserting the mass values obtained in section~\ref{3family}, 
as corrected in section~\ref{inftyV}, namely~\ref{mtop1} and~\ref{massbott},
and the value of the weak coupling \ref{awToday}, run at the 
bottom scale, 
$\alpha^{-1} \sim 24,1$~\footnote{\label{linearalpha}In principle,
also the weak coupling should undergo an effective beta-function
modification similar to the one of the electromagnetic coupling discussed
in section~\ref{fsc}. However, as discussed in section~\ref{fsc} and in the
footnote at page~\pageref{alphashift}, this is expected to be a local 
modification, that tends to vanish toward the upper end scale of the matter 
sector, the scale that at present time is around the TeV scale.
At the $W$-boson scale, $\alpha_{\rm w.i.}$ should have almost regained
its ``regular'' value. However, we cannot exclude a slight modification
toward a lower effective value, which could explain why we get
a boson mass slightly higher than the experimental one. If we assume a 
``linear'' decrease of the effect, from the MeV to the TeV scale,
we should find that, if at the MeV scale the weak coupling
undergoes a shift proportional to the one of the effective electromagnetic
coupling: $\alpha_{\rm w.i.} \vert_{\rm 1 \, MeV} \, \to \, 
\alpha_{\rm w.i.} \vert_{\rm 1 \, MeV} \times ( 132,8 / 137) $,
at the 80 GeV scale it should have lost $\sim 2/3$ of its effect,
leading to a $\sim 83,0 \, {\rm GeV}$ $W$-boson mass (see
Appendix~\ref{Ashift}).\label{aweakshift}}, 
we get:
\be
M_{W^{\pm}} ~ \sim \, 83,4  \, {\rm GeV} \, .
\label{mW+-}
\ee
In order to obtain this mass, we used for the top and bottom mass
the ``bare'' values of page~\pageref{improvmass}, not the values
after the correction that brings them to their actual experimental value.
Indeed, the relation~\ref{mtmbmW} involves in its ``bare'' formulation
bare particles. As it was for the quarks, also $W$ are unstable
and their mass is corrected by their interaction with the $m_{3/10}$ scale.
However, for gauge bosons things go differently than for matter states, and
their corrected mass cannot simply be obtained by plugging in~\ref{mtmbmW}
the corrected values of $m_t$ and $m_b$. Gauge bosons behave T-dually
with respect to particles; therefore, in their case, we must use an expression
like~\ref{Mam} in its T-dual form:
\be
{1 \over M^2_W} \, \to \, {1 \over M^2_W} \, \left( 1 \, + \, \alpha 
\times {1 \over M^2_W} \int {d^4  p \over (p + m_{3/10})^2} \right) \, ,
\label{mWpert1}
\ee  
where $m_{3/10}$ here basically stays for the neutron's mass, and the
integral is intended up to the $W$-boson energy. Since $M_W > m_{3/10}$,
$1 / M_W < 1 / m_{3/10}$, and, as in section~\ref{pi-k}, we correct the
lower (inverse) scale $1 / M_W$ with the higher (inverse) scale 
$1 / m_{3/10}$. Moreover, the effective $W$-boson contact interaction
is not suppressed by $W$-boson transfer propagators, and the strongest
interaction they have with the $m_{3/10}$ scale occurs through the weak 
coupling. Therefore, here $\alpha = \alpha_{\rm w.i.}$. Owing to the different 
type of effective loop correction to the boson interaction with matter, 
as compared to the one of matter with matter, the term that multiplies the 
coupling is of order 1. 
We have therefore:
\be
{1 \over M^2_W} \, \to \, {1 \over M^2_W} \, \left( 1 \, + \, 
\alpha_{\rm w.i.} \right) \, ,
\label{mWpert2}
\ee  
or, T-dualized back:
\be
{M^2_W} \, \to \, \approx \, {M^2_W} \, \left( 1 \, - \, \alpha_{\rm w.i.} 
\right) \, .
\label{mWpert3}
\ee 
Inserting the value of $\alpha_{\rm w.i.}$ at the $W$ mass scale, 
$\alpha^{-1}_{\rm w.i.} \vert_{M_W} \, \sim \, 23,46$\label{aweakW}, 
we obtain:
\be
M_{W^{\pm}} ~ \to ~ \approx \; 81,6 \, {\rm GeV} \, . 
\label{MWcorr1}
\ee
All the above expressions, \ref{mWpert1}, \ref{mWpert2} and \ref{mWpert3},
neglect terms of order ${\cal O}( \alpha^2 )$ (according to~\cite{pdb2010},
the fit of the current experimental
values of the $W^{\pm}$ mass is around $80,399 \pm 0,023$ GeV;
its difference with respect to our estimate is therefore of the order of
the corrections we are neglecting).

The mass of the $Z$ boson cannot be directly derived in a similar way,
by simply substituting $m_t$ to $m_b$ in \ref{mtmbmW}: when $m_1 = m_2$
the symmetry is not broken, and the boson is massless! 
In first approximation, we expect 
the $Z$ mass to be of the order of the mass of the $W$ bosons:
at the $SU(2)_{\rm w.i.}$ extended symmetry point the shift lifts 
all the three bosons by the same amount. However, as discussed in 
section~\ref{les}, at the point of minimal symmetry, a bit away from the orbifold
point, the group is broken to $U(1) = U(1)_Z$ through a parity-preserving,
left-right symmetric operation.
This means that what distinguishes the mass of the $Z$ boson from the one of
the chiral $W^{\pm}$ bosons is the fact that it acquires a ``right moving'' 
component: while the charged bosons interact only with a left-handed chiral 
current, the neutral boson has now a certain amount of coupling with a 
right-moving current \footnote{This is not a general property of string vacua: 
when going to the
$U(1)$ point, it is in general not true that vector fields get mixed up.
In the present case, these fields are massive, already lifted by
a shift, after which the left and right chiral components of fermions combine
to give rise to massive matter. 
From a ``pure'' string point of view the $SU(2)$ bosons don't exist as 
massless fields, and therefore not as gauge bosons. 
Here we are discussing about the properties of 
massive string excitations, that we account among the field degrees of freedom
only because their mass is small, lower than the Planck scale.
Since the parity-breaking shift reflects in a ``level-2'' realization
of the weak group, there is no surprise that a displacement toward
the $U(1)$ point indeed involves the breaking of two $U(1)$'s,
the left and right, as a matter of fact ``patched together''.}. 
This ``misbalance'' should be related to the mass difference between
the $Z$ and the $W^{\pm}$ bosons.  
In turn, as it is for all the other massive excitations,
also the $Z$ mass should be related to the volume it occupies 
in the phase space.
The disagreement between the $W$ and the $Z$ mass should then be tuned
by the strength of $SU(2)_{\rm w.i.}$ as compared to  $U(1)_Z$. 
In order to derive the mass of the $Z$ boson, consider therefore once again
the diagram~\ref{m1m2}, this time with $Z$, $W^-$ and $W^+$
replacing respectively the top, bottom quarks and the $W$ boson: in this case
we view the process as a transition between $W^-$ and $Z$, produced
by an element of the ``group'' $SU(2)_{\rm w.i.} / U(1)_Z$ 
(more precisely not a group but a coset)~\footnote{Notice that we are not 
defining here $Z$ as a linear combination of $W^0$ and the field $B_{\mu}$,
associated to the hypercharge.}. 
At the vertices, $g$ is now the ``coupling'' of $SU(2)_{\rm w.i.} / U(1)_Z$. 
More precisely, since, as we discussed in section~\ref{U1beta}, the 
relation between ``width'' in the phase space and mass, in the case of gauge
bosons, is the inverse with respect to the case of matter states (higher
probability = lower boson mass), the relation of diagram~\ref{m1m2}
has to be ``T-dualized'' in the space of couplings; namely, ``S-dualized''.
The coupling appearing at the vertices is therefore the inverse
of the ``coupling'' $g^{\ast}$ of $SU(2)_{\rm w.i.} / U(1)_Z$. 
This on the other hand is precisely what we should expect. If we set:
\be
\alpha_{SU(2)_{\rm w.i.}} \, = \, \alpha^{\ast}_{SU(2)_{\rm w.i.}/U(1)_Z}
\, \times \, \alpha_{U(1)_Z} \, ,
\label{aastar}
\ee 
being the $U(1)_Z$ coupling smaller than the one of the unbroken group,
we obtain that $\alpha^{\ast} > 1$, and the relation~\ref{m1m2} must be
dualized in order to reduce to the ordinary weak coupling diagram:
\vspace{1cm}
\be
\epsfxsize=6cm
\epsfbox{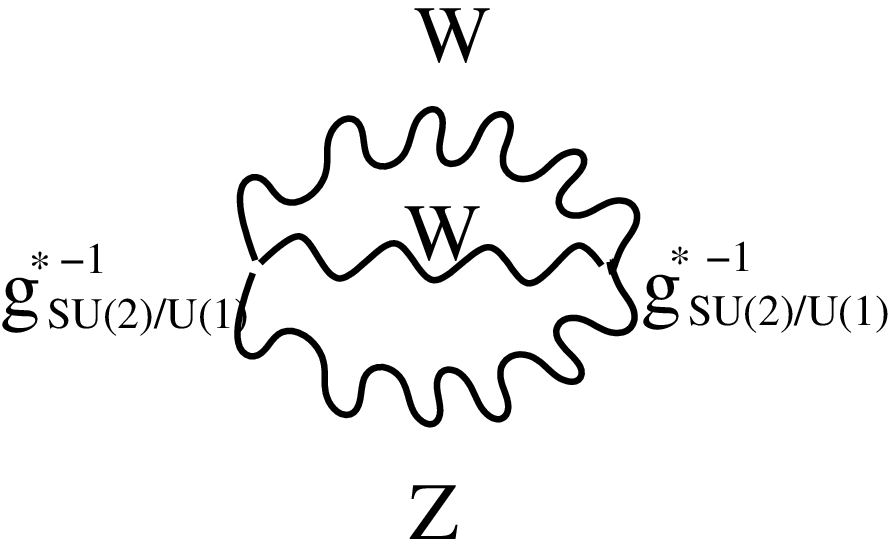}
\label{WWZ}
\ee
\vspace{0.3cm}

\noindent
Instead of a momentum to the fourth power, now the loop
integral pops out a momentum squared, and at the place of~\ref{p-cutoff}, 
the typical momentum of the space of the process is here:
\be
< p > ~ \sim ~ { M_Z \over M_W } \, .
\label{p-WZ}
\ee
The $W$ mass, the mass of the boson mediating
the process, appears in the denominator, as in~\ref{p-coffW},
T-dually to the case of~\ref{p-cutoff}.
From the diagram~\ref{WWZ} we obtain therefore:
\be
\left( {M_Z \over M_W} \right)^2 
~ \approx ~  \alpha^{\ast }_{SU(2)_{\rm w.i.}/U(1)_Z} ~ , 
\label{mWmZ}
\ee
and, using the relation \ref{aastar},
\be
M_Z ~ \sim ~ \sqrt{ \alpha_{SU(2)_{\rm w.i.}} \over \alpha_{U(1)_Z} } 
\; M_W \, .
\label{mZmW}
\ee
In order to obtain $ \alpha_{U(1)_Z}$ we can proceed as in 
section~\ref{U1beta},
this time by determining the fraction with respect to the volume occupied
by $SU(2)_{\rm w.i.}$ at the place of $SU(2)_{\Delta m}$. This means that
the coupling of $U(1)_Z$ should stay to the coupling of $U(1)_{\gamma}$
in the same ratio as the coupling of $SU(2)_{\rm w.i.}$ stays
to the one of $SU(2)_{\Delta m}$. Therefore, we expect:
\be
{\alpha_{U(1)_Z} \over \alpha_{SU(2)_{\rm w.i.}}} \, \approx \,
{\alpha_{U(1)_{\gamma}} \over \alpha_{SU(2)_{\Delta m}}} \, .
\label{U1/U2prop}
\ee 
At present time, \ref{aSU2} and \ref{bU1} and
the $W$-boson mass~\ref{MWcorr1} tell us that
the $Z$ boson mass should be approximately:
\be
M_Z \, \sim \, 1,127 \, M_W ~~ \approx 91,96 \, {\rm GeV} \, .
\label{mZvalue}
\ee
If we proceed as in the footnote at page~\pageref{aweakshift}, by assuming
a linear decrease of the local correction to the effective beta-function,
this time of the electromagnetic coupling discussed in section~\ref{fsc},
till its vanishing at the top scale of the charged matter phase space, 
\ref{mtop}, we get that at the $80 \, {\rm GeV}$ scale the shift
should have been reduced to around 1/4 of its size, producing a
relative modification of the electromagnetic coupling at this scale
of a factor $\sim 1,00791$, leading to a modification of the ratio~\ref{mZmW}
by a factor $1,00394553$, i.e. a $Z$ to $W^{\pm}$ mass ratio:
\be
{M_Z \over M_W} \, : \, 1,127 ~ \to ~ 1,132 ~ ,
\ee
a number that should be compared with the experimental ratio
of these masses, $\sim 1,134$ \cite{pdb2010}. Owing to the theoretical
uncertainties implicit in our derivation, it does not make sense to
refine the calculation, although it seems that the linear
approximation of the effective beta-function is not quite far from the
real behaviour.

\vspace{.5cm}

\noindent
Let's now come back to see how, for what concerns the neutral
and charged currents, the low-energy action looks like.
According to the results of section \ref{les}, the matter states feel:
\begin{enumerate} 
\item[i)] a long range force, mediated by
a massless field, with the strength of the non-anomalous traceless $U(1)$ group
discussed in section \ref{U1beta}, 
\item[ii)] a short range force, mediated by
the bosons obtained from the breaking of 
$SU(2)_{(\rm L)}$~($=SU(2)_{\rm w.i.}$), and 
\item[iii)] a would-be ``very short'' range 
force, ``mediated'' by the boson corresponding to $T_3^{\rm R}$.
As we said, this third is not an interaction in the sense of field theory, 
because the mass of this boson is higher than the Planck mass. 
On the other hand, its contribution is highly suppressed. 
\end{enumerate}
Our analysis tells us that 
\emph{$SU(2)$ singlets, both left and right moving}, 
couple to a massless boson in a diagonal way with the strength of the
group $U(1)_{\tilde{Y}}$, whose present time value is given by \ref{bU1}. 
The electric charge corresponds to a certain choice of charge distribution  
among the degrees of freedom constituting these singlets, as dictated
by minimization of symmetry. It can be viewed as a linear combination of the
three $U(1)$ charges $T_3^{\rm L}$, $T_3^{\rm R}$ and $\tilde{Y}$, resulting
in a traceless ``shift'' in the definition of the states.
The total charge remains of course the same as the ``hypercharge'' $\tilde{Y}$,
and therefore also the strength of the coupling, which is related to the 
``width'' of the interaction \footnote{This in first approximation:
mass differences lead in fact to a differentiation of the decay widths,
depending on the mass/charge assignments and distribution.}.
We conclude therefore that this is the strength of
the electromagnetic interaction, mediated by a massless boson that
we identify with the photon.  

The left-handed current couples in non-diagonal way to the $W^{+}$ and
$W^{-}$ bosons associated to the ``raising'' and ``lowering'' generators
of the $SU(2)_{\rm w.i.}$ group, and diagonally with the $U(1)_Z$ resulting
after the breaking of this group.
Owing to the fact that this symmetry is broken, the coupling and the mass of
this boson is not the same as the one of the $W^{\pm}$ bosons: they are related
as given in \ref{mZmW}.
If we now consider the values of the couplings of these terms, 
``run back'' to an infinite-volume effective action in order to make possible
a comparison with the standard description of low-energy physics,
we can see that approximately $\alpha_{\gamma}$, 
$\alpha_{\rm w.i.}$ and ``$\alpha_Z$'', the total coupling of the $Z$ boson, 
are related by ratios that can be written as:
\ba
\sqrt{\alpha_{\gamma}} & \approx & \sqrt{\alpha_{\rm w.i.}} \, \sin \theta \, ;
\label{sintheta} \\
&& \nn \\
\sqrt{\alpha_{Z}} & \approx & \sqrt{\alpha_{\rm w.i.}} \, \cos \theta \, , 
\label{costheta}
\ea 
for a certain angle $\theta$. The value of
$\alpha_{\gamma}$, the coupling of the electromagnetic current $J^{e.m.}$ 
to the photon, numerically approximately coincides with the one of the
fine structure constant. 
$\alpha_{\rm w.i.}$ is the coupling of the $W^{\pm}$ and $Z$ bosons to the
axial current $J^{\pm,0}$, and numerically approximately coincides with the 
standard weak coupling $\alpha_{w}$. Notice that in our framework also the 
$Z$ boson couples to this current by the same amount as the charged bosons.
In the Standard Model effective action, at the tree level the couplings of
the $J^{\pm,0}$ are instead:
\ba
W^{\pm} & \rightarrow & g \, J^{\pm }_{\mu} W^{\mp \, \mu} \, ,
\label{JpmStd} \\
& & \nn \\
Z & \rightarrow & {g \over \cos \vartheta_w } J_{\mu}^0 Z^{\mu} \, .
\label{ax0Std}
\ea
However, the effective amplitude is weighted also by the boson mass, 
in such a way that:
\ba
\alpha_{\rm eff}(W) & \sim & { \alpha_w \over M^2_W } \, , \label{aeffW}  \\
& & \nn \\
\alpha_{\rm eff}(Z) & \sim & { \alpha_w \over \cos^2 \vartheta_w } \times
{1 \over M^2_Z } \, = \, 
{ \alpha_w \over \cos^2 \vartheta_w } \times { \cos^2 \vartheta_w \over M^2_W }
\, = \, { \alpha_w \over M^2_W } \, . \label{aeffZ}
\ea 
Effectively, the two couplings are therefore the same. 
This agrees with the fact that, in our framework, couplings are related to 
the effective interaction/decay
width, not to the microscopical description in terms of gauge strengths
and connections. In the Standard Model,
the $Z$ boson couples also to the electromagnetic current,
in such a way that its total interaction with matter is:
\be
{g \over \cos \vartheta_w} \left( J_{\mu}^0 \, - \, 
\sin^2 \vartheta_w J^{e.m.}_{\mu}  \right) \, Z^{\mu} \, . \label{ZJstd} 
\ee 
This is a consequence of the fact that the boson mass eigenstates are 
obtained from the hypercharge and the Cartan of the $SU(2)_{(\rm L)}$ group
through an orthogonal transformation.
The ``total'' width of the $Z$ boson is therefore corrected by
the fact that it couples also to a left-right symmetric current, whose trace
is the same as the one of our ``hypercharge''. Effectively, the size of the 
coupling is:
\be
\alpha^{\rm eff}_Z \, \sim \, {1 \over 4 \pi} {g^2 \over \cos^2 \vartheta_w}
(1  -  \sin^2 \vartheta_w )^2 \,
= \, \alpha_w \cos^2 \vartheta_w \, . \label{aZtot}
\ee
In practice, this means that, for what concerns the axial coupling, the $Z$
boson behaves analogously to the charged bosons, namely, it couples with
the same strength and it ``propagates'' with the same mass.
However, it is also ``displaced'' by a left-right symmetrically interacting
term, in such a way that its total width corresponds to a higher mass and lower
coupling, as given in~\ref{aZtot}.
These relations should be compared with \ref{sintheta} and \ref{costheta}.
The empiric parametrization we gave in terms of the angle $\theta$ corresponds
then to the expressions in terms of the Weinberg angle $\vartheta_w$:
\be
\theta \, \cong \, \vartheta_w \, .
\label{theta-w}
\ee
Therefore, for what concerns the electromagnetic and the charged and neutral 
axial currents, we have effectively reproduced, although in a completely 
different way, the coupling terms of the effective
action of the Standard Model. Things turn out to correspond,
although the underlying basic description doesn't at all.
On the other hand, perhaps it is not so dramatic
that things don't exactly match with all the details of the Standard Model
description. It should be clear that our way of approaching all the issues
related to the low energy parameters is completely different
from the approach of the Standard Model, or its field theoretical
extensions, even ``string inspired'' ones.
In our case, there is no field theoretical Higgs mechanism at work;
masses are generated, and symmetries are broken, in a different way, and
consequently different is also the parametrization of the elementary
interactions. Our aim is not to show how in detail a Standard-Model-like
description of the ``microscopical physics'' is recovered.
The usual effective theory is anyway an approximation, a 
``choice of linearization'', which, although if appropriately restricted
to a certain region of the space of parameters it can nicely fit the 
experimental data, on a larger scale it seems to not work. 
Certainly, fitting of data shows a rather good agreement of many 
Standard Model predictions with the experiments. 
However, this agreement is achieved by advocating,
``predicting'', the existence of degrees of freedom
that continue to escape any attempt of detection.
We can almost perfectly fit the Weinberg angle and boson masses, but at the 
expense of introducing a Higgs field with a mass such that it should have been
already detected somewhere, and still some things don't match. 
We can further adjust the misprediction by enlarging
the number of degrees of freedom, which allow to ``improve the direction'',
pushing forward the problem. For instance,
detecting supersymmetric partners at low energy
etc. In this situation, what is the meaning of statements like ``the Standard
Model correctly predicts the one or the other quantity'', if at the end this
is the result of a fit which assumes the existence of non-detected degrees
of freedom? Of course, these questions are well known since a long time,
and the answer still open. 
We recall the problem here just to remind the reader 
that there is nothing illegal in the fact that we are 
not reproducing all the terms of the low-energy Standard Model action, 
apart from those aspects that one can
safely consider as ``experimentally detected'', and not just ``fitted''.

\vspace{.5cm}
\noindent
{\sl A remark on the meaning of ``massless'' in this framework}
\vspace{.3cm}

\noindent
In the usual field theoretical representation of particles and interactions 
in infinitely extended space-time,
a free photon can not be ``localized'' as can a particle,
and its energy can be as small as we want.
In our framework, this corresponds only to an asymptotic situation,
closer and closer approached but concretely never realized.
At present time, a photon can not be more extended than the distance from
us to the horizon of the universe.
Masslessness does not translate therefore in the energy beeing arbitrarily 
small, but just in the property that ``the minimal energy corresponds to the
inverse of the radius of the universe'':
\be
E_{\gamma} \, \sim \, {1 \over 2} \, {1 \over {\cal T}} \, .
\label{ephoton}
\ee
Massive particles and fields are characterized by a minimal energy
that departs from \ref{ephoton} in that it scales
as a lower power of the inverse of the universe, and therefore
they have an extension smaller than its size:
\be
E_{\rm massive} \, \sim \, {1 \over 2} {1 \over {\cal T}^{p}} \, , ~~~~~~~~
0  < \, p \, <  1 \, ,
\label{emassive}
\ee
so that the spread in space is:
\be
< \Delta X >_{\rm massive} ~ \sim ~  {\cal T}^p  ~ < ~ {\cal T} \, . 
\label{mass-ext}
\ee
For a massive object, the exponent $p$ is necessarily always smaller than 1.

\subsection{The Fermi coupling constant}
\label{GF}

We are now in a position to make contact with the experimental
value of the weak coupling. This is measured through the so-called
Fermi coupling constant $G_F$, a dimensional ($\left[ m^{-2} \right]$)
parameter defined as the effective coupling of the weak interaction
at low transferred momentum~\footnote{Low means here negligible when compared
to the $W$-boson mass.}:
\be
{G_F \over \sqrt{2}} ~ = ~ {g^2 \over 8 M^2_W} ~ = ~ { \pi \alpha_{w} \over
2 M^2_W} \, . 
\ee
From section~\ref{gmass} we know that we can identify $\alpha_{\rm w.i.}$
with the usual weak coupling $\alpha_w$ of the literature
Inserting our results for the $W$-boson mass, \ref{MWcorr1}, and the
value of the weak coupling at the $W$-boson scale, 
given at page~\pageref{aweakW}, we obtain:
\be
G_F \vert_{M_W} ~ = ~ 1,4221 \times 10^{-5} \, {\rm GeV}^{-2} \, .
\ee
As it was for the case of
the fine structure constant, once again we are faced with the problem of 
understanding what is the meaning of a physical quantity, whose value is
always related to a certain experimental process at a certain scale.
Here, from an experimental point of view the Fermi coupling
is obtained by inspecting the pion into muon decay.
The effective, infinite-volume renormalization of $G_F$ to the 
pion--muon scale is obtained in our framework in the same way
as for the other couplings, namely treating 
$G_F$ as a generic coupling, whose behaviour is
represented through an effective linearization as in~\ref{alphamueff}.
The relative variation from the $W$ to the 
$\mu \div \pi$ scale~\footnote{Within our degree
of approximation, it does not make such
a difference the choice of one scale or the other, between muon and pion.}
is of order:
\be
{\Delta G_F \over G_F} \vert_{M_W \to m_{\pi}} ~ \approx ~ 0,81 \, ,
\ee
and we get:
\be
G_F \vert_{\pi / \mu} ~ \approx ~
1,1519 \times 10^{-5} \, {\rm GeV}^{-2} \, ,
\ee
a value about $1 \, \%$ away from the effective experimental 
value~\cite{pdb2006}.
The percent is on the other hand the order of the precision we have
in our estimate of the $W$-boson mass, and as a consequence
we cannot hope to get something better for the Fermi coupling.

\begin{centering}
\centerline{$\star \star \star$}
\end{centering}

This brief survey does not pretend to exhaust the argument of masses.
In particular, certainly not the topic of meson and baryon masses.
However, we hope to have at least shed a bit of light onto the subject. 
As we mentioned in section \ref{GeomP}, for the
practical purpose of computing the fine corrections to masses and couplings
it may turn out convenient to map to an appropriate 
``linearized representation''
of the string space, in which the issue of the computation of these
quantities with the methods of geometric probability can be approached 
with a more evolved technology. What however is missing in these approaches is
a ``cosmological perspective'', which would give the possibility 
of ``fixing the scale''.
For instance, the agreement of (almost) all the quantities computed in 
Ref.~\cite{Smith:1997uw} with experiments is impressive, and,
according to what we discussed in section~\ref{masses} (in particular,
\ref{GeomP}), in many cases
it is not just mere coincidence or numerology; 
however, at least one input has to be 
supplied from outside; e.g. the electron's mass, or any other scale,
to serve as the ``measure'' to which to compare all the ratios of volumes.
Moreover, we expect that in order to refine the results such as those of
\cite{Smith:1997uw}, \cite{Castro:2006} or \cite{Smilga:2004uu},
a better understanding
of how the non-perturbative vacuum is mapped, and its degrees of freedom
are represented in such a linearized approximation, is in order.

\newpage

\section{\bf Mixing flavours}
\label{massmatrix}

As is well known, mass eigenstates are not weak-interaction eigenstates:
the weak currents cross off-diagonally the elementary particles, and, 
besides diagonal up/down
decays, there is a smaller fraction of non-diagonal, flavour changing
decays that mix up the three families. Experimentally, this phenomenon
is well known for what matter quarks. Still hypothetical is however 
whether it occurs also for leptons: its detection would go in pair with a
clear indication of non-vanishing neutrino masses. Indeed, in this case
one tries to go the other way around, looking for neutrino
oscillations as a signal of non-vanishing neutrino masses.
In the case of quarks,
the standard parametrization of these phenomena is made through the
introduction of a matrix $V_{\rm CKM}$, the Cabibbo-Kobayashi-Maskawa matrix,
which encodes all the information about the ``non-diagonal'' propagation
of elementary particles. It is defined as the matrix which rotates the base
of ``down'' quarks of the $SU(2)$ doublets, allowing to express
the currents eigenstates in terms of mass eigenstates:
\be
V_{\rm CKM} \, = \, 
\left(
\begin{array}{ccc}
V_{ud} & V_{us} & V_{ub} \\
V_{cd} & V_{cs} & V_{cb} \\ 
V_{td} & V_{ts} & V_{tb} \\
\end{array}
\right) \, .
\label{CKMdef}
\ee 
$V_{\rm CKM}$ accounts for the mixing among
different generations, as well as the CP violations.

Despite the elegance of the formal treatment, and the intriguing
relation between number of quarks and the existence of a phase,
from the point of view of the Standard Model 
the entries of the CKM matrix remain external inputs, chosen to fit
experimental data: there seems to be no deep reason why a mixing of
quark generations should exist at all, nor why there should be a phase responsible for
CP violations. The ordinary theoretical treatment simply provides
a parametrization of the quark mixing, for which the
number of quark families results to be precisely the
minimal one allowing the existence of a a phase giving rise to CP violations.   
In our approach, non-vanishing off-diagonal CKM entries, and CP violation,
are not simply allowed, but necessarily implied. As such, they come out
as a prediction of the theoretical framework. 
The reason is that any non-trivial mass ratio corresponds to the coupling of a 
broken symmetry; in particular, the CKM matrix expresses the mixing due to
what remains of the broken symmetry between quark families (the 
symmetry among the three orbifold planes of the internal string space). 
CP violations are related to the breaking of parity induced by
the shift along the extended space that gives origin to all sub-Planckian
masses. Therefore, both the family mixing and the parity violation have their
origin in the mechanism that produces masses, and is necessarily implied by
symmetry minimization/maximization of entropy. Nevertheless, from this point of view it
could seem that there is no relation between the existence of a CP violating 
phase, and the number of quarks, as is instead the case of the standard
CKM approach. However, even though apparently less tightly bound, in our
framework these aspects are related in an even stronger way, because
all this proceeds from a unique condition, the minimization of symmetry, necessary
to produce the configuration of highest entropy, i.e. the configuration 
that dominates in the physical world. In our case,
also the number of quarks is a consequence of 
symmetry minimization and the uniqueness of the representation of
the combinatorial scenario, encoded in~\ref{zsum}, in
terms of string theory.

\vspace{.5cm}

\subsection{The effective CKM matrix}
\label{ckm}

Although quark mixing and CP violations in our case are in principle not related to
a description in terms of a CKM matrix, in order to make contact with
the usual theoretical organization of experimental results, we will
derive the entries of an effective CKM within our theoretical framework.   
In our framework, masses have a different explanation than in
ordinary field theory. Nevertheless, it is
still possible to refer to an effective field theory description,
once accepted that this has to be considered only as a tool useful
for practical purpose, without any pretence of being a (self-)consistent 
theory. Although in our approach we directly consider 
decay amplitudes rather than effective terms of a Lagrangian,
in order to make possible an easy  comparison between our predictions
and the usual literature it is therefore 
convenient building an effective Lagrangian, that, 
within the rules of field theory, will reproduce, or ``mimic'', 
our amplitudes.

As is well known, the CKM matrix is a unitary matrix, in which 
all phases except one are reabsorbed into a redefinition 
of the quarks wave functions. Therefore, its nine entries are parametrized by
nine real coefficients and one phase, responsible for parity violation.  
In our framework, the analysis of the spectrum has been carried out
by classifying the degrees of freedom according to their charge.
This means that what we got are the ``current eigenstates'' 
(section~\ref{nps}). Subsequently,
we have considered a ``perturbation'' of this configuration, obtained
by switching-on so-far neglected parameters, in order to investigate
their masses (section~\ref{masses}).
In section~\ref{entromass} we have put mass ratios in relation with ratios
of sub-volumes of the phase space, which is divided into several sectors by
the breaking of the initial symmetry. Mass ratios are then related
to the couplings of the broken symmetries. As we anticipated, there are two
kinds of breaking: a ``strong breaking'', in which the would-be gauge
bosons acquire a mass above the Planck scale, and a 
``soft breaking'', in which the gauge bosons 
acquire a mass below the Planck scale.
Only in this second case the transition appears as an ordinary decay,
mediated by a propagating massive boson.
Otherwise, the boson of the broken symmetry works somehow like an external 
field: we don't see any boson propagating, and we 
interpret the phenomenon as a ``family mixing''. 
The off-diagonal entries of the CKM matrix precisely collect the
effect of this type of ``non-field-theory decay'': off-diagonal entries
account for transitions from one generation to another,
non mediated through gauge bosons as in the case of ordinary decays.

According to our previous discussion, the ratios between
entries of the CKM matrix should be of the same order of the mass ratios,
normalized to the full decay amplitude: mass ratios correspond in fact to
squares of ``elementary'' couplings: $m_f/m_i \sim \alpha_{i \to f} ~~ 
(\sim g^2_{i \to f})$. If $\alpha_{a b}$ is the coupling for the flip 
from family $a$ to family $b$, the decay amplitude
of a $a \to b$ flavour changing decay is expected to be proportional
to $\alpha^2_{a b}$.

Any decay amplitude depends on masses, both
of initial and final states of the process. With our non-perturbative methods
we have direct access to the full decay amplitudes. In order to make contact
with the ordinary description
of the mixing mechanism, we must consider that, as it is defined, 
the CKM matrix is unitary, and collects the
information about flavour changing, subtracted from any dependence on masses:
in expressions of amplitudes, this dependence is carried 
by other terms. This allows to normalize the matrix in such a way that, 
owing to the fact that off-diagonal elements are much smaller than diagonal 
ones, the diagonal elements are close to 1:
\be
|V_{ud}|, ~ |V_{cs}|, ~ |V_{tb}| \, \approx \, 1 \, ,
\label{CKMdiag}
\ee 
and, with a good approximation, 
\ba
|V_{us}| & \approx & |V_{cd}| \label{vus}\; \\
|V_{ub}| & \approx & |V_{td}| \; \label{vub}\\
|V_{cb}| & \approx & |V_{ts}| \; \label{vcb}
\ea
As for the computation of masses, a detailed evaluation of the CKM
matrix entries would require
taking into account all processes contributing to the determination
of the phase space. 
Here we want just to make a test of reliability of our scheme;
we are therefore only interested in a first approximation. 
To this purpose, it is 
reasonable to work within the framework of the simplifications 
\ref{CKMdiag}--\ref{vcb}. 
Owing to these simplifications, we can restrict our discussion
to the off-diagonal elements $|V_{ts}|$, $|V_{td}|$ and $|V_{cd}|$.
A direct, non-diagonal $t \to s$ decay should have an amplitude of order
$m_s / m_t$, normalized then through $m_b / m_t$ in order to reduce
to the scheme \ref{CKMdiag}. A rough prediction for $V_{ts}$ is therefore:
\be
V_{ts} \, \approx \, {m_s \over m_b} \, \sim \, 
{ 0,147 \, {\rm GeV} \over 3,6 \, {\rm GeV} } \, \sim \, 0,04 \, ,
\label{vts} 
\ee
where we have used the values \ref{mscorr1} and \ref{mbcorr}.
Similarly, we obtain:
\be
V_{td} \ \approx \, {m_d \over m_b} \, \sim \, 0,001 \, ,
\label{vtd}
\ee
and
\be
V_{cd} \, \approx \, {m_d \over m_s} \, \sim 0,027 \, .
\label{vcd}
\ee
While \ref{vts} basically agrees with the commonly expected value of this 
entry (see Ref.~\cite{pdb2006}), \ref{vtd} and \ref{vcd} are away by a factor
$\sim 4$ in the first case, and $ \sim 8 $ in the second. An adjustment
of the value is not a matter of ``second order'' corrections. 
Here the problem is that for these mixings, experimental results
are mostly obtained through branching ratios of meson ($\pi$, $K$) decays.
In these quark compounds, the strong non-perturbative resummation
is highly sensitive to the GeV scale. Indeed, an experimental value
$|V_{cd}| \sim 0,22$ seems to be much influenced not by the mass ratio
of the bare quarks, but of the $K$ and $\pi$ mesons:
\be
V_{cd} \, \approx \, { m_{\pi} \over m_K } ~ \sim \, {\cal O}(0,22) \, .
\label{vcdPi}
\ee
Although in a lighter way, the meson scale seems to modify also the ratio 
of the bottom to down quark transition. As we already said,
here it is not a matter of determining a physical quantity: only decay 
amplitudes are physical, the CKM matrix doesn't have a physical meaning 
in itself. It is therefore crucial to see how do we refer to this effective
tool: how much ``resummation'' we want to attribute to a correction
to be applied to ``bare'' decay amplitudes computed from a ``bare'' CKM matrix,
and how much of it we prefer to already include in the CKM
matrix. As long as the final products we consider are just meson amplitudes,
the two approaches are equivalent.

By comparing eqs.~\ref{vcd} and \ref{vcdPi}, we are faced with
something at the same time reasonable and which nevertheless sounds somehow
odd. On one hand, the fact that \ref{vcdPi} gives a higher ratio 
is not surprising:
it is in fact quite natural to think that a heavier particle has a larger
decay probability than a lighter one. On the other hand, when applied
to the $|V_{cd}|$ transition, this argument seems to lead to a 
contradiction: the basic degrees of freedom of a $K$- and $\pi$-mesons are the
quarks; nevertheless, the Kaon
has a larger decay probability than the quarks it is made of.
Indeed, it is not in this way that the enhancement of the
$V_{cd}$ entry due to the passage from quarks to mesons has to be interpreted.
The free quark ``does not exist'', Pions and Kaons are the lightest 
strong-interaction singlets containing the $d$ and $s$ quark. Once inserted
in the computation of a decay amplitude, the values we are proposing for the 
entries of the CKM matrix must be corrected by some overall
``form factor'', of the order of $m_K / m_s$ for the initial
state, and of $m_{\pi} / m_d$ for the final state. In practice, this
is equivalent to the introduction of an 
``effective'' CKM matrix entry, $V^{\rm eff}_{cd} \, \sim \,
(m_K / m_s) / (m_{\pi} / m_d) \times V_{cd}$. This rescaling eats the
factor $\sim 8$ of disagreement between our prediction and the usual
value of this entry, as reported in the literature.

Differently from the case of $|V_{cd}|$,
$|V_{ts}|$ turns out to be in agreement with what reported in the literature,
because the latter is derived by unitarity from $|V_{cb}|$, measured
through $B \to D$ decays. Both these mesons have a mass of the same 
order as the $b$ and $c$ quark respectively. To be more precise, 
in these cases the quark mass itself, as is given in the
literature, corresponds to the ``corrected mass'', basically coinciding
with the mass of the meson of which it constitutes the heaviest
component. The matrix entry is therefore ``by definition''
almost the same as the ``bare'' one.

The case of the $|V_{td}|$ (and $V_{ub}$) entries is even more involved,
being much higher the uncertainties in the experimental derivation
of the transition elements. 

A more detailed derivation of the CKM entries as a function of masses 
can be found in Ref.~\cite{Smith:1997uw}. As we said, this is perfectly fine
in a neighbourhood of our present time. 
In our case, the entries of \ref{CKMdef}
are however time-dependent, and the branching ratios of various decays
vary along the evolution of the universe. The various particles tend toward
a higher relative separation: although the curvature of space-time
tends to a ``flat'' limit, and the absolute value of masses decrease with time,
the ratios of the various masses increase, thereby lowering the probability
of mixing among families. This goes together with the T-dual increase with time
of the mass of the ``would be gauge bosons'' of the broken symmetry among
generations, the non-field-theoretical decay we discussed in 
section~\ref{entromass}.

In our framework, neutrinos are massive, and we expect that the CKM matrix
has a leptonic counterpart. The ``leptonic CKM'' entries should however
be more suppressed, as a consequence of the rearrangement of the phase space,
so that all the three neutrinos are lighter than the lightest charged lepton,
and their spaces have a higher separation.
According to the leptonic mass values derived in section \ref{neumass}, 
we expect at present time approximately:
\ba
V^{\rm leptons}_{\rm CKM} & \equiv & 
\left(
\begin{array}{ccc}
V_{e \nu_e} & V_{e \nu_{\mu}} & V_{e \nu_{\tau}} \\
V_{\mu \nu_e} & V_{\mu \nu_{\mu}} & V_{\mu \nu_{\tau}} \\ 
V_{\tau \nu_{e}} & V_{\tau \nu_{\mu}} & V_{\tau \nu_{\tau}} \\
\end{array}
\right) \nn \\
&& \nn \\
&& \nn \\
& \approx &
\left(
\begin{array}{ccc}
\sim 1 & \sim 0,007 & \sim 0,00005 \\
\sim 0,007 & \sim 1 & \sim 0,007 \\
\sim 0,00005 & \sim 0,007 & \sim 1 \\
\end{array}
\right) \,
\, .
\label{CKMlept}
\ea 
Non-diagonal lepton decays are therefore more difficult to observe
than those of quarks, perhaps more difficult to detect than neutrino
masses themselves.

\subsection{CP violations}
\label{cp}

In our framework, CP (and T) violation is already implemented
in the construction: it is a consequence of 
the shifts in the space and time coordinates, that
break parity (in the case of space), and time reversal symmetries. 
As we have seen, this is related to entropy, and to the fact
that also the second principle of thermodynamics is automatically implemented
in this scenario.

Let's consider the decay of a particle, e.g. $K \, \to \, \pi \, 
(+ \, \gamma)$, or $B \, \to K \, (+ \, \gamma)$. 
The decay probability is proportional to the square ``coupling'':
$\alpha^2_{K \to \pi \gamma}$, $\alpha^2_{B \to K \gamma}$.
The square effective couplings represent volumes in the phase space:
they are in fact proportional to mass ratios, and can be viewed
as the inverse of a proper time,
the proper time of the decaying particle measured in units of
the decay product, raised to the fourth power. The larger is this volume,
the higher is the decay probability.

However, if we consider the problem more in detail, we see that
not the entire length, given as the inverse of the mass, 
is at disposal for increasing the entropy 
through the decay process: part of this phase space is occupied by the rest 
energy of the product particle(s), that we collectively
indicate with $m_f$. As for all masses, the origin of $m_f$ is
a shift along space-time. 
We want now to see what happens if we invert the shifted coordinate.
Let's consider the simplified case of just one space-time coordinate, $t$.
In this case, this inversion is a time reversion.
Under $t \, \to \, - t$, the shift operation
turns out to act in the opposite direction, and results
in an expansion of the volume. In our framework, time is compact: 
$t \in [0, {\cal T}]$. A mass corresponds to a shift $t \, \to \, t + a$, 
$a \approx {1 \over m_f}$, such that now the zero point has been displaced
to $a$: $0 \, \to \, a$. If we perform a time reversal operation,
$t \, \to \, - t$, we obtain that the shift acts now as: 
$(-t) \, \to \, (- t) + a$.
By overall sign reversal, this is equivalent
to a mirror situation, $t \, \to \, t -a$, in which we have
a ``particle'' with mass $- m_f$. Roughly speaking, we can think
that the volume in the phase space occupied by this particle is ``stolen''
from the correct time evolution, and goes ``in the opposite direction''.
The asymmetry between a ``straight'' decay and its ``time-reversed'' one
is given by the fact that in the first case the phase space volume
is proportional to $[(m_i + m_f) / m_f]^4$, after
the inversion to  $[(m_i - m_f) / m_f]^4$.
The general result with four space-time coordinates
is then a simple consequence of the fact that a global inversion of 
all the space-time coordinates: $t \to -t, \vec{x} \to - \vec{x}$
is a symmetry of the system.

In decays involving transitions from neighbouring generations. e.g.
$b \, \to \, s$, $s \, \to \, d$, the mass ratio is at present time 
$m_i /m_f \, \sim \, 10$ and $\sim \, 3,8$ respectively 
(as in the case of the CKM angles,
the mass ratios we have to consider are not those of free quarks, but
of the particles effectively involved in the decays. In this specific
case, the $B$, $K$ and $\pi$ mesons). Therefore, at present we get an 
asymmetry of order $\sim (1/3,8)^4 \approx 
4,8 \times 10^{-3}$ for the K decays, and $\sim 10^{-4}$ for B decays.

In the case of D mesons ($(c \bar{d})$, $(c \bar{u})$ and conjugates), 
we get a ${\cal O}(10^{-2})$ decay asymmetry. This D result is of particular
relevance, because in this case there is no reliable prediction based
on the Standard Model effective parametrization. Indeed, while in our 
framework we obtain an asymmetry in the range of the experimental 
observations, the Standard Model prediction fails to account for
the magnitude of the observed effect (for a review, see for 
instance~\cite{pdb2006}). Because of this, 
CP violations in the D mesons decays 
are often considered a test for ``new physics''.
\begin{figure}
\vspace{.5cm}
\centerline{
\epsfxsize=12cm
\epsfbox{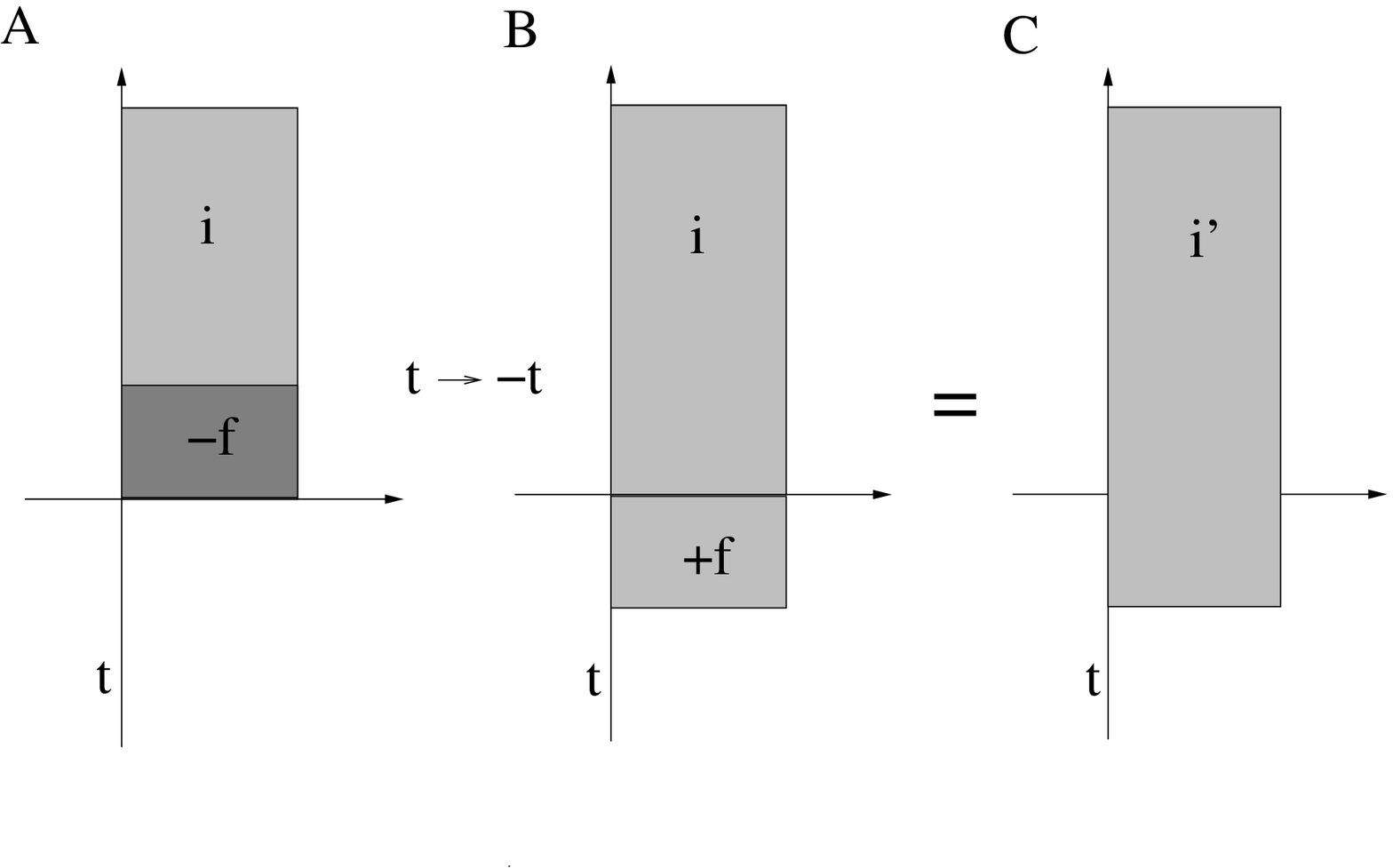}
}
\vspace{0.4cm}
\caption{The increase in the decay amplitude as produced by a time reversal.
While in picture A the phase space volume of the final state $f$ has to be 
subtracted from the volume at disposal of the initial state, after a time flip
it adds up (picture B), ending in an increased decay probability for the
initial state (picture C).}     
\label{t-reversal}
\vspace{.5cm}
\end{figure}

\newpage

\section{\bf Astrophysical implications}
\label{astropred}

Expression~\ref{zsum} contains in principle all the information about the
universe, at any time of its evolution. In section~\ref{ubh} we have already 
seen how it is possible to derive information also 
about ``astronomical'' data, such as the cosmological
constant and the energy density of the universe. 
We have then investigated the masses and interactions of the elementary
particles. Masses and couplings are of interest for the physics of
accelerators. However, a perhaps larger
potential source of stringent tests seems to come from astrophysics,
a field which appears the more and more as one of the most exciting
domains of present and future investigation. This is particularly true
for our framework, in which the present day physics 
is tightly related to the history and the evolution of the universe. 
Therefore, after having gained a better insight into the details of the 
elementary particles, here and in section \ref{exc} we come back
to consider some important issues addressed by astrophysical observations.

\subsection{The CMB radiation}
\label{CMB}

An important experimental cosmological 
observation is the detection of a ``ground'' cosmic electromagnetic
radiation with the typical spectrum of a black body
radiation, with a temperature of about 2,8 $^0 K$ \cite{penzias,Mather:1998gm}.
This phenomenon is often claimed to constitute a proof of the theory of the
Big Bang: this radiation would consist of photons cooled down during the 
expansion of the universe, and at the origin they should have possessed
an energy corresponding to a microwave length, as expected from
energy exchange due to Compton scattering through the plasma at the
origin of the universe. Here we want to discuss how these issues are 
addressed in our scenario. In the framework we are proposing, 
the history of the universe is already implied in the solution
of~\ref{zsum}; its expansion and
cooling down are already ``embedded'' in the framework, which is
in itself a cosmological scenario, and in principle don't need to be
added as separate inputs. As a consequence,
the 3 Kelvin temperature of 
this radiation can be justified by directly using
the present-day data of the universe. This does not mean that
the CMB radiation does not have any relation
with the universe at early times of its expansion: simply, the present 
configuration of the universe is in our framework
a particular case of a cosmological solution accounting for any era of
its evolution, and therefore already ``contains the information''
about the earlier times. In particular, all energies and mass scales of our
present time are the primordial ones, cooled down.   

According to the discussion of section~\ref{les}, the spectrum of the universe
``stabilizes'' into neutrons/protons+electrons (the visible matter) plus
free neutrinos $\nu_e$, photons and, of course, gravitons. The CMB radiation
consists of photons of ``ground energy'', i.e. they are not produced
by interactions and decays of visible matter. This background radiation can
be viewed as a bath of photons in contact with a ``thermal reservoir'',
a neutral vacuum, whose energy corresponds to the inverse square root
scale of the age of the universe, the ground energy scale of 
matter~\footnote{Notice that we must use the ground energy scale,
not the ``mean'' mass scale, roughly corresponding to the neutron mass scale.
Here we are not interested in the non-perturbative mass eigenstate 
at any finite time, but in the ground energy step of a universe evolving
toward a ``flat'' limit at infinity (flat in the sense of the non-compact
orbifolds discussed in the previous sections).}.
Particle's pair production out of this vacuum leads eventually to
photon production. In order to understand what the average energy of
this ``photon sea'', the ``temperature'', is,
let's look at a typical photon/vacuum transition 
process, taking into account that at equilibrium the photon
propagator, intended as ``the mean propagator'', is not renormalized.
As we already discussed several times, in the physical picture
Feynman-like diagram corrections act multiplicatively, a passage
from logarithms of quantities such as couplings and coordinates
to their exponential being subtended. The traditional series expansion
corresponds to a linearization, to working with algebras,
on the ``tangent space'', of a phenomenon
that, once resummed, has to be viewed as working at the level of the group. 
The diagram representing the class of processes of interest for us is
sketched in figure~\ref{gammavac}.
\vspace{1cm}
\be
\epsfxsize=7cm
\epsfbox{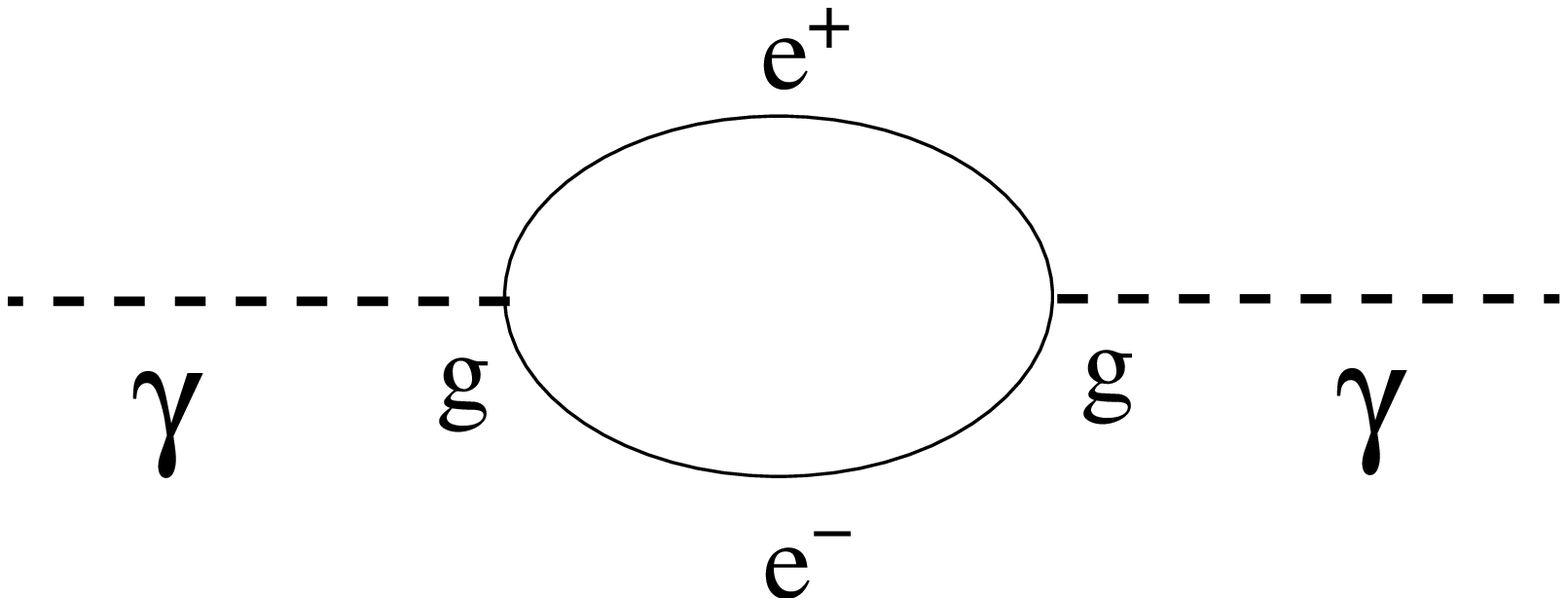}
\label{gammavac}
\ee
\vspace{0.1cm}

\noindent
As seen from the ``exponential picture'', diagrams like this one
contribute to a multiplicative correction to the photon propagator, such that:
\be
{1 \over p^2_{\gamma}} \, \to \, {1 \over p^2_{\gamma}} \, \times \,
\left[ {\alpha \, \int ({\rm loop}) \over p^2_{\gamma}} \right] \, ,
\label{pgamma}
\ee 
where $\alpha \, = \, \alpha_{\rm e.m.} \, \equiv \, \alpha_{\gamma}$ 
and the loop integral pops out a $[m^2]$, corresponding to the energy of our
black body reservoir $E_0$: 
\be
\int ({\rm loop}) ~ \approx ~ E^2_0 ~ \approx ~ 
\left(   { 1 \over  2 \sqrt{\cal T} } \right)^2  \, .
\label{loopE}
\ee
In other words, we can say that the ground electromagnetic radiation
consist of a gas of photons at equilibrium, interacting with a matter vacuum,
whose ground energy fluctuations are, according to the Uncertainty Principle,
of the order of 1/2 the inverse of the typical length of the matter sector,
the square root scale of the age of the universe.
Since at equilibrium, as we have said, 
the average photon propagator is not renormalized, we conclude that
the term under square brackets in \ref{pgamma} must contribute by
a factor~1, and therefore:
\be
< p_{\gamma} > ~ \sim ~ \sqrt{\alpha_{\gamma}} ~
{ 1 \over 2 } \,
{1 \over \sqrt{\cal T} } \, .
\label{pgT}
\ee 
Using the value of $\alpha_{\gamma}$ at the electron's scale,
derived through an effective running of the type~\ref{alphamueff} from
the initial value~\ref{bU1} at the ${\cal T}^{-1/2}$ scale
to the $0,5 \, {\rm MeV}$ scale, $\alpha^{-1}_{\gamma} \vert_{m_e} \sim
132,3$, \ref{calt0} for the present age
of the universe, and converting energy into temperature through the Boltzmann 
constant, we obtain:
\be
T_{\gamma} \, \equiv  \, k^{-1} < p_{\gamma} > ~ =
\, k^{-1} E^0_{\gamma} \, \sim \, 2,72 \, ^0 K \, .
\label{Tcmb} 
\ee  
A signal that the present day phenomenon results from the cooling down
of a primordial situation comes from the spatial inhomogeneity of 
this radiation over a solid angle of observation~\cite{Smoot:1992td}.
We have seen that in our framework, that describes the physics ``on shell'',
the invariance under space rotations is broken by the same mechanism that
gives rise to non vanishing masses for the matter states. 
The scaling of masses, that, going backwards in time, 
tend to the Planck scale, tells us that in the primordial universe also
the space inhomogeneity must have been higher than what it is today.
Toward the Planck scale, the spatial inhomogeneities are expected to 
become of a size comparable with the size of the universe itself.

\subsection{The fate of dark matter and the Chandra observations}
\label{darkm}

A discrepancy between our framework and the common
expectations is the absence in our scenario of dark matter.
According to our analysis, the universe consists only of the already
known and detected particles.
Of course, there can be regions of the space in which a high concentration
of neutrinos, which for us are massive, increases the curvature
without being electromagnetically detected. But this is not going to
change dramatically the scenario: there is no hidden matter acting as
an extra source able to increase the gravitational force by around a factor
ten over what is produced by visible matter, as it seems to be required
in order to explain a gravitational attraction among galaxies much higher
than expected on the base of the estimated mass of the visible stars. 
The problem arises in several contexts: Big-Bang nucleosynthesis,
rotational speed of galaxies, gravitational lensing.
All these points would require a detailed investigation, beyond the scope
of this work. We will also not attempt to rediscuss a huge literature,
and limit ourselves here to mention some hypotheses. The first remark
is that the discrepancies between theoretical expectations and the 
observed effects, which are found in so different
issues as primordial universe, nucleosynthesis and galaxy phenomenology,
don't need necessarily to be explained all in the same way.

About nucleosynthesis we will spend some words in section~\ref{nsyn}.
Let's consider here the problems related to the motion of external stars
in spiral galaxies, where for the first time the issue of dark matter
has been addressed, and the ``anomalous'' gravitational lensing, 
with reference to the
recently observed effect in the 1E0657-558 cluster~\cite{Clowe:2006eq}. 

It is since 1933 (Fritz Zwicky) that, by looking at the
amount of red-shift in the light emitted by the stars in the wings of
a spiral galaxy, it has been noticed
how, differently from what expected, the rotation speed
does not decrease with the inverse of the square root of the radius: it is a 
constant~\cite{rubin1970,rubin1980}. 
Presence of invisible matter has been advocated, in order
to fill the gap between the mass of the observed matter and the amount
necessary to increase the gravitational force. Indeed, the expectation that
the rotation speed
of stars in the external legs should decrease is based on the assumption
that almost the entire mass of the galaxy is concentrated in the bulge
at the center of the spiral. Any star on the wings would therefore feel
the typical gravitational field due to a fixed, central mass. 

\begin{figure}
\vspace{.5cm}
\centerline{
\epsfxsize=10cm
\epsfbox{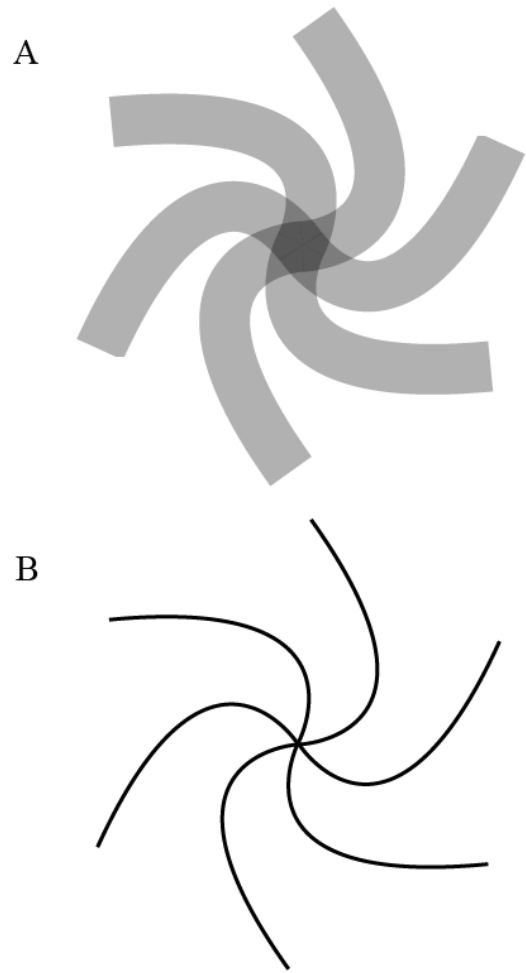} 
}
\vspace{.8cm}
\caption{Picture A is the rough sketch of a spiral galaxy, in which the arms
are broad and shadowed in a way to highlight the increasing mass density
due to their superposition at the center. Figure B represents the same
object, with the arms narrowed down, in order to highlight the 
one-dimensional nature of the physical problem, for what concerns the mass 
density.}     
\label{spiral}
\end{figure}

In the framework of our scenario, masses have been in the past higher than
what they are now. Moreover, owing to the fact that, as we discuss 
in~\cite{assiom-2011}, the universe ``closes up'', in such a way that the
horizon we observe corresponds to a ``point'', the space separation 
between objects located at a certain cosmic distance from us appears to
be larger than what actually is. All this could mean that the mass of
the center of a galaxy, as compared to the wings, 
has been systematically overestimated. It would be interesting to see, by
carrying out a detailed re-examination of the astronomical observations,
whether the behaviour of the center of a galaxy still requires
to advocate the presence of a heavy black hole, in order to explain 
a gravitational force higher than what expected on the base of 
the estimated mass of the visible stars. 
In any case, it is possible that, once the downscaling of length 
and upscaling of masses has been appropriately taken into account,
a better approximation of a spiral galaxy is the one sketched in 
figure~\ref{spiral}. In part A of the picture the galaxy is (very roughly)
represented with wide wings, with stars relatively ``broadened'' on 
the plane of the galaxy. Part B shows the same figure, simply with much 
narrower arms. In picture A the broad lines have been shadowed in a way
to make evident that the higher star density of the bulge is
largely due to the ``superposition'' of the various arms. 
Nevertheless, as it is clear from picture B,
the problem remains basically ``one-dimensional'':
the wings are one-dimensional lines coming out of the center of the galaxy.
Under the hypothesis that all the stars have the same mass, the linear density
of a wing is constant, and, once integrated from the center up to a certain 
radius $R$, the total mass $M_R$ of the portion
of galaxy enclosed within a distance $R$ from the center is roughly
proportional to $R$:
\be
\rho \, = \, {d M \over d r} \, \sim \, {\rm const.} ~~
\Rightarrow ~
M_R \, \sim \, {\rm const} \, \times \, R \, .
\label{Mspiral}
\ee
In the expression of the
gravitational potential, the linear $R$ dependence of the mass cancels 
against the $R$ appearing in the denominator (the potential remains the 
one of a Coulomb force). The gravitational
potential energy is therefore a constant times the mass of the star in the 
wing. Conservation of energy implies therefore that also the velocity of the
star does not depend on the radius $R$.
\noindent
We stress that this is only an approximation: it would be exact if
the arms were not those of a spiral but straight legs coming
out radially from the center, and under the assumption that all the stars
of the bulge correspond to the superposition of the arms.

In the case of the 1E0657-558 cluster, the Chandra observatory has
detected a gravitational lensing higher than what expected on the base
of the amount of luminous matter. Moreover, the highest effect corresponds 
to two dark regions close to the cluster, rather than to places where
the visible matter is more dense. In the framework of our scenario,
a possible explanation could be that what is observed is the effect
of a ``solitonic'' gravitational wave, produced as a consequence of
the separation of one sub-cluster from the other one. This could increase
the gravitational force by an amount equivalent to the displaced
cluster mass, for a length/time comparable to the cluster size,
therefore a time much higher than the few hours during which the effect
has been measured ($\sim$ 140 hours). It remains that the lensing
is around 8-9 times higher than what expected on the base of the amount of 
visible mass. However, the cluster under consideration is at about
4 billion light years away from us. This is around 1/3 of the age of the
universe. This time distance is large enough to make relevant the effects
due to a change of the curvature of space-time along the evolution
of the universe, as well as a change of masses, according to 
\ref{rholambda}--\ref{rhor} 
and \ref{maverage}. Furthermore, as we discussed above,
the apparent space separation 
between objects located at a certain cosmic distance from us must be
appropriately downscaled, in order to account for the curving up of space-time
into a sphere, with the horizon ``identified'' with the origin.
Putting all this together, we obtain that
the effective gravitational force experienced on the 1E0657-558 cluster
is (or, better, it was) indeed 8-9 times higher than what it appears to us
on the base of the expected mass of the objects in the cluster,
i.e. precisely the amount otherwise referred to dark matter.

\newpage

\section{\bf Cosmological constraints}
\label{exc}

Cosmology addresses two kinds of problems for what concerns
the ``running back'' of a theory, or an ``early time'' model. 
Namely, 1) the possible non-constancy of what are commonly called 
``constants'', and 2) the agreement
with the expected origin/evolution of the early universe
(baryogenesis, nucleosynthesis etc...).
In our framework, these issues are put in a light quite different from
the usual perspective: there are in fact indeed no constants; therefore,
a variation of couplings, masses, cosmological parameters, and, as a 
consequence, energy spectra, is naturally implemented. However, 
there is a peculiarity: all these parameters scale as appropriate
powers of the age of the universe. As a consequence, a ``number'' close to one
at present day has a very mild time dependence:
\be
{\cal O}(1) \, \approx \, {\cal T}^{\epsilon} ~ \Rightarrow ~ 
| \epsilon | \ll 1 \, ,
\label{1Te}
\ee 
and therefore varies quite a little with time. Oklo and nucleosynthesis
bounds, being given as ratios of masses and couplings that cancel each other 
to an almost ``adimensional'' quantity, are precisely of this kind. 
In our case they don't provide therefore any dangerous constraint.

For what concerns the non-constancy of ``constants'', 
there are not enough data enabling to test our prediction about a time 
variation of the cosmological constant, whose measurement
is still too imprecise. A more stringent test of the variation of parameters
comes from the observations on the light emitted by ancient Quasars. 
In this case, the spectrum shows
an ``anomalous'' red-shifted spectrum. This shift
should not be confused with the usual red-shift, of which we have discussed
in section~\ref{frw}. The effect we consider here persists once the
``universal'' red-shift effect has been subtracted. As an explanation,
it is often advocated a possible time variation of the fine structure 
constant $\alpha$. We already devoted a paper to this subject~\cite{alpha},
at a time in which many issues concerning our framework were not enough clear.
We therefore rediscuss here the argument, in the light of a better
understanding of the theoretical framework we are proposing.

\subsection{The ``time dependence of $\alpha$''}
\label{talpha}

The question of the possible time variation of the fine structure ``constant''
arises in the framework of string theory derived effective
models for cosmology and elementary particles. Various
investigations have considered the possibility of producing some evidence
of this variation, or at least a bound on its size. To this regard,
astrophysics is certainly a favoured field of research, in that it naturally
provides us with data about earlier ages of the universe. 
A possible signal for such a time variation could be
an observed deviation in the absorption spectra of ancient 
Quasars~\cite{Webb:1998cq,dfw,wetal,metal}. This effect consists
is a deviation in the energies corresponding to some electron transitions, 
which remains after subtraction of the background effect of the red-shift,
and is obtained with interpolations and fitting of data.

What is observed is a decrease
of the relativistic effects in the energies of the electrons cloud,
with respect to what expected on the base of present-day parameters 
(in particular, the fine structure constant).
Indeed, while the atomic spectra are universally proportional to the
atomic unit $m e^2 \, \propto \, m \alpha^2$,
the relativistic corrections depend on the coupling $\alpha$. 
After subtraction of the ``universal'' red-shift effects, their variation
should then be directly related to a variation of $\alpha$. 
In our framework,
the explanation comes from considering both the scaling of $\alpha$
and the one of masses at the same time:
going backwards in time $\alpha$ 
increases, as also the proton and the electron mass do, but
\emph{the ratio of $\alpha$ to the mass scales \underline{decreases}}.
This is the ``variation of $\alpha$ after subtracting the universal red-shift''
which is usually considered in the discussions of the literature. 
Namely, if we measure the
variation of $\alpha$ with respect to the electron's mass scale (whether the 
true electron mass or the ``reduced'' mass doesn't make a relevant 
difference~\footnote{In the hydrogen
atom this is given by $m_{e} = {m_{e} m_{\rm p} \over m_{e} 
+ m_{\rm p}}$. The possibility of referring to a change of this quantity 
the effect measured in Ref.~\cite{dfw} can be found in 
Ref.~\cite{cf,cf2,fshu}.}), 
i.e. if we rescale quantities in the frame in which masses are
considered fixed, we indeed observe a decrease of the coupling $\alpha$.
Indeed, what is done in the literature (see Refs.~\cite{dfw}) is not
only to consider masses fixed, but to exclude from the evaluation 
also the effect of the red-shift. 
With current experimental methods, based on the interpolation of spectral data
in order to find out the ``background'' and the variations out of it,
this subtraction is somewhat unavoidable.

In order to obtain what the prediction in our
scenario is, and how it compares with the literature, 
let's first see how the decrease of 
the relativistic effects, when going backwards in time, turns out to be 
a prediction of our framework. Consider the effective scaling 
of $\alpha$ in terms of $m \alpha^2$ units, the ``universal'' scaling of 
emission/absorption atomic energies. We have that:
\be
\bar{\alpha} ~ \stackrel{\rm def}{\equiv} ~
{ \alpha \over m \alpha^2 } ~ \approx ~  {\cal T}^{{1 \over 3} + {1 \over 28}}
\, .
\label{abar}
\ee 
The ``effective'' coupling $\bar{\alpha}$ scales as a positive power
of the age of the universe: going backwards in time, it decreases.
According to the literature, atomic energies have an approximate
scaling of the type~\footnote{See for instance Ref.~\cite{dfw}}:
\be
E_n \, \approx \, K_n \, ( m \, \alpha^2 ) \, + \, \Gamma_n \, \alpha^2 \, 
( m \, \alpha^2 ) \, ,
\label{E-n}
\ee 
where $K_n$ and $\Gamma_n$ are constants and the second term, 
of order $\alpha^2$ with respect to the first one, accounts for the
relativistic corrections. Investigations on the possible variation
of $\alpha$ use interpolation methods to disentangle the second term
from the first one. Since the universal part is reabsorbed into the red-shift,
the relative variation should give information on just the variation
of $\alpha$. Expression \ref{E-n} is of the form:
\be
E_n \, \approx \, E^0_n \, ( 1 \, + \, a_1 \, \alpha^2 ) \, .
\label{E-nx}
\ee   
It is derived by considering the first terms of a field theory expansion
around the fine structure constant (the electric coupling).
Indeed, since we are interested in the correction subtracted of
the universal part reabsorbable in the red shift, we can separate
the ${\cal O} ( \alpha^2 )$ term in \ref{E-n} as:
\be
K_n ~ = ~ ( K_n \, - \, \Gamma_n ) \, + \, \Gamma_n \, . 
\label{knGn}
\ee
This allows to reduce the part of interest for us to:
\be
E^{\rm eff}_n \, \approx \, E^0_n \, ( 1 \, + \, \alpha^2 ) \, .
\label{E-nexp}
\ee  
As we already observed 
several times along this work, perturbative expressions involving
elementary particles are naturally defined and carried out in a
logarithmic representation of the physical vacuum. In particular,
when writing expressions like \ref{E-n} it is intended that the coupling
$\alpha$ scales logarithmically.
An expression like \ref{E-nexp} should be better viewed
as accounting for the first terms of a series that sums up
to an expression scaling as a certain power of the age of the universe:
\be
\alpha_{\rm eff} \; \equiv \; 
1 \, + \, \alpha^2 ~ \approx  ~ 1 \, + \, \alpha^2 \, +
\, {\cal O}(\alpha^4) ~ \leadsto ~ \exp \alpha ~~ \sim \, {\cal T}^{\beta} \, ,
\label{aeff}
\ee 
where $\alpha$ is then not the full coupling, intended in the non-perturbative
sense of \ref{mTbeta}, but its logarithm.
According to \ref{abar}, in the hypothesis of keeping masses fixed,
this term should then effectively scale as a \underline{positive}
power of the age of the universe: $\beta > 0$.
The exponent $\beta$ can be fixed by comparing values at present time:
\be
\alpha \vert_{\rm today} 
\, \approx \, \sqrt{5 \, \times \, 10^{-5}} \, .
\label{alphatd}
\ee
We obtain therefore:
\be
1 \, + \, \alpha^2 \, \approx \,
{\cal O}(1 \, + \, 5 \, \times \, 10^{-5}) \; \approx \, {\cal T}^{\beta} ~~
\Rightarrow ~~ \beta \; \sim \, {\cal O}(10^{-6}) \, ,
\label{b-value}
\ee
and a relative time variation:
\be
{\dot{\alpha}_{\rm eff} \over \alpha_{\rm eff}} ~ \approx ~
\beta \, {\cal T}^{-1} ~  \approx ~ {\cal O} \left( 
10^{-16 } \, {\rm yr}^{-1} \right) \, .
\label{alphadot}
\ee
This is the relative variation of the relativistic correction
subtracted of the universal part (reabsorbed in the red-shift),
to be compared with the results of~\cite{Webb:1998cq},
as reported also in~\cite{dfw}:
\be
{ \langle \dot{\alpha} \rangle \over \alpha}
\, = \, - 2.2 \, \pm \, 5.1 \, \times \, 10^{-16} ~ {\rm yr}^{-1} \, .
\ee
Since the deviation of the resummed function \ref{E-nexp} from
a pure exponential is of order $\alpha^4 \, \sim \, 2 \, \times \,
10^{-9 }$, four orders of magnitude smaller than the dominant term,
the inaccuracy in our
computation is much lower than the order of magnitude of the result.

\subsection{The Oklo bound}
\label{oklo}

Data from the natural fission reactor, active in Oklo around two billions
years ago, are today considered one of the most important sources of
constraints on the time variation of the fundamental constants. By comparing
the cross section for the neutron capture by Samarium at present time with
the one estimated at the time of the reactor's activity, one derives a bound
on the possible variation of the fine structure constant, and on the ratio
$G_F m^2_{\rm p}$, in the corresponding time interval. 
The interpretation of the experimental measurements and their translation
into a bound on the variation of the capture energy resonance
is not so straightforward, and depends on several hypotheses.
In any case, all these steps are sufficiently under control.
More uncertain is the translation of this bound on the energy 
variation into a bound on the variation of the fine structure constant 
and other parameters: this passage requires strong assumptions about 
what is going to contribute to the atomic energies. 
This analysis was carried out in Ref.~\cite{oklo}, 
basically  on the hypothesis that the main contribution to the resonance energy
comes from the Coulomb potential of the electric interaction among
the various protons of which the nucleus of Samarium consists. According 
to~\cite{oklo}, after a certain amount of reasonable approximations, the 
energy bound translates into a bound on the variation of the 
electromagnetic coupling. A simple look at expression~\ref{betaSU2}
shows that, in our scenario, the variation of this coupling over the time 
interval under consideration violates the Oklo bound. 
This bound seems therefore to rule out our theoretical framework. 
However, things are not so simple: the derivation of a bound on a coupling
out of a bound on energies works much differently in our framework, and
we cannot simply use for our purpose the results of~\cite{oklo}.
Indeed, in our framework what varies with time
is not only the fine structure constant, but also the nuclear force,
and the proton and neutron mass as well. 
Of relevance for us is therefore not a bound on a coupling, derived
under the hypothesis of keeping everything else fixed, but
the bound on the energy itself~\cite{oklo}:
\be
-0,12 \; {\rm eV} \,  < \, \Delta E \, < \, 0,09 \; {\rm eV} \, .
\label{obound}
\ee
In order to give an estimate of the amount of the energy  
variation over time, as expected in our framework, we don't need to know
the details of the evaluation of the resonance energy starting from the 
fundamental parameters of the theory.
To this purpose, it is enough to consider that, whatever the expression of 
this energy is, it must be built out of i) masses, ii) couplings 
(electro-weak and strong)
and iii) the true fundamental constants (the speed of light $c$, 
the Planck constant $\hbar$, and the Planck mass ${\rm M}_{\rm p}$).
Working in units in which the latter are set to 1 (reduced Planck units), 
all parameters of points i) and ii) scale as a certain power of the age of 
the universe. As a consequence, the resonance energy itself mainly scales as 
a power of the age of the universe: 
\be
E \, \sim \, a {\cal T}^{-b} \, .
\label{eaTb}
\ee
(More generically, it could be a polynomial: $E \, \sim \, a_1 {\cal T}^{-b_1}
+ a_2 {\cal T}^{-b_2} + \ldots + a_n {\cal T}^{-b_n}$.
In this case, to the purpose of checking the agreement with a bound, it is
enough to look at the dominant term).
We can fix the exponent $b$ by comparing the expression, evaluated
using the present-day age of the universe, with the value of the resonance,
that we take from~\cite{oklo}:
\be
E \, \sim \, a {\cal T}^{-b} \, = \, 0,0973 \, {\rm eV} 
~ \times ~ 1,2 \, \times \, 10^{- 28} ~
= \; 1,2 \, \times \, 10^{- 29} \, {\rm M}_{\rm P} \, .  
\label{eres}
\ee
In order to solve the equation, we would need to know the coefficient $a$,
something we don't. However, as long as we are just interested in a rough
estimate, it is reasonable to assume that, 
since this coefficient mostly accounts for possible 
symmetry factors, it may affect the value of the result for about no
more than one order of magnitude. Inserting the value
${\cal T} \, \sim \, 5 \, \times \, 10^{60} 
{\rm M}^{-1}_{\rm P}$ for the age of the universe, we obtain:
\be
b \, \sim \, {1 \over 2} \, ,
\label{bres}
\ee
and finally:
\be
\left| \Delta E \right| 
\, \sim \, {1 \over 10 } E ~ \sim \; 0,01 \; {\rm eV}\, .
\label{deltares}
\ee 
over a time of two billion years.
This is compatible with the Oklo bound, eq.~\ref{obound}.

\
\\
From the Oklo data one tries also to derive a bound on the adimensional 
quantity
\be
\beta \, \equiv \, G_F m^2_{\rm p}(c / \bar{h}^3) \, .
\label{betaGdef}
\ee 
In this case, our discussion
is easier, because we know the scaling of all the quantities 
involved~\footnote{We recall that $G_F / \sqrt{2} = g^2 / 8 M^2_W$.
Therefore, $\beta = \pi \alpha m^2_{\rm p}   / \sqrt{2} M^2_W$.
For times much higher than 1 in reduced Planck units, 
the proton mass can be assumed to scale approximately like the mean mass 
scale~\ref{maverage}.}.
Once again, we have to deal with a quantity that scales as a power of the age
of the universe. At present time, we have:
\be
\beta \, \sim \, {\cal T}^{-b_{\beta}} \, = \, 1,03 \, \times \, 10^{-5} \, .
\label{betaT}
\ee
Inserting the actual value of the age of the universe,
we obtain $b_{\beta} \, \sim \, {1 \over 12}$. 
Over a time interval of around 1/5 of the age of the universe, this
gives a relative variation:
\be
{\Delta \beta \over \beta } \, \sim \, 0,017 \, , 
\label{beta-bound}
\ee
to be compared with the one quoted in Ref.~\cite{oklo}:
\be
{\vert \beta^{\rm Oklo} - \beta^{\rm now} \vert \over \beta } \, < \, 0,02 \, .
\label{dbOklo}
\ee
Both results \ref{deltares} and \ref{beta-bound}, although still
within the allowed range of values, seem to be quite close to
the threshold, beyond which the model is ruled out. One would therefore
think that a slight refinement on the measurement and derivation
of these bounds could in a near future decide whether it is
still acceptable or definitely ruled out. Things are not like that.
Indeed, as we already stressed in several similar cases, the \emph{entire}
derivation of bounds and constraints, involving at any level
various assumptions about the history of the universe and therefore
of its fundamental parameters, should be rediscussed within the new
theoretical framework: it doesn't make much sense to compare
pieces of an argument, extracted from an analysis carried out in a different
theoretical framework, with different phenomenological implications.
To be explicit, in the case of the derivation of the Oklo bounds,
one should reconsider all the derivation of absorption thresholds
and resonances. We should therefore better take into account from the 
beginning the time variation of all masses, and in particular the neutron
and proton masses, as well as couplings. Perhaps a more meaningful
quantity is then not anymore the pure resonance shift, but this
quantity rescaled by the neutron mass.
In this case, the effective variation of interest for our test is not
\ref{deltares}, but:
\be
{ \Delta ( E / m_n ) \over ( E / m_n ) }   \, \approx  \, 
{ \Delta {\cal T}^{- {1 \over 9}} \over {\cal T}^{- {1 \over 9}} } 
~ \sim ~ 0,02
\, ,
\label{demn}
\ee
a variation one order of magnitude smaller than \ref{deltares} 
($ \Delta E / E \sim 0,1$).
Analogous considerations apply also to the case of the second bound 
\ref{beta-bound}, basically equivalent to the nucleosynthesis bound.

\subsection{The nucleosynthesis bound}
\label{nsyn}

Bounds derived from nucleosynthesis models are even more questionable:
they certainly make sense within a certain cosmological model, but,
precisely because of that, they cannot be simply translated into a 
framework implying a rather different cosmological scenario.
Once again, the only anchor points on which we can rely are the few
``pure'' experimental observations, to be interpreted in a consistent way
in the light of a different theory. 
The point of nucleosynthesis is that there is a very narrow ``window''
of favourable conditions under which, out of the initial hot plasma,
our universe, with the known matter content, has been formed. 
Of interest for us is the very stringent condition about the temperature
(and age of the universe) at which the amount of neutrons in baryonic matter
have been fixed. As soon as, owing to a cooling down of the temperature,
the weak interactions are no more at 
equilibrium, the probability for a proton to transform into a neutron
is suppressed with respect to the probability of a neutron to decay into a 
proton. Owing to their short life time, comparable
with the age of the universe at which the equilibrium is broken,
basically almost all neutrons rapidly decay into protons, except for those
that bound into $^4$He. Nucleosynthesis predicts a  
fraction of $^4$Helium and Hydrogen baryon numbers ($\sim 1/4$) 
in the primordial universe, which is in good agreement with 
experimental observations. The formula for the equilibrium of neutron/proton
transitions is given by:
\be
{ {\rm n} \over {\rm p}} \, = \, {\rm e}^{- {\Delta m \over k T}} \, 
\sim \, 1 \, ,
\label{nucleq}
\ee
where $\Delta m = m_{\rm n} - m_{\rm p}$.
In the standard scenario, this mass difference is a constant, and
the temperature runs as the inverse of the age of the universe.
The equilibrium is broken at a temperature of around $0,8$ MeV,
when $({\rm n} / {\rm p}) \simeq 1 / 7$. 
In our framework too the temperature runs as the inverse of the age of the 
universe, but the mass difference $\Delta m$ is not a constant:
all masses run with time. At large times (${\cal T} \gg 1 $ in Planck 
units), we are in a regime in which we can use the arguments
of section~\ref{hmass}, in order to conclude that, being the $u$ and $d$ quark 
masses much lighter than the neutron mass scale (which is related
to the ``$m_{11/36}$'' mass scale), we can consider $\Delta m$ as a
perturbation of $m \simeq m_{\rm n}$. In this regime,
the neutron-proton mass difference is basically of the order of the
constituent quark mass difference, and we have reasons to expect that it 
also runs accordingly. It would therefore seem that, in our case,
going backwards in time, the ratio (n/p) remains lower than in the standard
case, and the equilibrium \ref{nucleq} is attained at a temperature much 
higher. However, to the purpose of determining the processes of the 
nucleosynthesis, essential is not just the scaling of the equilibrium law
of the neutron-to-proton ratio, but also that of the mean life of the neutron.
It is the combined effect of these two quantities what determines the
primordial baryon composition. In the standard case, the neutron mean
life is assumed to be constant. Being related to the neutron decay amplitude,
i.e. to the volume occupied by the neutron in the phase space, in our 
framework this quantity too is not constant. In order to see what
in practice changes in our scenario with respect to the standard one,
instead of attempting to guess what the scaling behaviour of the neutron mean
life could be, we can proceed by considering, instead of the pure running
of the equilibrium equation, the \emph{reduced running at fixed neutron mean
life}. Certainly the mean life is constant if the neutron mass is constant.
The quantity of interest for us is therefore the scaling of the mass 
difference, as measured in units of the neutron mass itself.
According to our considerations of above, we have:
\be
\Delta m_{\rm red} ({\cal T}) \, \equiv \,     
{\Delta m \over m_{\rm n}} \, \sim \, 
{{\cal T}^{p_{(u-d)}} \over {\cal T}^{p_{\rm n}}} \, ,
\label{dm/mT}
\ee 
where $p_{(u-d)}$ and $p_{\rm n}$ are exponents corresponding to the up-down 
quark mass difference and to the neutron mass respectively.
This running is expected to hold not only at present time
but also at a temperature of $\sim 1$ MeV, which is anyway much lower than the 
Planck scale. We can therefore compare our prediction with the standard
one by simply considering the relative deviation of equation~\ref{nucleq}
from its standard value,
as obtained by replacing the constant mass difference $\Delta m$ 
with $\Delta m_{\rm red} ({\cal T})$:
\be
{ {\rm n} \over {\rm p}} \, = 
\, {\rm e}^{- { \Delta m \over k T}} ~~
\to ~ \left( { {\rm n} \over {\rm p}} \right)_{\rm red} \,
\equiv \, 
{\rm e}^{- { \bar{m}_{\rm n} \Delta m_{\rm red}({\cal T}) \over k T}}
\, ,
\label{nucleqRed}
\ee
where $\bar{m}_{\rm n}$ is the \emph{fixed}, time-independent present-day
value of the neutron mass. Therefore, in the standard case
$( { {\rm n} / {\rm p}} )_{\rm red}$ coincides with (n/p).
According to the mass values given in section~\ref{masses}, we have:
\be
\Delta m_{\rm red} ({\cal T}) \, \approx \, {\cal T}^{- {1 \over 24}} \, .
\label{dmred}
\ee
Considering that the time variation between the point ${\cal T}_f$
of the breaking of equilibrium and the present day is of the order of the
age of the universe itself, 
$\Delta T \equiv {\cal T} - {\cal T}_f \sim {\cal T}$, 
we obtain approximately that the integral variation of 
$x \, \equiv \, \Delta m_{\rm red} ({\cal T})$ over this time interval is:
\be
\Delta x \, \sim \, {1 \over 24} x \, .
\label{dx/x}
\ee
The ``reduced value'' of (n/p), $({\rm n} / {\rm p})_{\rm red}$,
is now modified to:
\be
\left( { {\rm n} \over {\rm p}} \right)_{\rm red.}: \, {1 \over 7} 
~ \to ~~ \sim \, 
{1 \over 7} \left( 1 \, - \, {\ln 7 \over 24}  \right) ~ \approx \, 0,131
\, . 
\label{n/pnew}
\ee
This value leads to a ratio $X_4$ of helium to Hydrogen of around:
\be
X_4 \, \sim \, 0,232 \, ,
\label{X4}
\ee 
still in excellent agreement with what expect from today's most
precise determinations (for a list of results and references,
see Ref.~\cite{pdb2006}).

As mentioned above, there is here no reason to push the discussion into 
further detail, because the entire issue,
as well as all the extrapolations from experimental observations, should be
rediscussed within the framework of this scenario, something well beyond the
scope of this work.
We want however to point out another aspect of the problem, which
arises in our theoretical framework.
A peculiarity of our scenario is that, at any time, there is
a bound on the number of particles that can exist in the universe.
At any time the volume of space-time is finite;
the maximal amount of energy is fixed by the Schwarzschild black hole relation,
$2M = R$, where $R$ is here identified with ${\cal T}$,
and the lowest particle's mass is the one of the electron's neutrino, given in
~\ref{mnueV}. The number of matter states is therefore finite, and cannot 
exceed the Schwarzschild mass of the universe divided by the lowest neutrino
mass. On the other hand,
this statement has the ``mean value'' validity of any other statement
related to \ref{zsum}.
As we already discussed,
only in an ``average'' sense we can in fact talk of geometry of space-time,
and make contact with ``classical'' objects such as black holes, and in
general with the Einstein's equations and their Scharzschild's solution.  
One may then ask if the ``black hole bound'' on the total
mass of the universe can be ``violated'' in a quantum way, 
for a time $\Delta t \sim 1 / \Delta m$, also along the 
minimal entropy solutions: although they are the configurations with
a space-time more close to a classical geometry description,
they are nevertheless string vacua and, as such, quantum vacua.
The point is however how do we observe such a violation:
in order to measure a mass fluctuation $\Delta m \sim 1 / \Delta t$,
we need precisely a time $\Delta t > 1$ in Planck units.
According to the black hole bound, in a time $\Delta t$ the universe
increases its energy from $E \sim {\cal T}$ to 
$E^{\prime} \sim {\cal T} + \Delta t $, i.e. $\Delta E \sim \Delta t$.
Since $\Delta t > 1$, we have $\Delta E > \Delta m$. In other words,
the mass/energy fluctuation implied by the Uncertainty Principle is always
lower than the total energy increase due to the time evolution
of the universe: in the time we need in order to measure a possible
quantum fluctuation out of the ground value of the energy, this value
changes by a much larger amount. Quantum fluctuations are smaller
than the ``classical'' increase due to the 
expansion.

\newpage

\centerline{\bf Appendix}

\appendix

\section{Conversion units for the age of the universe}
\label{A1}

We give here some conversion factors from time units to Planck mass units.
\[
1~ {\rm year} ~({\rm yr}) \, = \, 3,1536 \, \times \, 10^{7} ~ {\rm s} 
\]

\noindent
In order to convert this value to eV units we divide by 
$\hbar = 6,582122 \, \times \, 10^{-22}$ MeV s. We obtain:

$$
1 ~ {\rm yr} \, = \, 4,791160054 \, \times \, 10^{28} ~ {\rm MeV}^{-1} 
$$

\noindent
Considering that the Planck mass ${\rm M}_{\rm P} \, = \, 1,2 \, \times \, 
10^{19}$~GeV, we have also the relation:

$$
1 ~ {\rm yr} \, = \,3,992633379 \, \times \, 
10^{50} \, {\rm M}^{-1}_{\rm P} \, .
$$

\noindent
The age of the universe ${\cal T}$, 
estimated to be around 11,5 to 14 billion years, reads therefore:

\[
{\cal T} \, \approx \, \left\{ \begin{array}{l}
                   ~ 4,59152839 \\
                   ~ 5,58968673 
                               \end{array}
                              \right.
\, \times \, 10^{60} ~ {\rm M}^{-1}_{\rm P}
\]
If instead we take the neutron mass as the most precise way of
deriving the age of the universe,
from expression \ref{mneutron} and the present-day measured neutron mass, 
we obtain: 
\[
{\cal T} \, \approx \, ~ 5.038816199 
\, \times \, 10^{60} ~ {\rm M}^{-1}_{\rm P} ~~~ 
(\, = \, 12,6202827 \times 10^9 ~ {\rm yr})
\]

\section{The type II dual of the ${\cal N}_4 = 1$ orbifold}
\label{N=0}

We discuss here the type II dual
construction of the ${\cal N}_4 = 1$ orbifold vacuum of 
section~\ref{breaksusy}.
On the heterotic side, this appears as a supersymmetric
construction. We claimed that ${\cal N}_4 = 1$ supersymmetry
exists only perturbatively, but when the full, non-perturbative
construction is considered, one sees that this symmetry is broken.
From the heterotic point of view, the breaking is non-perturbative, being
produced by a ``twist'' along the coupling-coordinate around which
the perturbative expansion is built. The only signal of the supersymmetry
breaking is then indirectly provided by the way couplings of non-perturbative
matter and gauge sectors (parametrized by perturbative fields of the heterotic
string) enter in the expressions of threshold corrections of effective
couplings. Namely, with the ``wrong'' power, as if these couplings
were ``inverted'', from a $< 1$ to a $> 1$ value.
Indeed, these couplings are parametrized by moduli only at the ${\cal N}_4 = 2$
level (see Ref.~\cite{striality}). When the perturbative supersymmetry
is reduced to ${\cal N}_4 = 1$, these fields are twisted. This however only 
means that the expectation value is not anymore a running parameter, but is
fixed. We can nevertheless trace the fate of the couplings by investigating
the so-called ``${\cal N} = 2$'' sectors.

In order to follow the operation of supersymmetry breaking from the
the type II side, let's first consider the starting point, the
${\cal N}_4 = 2$ construction. In order to make easier the investigation of
the projections, it is convenient to express the degrees of freedom
in terms of free fermions (Ref.~\cite{abk}). 
In the case of type II strings, these constructions have been extensively
analysed in Ref.~\cite{gkr}. Indeed, the cases we are referring to 
are embedded in a infinitely extended space-time, a situation 
deeply different from the one considered in this paper, where space-time is
compact. However, as we have 
seen, in practice this reflects in a different interpretation of the results
(e.g. the fact that densities become global quantities), whereas from a 
technical point of view the usual analysis carries over from a scenario to 
the other one with minor, obvious changes (the substitution
of a continuum of modes along the space-time coordinates with a 
discrete lattice of momenta/energies). For simplicity, we use therefore here
the same notation for the string modes as in the cited works. 
The set of all fermions is therefore: 
\be 
F=\left\{ \begin{array}{l} \psi_{\mu}^L, \chi_I^L, y_I^L, \omega_I^L \\
        \psi_{\mu}^R, \chi_I^R, y_I^R, \omega_I^R \end{array}
              \right\}, 
~~~(\mu=1,2;~I=1,...,6),
\ee
where $\psi_{\mu}^{L,R}$ indicate the left and right
moving fermion degrees of freedom along the transverse space-time coordinates,
while $\chi_I^{L,R}$ those along the internal coordinates.
$y_I^{L,R}\omega_I^{L,R}$ correspond instead to the internal fermionized
bosons. The basic sets of boundary conditions are
$S$ and $\bar{S}$, which contain only eight left- or
right-moving fermions, and distinguish the boundary
conditions of the left- and right- moving world-sheet superpartners:
\be
S=\left\{ \psi_{\mu}^L, \chi_1^L, \ldots,\chi_6^L \right\},~~~~~  
\bar{S}=\left\{ \psi_{\mu}^R, \chi_1^R,\ldots,\chi_6^R \right\}.
\ee
In order to obtain a $Z_2 \times Z_2$ symmetric orbifold, we need 
then the two sets $b_1$ and $b_2$:
\ba
b_{1} & = & 
      \left\{ \begin{array}{l} \psi_{\mu}^L,\chi_{1,2}^L,y_{3,\ldots,6}^L \\
       \psi_{\mu}^R,\chi_{1,2}^R,y_{3,\ldots,6}^R \end{array}
              \right\}~, 
\label{b1} \\
&& \nn \\
b_{2}& = & 
  \left\{ \begin{array}{l} \psi_{\mu}^L,\chi_{3,4}^L,y_{1,2}^L,y_{5,6}^L \\
       \psi_{\mu}^R,\chi_{3,4}^R,y_{1,2}^R,y_{5,6}^R \end{array}
              \right\}~. 
\label{b2}
\ea
These sets assign $Z_2$ boundary conditions and break the ${\cal N}_4=8$ 
supersymmetry to ${\cal N}_4=2$. The lowest entropy configuration
is then obtained by further partial shifting of some states of the
twisted sectors. We will not consider these further operations: they
commute with the projection we want to consider in the following, 
namely the one that leads to the breaking of supersymmetry; considering them 
complicates the construction without altering the conclusions. 
As discussed in Ref.~\cite{gkr} and~\cite{striality},
depending on the relative phase of the projections introduced 
by $b_1$ and $b_2$, we obtain two mirror configurations which, according
to~\cite{striality}, are two slices of the same model: in one we see
only the vector multiplets, in the other only the hypermultiplets,
of the same $U(16)$ model.

We want now to introduce another
projection, dual to the $Z_2^{(2)}$. In order to understand what we have
to expect from this further operation, we must take into account that
1) in order to preserve the pattern of duality with the heterotic and
type I string established at the ${\cal N}_4 = 2$ level, i.e.
the identification of the geometric moduli of the type II space
with those of the heterotic space and the type I coupling moduli,
also this third projection must act symmetrically on left and right movers;
2) it must twist all these moduli. On the other hand, we cannot pretend  
to see the extended space-time represented in a similar way in both the 
heterotic/type I and the type II dual: a further symmetric, independent
twist on the type II space must necessarily act also on the coordinates with 
index ``$\mu$''. This means that, in order to see the action of the
heterotic $Z_2^{N=2 \to 1}$ projection, on the type~II side we must trade
the space-time coordinates for internal ones. From the type~II point of view
the heterotic space-time will therefore be entirely non-perturbative,
and the type II construction will look perturbatively 
compactified to two dimensions. Being two coordinates
hidden in the light-cone gauge, we see therefore no transverse
non-compact coordinates. As we discussed in section~\ref{dimString}, 
representing the ``11-th coordinate''
of string theory in orbifolds entails its linear realization through an
embedding in a two-dimensional toroidal space. The space gives therefore
the fake impression of being ``12-dimensional''. This is however
an artifact of the perturbative representation.

Compactifying the ``$\mu$'' indices implies that we can now
fermionize the bosons also along these coordinates.
The boson degrees of freedom 
$\partial X_{\mu}$ and $\bar{\partial} X_{\mu}$ will
be now represented as $y^L_{\mu} \omega^L_{\mu}$ and
$y^R_{\mu} \omega^R_{\mu}$. 
(By the way, we remark that in the scenario discussed in this work,
all string coordinates are always compactified. Therefore, in principle
fermionization of the space-time degrees of freedom is always possible.
On the other hand,
when considering explicit string constructions, perturbation is always
possible only around a decompactified coordinate, that works as the
vanishing coupling around which to perturb. The very fact of
writing a perturbative representation of a string vacuum
implies the assumption that a certain limiting procedure toward
a non-fermionizable point of some coordinates has been taken.)

From this two-dimensional point of view, the ${\cal N}_4 = 2$ type II
construction contains only scalar fields: the space-time is non-perturbative, 
and therefore so are all indices (vector, spinor and tensor) 
running along space-time coordinates. The type II construction is therefore
blind to the distinction between gauge and matter, whose degrees of 
freedom have a space-time
vector or spinor index, and an internal, scalar index: only this last index
is visible on the type II, and these states appear all as scalars.
There is no trace of the graviton, because it bears only space-time indices.
Moreover, the fields $T^{i}$ and $U^{i}$, $i = 1,2,3$, usually
appearing in one-loop expressions of threshold corrections,
don't correspond now to geometric moduli of two-tori.
Indeed, for any twist what remains untwisted is a four-torus. 
In practice, we have added a two-torus. However, as we discussed, this is
an artifact of the linearization of the space; there is indeed no
twelve-dimensional theory, and the appearance of the
two-torus is due to an ``over-dimensional''
representation of a curved space with just one more coordinate,
the one that served as the coupling on which to expand in the
four-dimensional vacuum. The 12-th coordinate
is instead a curvature. There is no surprise that, in this representation,
the former moduli $T^{i}$ and $U^{i}$ are now multiplied by what was the
coupling of the theory: its dependence was simply ``frozen'' by construction.
For what matters duality with the heterotic construction, nothing changes,
because the value of these fields was not fixed. 
We can recover a description in terms of moduli of two-tori by introducing
independent boundary conditions for the ``complex planes'' (1,2), (3,4),
(5,6), (7,8) (see Ref.~\cite{gkr} for a detailed discussion of these sets).
This allows to disentangle the two-torus moduli, by factorizing the space
in four two-tori. On the type~II side we see then that, besides
the  $T^{i}$ and $U^{i}$, $i = 1,2,3$, we have now  
one more field, corresponding to what was the (hidden) coupling
of the four-dimensional construction. It misleadingly appears as a pair of
torus moduli, $T^4$, $U^4$, respectively corresponding to the volume form
and the complex structure.
Owing to the symmetry of the construction under exchange of the
three tori with the fourth one, a $T^{4} \leftrightarrow U^{4}$ reflection
exchanges the two ${\cal N} = 4$ mirror constructions (the one with
only vectors with the one with only hyper multiplets). It is worth to
consider more in detail this property. The ``fourth torus'' volume
form is the product of two radii, that we call $R_{11}$ and $R_{12}$
for obvious reasons. The moduli $T^4$ and $U^4$ are related to these
radii by: $\Im T^4 = R_{11} R_{12}$, $\Im U^4 = R_{11}/ R_{12}$.
As we said, one of the two radii is indeed 
not a real further coordinate, but a curvature. When seen from the 
``four dimensional point of view'', an inversion of this radius
corresponds to an inversion of the full string coupling. Therefore,
the $T^4 \leftrightarrow U^4$ mirror exchange that relates the two
constructions is an ``S-duality'' of the ``normal representation'' of
the type II vacuum.

We already discussed in Ref.~\cite{striality} how
the heterotic construction, containing both vector and hyper multiplets,
corresponds to a slice, built around a corner of the moduli space, 
of the ``union'' of both the type II
mirror models. From this point of view it is therefore ``self-mirror''.
Here we understand that this mirror symmetry is indeed a strong-weak coupling
duality of the type II string, an operation which
is perturbative on the heterotic dual~\footnote{On the heterotic side,
matter and gauge sectors are exchanged by an exchange of the twisted and
the untwisted sectors. This corresponds to an inversion of the world-sheet
parameter $\tau$: $\tau \to - 1 / \tau$. This parameter is integrated
out, and it never appears explicitly in the effective theory.
On the other hand, we have seen that the world-sheet coordinates are roughly
``identified'' with the two longitudinal coordinates of the light-cone gauge.
Any trace of the moduli of this symmetry 
is therefore hidden by the gauge fixing.}.
For the rest, it is important to observe that, although we cannot
explicitly verify it on the base of the carried space-time indices,
all hidden, the identification of the degrees of freedom allows anyway to 
see the $S$ and $\bar{S}$ as the generators of space-time supersymmetry.
This time they are to be intended as a representation of
the ``internal part'' of the supersymmetry sets.

From the above considerations, we conclude that, on the type II side,
the new projection, corresponding to the step ${\cal N}_4 = 2 \, \to \,
{\cal N}_4 = 1$, must be represented by a set $b_3$ given, up
to a permutation of the three complex planes corresponding to the indices
$I = 1, \ldots, 6$, by:
\be
b_{3} \, = \, 
      \left\{ \begin{array}{l} \chi_{3,\ldots,6},y_{\mu}^L,y_{1,2}^L \\
       \chi_{3,\ldots,6}^R,y_{\mu}^R,y_{1,2}^R \end{array}
              \right\}~. 
\label{b3} 
\ee
The condition 2) of above tells us however that,
differently from the case of $b_1$ and $b_2$, the ``GSO phase''
of this set must be \footnote{We refer the reader to \cite{abk} for
an explanation of this coefficient and its role.}:
\be
\delta_{b_3} = -1 \, ,
\ee
(we recall that $\delta_{b_1} = \delta_{b_2} = 1$ and 
$\delta_S = \delta_{\bar{S}} = -1$). This condition projects out
all the states of the type $\phi^L \otimes \phi^{R}$, for whatever
indices and $\phi \in \{ \psi, \chi, y, \omega \}$, i.e. all
the states of the untwisted sector. 
The moduli ``$T$'' and ``$U$'' are now ``twisted'', and  
the only massless states come from the twisted sectors.
The projection coefficients of the fermionic construction are given
in the following table:
\be
\begin{tabular} {| l |r|r|r|r|r|r|} \hline 
 & $F$ & $S$ & $\bar{S}$ & $b_1$ & $b_2$ & $b_3$
\rule[-.2cm]{0cm}{.7cm} \\ \hline
$F$    & 1 & $-1$ & $-1$ & 1 & 1 & 1 \rule[-.2cm]{0cm}{.7cm} \\ \hline
$S$    & $-1$ & 1 &  1 & $-1$ & $-1$ & $-1$ \rule[-.2cm]{0cm}{.7cm} \\ \hline
$\bar{S}$ & $-1$ & 1 & 1 & $-1$ & $-1$ & $-1$ \rule[-.2cm]{0cm}{.7cm} \\ \hline
$b_1$ & 1 & 1 & 1 & 1 & $ 1$ & $1$ \rule[-.2cm]{0cm}{.7cm} \\ \hline
$b_2$ & 1 & 1 & 1 & $ 1$ & $1$ & $1$ \rule[-.2cm]{0cm}{.7cm} \\ \hline
$b_3$ & 1 & 1 & 1 & $ 1$ & $1$ & $1$ \rule[-.2cm]{0cm}{.7cm} \\ \hline
\end{tabular}
\label{cfermion}
\ee
together with the conditions: 
$\delta_S = \delta_{\bar{S}} = \delta_{b_3} = -1$,
$\delta_{\phi} = \delta_{b_1} = \delta_{b_2} = 1$.
Observe that, with this choice, $b_3$, although a type II symmetric twist
as $b_1$ and $b_2$, projects the states with the same phase as
a heterotic $Z_2$ orbifold projection, as we precisely wanted.
Notice also that, differently from how it appears on the heterotic side,
the projection introduced by $b_3$ is not
exactly symmetrical to the one introduced by $b_2$.
For instance, it seems that it would project out all the $T$ and $U$
fields even when acting alone, i.e. before the introduction of $b_2$.
This impression is however misleading, in that it neglects that, as we
have seen, from the point of view of this two-dimensional compactification,
these fields are no more moduli of a torus, but have a more complicate
expression as functions also of the former coupling coordinate,
here ``embedded'' in the further, fourth torus.
And indeed, if we want to introduce the ``planes'' as in 
Ref.~\cite{gkr,striality} in order to lower the rank of the twisted sectors,
the sets which introduce separate boundary conditions for the coordinates
must be defined in order to include more than one bosonic coordinate.
Namely, they must contain also the ``coupling plane''.
In the ${\cal N}_4 = 2$ model constructed with just $\{b_3, b_1 \}$ (or
$\{b_3, b_2 \}$) the moduli $T^{i}$, $U^{i}$ are no more built 
from the states:
\be
\delta_{ij} \, x_i \bar{x}_j | 0 >  
\ee
but as combinations of states of the type:
\be
x_i \bar{x}_j | 0 > ~~~ i \neq j \, , ~ \{i,j \} \in 
\left( \{ 3,4  \}, \, \{ 5,6  \}, \{ 7,8  \} \right) ~ \cup \{ 11,12  \} \, .
\ee
The partition function of this orbifold 
is given by the integral over the modular
parameter $\tau$, with modular-invariant measure 
$(\Im \, \tau)^{-2} d \tau d \bar{\tau} $, of:
\be
Z^{{\rm string}} =   
\left( {1 \over 2} \right)^3 \sum_{(H_1,G_1,H_2,G_2,H_3,G_3)}
~Z^F_{\rm L}~  Z^F_{\rm R} ~
\sum_{(\gamma,\delta)} ~ Z_{8,8} {\ar{\gamma}{\delta}}~,
\label{z}
\ee
where $Z^F_{L,R}$ contain the contribution of the world-sheet 
fields $\psi_{\mu}^{L,R}$, $\chi_{a}^{L,R}$ (the sets $S$ and $\bar{S}$); 
$Z_{8,8}$ substitutes what in four dimensional constructions is
$Z_{6,6}$, the $c=(6,6)$ internal space. Now this space spans all bosonic
degrees of freedom and has $c=(8,8)$, 
corresponding to the fields $\omega_I^{L,R}$, $y_I^{L,R}$, $I = 1, \ldots, 8$. 
Notice that we don't have now the factor 
$ 1 / (\Im \tau | \eta(\tau)|^4)$, the contribution of the 
space-time transverse bosonic degrees of freedom, now accounted in $Z_{8,8}$.
We have:
\be
Z^F_{\rm L}={1 \over 2} \sum_{(a,b)} {{\rm e}^{i \pi \varphi_{\rm L} 
\left( a,b,\vec H,\vec G \right)} \over \eta^4}
\vartheta {\ar{a+H_3}{b+G_3}}
\vartheta {\ar{a+H_2-H_3}{b+G_2-G_3}}
\vartheta {\ar{a+H_1}{b+G_1}}\vartheta {\ar{a-H_1-H_2}{b-G_1-G_2}}
~, \label{fl}
\ee
\be
Z^F_{\rm R}={1 \over 2} \sum_{(\bar{a},\bar{b})} 
{{\rm e}^{i \pi \varphi_{\rm R} 
\left( \bar{a},\bar{b}, \vec H, \vec G \right)}
 \over \bar{\eta}^4}
\vartheta {\ar{\bar{a}+H_3}{\bar{b}+G_3}}
\vartheta {\ar{\bar{a}+H_1-H_3}{\bar{b}+G_1-G_3}}
\vartheta {\ar{\bar{a}+H_2}{\bar{b}+G_2}}\vartheta {\ar{\bar{a}-
H_1-H_2}{\bar{b}-G_1-G_2}}~,
\label{fr}
\ee
with:
\ba
\varphi_{\rm L} & = & a+b+a b ~, \\
\varphi_{\rm R} & = & \bar{a}+\bar{b}+\bar{a} \bar{b}~.
\ea
The contribution of the compact bosons is:
\ba
Z_{8,8} \ar{\gamma}{\delta} & = & {\rm e}^{i \pi (H_3 + G_3 + H_3 G_3)}
\nonumber \\ && \nonumber \\
& \times & 
{1 \over  |\eta|^4}\, \left\vert \vartheta \ar{\gamma}{\delta}
\vartheta \ar{\gamma +H_3}{\delta+G_3} 
 \right\vert^2 
 \nonumber \\ && \nonumber \\
& \times & 
{1 \over  |\eta|^4}\, \left\vert \vartheta \ar{\gamma}{\delta}
\vartheta \ar{\gamma + H_2+ H_3}{\delta + G_2+ G_3} 
\right\vert^2 
 \nonumber \\ && \nonumber \\
& \times & 
{1\over |\eta|^4}\, \left\vert \vartheta \ar{\gamma}{\delta}
\vartheta \ar{\gamma+H_1}{\delta+G_1}
\right\vert^2
 \label{zb}\\ && \nonumber \\
& \times &
{1\over |\eta|^4}\, \left\vert \vartheta \ar{\gamma}{\delta}
\vartheta \ar{\gamma+H_1+H_2}{\delta+G_1+G_2}
\right\vert^2 ~. \nonumber
\ea
The pairs $(a,b)$ and $(\bar{a},\bar{b})$ specify the
boundary conditions, in the directions ${\bf 1}$ and $\tau$ of the
world-sheet torus, of the sets $S$ and $\bar{S}$, while
$(\gamma,\delta)$ refer to the set of all fermionized bosons; 
$(H_1,G_1)$, $(H_2,G_2)$ and $(H_3,G_3)$
refer to the sets $b_1$, $b_2$ and $b_3$.
Notice the presence of the phase ${\rm e}^{i \pi (H_3 + G_3 + H_3 G_3)}$,
corresponding to the choice $\delta_{b_3} = -1$.

In this model there are nine massless sectors, corresponding to the
previous $b_1$, $b_2$, $b_1b_2$, the new ones, $b_3$, $b_3 b_1$,
$F b_3 b_2$, $b_3 b_1 b_2$, $S \bar{S} b_3 b_2$, and the $S \bar{S}$
sector. Only three sectors have a perturbative dual on the heterotic side,
and correspond to a tern generated by a pair of intersecting projections.
Here $b_3 \cap b_1 \neq \emptyset$ and $b_2 \cap b_1 \neq \emptyset$,
while $b_3 \cap b_2 = \emptyset$, therefore the pair is either
$\{b_3,b_1  \}$ or $\{b_2,b_1  \}$. On the sets generated by one of these 
pairs, the third independent projection doesn't impose any further constraint.
The third projection is already ``built-in'' by construction in the heterotic
string, which starts with half the maximal supersymmetry of the
type II string. Therefore, apart from the supersymmetry reduction,
from the heterotic point of view the further projection triplicates
the structure of the ${\cal N}_4 = 2$ model. However,
on the type II side, where we have access to all the sectors, 
we can see that some of the sectors hidden for the
heterotic string are not supersymmetric: owing to the $\delta_{b_3}$
GSO torsion, the $S \bar{S}$ states are here supersymmetric to nothing,
and the same is true for the states of the $F b_3 b_2$ and  
$S \bar{S} b_3 b_2$ sectors: their superpartners are massive.
This is a representation in terms of free fermions of
what more generally is a mass shift (see Ref.~\cite{gkr} for
a discussion of the translation of the fermionic language in terms
of orbifold operations).

With different choices of the relative GSO projections of one sector
to the other one, the coefficients $(b_3| b_j) $ in table~\ref{cfermion},
we obtain mirror configurations in which supersymmetry is broken in
a different way: a negative projection of $b_3$ to $b_1$ and $b_2$
implies that all the twisted sectors are projected out.
Some of them, not as a consequence of a shift, but due to
incompatibility of the selected chiralities of the spinors
of the twisted sectors. It seems therefore that the model is empty
unless the $S$ and $\bar{S}$ projections are removed from the definition
of the basis: only the pure Ramond-Ramond sector survives (the projections
$(b_3 | S)$ and $(b_3 | \bar{S})$ remain unchanged). 
These mirror models seem to exist only at a ``delta-function'' point
in the string moduli space.

\section{The supersymmetry-breaking scale}
\label{susybreaking}

The string vacuum whose type~II dual has been
discussed in appendix~\ref{N=0} shows
that, once the string space attains the maximal amount of twisted
coordinates, supersymmetry is broken and the space is necessarily curved.
In section~\ref{breaksusy} we pointed out that, in a 
compact space, supersymmetry is always broken, as a consequence of the
missing invariance under space-time translations, which are part of
the super-Poincar\'{e} group ($\{ Q , \bar{Q} \} \sim P$). Indeed,
minimization of entropy requires that all coupling moduli, and in particular
the heterotic dilaton field, are twisted, and frozen at the Planck scale.
All this means that the heterotic construction discussed in 
sections~\ref{spectrumZ2}, \ref{breaksusy} doesn't correspond to a
true perturbation around a decompactified coordinate, but is a kind of
non-compact orbifold, in which, in order to be able to build the states
around a small/vanishing value of a coordinate, we artificially neglect
its being twisted, and proceed as if along this direction we would not
have fixed points. In some sense, this is also what is done in 
Ref.~\cite{hw1,hw2}. 
However, in that case, as long as one is not interested in 
further curling of the string space, and remains at the level of maximal 
supersymmetry, the game may appear basically innocuous: it works because
there are many other de-compactifiable coordinates. Problems arise when
proceeding to further twisting, as we observed in Ref~\cite{striality}. 

The result is that there are ``hidden'' sectors, whose origin has to be
traced in the original, neglected T-duality of the theory, which are
non-perturbative, and where all the breaking of supersymmetry is relegated.
All this is the misleading consequence of an artificial, ``illegal'' 
flattening of an intrinsically curved space. The coordinate, or better, the
curvature, which is related to the size of the flattened coordinate,
works as ``order parameter'' for the breaking of supersymmetry.
Some aspects of this phenomenon are precisely responsible for what 
we observe on the type~II side.

As we have seen in appendix~\ref{N=0}, on the type~II side we have 
two mirror situations, in which the breaking of supersymmetry manifests itself 
in a different way. However, both of them can be related to the same
mechanism; they must be seen as two aspects of the same phenomenon.
The key point is that non-freely acting projections can be viewed as obtained 
at the corner of the moduli space of freely acting constructions. In 
this case, at a generic point in the orbifold moduli space, projected states
receive a non-vanishing mass as a consequence of a coordinate shift associated
to the orbifold twist. In the decompactification limit of this coordinate,
masses become infinite and the projected states disappear from the spectrum, as
they usually do in ordinary non-freely acting orbifolds. 
This allows us to get an idea of the scale at which supersymmetry is broken:
the supersymmetric partners are lifted by a shift picked along an
internal coordinate $X$, which is also twisted.
In order to describe the GSO projection
process in terms of freely acting shifts, we 
must look for a dual configuration in which $X$, that 
we know to be fixed by minimization of symmetry
to a value around the Planck scale, 
reaches this value as a limit at the corner of the moduli space. 
The map between the two descriptions, 
$X \, \to \, R (X)$, must therefore be some kind of logarithm, so that:
\be
X \, \to \, 1 ~~ \leftrightarrow ~~ R \, \to \, 0/\infty \, .
\label{R(X)}
\ee   
This implies that either $R \, \approx \, \ln X$ or $1/R \, \approx \, \ln X$.
As discussed in ref.~\cite{striality,gkr},
a change in sign of the $( b_i | b_j)$, projections corresponds to the
inversion of some internal radius. The mirror constructions of 
appendix~\ref{N=0} correspond therefore precisely to the one or the other of
these two possibilities, for some of the internal coordinates.
According to the mechanism of freely acting projections, in the dual picture
the mass of the projected states reads:
\be
\tilde{m}~ (\sim \ln m ) ~ \sim ~ R \, .
\label{mR}
\ee
Pulled back to the physical picture, we obtain that, at the
``twisting point'', the mass is of order one in Planck units.
This is the mass gap between the observed particles (and fields),
with sub-Planckian mass, vanishing at first order, and their
superpartners.   

\vspace{1cm}
\section{Local correction to effective beta-functions}
\label{Ashift}

The running of the electromagnetic and
weak couplings in the representation in which they are going to
be compared with experimental data is logarithmic, with a slope
determined by an effective beta-function coefficient.
However, as discussed in section~\ref{fsc}, around the scale $\sim m_e$,
the volumes of the matter phase space are expanded (or, logarithmically,
shifted), in such a way that for instance
the electromagnetic coupling at the scale $m_e$ (i.e. the fine structure
constant) effectively
corresponds to the value of the coupling \emph{without correction} at a 
run-back scale, $m^{\rm eff.}_e$. The amount of running-back in the scale
of the logarithmic effective coupling is equivalent to the
amount of the forward shift in the logarithmic representation 
of the volumes of particles in the phase space. If volumes get
multiplied by a factor, their logarithm gets shifted, and so 
gets shifted back the scale at which the coupling in its
logarithmic representation is effectively evaluated. 
This deviation can be considered as a perturbation of the logarithmic
running, that we illustrate here. In the figure,
$\mu_0$ stays for the 
starting scale of the running: $\mu_0 = (1 /2) \, {\cal T}^{- 1 / 2}$,
$\mu_{\sim 1/4}$ for the upper end scale of the matter sector,
the thick solid line shows the approximate expected behaviour of 
the inverse coupling $\alpha^{-1}$, including the correction to the
shape, while the thin solid line indicates the original logarithmic behaviour.
The dashed segments indicate the linear approximation of the curve
we considered in the footnote
at page~\pageref{linearalpha} in order to compute the effective weak
coupling at the $W$-boson scale:  

\vspace{1cm}
\centerline{
\epsfxsize=12cm
\epsfbox{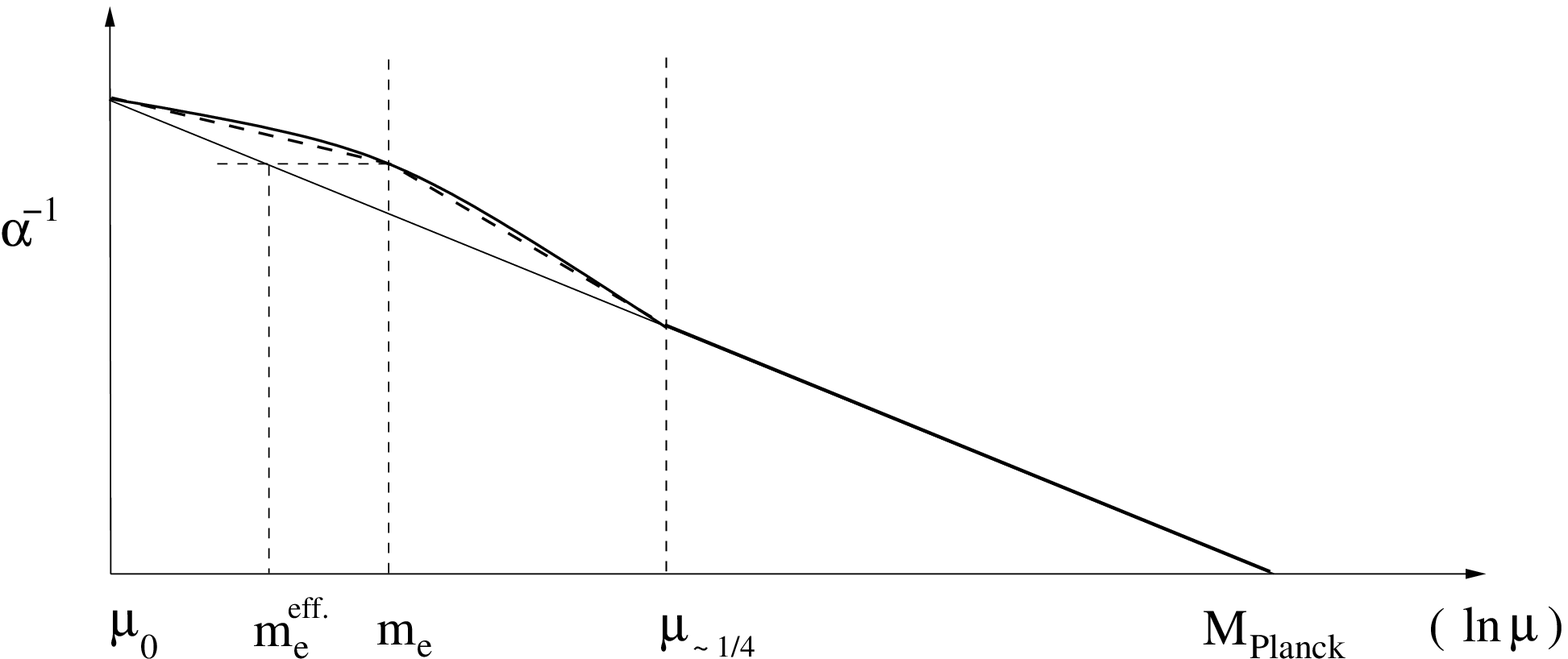}
}

\vspace{1cm}
\section{A note on the string partition function}
\label{nspf}

The string partition function 
is defined in the perturbative constructions at the one-loop
(genus one) level of expansion over the string coupling, as the
integral over the ``modular parameter'' $\tau$ of the sum over all
string excitations at any energy level.
In the closed superstring, it is an expression of the type \footnote{We skeep here to
consider the type I string, in which the parameter $\tau$ has a slightly
different definition \cite{Sagnotti:1987tw}.}:
\be
Z ~ = ~ \int d^2 \tau \sum q^{{1 \over 2}P^2_{\rm L}} 
\bar{q}^{{1 \over 2}P^2_{\rm R}} \times 
F \left[\tau, \eta(q), \vartheta(q)) \right]
\, ,
\ee
where $q = \exp {\rm i} \tau$, and we separated the sum over momenta 
(and windings) of left and right movers, from the rest of the contribution,
here collected in the function $F \left[\tau, \eta(q), \vartheta(q)) \right]$:
The details of this expression are of no interest for our present 
discussion~\footnote{See for instance Ref.~\cite{Kiritsis:1997hj}.};
what we want here to remark is the appearance in
the summation of the \emph{square}
of the momenta, i.e. of energies. 
This is due to the fact that these momenta come from the expansion 
over the modes of bosonic target space coordinates, for which
the Hamiltonian, i.e. the energy operator, is quadratic in the momenta.
The bosonic relation between momentum and energy is in fact of the
type $E \sim P^2 / 2m$. In a string construction, the mass is assumed to be
one (the string mass, defined as the inverse of the string length),
and one obtains an identification of energy with the square of a momentum.
All this is perfectly fine and legal. However, in our context it sounds a bit
strange, because we are going to eventually identify the lowest level
of momentum ($m \sim p_0 \sim 1/R$) with a mass, and we would like to see
this to be linearly related to an energy: $E_0 \sim m$.   
The point is that we started with bosonic coordinates, and we
and up with the energies of a physical system in which as elementary
matter particles we have fermions (leptons and quarks). 
From a technical point of view, things are adjusted by the fact that 
the orbifold operation which lifts all matter states to a non-zero mass,
producing the string vacuum of highest entropy, acts perturbatively
as $R \to {1 \over 2}R$, therefore as $R \to \sqrt{R}$ in the physical
coordinates (see section~\ref{entromass}, page~\pageref{logpict}). 
As a consequence, also \emph{physical} momenta
are ``square-root'' contracted:
\be
P^2 ~ \sim ~ \left( {1 \over R} \right)^2 
~~~ \longrightarrow ~~ P ~ = ~ \left( {1 \over \sqrt{R}} \right)^2 \, .  
\ee
Owing to this shift, the string partition function bears therefore the correct dependence
on the physical momenta and masses, and can be matched with
the partition function of a statistical system, as implied by the
combinatorial approach discussed in~\cite{assiom-2011}, 
and encoded in~\ref{zsum}.
From a physical point of view, 
this contraction can be viewed as follows:
in order to produce massive matter, the shift must couple fermions
two by two. Indeed, the
entire perception of the physical world an observer can have is mediated
by boson fields (graviton, photon), which couples to fermion pairs (not only
the photon but also
the graviton couples to fermions through terms of the effective action in which
fermions are paired, as are also all mass terms), and so, i.e. of boson type, are
also all the energy levels appearing in the partition function of the universe.
Physical momenta are made out of the 
pairing of two momenta of the elementary description.

\newpage

\providecommand{\href}[2]{#2}\begingroup\raggedright\endgroup

\end{document}